\documentclass[12pt]{article}%
\pdfoutput=1

\usepackage{cite}
\usepackage{mathtools}
\usepackage{amsfonts}
\usepackage{amssymb}
\usepackage{authblk}
\usepackage{float}
\usepackage{bbm}
\usepackage{bm}
\usepackage{caption}
\usepackage{epsfig}
\usepackage{epstopdf}
\usepackage{heck}
\usepackage[margin=1in]{geometry}
\usepackage{tabularx}
\usepackage{xcolor}
\usepackage{graphicx}

\usepackage{caption}
\usepackage{tikz}
\usetikzlibrary{decorations.pathreplacing}
\usetikzlibrary{calc}

\usepackage{hyperref}
\usepackage[nameinlink,capitalize]{cleveref}

\usepackage{tikz}
\usetikzlibrary{arrows}
\usetikzlibrary{arrows.meta}
\usetikzlibrary{arrows,decorations.pathmorphing}
\usetikzlibrary{shapes.geometric,calc,arrows, positioning,shapes.misc,decorations.markings}
\tikzset{
  big arrow/.style={
    decoration={markings,mark=at position 1 with {\arrow[scale=2,#1]{>}}},
    postaction={decorate},
    shorten >=0.4pt},
  big arrow/.default=black}

\pgfdeclarelayer{edgelayer}
\pgfdeclarelayer{nodelayer}
\pgfsetlayers{edgelayer,nodelayer,main}
\tikzstyle{none}=[inner sep=0pt]

\tikzstyle{NodeCross}=[draw, shape=circle, cross out, inner sep=0pt, minimum size=10pt,line width=0.25mm]
\tikzstyle{Circle}=[draw, shape=circle, black, fill=black, inner sep=0pt, minimum size=4pt]
\tikzstyle{circle}=[draw, shape=circle, black, fill=black, inner sep=0pt, minimum size=16pt]
\tikzstyle{CircleRed}=[draw, shape=circle, fill={rgb,255: red,191; green,0; blue,0}, inner sep=0pt, minimum size=8pt]
\tikzstyle{Star}=[draw, shape=star, fill=red, star points=8, inner sep=0pt, minimum size=12pt]
\tikzstyle{CircleBlue}=[draw, shape=circle, fill=blue, inner sep=0pt, minimum size=8pt]
\tikzstyle{CirclePurple}=[draw, shape=circle, fill={rgb,255: red,191; green,0; blue,191}, inner sep=0pt, minimum size=8pt]
\tikzstyle{EmptyCircle}=[draw, shape=circle, inner sep=0pt, minimum size=4pt]
\tikzstyle{GreenCircle}=[draw, shape=circle,  fill={rgb,255: red,80; green,200; blue,120}, inner sep=0pt, minimum size=8pt]
\tikzstyle{BrownCircle}=[draw, shape=circle,  fill={rgb,255: red,115; green,115; blue,115}, inner sep=0pt, minimum size=8pt]
\tikzstyle{CirclePurpleSmall}=[draw, shape=circle, fill={rgb,255: red,191; green,0; blue,191}, inner sep=0pt, minimum size=4pt]

\tikzstyle{DashedLine}=[-, densely dashed, line width=0.25mm]
\tikzstyle{DottedLine}=[-, dotted, line width=0.25mm]
\tikzstyle{ThickLine}=[-, line width=0.25mm]
\tikzstyle{ArrowLineRight}=[-, -{Stealth[scale=1.75]}, line width=0.15mm, scale=5]
\tikzstyle{RedLine}=[-, draw={rgb,255: red,191; green,0; blue,0}, fill=none, line width=0.5mm]
\tikzstyle{DashedLineThin}=[-, densely dashed, line width=0.125mm, fill=none, draw=black]
\tikzstyle{ArrowLineRed}=[-, draw={rgb,255: red,191; green,0; blue,0}, -{Stealth[scale=1.75]}, line width=0.15mm, scale=5]
\tikzstyle{BlueLine}=[-, draw={rgb,255: red,0; green,0; blue,191}, fill=none, line width=0.5mm]
\tikzstyle{DashedRed}=[-, densely dashed, draw={rgb,255: red,191; green,0; blue,0}, fill=none, line width=0.5mm]
\tikzstyle{DottedRed}=[-, dotted, draw={rgb,255: red,191; green,0; blue,0}, fill=none, dotted, line width=0.5mm]
\tikzstyle{BlueDottedLight}=[-, dotted, draw={rgb,255: red,0; green,0; blue,191}, fill=none, line width=0.5mm]
\tikzstyle{ArrowLinePurple}=[-, draw={rgb,255: red,191; green,0; blue,191}, -{Stealth[scale=1.75]}, line width=0.15mm, scale=5]
\tikzstyle{DashedLineGreen}=[-, densely dashed, draw={rgb,255: red,74; green,103; blue,65}, line width=0.25mm]
\tikzstyle{LineGreen}=[-, draw={rgb,255: red, 74; green,200; blue,65}, line width=0.5mm]

\tikzset{snake it/.style={decorate, decoration=snake}}

\tikzstyle{PurpleLine}=[-, draw={rgb,255: red,191; green,0; blue,191}, fill=none, line width=0.5mm]
\tikzstyle{BrownLine}=[-, draw={rgb,255: red,115; green,115; blue,115}, fill=none, line width=0.5mm]

\usetikzlibrary{decorations.markings}

\tikzset{
dashstar/.style={
 dash pattern=on 5pt off 5pt,
 postaction={
  decorate,
  decoration={
   markings,
   mark=between positions 9pt and 1 step 10pt with {
     \node[color=red] {*};
   }
  }
 }
},
dashstarstar/.style={ 
 dash pattern=on 5pt off 10pt,
 postaction={
   decorate,
   decoration={
     markings,
     mark=between positions 10pt and 1
          step 15pt
           with {
            \node at (-2pt,0pt) {\pgfuseplotmark{star}};
            \node at (2pt,0pt) {\pgfuseplotmark{star}};
           }
   }
 }
}
}
\usepackage{pgfplots}
\pgfplotsset{compat=1.16}

\DeclarePairedDelimiter\bra{\langle}{\rvert}
\DeclarePairedDelimiter\ket{\lvert}{\rangle}

\newcommand*{\diff}{\mathop{}\!\mathrm{d}}

\DeclareMathOperator{\AdS}{AdS}
\DeclareMathOperator{\CFT}{CFT}

\DeclareMathOperator{\SO}{SO}
\DeclareMathOperator{\SU}{SU}

\DeclareMathOperator{\U}{U}

\newcommand*{\bO}{\ensuremath\mathbb{O}}

\newcommand*{\cH}{\ensuremath\mathcal{H}}

\newcommand*{\cO}{\ensuremath\mathcal{O}}

\newcommand{\lb}{\left(}
\newcommand{\rb}{\right)}
\newcommand{\lbb}{\left[}
\newcommand{\rbb}{\right]}

\newcommand{\be}{\begin{equation}}
\newcommand{\ee}{\end{equation}}
\newcommand{\ba}{\begin{aligned}}
\newcommand{\ea}{\end{aligned}}
\newcommand{\Z}{{\mathbb Z}}
\newcommand{\R}{{\mathbb R}}
\newcommand{\C}{{\mathbb C}}

\definecolor{rred}{RGB}{191,0,0}

\begin{document}

\date{October 2023}

\title{SymTrees and Multi-Sector QFTs}

\institution{HAMBURG}{\centerline{$^{1}$II. Institut fur Theoretische Physik, Universitat Hamburg, Hamburg 22607, Germany}}
\institution{PENN}{\centerline{$^{2}$Department of Physics and Astronomy, University of Pennsylvania, Philadelphia, PA 19104, USA}}
\institution{PENNmath}{\centerline{${}^{3}$Department of Mathematics, University of Pennsylvania, Philadelphia, PA 19104, USA}}
\institution{CERN}{\centerline{${}^{4}$CERN Theory Department, CH-1211 Geneva, Switzerland}}
\institution{NYU}{\centerline{$^{5}$Center for Cosmology and Particle Physics, New York University, New York, NY 10003, USA}}
\institution{VT}{\centerline{$^{6}$Physics Department, Robeson Hall, Virginia Tech, Blacksburg, VA 24061, USA}}

\authors{Florent Baume\worksat{\HAMBURG}\footnote{e-mail: \texttt{florent.baume@desy.de}},
Jonathan J. Heckman\worksat{\PENN,\PENNmath}\footnote{e-mail: \texttt{jheckman@sas.upenn.edu}},
Max H\"{u}bner\worksat{\PENN}\footnote{e-mail: \texttt{hmax@sas.upenn.edu}}, \\[4mm]
Ethan Torres\worksat{\PENN, \CERN}\footnote{e-mail: \texttt{ethan.martin.torres@cern.ch}},
Andrew P. Turner\worksat{\PENN}\footnote{e-mail: \texttt{turnerap@sas.upenn.edu}}, and
Xingyang Yu\worksat{\NYU, \VT}\footnote{e-mail: \texttt{xingyangy@vt.edu}}
}

\abstract{The global symmetries of a $D$-dimensional QFT can, in many cases, be captured in terms of a
$(D+1)$-dimensional symmetry topological field theory (SymTFT).
In this work we construct a $(D+1)$-dimensional theory which governs the symmetries of QFTs with
multiple sectors which have connected correlators that admit a decoupling limit.
The associated symmetry field theory decomposes into a SymTree, namely a treelike structure of SymTFTs
fused along possibly non-topological junctions. In string-realized multi-sector QFTs, these junctions
are smoothed out in the extra-dimensional geometry, as we demonstrate in examples.
We further use this perspective to study the fate of higher-form symmetries in the
context of holographic large $M$ averaging where the topological sectors of
different large $M$ replicas become dressed by additional extended operators associated with the SymTree.}

{\small \texttt{\hfill ZMP-HH/23-13}}

{\small \texttt{\hfill CERN-TH-2023-183}}

\maketitle

\setcounter{tocdepth}{2}

\tableofcontents

\newpage

\section{Introduction}

Global symmetries play an important role in constraining the dynamics
of quantum field theories (QFTs). It has recently been shown that global
symmetries are also associated with deep topological structures \cite{Gaiotto:2014kfa}.
This has led to many generalizations, which now encompass the standard textbook 0-form symmetries, as well as various higher-form,
higher-group, and higher categorical structures.\footnote{For some recent reviews, see e.g., \cite{Cordova:2022ruw, Schafer-Nameki:2023jdn, Bhardwaj:2023kri, Luo:2023ive, Shao:2023gho}.}

For a $D$-dimensional QFT,\footnote{We assume throughout that our QFT\ is
Lorentz invariant when formulated on $\mathbb{R}^{D-1,1}$.} this topological
structure can often be captured in terms of a $(D+1)$-dimensional topological
field theory, often referred to as a symmetry
topological field theory (SymTFT).\footnote{See e.g., \cite{Reshetikhin:1991tc, Turaev:1992hq, Barrett:1993ab, Witten:1998wy, Fuchs:2002cm, Kirillov:2010nh, Kapustin:2010if, Kitaev:2011dxc, Fuchs:2012dt, Freed:2012bs, Freed:2018cec, Freed:2022qnc, Kaidi:2022cpf}.}
In this framework, the structure of the
$D$-dimensional QFT\ is specified by fixing appropriate boundary conditions in
the SymTFT: we have a non-topological physical boundary condition specifying a choice of relative QFT,
as well as a topological boundary condition which fixes the global form of the QFT.
This approach is quite powerful, and immediately provides a framework for
extracting higher-categorical symmetries as captured by topological fusion
rules of the SymTFT. While the existence of this SymTFT can be formulated in
purely bottom up terms, it is helpful to note that for QFTs with a top down
(i.e., stringy) realization, the SymTFT naturally appears via a topological
reduction of the associated extra-dimensional non-compact geometry \cite{Apruzzi:2021nmk} (see also \cite{Aharony:1998qu, Heckman:2017uxe, Apruzzi:2021phx, vanBeest:2022fss}).
This is especially helpful in the context of
intrinsically strongly coupled QFTs, e.g., $D>4$ conformal
fixed points.\footnote{See \cite{Heckman:2018jxk, Argyres:2022mnu} for recent reviews.}

But QFTs can exhibit a range of possible phenomena and it is
natural to ask whether the current paradigm of SymTFTs is flexible enough to
accommodate all these possibilities. In this work we study the
structure of SymTFTs for QFTs with multiple
\textquotedblleft decoupled\textquotedblright\ sectors in which all connected correlators involving
non-topological operators admit a decoupling limit. In practice, this means we have parameters / mass scales such that for
connected correlators between different sectors:
\begin{equation}
\langle \mathcal{O}_1 \mathcal{O}_2 \rangle_{\mathrm{conn}} \rightarrow 0,
\end{equation}
in a suitable decoupling limit. We refer to these as multi-sector QFTs. In our terminology,
each sector is itself a non-trivial interacting relative QFT.

Even though the different sectors have decoupled local dynamics, their global form can still be non-trivially
coupled topologically.\footnote{This is somewhat distinct from
the case of coupling a QFT to a TFT studied in \cite{Kapustin:2014gua},
but we explain the relation to the present work in Appendix \ref{app:KapustinSeiberg}.}
For example, the global form of the gauge group in a multi-sector model can impose non-trivial
constraints on the spectrum of Wilson lines.

To frame the discussion, an example which we repeatedly return to is that of a UV $\mathfrak{su}_{N+M}$ gauge theory
which undergoes adjoint Higgsing to $\mathfrak{su}_{N}\times\mathfrak{su}_{M}\times\mathfrak{u}(1)$ gauge theory in the IR.\ While there is no issue in
defining Wilson lines in the parent $\mathfrak{su}_{N+M}$ theory, the
construction of Wilson lines for just the $\mathfrak{su}_{N}$ or $\mathfrak{su}_{M}$
gauge theory sector meets with immediate subtleties such as the proper
treatment of 1-form symmetries in a given sector.

In general, we can construct the corresponding SymTFT for each individual sector,
and then construct a junction with a SymTFT which captures the symmetries of
the parent UV theory. Our main claim is that this procedure can indeed be
carried out, but there is in general no guarantee that the theory living at
the junction is topological. Carrying this out for multiple SymTFTs fused by junctions, we arrive at a treelike structure.

We refer to this tree as a ``SymTree''. A SymTree consists of branches, which merge at junctions. Each branch is associated with a SymTFT, and each junction specifies gluing / compatibility conditions for these bulk TFTs. The junctions themselves need not be topological, and often support additional degrees of freedom. This also leads to a natural categorical structure where collections of SymTFTs are the objects and the junctions serve as morphisms. Rearrangements of the branches of a SymTree amount to compatibility conditions for the morphisms,
i.e., homotopy equivalences.\footnote{In principle there can be anomalies / obstructions in carrying out such moves.}

To establish this, we begin with the $(D+1)$-dimensional SymTFT for the full multi-sector QFT. Treating the topological couplings between sectors as supported in $D$ dimensions, we can pull these into the bulk. This results in junctions of SymTFTs for the different relative QFTs. Heavy defects defined in one relative theory are dressed by extended operators in the bulk SymTFTs, which can split / attach to other relative QFT sectors by passing through the junctions. Similarly, symmetry operators attached to the topological boundary of the full SymTFT can be pushed through the junctions, resulting in symmetry operators which are dressed by defects possibly attached to the junction, and can also be shared between multiple sectors.

SymTrees have a direct geometric interpretation in string theory.
To begin, recall that string-realized QFTs decoupled from gravity
naturally arise from local geometries with various singularities. In stringy terms, a
multi-sector QFT simply amounts to having a non-compact geometry with more than one
such singularity. Near each singularity we get a collection of local
operators, and heavy states which stretch across the different sectors are
integrated out, leaving their imprint in the low energy effective field
theory via higher dimension operators suppressed by a scale $\Lambda$. Taking $\Lambda$ very large and possibly tuning other moduli then results in seemingly decoupled QFT sectors.\footnote{This further tuning of moduli is sometimes necessary to truly decouple the sectors. For example, another contribution which
often does not decouple are kinetic mixing terms between $\U(1)$ gauge fields \cite{Holdom:1985ag}.
These arise from integrating out charged states which
could in principle be very heavy. This leads to a more ``obvious'' non-decoupling effect,
but one which is somewhat orthogonal to the considerations of the present work. This contribution can also be switched off via suitable tuning
of moduli.
}

While this provides a way to partially sequester the contributions
from different sectors, there are still residual topological couplings which
persist, even into the deep infrared. Extending the picture of reference \cite{Apruzzi:2021nmk}
to cover this case, we observe that for a $D$-dimensional QFT, we can
indeed locally construct a $(D+1)$-dimensional SymTFT by extending in the
\textquotedblleft radial direction\textquotedblright\ emanating out from a
given singularity. But with multiple QFT\ sectors, this radial direction
inevitably fuses with \textit{other} locally defined SymTFTs. The resulting structure
is thus really a junction of individual SymTFTs which fuse along a possibly
non-topological $D$-dimensional interface. While the $(D+1)$-dimensional
description is somewhat singular, it is clear that this is smoothed out in the extra-dimensional geometry of string theory.
From this perspective, the stringy construction (when
available) leads to a systematic method for constructing a SymTree. Any ambiguities in reading off the SymTree
amount to dualities / homotopy moves which rearrange the branches of the SymTree.


The string theory characterization of heavy defects and topological symmetry
operators exactly fits with these general considerations. Much
as in \cite{DelZotto:2015isa} (see also \cite{GarciaEtxebarria:2019caf, Albertini:2020mdx, Morrison:2020ool}), heavy defects arise from branes wrapped on non-compact cycles which extend from a given QFT\ sector out to \textquotedblleft
infinity\textquotedblright\ (i.e., where we impose the topological boundary
conditions), which are then partially screened by branes wrapped on collapsing cycles.
On the other hand, symmetry operators arise from branes wrapped
\textquotedblleft at infinity\textquotedblright\ \cite{Apruzzi:2022rei, GarciaEtxebarria:2022vzq, Heckman:2022muc} (see also \cite{Heckman:2022xgu, Acharya:2023bth, Dierigl:2023jdp, Cvetic:2023plv, Apruzzi:2023uma, Cvetic:2023pgm}).
Pushing these branes in from infinity so that they are shared across multiple
QFT sectors exactly matches with the bottom up description in terms of
junctions of SymTFTs.

One of the general lessons from this sort of analysis is that trying to
characterize all categorical symmetry structures in terms of a single bulk
SymTFT can obscure some important features (though of course they are still present).
For example, there have been recent proposals that many categorical structures are captured by a
suitable fusion $(D-1)$ Category (see e.g., \cite{Bartsch:2023pzl, Bhardwaj:2023wzd, Bartsch:2023wvv, Bhardwaj:2023ayw}).
Our present considerations illustrate that
both the objects, as well as morphisms of the correct symmetry category for a
general QFT will inevitably be somewhat broader.\footnote{At the very least, the presence of non-topological
interfaces suggests that the collection of $k$-morphisms must be enriched.}

We illustrate these general features with examples, many of which also admit a
top down construction. As an illustrative example, we consider 7D\ gauge
theories engineered from M-theory on an ADE\ singularity. In this case, the
local geometry takes the form $\mathbb{C}^{2}/\Gamma$ for $\Gamma$ a finite
subgroup of $\SU(2)$. There is an ADE\ classification of such singularities,
and this fixes the Lie algebra type of the corresponding gauge theory, i.e., the relative QFT.
The global form of the gauge group is fixed by a choice of boundary conditions
\textquotedblleft at infinity\textquotedblright\ on the generalized lens space
$S^{3}/\Gamma$ in the asymptotic conical geometry. Complex structure
deformations of the singularity correspond to adjoint Higgsing of the
singularity, and can result in multiple QFT\ sectors where all connected
correlators for local operators in different sectors decouple below the Higgsing scale.
Even so, there can still be topological couplings between these sectors which correlate the
structure of heavy defects and topological symmetry operators. Focusing on the
local radial profile for these geometries, we uncover a junction of symmetry
TFTs with a non-topological interface theory, supported on the junction, setting boundary conditions for the TFTs.

This basic geometric example generalizes in a number of ways. For example, we can produce similar SymTree structures for
6D superconformal field theories (SCFTs), as well as their compactifications to lower-dimensional systems.
Similar considerations also apply in QFTs engineered via D-branes probing
singularities. For example, we can also realize 4D $\mathcal{N}=4$ Super
Yang-Mills theory with an A-type gauge group via spacetime filling D3-branes
sitting at a common point of $\mathbb{C}^{3}$. Partitioning up these D3-branes
to multiple stacks, we observe that these sectors decouple in the deep IR, but
that there are also massive strings which are integrated out in taking this
limit. The associated bulk SymTree exhibits the same structure of
SymTFTs fused along a non-topological junction. One can also
apply the same reasoning in hybrid situations where we have branes probing
singularities; we can deform the singularities and at the same time also
separate the stacks of D-branes in the extra dimensions, much as in \cite{Heckman:2021vzx}.

The unifying theme in all of these examples is that we start with a single ``parent theory'' and then consider a flow in the moduli space of vacua to a multi-sector QFT. The SymTree encodes a topological treelike structure associated with this flow.

In addition to these examples, we also present examples where the multi-sector model is not obtained from a flow in moduli space.
Such multi-sector models are ubiquitous in string compactifications which typically have other sequestered sectors anyway. From a bottom up perspective, these sectors can be viewed as always being at infinite distance in moduli space.

In all of these cases, we can use the \textquotedblleft branes at
infinity\textquotedblright\ perspective to construct heavy defects as well as
topological symmetry operators. Moving these objects into the bulk and passing
them to another sector explicitly illustrates that defects of one theory
inevitably need to be dressed by additional operators.

We anticipate that these considerations can be used to study the structure of
a wide variety of multi-sector QFTs. Indeed,
while our examples mainly focus on supersymmetric multi-sector QFTs, the structure of a SymTree is largely complementary data.
Along these lines we also give an example of a non-supersymmetric Yang-Mills theory coupled to a complex adjoint-valued scalar which has precisely the same sort of SymTree as found in the supersymmetric setting.

As another application, we use this perspective to study large $M$ ensemble averaging in the
context of the AdS/CFT correspondence \cite{Schlenker:2022dyo}. At a practical
level, this is expected for any ``self-averaging'' observable which is smooth
in the value of Newton's constant.\footnote{We review some aspects of self-averaging
observables in Appendix \ref{app:ENSEMBLE}.} On the other hand, phenomena such as the
confinement / deconfinement transition are observable in semi-classical
gravity, but are also quite sensitive to the specific higher-form symmetries
of the boundary theory \cite{Witten:1998zw, Aharony:1998qu}. One would presumably
still like to assert that even with large $M$ averaging, the Wilson lines of
$\SU(M)$ gauge theory serve as an order parameter for confinement /
deconfinement. Reconciling these two points of view, we can consider a
collection of large $M$ replicas with extended operators dressed by additional
extended operators. Dressing the Wilson lines of an individual replica by
operators shared across all the sectors yields a general procedure for
producing an order parameter which is still sensitive to the higher-form
symmetries of the original large $M$ gauge theory with no averaging. This sort
of construction also lifts to a top down proposal for implementing disorder
averaging \cite{Heckman:2021vzx}.\footnote{We caution that while this top down
procedure is designed to produce the same answers \textquotedblleft in the IR\textquotedblright, it will inevitably
depart from the single throat large $M$ answer at short distances / high energies.
See also reference \cite{Antinucci:2023uzq} for other aspects of generalized
symmetries in disorder averaged systems.}

The rest of this paper is organized as follows. In Section \ref{sec:SYMTREE}
we analyze in general terms the symmetry field theory associated with a
multi-sector QFT. In particular, we explain how junctions of SymTFTs arise in
this setting. In Section \ref{sec:TOPDOWN} we show how this treelike structure
is smoothed out in the extra dimensions of string constructions. Section \ref{sec:7DSYM}
presents an illustrative example of SymTrees for 7D Super Yang-Mills theory. We present additional
examples constructed via vacuum moduli space flows in Section \ref{sec:FLOW}, and in Section \ref{sec:ISOLATED} we
construct examples where each sector is at infinite distance in moduli space from its counterpart. Section \ref{sec:NONSUSY}
presents a non-supersymmetric example of SymTrees for Yang-Mills theory coupled to a complex adjoint-valued scalar.
In Section \ref{sec:LARGEN} we use this structure to study higher-form symmetries in large $M$ ensemble averaging. Section
\ref{sec:CONC} contains our conclusions and future directions. In Appendix \ref{app:KapustinSeiberg} we study the SymTree of a gauge theory coupled to a TFT. Appendix \ref{app:Kineticterms} gives additional details on a top down derivation of single derivative terms of a SymTFT.
In Appendices \ref{app:FiltrationsALE} and \ref{app:SEQISOLATED} we present more details on some of the (co)homology calculations used in the main body. Appendix \ref{app:ENSEMBLE} reviews some additional details on ensemble averaging in holography.

\section{SymTrees} \label{sec:SYMTREE}

\begin{figure}
\centering
\scalebox{0.8}{
\begin{tikzpicture}
\begin{pgfonlayer}{nodelayer}
		\node [style=none] (0) at (-6, -2) {};
		\node [style=none] (1) at (-6, 2) {};
		\node [style=none] (2) at (-2, 2) {};
		\node [style=none] (3) at (-2, -2) {};
		\node [style=none] (8) at (-1.25, 0) {$\mathcal{B}_{\textnormal{phys}}$};
		\node [style=none] (9) at (-6.75, 0) {$\mathcal{B}_{\textnormal{top}}$};
		\node [style=none] (11) at (-6, 0) {};
		\node [style=none] (12) at (-2, 0) {};
		\node [style=none] (13) at (-4.5, 1.5) {};
		\node [style=none] (14) at (-3.5, 1.5) {};
		\node [style=none] (15) at (-4.18, 0.18) {};
		\node [style=none] (16) at (-4.15, -0.05) {};
		\node [style=none] (17) at (-4.75, 0) {};
		\node [style=none] (18) at (-4.625, 0) {};
		\node [style=none] (19) at (-4.875, 0) {};
		\node [style=none] (20) at (-4, -2.5) {SymTFT Slab};
		\node [style=none] (21) at (-4.4, -0.75) {$\mathcal{U}$};
		\node [style=none] (22) at (-3, 0.5) {$D$};
	\end{pgfonlayer}
	\begin{pgfonlayer}{edgelayer}
		\draw [style=BlueLine] (1.center) to (0.center);
		\draw [style=RedLine] (2.center) to (3.center);
		\draw [style=ThickLine] (1.center) to (2.center);
		\draw [style=ThickLine] (3.center) to (0.center);
		\draw [style=LineGreen, in=90, out=90, looseness=1.75] (17.center) to (15.center);
		\draw [style=LineGreen, in=-90, out=-90, looseness=2] (16.center) to (17.center);
		\draw [style=BrownLine, snake it] (11.center) to (19.center);
		\draw [style=BrownLine, snake it] (18.center) to (12.center);
	\end{pgfonlayer}
\end{tikzpicture}}
\caption{Standard SymTFT setup. Topological symmetry operators (green, $\mathcal{U}$) link heavy defect operators (grey, $D$) in the $(D+1)$-dimensional slab. The defects stretch from the topological boundary (blue, $\mathcal{B}_{\mathrm{top}}$) to the physical boundary (red, $\mathcal{B}_{\mathrm{phys}}$).}
\label{fig:Setup}
\end{figure}
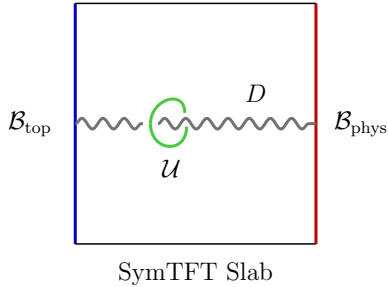

In this paper we shall be interested in the structure of $D$-dimensional multi-sector QFTs, and a $(D+1)$-dimensional bulk
theory which governs its symmetries. While our considerations are motivated by string-theoretic constructions, they can be stated in purely field theoretic terms, and so in this Section we opt to give a bottom up characterization of these structures. We defer a top down string-theoretic approach to Section \ref{sec:TOPDOWN}.\footnote{There are many excellent resources for learning more about string theory. See, e.g., reference \cite{StringDummies}.}

To begin, we recall that the symmetries of a (relative) $D$-dimensional QFT can be encapsulated in terms of a corresponding $(D+1)$-dimensional symmetry topological field theory (SymTFT) \cite{Freed:2012bs, Freed:2018cec, Freed:2022qnc, Kaidi:2022cpf}. In this symmetry TFT, the global form of the QFT is specified by suitable boundary conditions. More precisely, we have a state / physical boundary condition $\vert \mathcal{B}_{\mathrm{phys}} \rangle$, or simply $\mathcal{B}_{\mathrm{phys}}$, as well as a topological boundary condition $\langle \mathcal{B}_{\mathrm{top}} \vert$, or simply $ \mathcal{B}_{\mathrm{top}}$, which amounts to a choice of Dirichlet boundary conditions for some of the fields of the SymTFT and Neumann boundary conditions for others. The partition function of the absolute QFT is then given by evaluation of the overlap of states:
\begin{equation}
\mathcal{Z}_{\mathcal{B}_{\mathrm{top}} , \mathcal{B}_{\mathrm{phys}}} = \langle \mathcal{B}_{\mathrm{top}} \vert \mathcal{B}_{\mathrm{phys}}\rangle,
\end{equation}
in the obvious notation.

Heavy defects and topological symmetry operators can be introduced in this framework in a straightforward manner. First of all, we can consider symmetry operators localized near the topological boundary condition. We can of course move this operator into the $(D+1)$-dimensional bulk and over to the physical boundary. These symmetry operators acts on the heavy defects of the QFT. In the SymTFT, these heavy defects lift to defects which fill out one more direction in the bulk, and stretch from $\mathcal{B}_{\mathrm{top}}$ to $\mathcal{B}_{\mathrm{phys}}$ (see figure \ref{fig:Setup}). Observe that symmetry operators and physical operators now link both in the $D$-dimensional physical boundary, as well as in the $(D+1)$-dimensional bulk.

Our interest here will be in QFTs with multiple sectors. Our definition of this is to begin with distinct relative theories $\mathcal{T}_1$ and $\mathcal{T}_2$. These theories might be coupled via operator mixing terms. We demand, however, that there is a limit of various mass scales and / or parameters in which they decouple:
\begin{equation}\label{eq:decoup}
\langle \mathcal{O}_1 \mathcal{O}_2 \rangle_{\mathrm{conn}} \rightarrow 0\,.
\end{equation}
A typical situation is operator mixing via higher-dimension operators. In the limit where the suppression scale $\Lambda \rightarrow \infty$, this mixing term vanishes. While our definition also allows for possible mixing by marginal operators (e.g., as would occur in models with kinetic mixing) the essential points are already covered by cases with just higher-dimension operator mixing.

In any case, when the conditions leading to line \eqref{eq:decoup} are satisfied
we refer to this as a multi-sector QFT. Clearly, we can extend this to include any number of theories $\mathcal{T}_i$.
We then introduce another relative theory $\mathcal{J}$ (for ``junction'') which has only topological couplings to the original relative theories and thus mixes the different sectors via terms which do not fully decouple.

For each such sector $\mathcal{T}_i$, we can therefore speak of a symmetry TFT $\mathcal{S}_{i}$ which lives in $D+1$ dimensions, and has a physical boundary condition $ \mathcal{B}_{\mathrm{phys}}^{(i)}$ specifying a relative QFT. For each such SymTFT, we can also speak of the associated boundary conditions $ \mathcal{B}_{\mathrm{top}}^{(i)} $ which fixes an absolute theory.

We now glue the theories together.
We start with the original decoupled theories, $\mathcal{T}_{1}$ and $\mathcal{T}_{2}$, and assume these admit descriptions by actions $S_1$ and $S_2$ respectively. Then introduce the junction theory $\mathcal{J}_{12}$, again assumed to have an associated action $S_{\mathcal{J}_{12}}$. In these cases the topological couplings captured by the SymTree are given by an overall $D$-dimensional action
\be
S_{\rm full}=S_1+S_2+S_{\mathcal{J}_{12}}+S_{\rm Tmix}
\ee
where $S_{\rm Tmix}$ describes topological mixing terms.\footnote{This is distinct from the procedure of coupling a QFT to a TFT discussed in reference \cite{Kapustin:2014gua}. In that case, the TFT is coupled in a way such that overall no degrees of freedom are added and only a change in polarization is achieved. We comment on the relation to our construction further in Appendix \ref{app:KapustinSeiberg}.} The full SymTree then further supplements $S_{\rm full}$ by specifying the path-integral. We refer to the relative theory associated with the action $S_{\rm full}$ as $\mathcal{T}_{\mathrm{full}}$.

For non-Lagrangian theories the SymTree should be considered as the definition of the topological couplings we consider. More precisely, $\mathcal{T}_{\mathrm{full}}$ is schematically presented as
\be\label{eq:ABC}
\begin{tikzpicture}
	\begin{pgfonlayer}{nodelayer}
		\node [style=none] (0) at (-5, 0) {$\mathcal{T}_{1} $};
		\node [style=none] (1) at (-4.375, 0) {};
		\node [style=none] (2) at (-2.625, 0) {};
		\node [style=none] (3) at (-2, 0) {$ \mathcal{J}_{12} $};
		\node [style=none] (4) at (-1.375, 0) {};
		\node [style=none] (5) at (0.375, 0) {};
		\node [style=none] (6) at (1, 0) {$\mathcal{T}_{2} $};
		\node [style=none] (7) at (-0.5, 0.35) {\small $\mathsf{TFT}_{\mathcal{J}_{12},\mathcal{T}_{2}}$};
		\node [style=none] (8) at (-3.5, 0.35) {\small $\mathsf{TFT}_{\mathcal{T}_{1},\mathcal{J}_{12}}$};
	\end{pgfonlayer}
	\begin{pgfonlayer}{edgelayer}
		\draw [style=ThickLine] (1.center) to (2.center);
		\draw [style=ThickLine] (4.center) to (5.center);
	\end{pgfonlayer}
\end{tikzpicture}
\ee
with TFTs $\mathsf{TFT}_{\mathcal{T}_1,\mathcal{J}_{12}}$ and $\mathsf{TFT}_{\mathcal{J}_{12},\mathcal{T}_{2}}$ in one higher dimension, which have edge mode theories as indicated by the subscripts. In the end, the original sectors $\mathcal{T}_1$ and $\mathcal{T}_2$ now interact via topological terms, as well as with an intermediate gluing theory $\mathcal{J}_{12}$. Clearly this same structure extends to QFTs with many sectors, and so we can label the original decoupled sectors as $\mathcal{T}_{i}$, with $i \in I$ an index. From this, we can fuse together multiple decoupled sectors by picking a subset $J \subset I$, with an associated $\mathcal{J}_{J}$ of gluing theories and a topological field theory $\mathsf{TFT}_{J}$ which couples the different sectors together. We refer to the full (relative) $D$-dimensional theory obtained in this way as $\mathcal{T}_{\mathrm{full}}$. Let us note that this sort of structure naturally appears in a number of contexts, for example in adjoint Higgsing of a gauge theory where the IR theory separates into sectors which are decoupled (up to topological terms). It is also quite common in stringy realizations of QFTs where there is a clear notion of geometric localization of operators, including geometrically delocalized sectors (the $\mathcal{J}$'s) which are shared across multiple sectors. We turn to examples of this sort in later sections.

\begin{figure}
\centering
\scalebox{0.8}{
\begin{tikzpicture}
\begin{pgfonlayer}{nodelayer}
		\node [style=none] (0) at (-4, -2) {};
		\node [style=none] (1) at (-4, 2) {};
		\node [style=none] (2) at (0, 2) {};
		\node [style=none] (3) at (0, -2) {};
		\node [style=none] (4) at (3.75, 3.25) {};
		\node [style=none] (5) at (3, 0.75) {};
		\node [style=none] (6) at (3.75, -0.75) {};
		\node [style=none] (7) at (3, -3.25) {};
		\node [style=none] (8) at (3, -1) {};
		\node [style=none] (9) at (-2, 0) {$\mathcal{S}_{\textnormal{full}}$};
		\node [style=none] (10) at (-5, 0) {$\mathcal{B}_{\textnormal{top}}$};
		\node [style=none] (11) at (3.75, -2.5) {$\mathcal{B}_{\textnormal{phys}}^{(1)}$};
		\node [style=none] (12) at (4.5, 2.5) {$\mathcal{B}_{\textnormal{phys}}^{(2)}$};
		\node [style=none] (13) at (0, 2.75) {$\mathcal{J}$};
		\node [style=none] (14) at (2.25, -2.25) {$\mathcal{S}_1$};
		\node [style=none] (15) at (3, 2.25) {};
		\node [style=none] (16) at (3, 2.25) {$\mathcal{S}_2$};
	\end{pgfonlayer}
	\begin{pgfonlayer}{edgelayer}
		\draw [style=BlueLine] (1.center) to (0.center);
		\draw [style=PurpleLine] (2.center) to (3.center);
		\draw [style=RedLine] (4.center) to (6.center);
		\draw [style=RedLine] (5.center) to (7.center);
		\draw [style=ThickLine] (1.center) to (2.center);
		\draw [style=ThickLine] (2.center) to (4.center);
		\draw [style=ThickLine] (2.center) to (5.center);
		\draw [style=ThickLine] (7.center) to (3.center);
		\draw [style=ThickLine] (3.center) to (0.center);
		\draw [style=ThickLine] (8.center) to (6.center);
		\draw [style=DottedLine] (3.center) to (8.center);
	\end{pgfonlayer}
\end{tikzpicture}}
\caption{We depict a trivalent junction $\mathcal{J}$ of symmetry TFTs. The junctions supports the $D$-dimensional theory $\mathcal{G}_{J} \otimes \mathsf{TFT}_{J}$. Color conventions: Junctions are purple.}
\label{fig:SymTree}
\end{figure}
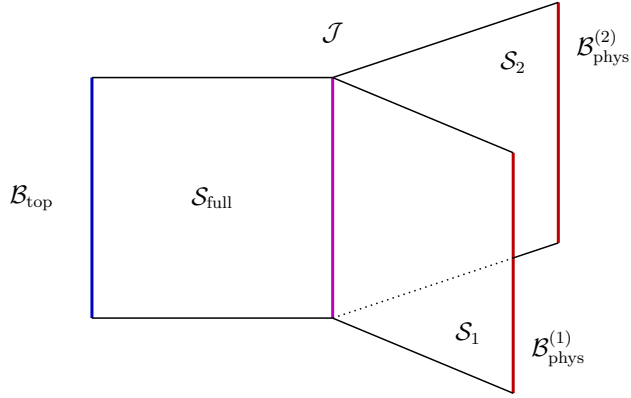

Now, on general grounds, the $D$-dimensional theory $\mathcal{T}_{\mathrm{full}}$ has its own SymTFT, which we refer to as $\mathcal{S}_{\mathrm{full}}$. That being said, there is clearly more fine-grained structure available from decomposing the boundary $D$-dimensional theory into its constituent parts, where each individual sector has its own SymTFT. Indeed, for a given sector $\mathcal{T}_{i}$, we have an associated SymTFT $\mathcal{S}_{i}$ and in the discussion above we make the identification:
\begin{equation}
\mathcal{S}_{i}= \mathsf{TFT}_{\mathcal{J}_{J},\mathcal{T}_{i}}.
\end{equation}
Here, $\mathcal{J}_J$ refers to a junction theory which couples together some collection of theories as obtained from a subset $J \subset I$.
The process of coupling the $\mathcal{T}_{i}$ sectors together can be visualized in terms of a treelike structure $\Upsilon$: Along each terminating branch, we have a SymTFT $\mathcal{S}_i$ as associated with the theory $\mathcal{T}_i$. In this theory, we have a physical boundary condition $ \mathcal{B}_{\mathrm{phys}}^{(i)}$. At the other end of $\mathcal{S}_i$, we can fuse it to a collection of other SymTFTs. The fusion in question involves a collection of theories indexed by $J \subset I$. At this junction, we have the $D$-dimensional theory,
and emanating out from it, we have the other SymTFTs (see figure \ref{fig:SymTree}).

\begin{figure}
\centering
\scalebox{0.55}{
\begin{tikzpicture}
	\begin{pgfonlayer}{nodelayer}
		\node [style=none] (0) at (2.25, 1.5) {};
		\node [style=none] (1) at (2.25, -2.5) {};
		\node [style=none] (2) at (5.75, 4) {};
		\node [style=none] (3) at (5.75, 0) {};
		\node [style=none] (4) at (5.75, -1.25) {};
		\node [style=none] (5) at (5.75, -5.25) {};
		\node [style=none] (6) at (-1.75, 1.5) {};
		\node [style=none] (7) at (-1.75, -2.5) {};
		\node [style=none] (8) at (5.75, 0) {};
		\node [style=none] (9) at (2.25, -2.5) {};
		\node [style=none] (10) at (2.25, 1.5) {};
		\node [style=none] (13) at (-2.75, -0.5) {\Large $\mathcal{B}_{\textnormal{top}}$};
		\node [style=none] (14) at (5.75, 4.75) {\Large $\mathcal{B}_{\textnormal{phys}}^{(N)}$};
		\node [style=none] (15) at (6.5, 6.5) {\Large $\mathcal{B}_{\textnormal{phys}}^{(1)}$};
		\node [style=none] (22) at (5.75, -6) {\Large $\mathcal{B}_{\textnormal{phys}}^{(1)}$};
		\node [style=none] (23) at (0, 13) {};
		\node [style=none] (24) at (0, 9) {};
		\node [style=none] (25) at (-4, 13) {};
		\node [style=none] (26) at (-4, 9) {};
		\node [style=none] (27) at (0, 9) {};
		\node [style=none] (28) at (0, 13) {};
		\node [style=none] (29) at (-5, 11) {\Large $\mathcal{B}_{\textnormal{top}}$};
		\node [style=none] (30) at (7, 12) {};
		\node [style=none] (31) at (7, 8) {};
		\node [style=none] (32) at (8, 13) {};
		\node [style=none] (33) at (8, 9) {};
		\node [style=none] (34) at (9, 14) {};
		\node [style=none] (35) at (9, 10) {};
		\node [style=none] (36) at (6, 11) {};
		\node [style=none] (37) at (6, 7) {};
		\node [style=none] (38) at (3.5, 12.5) {};
		\node [style=none] (39) at (3.5, 8.5) {};
		\node [style=none] (40) at (1.75, 12.75) {};
		\node [style=none] (41) at (1.75, 8.75) {};
		\node [style=none] (42) at (5.5, 12.9) {};
		\node [style=none] (43) at (5.5, 8.9) {};
		\node [style=none] (44) at (6, 8.125) {};
		\node [style=none] (45) at (7, 8.95) {};
		\node [style=none] (46) at (8, 9.875) {};
		\node [style=none] (54) at (7.5, 7.5) {\Large $\mathcal{B}_{\textnormal{phys}}^{(2)}$};
		\node [style=none] (55) at (8.5, 8.5) {\Large $\mathcal{B}_{\textnormal{phys}}^{(3)}$};
		\node [style=none] (56) at (9.5, 9.5) {\Large $\mathcal{B}_{\textnormal{phys}}^{(4)}$};
		\node [style=none] (57) at (6.75, 1.5) {};
		\node [style=none] (58) at (6.75, -2.5) {};
		\node [style=none] (59) at (5.75, -2.5) {};
		\node [style=none] (60) at (7.75, -0.5) {\Large $\mathcal{B}_{\textnormal{phys}}^{(i)}$};
		\node [style=none] (61) at (6.75, -3.25) {};
		\node [style=none] (62) at (6.25, -4.75) {};
		\node [style=none] (63) at (6.75, 2.25) {};
		\node [style=none] (64) at (6.25, 3.5) {};
		\node [style=none] (65) at (0, 7) {\Large (i)};
		\node [style=none] (66) at (2.25, -4.5) {\Large (ii)};
		\node [style=none] (67) at (2.25, -6.75) {};
	\end{pgfonlayer}
	\begin{pgfonlayer}{edgelayer}
		\draw [style=ThickLine] (6.center) to (0.center);
		\draw [style=ThickLine] (0.center) to (2.center);
		\draw [style=ThickLine] (1.center) to (7.center);
		\draw [style=BlueLine] (6.center) to (7.center);
		\draw [style=RedLine] (4.center) to (5.center);
		\draw [style=RedLine] (2.center) to (3.center);
		\draw [style=ThickLine] (4.center) to (0.center);
		\draw [style=ThickLine] (5.center) to (1.center);
		\draw [style=DottedLine] (1.center) to (8.center);
		\draw [style=RedLine] (10.center) to (9.center);
		\draw [style=PurpleLine] (10.center) to (9.center);
		\draw [style=ThickLine] (25.center) to (23.center);
		\draw [style=ThickLine] (24.center) to (26.center);
		\draw [style=BlueLine] (25.center) to (26.center);
		\draw [style=RedLine] (28.center) to (27.center);
		\draw [style=PurpleLine] (28.center) to (27.center);
		\draw [style=RedLine] (30.center) to (31.center);
		\draw [style=ThickLine] (28.center) to (30.center);
		\draw [style=RedLine] (32.center) to (33.center);
		\draw [style=RedLine] (34.center) to (35.center);
		\draw [style=RedLine] (36.center) to (37.center);
		\draw [style=RedLine] (38.center) to (39.center);
		\draw [style=PurpleLine] (38.center) to (39.center);
		\draw [style=RedLine] (40.center) to (41.center);
		\draw [style=PurpleLine] (40.center) to (41.center);
		\draw [style=ThickLine] (40.center) to (32.center);
		\draw [style=ThickLine] (39.center) to (37.center);
		\draw [style=ThickLine] (38.center) to (36.center);
		\draw [style=ThickLine] (34.center) to (28.center);
		\draw [style=ThickLine] (27.center) to (39.center);
		\draw [style=ThickLine] (44.center) to (31.center);
		\draw [style=ThickLine] (45.center) to (33.center);
		\draw [style=ThickLine] (46.center) to (35.center);
		\draw [style=DottedLine] (39.center) to (44.center);
		\draw [style=DottedLine] (41.center) to (45.center);
		\draw [style=DottedLine] (27.center) to (46.center);
		\draw [style=RedLine] (57.center) to (58.center);
		\draw [style=ThickLine] (10.center) to (57.center);
		\draw [style=ThickLine] (59.center) to (58.center);
		\draw [style=DottedLine] (9.center) to (59.center);
		\draw [style=DottedRed, in=90, out=-45, looseness=0.75] (64.center) to (63.center);
		\draw [style=DottedRed, in=45, out=-90, looseness=0.75] (61.center) to (62.center);
	\end{pgfonlayer}
\end{tikzpicture}}
\caption{Junctions can be assembled into trees (i). The tree $\Upsilon$ can be visualized as a horizontal cross-section. Junctions can have arbitrary valency (ii).
}
\label{fig:Junctions}
\end{figure}
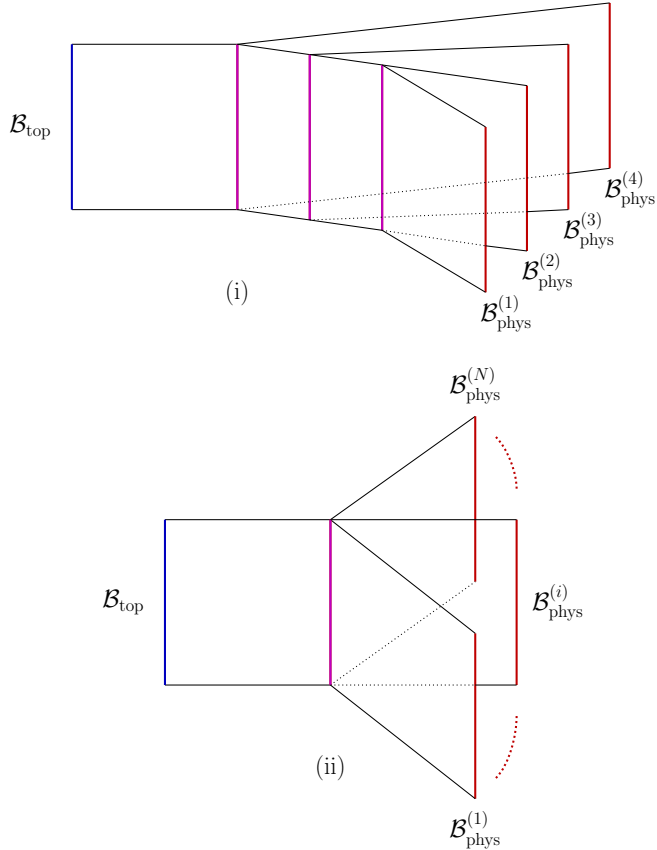

Clearly, there are many ways to construct such a junction, and each of them leads to a different treelike structure (see figure \ref{fig:Junctions}). That being said, for each choice of tree, we get a notion of a $(D+1)$-dimensional bulk. Away from all of these junctions, we can also speak of the topological boundary for $\mathcal{S}_{\mathrm{full}}$. Indeed, pushing all of the junctions into the physical boundary conditions results in the SymTree reducing to a single slab filled by $\mathcal{S}_{\mathrm{full}}$. We denote the boundary condition, obtained by stacking all junctions and physical boundary conditions of the multi-sector QFT, by $\mathcal{B}_{\textnormal{phys}}^{(\textnormal{retract})}$.

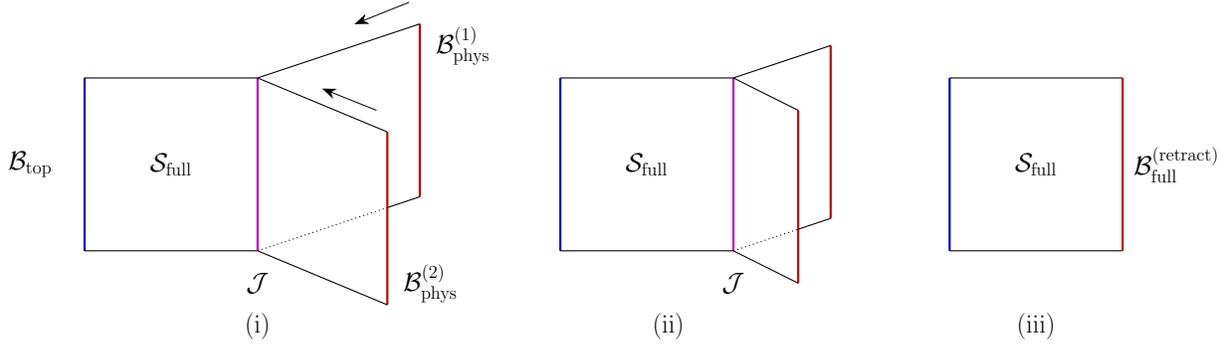
\begin{figure}
\centering
\scalebox{0.575}{
\begin{tikzpicture}
	\begin{pgfonlayer}{nodelayer}
		\node [style=none] (0) at (-11.5, -1.25) {};
		\node [style=none] (1) at (-11.5, 2.75) {};
		\node [style=none] (2) at (-7.5, 2.75) {};
		\node [style=none] (3) at (-7.5, -1.25) {};
		\node [style=none] (4) at (-3.75, 4) {};
		\node [style=none] (5) at (-4.5, 1.5) {};
		\node [style=none] (6) at (-3.75, 0) {};
		\node [style=none] (7) at (-4.5, -2.5) {};
		\node [style=none] (8) at (-4.5, -0.25) {};
		\node [style=none] (9) at (-0.5, -1.25) {};
		\node [style=none] (10) at (-0.5, 2.75) {};
		\node [style=none] (11) at (3.5, 2.75) {};
		\node [style=none] (12) at (3.5, -1.25) {};
		\node [style=none] (13) at (5.75, 3.5) {};
		\node [style=none] (14) at (5, 2) {};
		\node [style=none] (15) at (5.75, -0.5) {};
		\node [style=none] (16) at (5, -2) {};
		\node [style=none] (17) at (5, -0.75) {};
		\node [style=none] (18) at (-4, 4.5) {};
		\node [style=none] (19) at (-5.25, 4) {};
		\node [style=none] (22) at (8.5, -1.25) {};
		\node [style=none] (23) at (8.5, 2.75) {};
		\node [style=none] (24) at (12.5, 2.75) {};
		\node [style=none] (25) at (12.5, -1.25) {};
		\node [style=none] (26) at (-2.75, 3.5) {\Large $\mathcal{B}^{(1)}_{\rm phys}$};
		\node [style=none] (27) at (-3.5, -2) {\Large $\mathcal{B}^{(2)}_{\rm phys}$};
		\node [style=none] (28) at (-12.75, 0.75) {\Large $\mathcal{B}_{\rm top}$};
		\node [style=none] (29) at (-7.5, -2) {\Large $\mathcal{J}$};
		\node [style=none] (30) at (3.5, -2) {\Large $\mathcal{J}$};
		\node [style=none] (31) at (13.75, 0.75) {\Large $\mathcal{B}_{\rm full}^{\rm (retract)}$};
		\node [style=none] (32) at (-7.5, -3) {\Large (i)};
		\node [style=none] (33) at (2, -3) {\Large (ii)};
		\node [style=none] (34) at (10.5, -3) {\Large (iii)};
		\node [style=none] (35) at (2.25, -3.75) {};
		\node [style=none] (36) at (-9.5, 0.75) {\Large $\mathcal{S}_{\rm full}$};
		\node [style=none] (37) at (1.5, 0.75) {\Large $\mathcal{S}_{\rm full}$};
		\node [style=none] (38) at (10.5, 0.75) {\Large $\mathcal{S}_{\rm full}$};
		\node [style=none] (39) at (-4.75, 2) {};
		\node [style=none] (40) at (-6, 2.5) {};
	\end{pgfonlayer}
	\begin{pgfonlayer}{edgelayer}
		\draw [style=BlueLine] (1.center) to (0.center);
		\draw [style=PurpleLine] (2.center) to (3.center);
		\draw [style=RedLine] (4.center) to (6.center);
		\draw [style=RedLine] (5.center) to (7.center);
		\draw [style=ThickLine] (1.center) to (2.center);
		\draw [style=ThickLine] (2.center) to (4.center);
		\draw [style=ThickLine] (2.center) to (5.center);
		\draw [style=ThickLine] (7.center) to (3.center);
		\draw [style=ThickLine] (3.center) to (0.center);
		\draw [style=ThickLine] (8.center) to (6.center);
		\draw [style=DottedLine] (3.center) to (8.center);
		\draw [style=BlueLine] (10.center) to (9.center);
		\draw [style=PurpleLine] (11.center) to (12.center);
		\draw [style=RedLine] (13.center) to (15.center);
		\draw [style=RedLine] (14.center) to (16.center);
		\draw [style=ThickLine] (10.center) to (11.center);
		\draw [style=ThickLine] (11.center) to (13.center);
		\draw [style=ThickLine] (11.center) to (14.center);
		\draw [style=ThickLine] (16.center) to (12.center);
		\draw [style=ThickLine] (12.center) to (9.center);
		\draw [style=ThickLine] (17.center) to (15.center);
		\draw [style=DottedLine] (12.center) to (17.center);
		\draw [style=ArrowLineRight] (18.center) to (19.center);
		\draw [style=BlueLine] (23.center) to (22.center);
		\draw [style=ThickLine] (23.center) to (24.center);
		\draw [style=ThickLine] (25.center) to (22.center);
		\draw [style=RedLine] (24.center) to (25.center);
		\draw [style=ArrowLineRight] (39.center) to (40.center);
	\end{pgfonlayer}
\end{tikzpicture}}
\caption{Depiction of retracting a SymTree to produce the corresponding SymTFT $\mathcal{S}_{\mathrm{full}}$ for the multi-sector QFT with topological couplings between the different sectors. In terms of the SymTree, this amounts to pulling in the different branches into the physical boundaries.}
\label{fig:Tuck}
\end{figure}

For a depiction of ``retracting'' a SymTree, see figure \ref{fig:Tuck}. The physical boundary condition after retracting, $\mathcal{B}_{\textnormal{phys}}^{(\textnormal{retract})}$, is equivalent to the $D$-dimensional theory given by dimensionally reducing $\mathcal{S}_1\otimes \mathcal{S}_2$ along the interval with boundary conditions on one end given by the junction $\mathcal{J}$, and on the other end by $\mathcal{B}^{(1)}_{\mathrm{phys}}\otimes \mathcal{B}^{(2)}_{\mathrm{phys}}$.

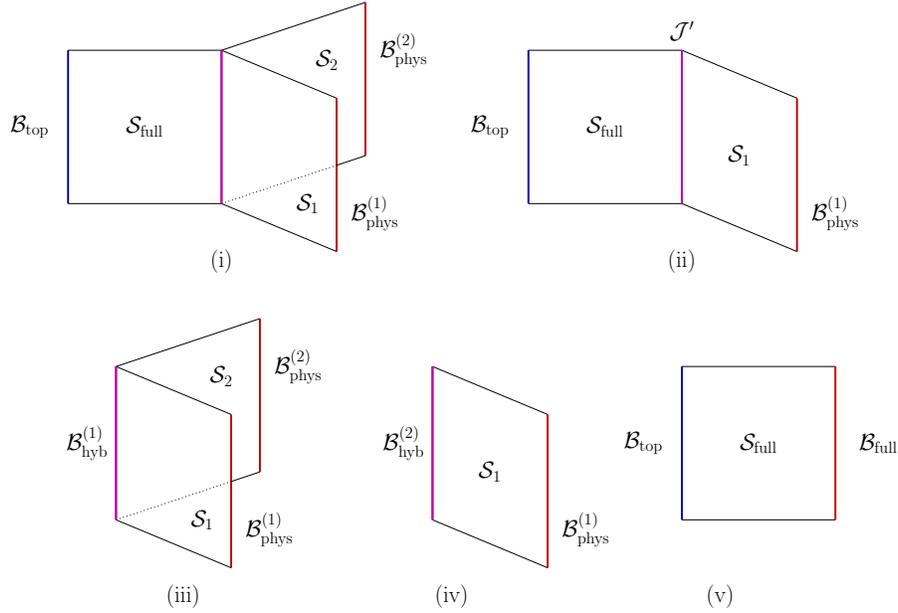
\begin{figure}
    \centering
    \scalebox{0.51}{
    \begin{tikzpicture}
	\begin{pgfonlayer}{nodelayer}
		\node [style=none] (0) at (-6, 2) {};
		\node [style=none] (1) at (-6, -2) {};
		\node [style=none] (2) at (-2.25, 3.25) {};
		\node [style=none] (3) at (-2.25, -0.75) {};
		\node [style=none] (4) at (-3, 0.75) {};
		\node [style=none] (5) at (-3, -3.25) {};
		\node [style=none] (6) at (-10, 2) {};
		\node [style=none] (7) at (-10, -2) {};
		\node [style=none] (8) at (-3, -1) {};
		\node [style=none] (9) at (-6, -2) {};
		\node [style=none] (10) at (-6, 2) {};
		\node [style=none] (13) at (-11, 0) {\Large $\mathcal{B}_{\textnormal{top}}$};
		\node [style=none] (14) at (-1.25, 2) {\Large $\mathcal{B}_{\textnormal{phys}}^{(2)}$};
		\node [style=none] (15) at (-2, -2.25) {\Large $\mathcal{B}_{\textnormal{phys}}^{(1)}$};
		\node [style=none] (16) at (6, 2) {};
		\node [style=none] (17) at (6, -2) {};
		\node [style=none] (20) at (9, 0.75) {};
		\node [style=none] (21) at (9, -3.25) {};
		\node [style=none] (22) at (2, 2) {};
		\node [style=none] (23) at (2, -2) {};
		\node [style=none] (25) at (6, -2) {};
		\node [style=none] (26) at (6, 2) {};
		\node [style=none] (27) at (1, 0) {\Large $\mathcal{B}_{\textnormal{top}}$};
		\node [style=none] (28) at (6, 2.5) {\Large $\mathcal{J}'$};
		\node [style=none] (29) at (10, -2.25) {\Large $\mathcal{B}_{\textnormal{phys}}^{(1)}$};
		\node [style=none] (30) at (-8.75, -6.25) {};
		\node [style=none] (31) at (-8.75, -10.25) {};
		\node [style=none] (32) at (-5, -5) {};
		\node [style=none] (33) at (-5, -9) {};
		\node [style=none] (34) at (-5.75, -7.5) {};
		\node [style=none] (35) at (-5.75, -11.5) {};
		\node [style=none] (38) at (-5.75, -9.25) {};
		\node [style=none] (39) at (-8.75, -10.25) {};
		\node [style=none] (40) at (-8.75, -6.25) {};
		\node [style=none] (41) at (-9.5, -8.25) {\Large $\mathcal{B}_{\textnormal{hyb}}^{(1)}$};
		\node [style=none] (42) at (-4, -6.25) {\Large $\mathcal{B}_{\textnormal{phys}}^{(2)}$};
		\node [style=none] (43) at (-4.75, -10.5) {\Large $\mathcal{B}_{\textnormal{phys}}^{(1)}$};
		\node [style=none] (44) at (-0.5, -6.25) {};
		\node [style=none] (45) at (-0.5, -10.25) {};
		\node [style=none] (48) at (2.5, -7.5) {};
		\node [style=none] (49) at (2.5, -11.5) {};
		\node [style=none] (51) at (-0.5, -10.25) {};
		\node [style=none] (52) at (-0.5, -6.25) {};
		\node [style=none] (53) at (-1.25, -8.25) {\Large $\mathcal{B}^{(2)}_{\textnormal{hyb}}$};
		\node [style=none] (55) at (3.5, -10.5) {\Large $\mathcal{B}_{\textnormal{phys}}^{(1)}$};
		\node [style=none] (56) at (-6, -3.5) {\Large (i)};
		\node [style=none] (57) at (6, -3.5) {\Large (ii)};
		\node [style=none] (58) at (-7, -12.25) {\Large (iii)};
		\node [style=none] (59) at (0, -12.25) {\Large (iv)};
		\node [style=none] (60) at (1, -13) {};
		\node [style=none] (61) at (10, -6.25) {};
		\node [style=none] (62) at (10, -10.25) {};
		\node [style=none] (63) at (6, -6.25) {};
		\node [style=none] (64) at (6, -10.25) {};
		\node [style=none] (65) at (10, -10.25) {};
		\node [style=none] (66) at (10, -6.25) {};
		\node [style=none] (67) at (5, -8.25) {\Large $\mathcal{B}_{\textnormal{top}}$};
		\node [style=none] (68) at (11.125, -8.25) {\Large $\mathcal{B}_{\textnormal{full}}$};
		\node [style=none] (69) at (7, -12.25) {\Large (v)};
		\node [style=none] (70) at (-3.75, -2) {\Large$\mathcal{S}_1$};
		\node [style=none] (71) at (-3.25, 1.75) {\Large $\mathcal{S}_2$};
		\node [style=none] (72) at (-8, 0) {\Large $\mathcal{S}_{\textnormal{full}}$};
		\node [style=none] (73) at (4, 0) {\Large $\mathcal{S}_{\textnormal{full}}$};
		\node [style=none] (74) at (7.5, -0.75) {\Large $\mathcal{S}_1$};
		\node [style=none] (75) at (1, -9) {\Large $\mathcal{S}_1$};
		\node [style=none] (76) at (-6.5, -10.25) {\Large $\mathcal{S}_1$};
		\node [style=none] (77) at (-6, -6.5) {\Large $\mathcal{S}_2$};
		\node [style=none] (78) at (8, -8.25) {\Large$\mathcal{S}_{\textnormal{full}}$};
	\end{pgfonlayer}
	\begin{pgfonlayer}{edgelayer}
		\draw [style=ThickLine] (6.center) to (0.center);
		\draw [style=ThickLine] (0.center) to (2.center);
		\draw [style=ThickLine] (1.center) to (7.center);
		\draw [style=BlueLine] (6.center) to (7.center);
		\draw [style=RedLine] (4.center) to (5.center);
		\draw [style=RedLine] (2.center) to (3.center);
		\draw [style=ThickLine] (4.center) to (0.center);
		\draw [style=ThickLine] (5.center) to (1.center);
		\draw [style=ThickLine] (8.center) to (3.center);
		\draw [style=DottedLine] (1.center) to (8.center);
		\draw [style=RedLine] (10.center) to (9.center);
		\draw [style=PurpleLine] (10.center) to (9.center);
		\draw [style=ThickLine] (22.center) to (16.center);
		\draw [style=ThickLine] (17.center) to (23.center);
		\draw [style=BlueLine] (22.center) to (23.center);
		\draw [style=RedLine] (20.center) to (21.center);
		\draw [style=ThickLine] (20.center) to (16.center);
		\draw [style=ThickLine] (21.center) to (17.center);
		\draw [style=RedLine] (26.center) to (25.center);
		\draw [style=PurpleLine] (26.center) to (25.center);
		\draw [style=ThickLine] (30.center) to (32.center);
		\draw [style=RedLine] (34.center) to (35.center);
		\draw [style=RedLine] (32.center) to (33.center);
		\draw [style=ThickLine] (34.center) to (30.center);
		\draw [style=ThickLine] (35.center) to (31.center);
		\draw [style=ThickLine] (38.center) to (33.center);
		\draw [style=DottedLine] (31.center) to (38.center);
		\draw [style=RedLine] (40.center) to (39.center);
		\draw [style=PurpleLine] (40.center) to (39.center);
		\draw [style=RedLine] (48.center) to (49.center);
		\draw [style=ThickLine] (48.center) to (44.center);
		\draw [style=ThickLine] (49.center) to (45.center);
		\draw [style=RedLine] (52.center) to (51.center);
		\draw [style=PurpleLine] (52.center) to (51.center);
		\draw [style=ThickLine] (63.center) to (61.center);
		\draw [style=ThickLine] (62.center) to (64.center);
		\draw [style=BlueLine] (63.center) to (64.center);
		\draw [style=RedLine] (66.center) to (65.center);
	\end{pgfonlayer}
\end{tikzpicture}
    }
    \caption{Subfigures (ii),(iii),(iv),(v) show various degenerations of (i) achieved by contracting one or more of the three symmetry theory slabs. Here $\mathcal{B}_{\textnormal{hyb}}$ denotes a hybrid boundary condition which occurs whenever the branch of the SymTree connecting to $\mathcal{B}_{\textnormal{top}}$ is contracted. Hybrid boundary are generically not purely topological. When a physical boundary condition is fused with a junction a new junction $\mathcal{J}'$ emerges.}
    \label{fig:ZIPPER}
\end{figure}

We also have a operation related to retraction, which we will refer to as ``unzipping". Whenever the multi-sector QFT emerges via limits of various mass scales and / or parameters, as discussed around \eqref{eq:decoup}, we are also handed the initial, single sector, parent QFT with associated boundary $\mathcal{B}_{\textnormal{phys}}^{(\textnormal{full})}$. Taking the discussed limits $\mathcal{B}_{\textnormal{phys}}^{(\textnormal{full})}$ reduces to $\mathcal{B}_{\textnormal{phys}}^{(\textnormal{retract})}$, i.e., the difference between these boundary conditions are precisely the states which decouple in the limit. Here we can immediately anticipate a convenient feature of the SymTree: it often happens that the decoupled states emerge as defects in the effective description, i.e., the spectrum of defects enhances. These are then manifest in the SymTree as we momentarily discuss.
Retractions of SymTrees are always allowed, these are a field theory manipulation.
There are of course other degenerations in which some other subset of the edges of the SymTree are contracted. For a trivalent SymTree we show the possible configurations in figure \ref{fig:ZIPPER}. We will also see examples in top-down approaches where, using the string theory construction, we are able to embed a given multi-sector QFT into a moduli space which has loci described by a single sector QFT. In this case, moving between the different moduli, we can also ``zip up" the Symtree mapping $\mathcal{B}_{\textnormal{phys}}^{(i)}$ and the junctions to $\mathcal{B}_{\textnormal{phys}}^{(\textnormal{full})}$.

Summarizing, the local subsectors $\mathcal{T}_i$ contribute relative theories \cite{Freed:2012bs}. Practically, this means that they each determine physical boundary conditions $\mathcal{B}^{(i)}_{\textnormal{phys}}$ of a symmetry topological field theory \cite{Freed:2022qnc, Kaidi:2023maf}. We have argued that topological non-decoupling between a collection of such relative theories amounts to interactions between their symmetry theories, which we formalize via junctions. Such junctions arise at the fusion of symmetry theories. These boundary conditions are not necessarily purely topological, rather they can be partially topological and partially physical. As such, the junctions may themselves support relative theories. Overall this results in a SymTree of symmetry theories with internal junctions and external boundaries (see figure \ref{fig:Junctions}). The data entering a SymTree includes:
\begin{equation}
    \begin{split}
        \textcolor{blue}{\mathcal{B}_{\text{top}}}: &~\text{Topological (i.e., gapped) boundary conditions},\\
        \textcolor{red}{\mathcal{B}_{\text{phys}}^{(i)}}: &~ \text{Physical boundary conditions for relative theory $\mathcal{T}_i$},\\
        \textcolor{purple}{\mathcal{J}}: &~\text{Junction with partially topological and partially physical boundary conditions,}\\
        \Upsilon: &~ \text{Tree built from SymTFTs and their junctions.}
    \end{split}
\end{equation}
Evaluation of the partition function for the SymTree theory depends on all this data, which we write as $\mathcal{Z}(\mathcal{B}_{\mathrm{top}}, \{\mathcal{B}^{(i)}_{\mathrm{phys}} \}, \Upsilon)$. Instead of trees one could of course consider arbitrary graphs, however, we find trees to arise in examples throughout and therefore restrict to these.

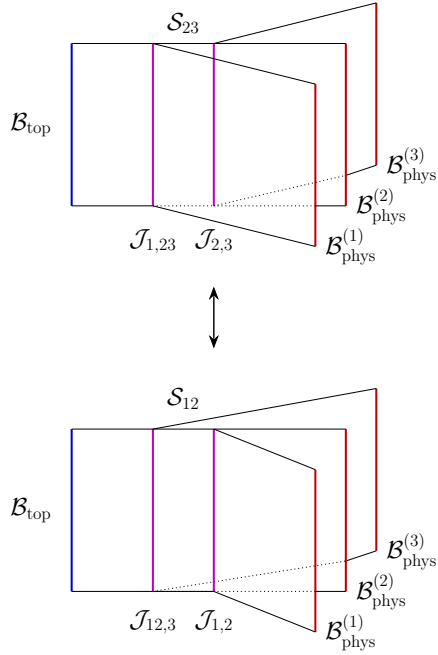
\begin{figure}
\centering
\scalebox{0.54}{
\begin{tikzpicture}
\begin{pgfonlayer}{nodelayer}
		\node [style=none] (0) at (-2.25, 0) {};
		\node [style=none] (1) at (-2.25, 4) {};
		\node [style=none] (2) at (1.25, 4) {};
		\node [style=none] (3) at (1.25, 0) {};
		\node [style=none] (4) at (5.25, 5) {};
		\node [style=none] (5) at (3.75, 3) {};
		\node [style=none] (6) at (5.25, 1) {};
		\node [style=none] (7) at (3.75, -1) {};
		\node [style=none] (8) at (4.5, 0.75) {};
		\node [style=none] (9) at (-0.25, 4) {};
		\node [style=none] (10) at (-0.25, 0) {};
		\node [style=none] (11) at (4.5, 4) {};
		\node [style=none] (12) at (4.5, 0) {};
		\node [style=none] (13) at (3.75, 0) {};
		\node [style=none] (14) at (-2.25, -9.5) {};
		\node [style=none] (15) at (-2.25, -5.5) {};
		\node [style=none] (16) at (-0.25, -5.5) {};
		\node [style=none] (17) at (-0.25, -9.5) {};
		\node [style=none] (18) at (5.25, -4.5) {};
		\node [style=none] (19) at (3.75, -6.5) {};
		\node [style=none] (20) at (5.25, -8.5) {};
		\node [style=none] (21) at (3.75, -10.5) {};
		\node [style=none] (22) at (4.5, -8.75) {};
		\node [style=none] (23) at (1.25, -5.5) {};
		\node [style=none] (24) at (1.25, -9.5) {};
		\node [style=none] (25) at (4.5, -5.5) {};
		\node [style=none] (26) at (4.5, -9.5) {};
		\node [style=none] (27) at (3.75, -9.5) {};
		\node [style=none] (28) at (1.25, -2.25) {};
		\node [style=none] (29) at (1.25, -3.5) {};
		\node [style=none] (30) at (4.625, -1) {\Large $\mathcal{B}^{(1)}_{\textnormal{phys}}$};
		\node [style=none] (31) at (5.375, 0) {\Large$\mathcal{B}^{(2)}_{\textnormal{phys}}$};
		\node [style=none] (32) at (6.125, 1) {\Large$\mathcal{B}^{(3)}_{\textnormal{phys}}$};
		\node [style=none] (33) at (-3.25, 2) {\Large $\mathcal{B}_{\textnormal{top}}$};
		\node [style=none] (34) at (4.625, -10.5) {\Large $\mathcal{B}^{(1)}_{\textnormal{phys}}$};
		\node [style=none] (35) at (5.375, -9.5) {\Large $\mathcal{B}^{(2)}_{\textnormal{phys}}$};
		\node [style=none] (36) at (6.125, -8.5) {\Large $\mathcal{B}^{(3)}_{\textnormal{phys}}$};
		\node [style=none] (37) at (-3.25, -7.5) {\Large $\mathcal{B}_{\textnormal{top}}$};
		\node [style=none] (39) at (-0.25, -0.875) {\Large $\mathcal{J}_{1,23}$};
		\node [style=none] (40) at (1.25, -0.875) {\Large $\mathcal{J}_{2,3}$};
		\node [style=none] (41) at (-0.25, -10.25) {\Large $\mathcal{J}_{12,3}$};
		\node [style=none] (42) at (1.25, -10.25) {\Large $\mathcal{J}_{1,2}$};
		\node [style=none] (43) at (0.5, 4.5) {\Large $\mathcal{S}_{23}$};
		\node [style=none] (44) at (0.5, -4.75) {\Large $\mathcal{S}_{12}$};
		\node [style=none] (45) at (1.25, -11.5) {};
		\node [style=none] (46) at (1.25, -3.25) {};
		\node [style=none] (47) at (1.25, -2) {};
	\end{pgfonlayer}
	\begin{pgfonlayer}{edgelayer}
		\draw [style=BlueLine] (1.center) to (0.center);
		\draw [style=PurpleLine] (2.center) to (3.center);
		\draw [style=RedLine] (4.center) to (6.center);
		\draw [style=RedLine] (5.center) to (7.center);
		\draw [style=ThickLine] (1.center) to (2.center);
		\draw [style=ThickLine] (2.center) to (4.center);
		\draw [style=ThickLine] (8.center) to (6.center);
		\draw [style=DottedLine] (3.center) to (8.center);
		\draw [style=PurpleLine] (9.center) to (10.center);
		\draw [style=ThickLine] (9.center) to (5.center);
		\draw [style=ThickLine] (10.center) to (7.center);
		\draw [style=ThickLine] (0.center) to (10.center);
		\draw [style=DottedLine] (10.center) to (3.center);
		\draw [style=RedLine] (11.center) to (12.center);
		\draw [style=ThickLine] (2.center) to (11.center);
		\draw [style=ThickLine] (13.center) to (12.center);
		\draw [style=DottedLine] (13.center) to (3.center);
		\draw [style=BlueLine] (15.center) to (14.center);
		\draw [style=PurpleLine] (16.center) to (17.center);
		\draw [style=RedLine] (18.center) to (20.center);
		\draw [style=RedLine] (19.center) to (21.center);
		\draw [style=ThickLine] (15.center) to (16.center);
		\draw [style=ThickLine] (16.center) to (18.center);
		\draw [style=ThickLine] (22.center) to (20.center);
		\draw [style=DottedLine] (17.center) to (22.center);
		\draw [style=PurpleLine] (23.center) to (24.center);
		\draw [style=ThickLine] (23.center) to (19.center);
		\draw [style=ThickLine] (24.center) to (21.center);
		\draw [style=ThickLine] (14.center) to (24.center);
		\draw [style=DottedLine] (24.center) to (17.center);
		\draw [style=RedLine] (25.center) to (26.center);
		\draw [style=ThickLine] (16.center) to (25.center);
		\draw [style=ThickLine] (27.center) to (26.center);
		\draw [style=DottedLine] (27.center) to (17.center);
		\draw [style=ArrowLineRight] (28.center) to (29.center);
		\draw [style=ArrowLineRight] (46.center) to (47.center);
	\end{pgfonlayer}
\end{tikzpicture}}
\caption{We depict two SymTrees related by an associator move $\Upsilon \leftrightarrow \Upsilon'$. There is an anomaly whenever fusion of trivalent junctions produces distinct tetravalent junctions. Generalizations are immediate. This can be accompanied by a non-zero obstruction class / anomaly $\alpha_{\Upsilon, \Upsilon^{\prime}}$, although the examples considered in this paper have no such obstruction.}
\label{fig:2gp}
\end{figure}

It is natural to ask what happens if we rearrange the branches of the tree, i.e., via an ``associator move''. As an example, consider the trees in figure \ref{fig:2gp}. In passing from one theory to the next, we get a possibly non-topological junction, and we are stacking and unstacking it with other junctions. This results in a new tree (and implicitly a new set of junctions) $\Upsilon^{\prime}$. For each such tree, there is a well-defined $(D+1)$-dimensional bulk. There can in principle be an anomaly in performing this maneuver, and this just amounts to the evaluation of the partition function for the fused theory at the junction:
\begin{equation}
\mathcal{Z}(\mathcal{B}_{\mathrm{top}}, \{\mathcal{B}^{(i)}_{\mathrm{phys}} \}, \Upsilon) = \exp(i \alpha_{\Upsilon,\Upsilon^{\prime}}) \mathcal{Z}(\mathcal{B}_{\mathrm{top}}, \{\mathcal{B}^{(i)}_{\mathrm{phys}} \}, \Upsilon^{\prime}),
\end{equation}
where the factor $\exp(i \alpha_{\Upsilon,\Upsilon^{\prime}})$ is a possible ``anomaly'' associated with the branch rearrangement.\footnote{A priori, it could happen that the discrepancy between the two theories is captured by more than just a complex phase. In such cases, we anticipate that the ``anomaly'' is captured by the associator $\alpha$ of the $(D+1)$-category whose objects are SymTFTs, where $\alpha$ is a  natural collection of isomorphisms
\begin{equation}
    \alpha_{1,2,3}: \mathcal{S}_1\otimes (\mathcal{S}_2\otimes \mathcal{S}_3)\stackrel{\cong}\rightarrow (\mathcal{S}_1\otimes \mathcal{S}_2)\otimes \mathcal{S}_3.
\end{equation}} In the cases we study in this paper, we typically have $\alpha_{\Upsilon, \Upsilon^{\prime}} = 0$, but in principle it can be non-zero.\footnote{For example, in stringy models with a bulk flavor brane or other gapless QFT it is quite likely that the obstruction class is non-trivial.}

Implicit here is a categorical structure which accompanies our SymTree. While we defer a full analysis of this to future work, let us sketch some of its structure. Since we are able to fuse more than two SymTFTs into a single SymTFT, and since such manipulations take place in the category of $(D+1)$-dimensional TFTs, we have a $(D+1)$-functor category between $\mathsf{Bord}_{D+1}$ and a suitably defined $(D+1)$-category $\mathcal{C}$. This naturally suggests a multi-fusion $(D+1)$-category whose objects are SymTFTs. In this language, the zipping and retracting procedures can also be phrased naturally. First, notice that any boundary condition $\mathcal{B}_\mathrm{phys}$ can be thought of as an element in $\mathrm{Hom}(\mathcal{S}_{\mathrm{full}}, \mathbf{1}_{D+1})$ where $\mathbf{1}_{D+1}$ is the trivial $(D+1)$-theory. We have seen that the retraction of two branches with boundary conditions $\mathcal{B}^{(1)}_\mathrm{phys}$ and $\mathcal{B}^{(2)}_\mathrm{phys}$ to some $\mathcal{B}^{(\mathrm{retract})}_\mathrm{full}$ always exists by dimensionally reducing $\mathcal{S}_1$ and $\mathcal{S}_2$. This means we have a product $\otimes_{\mathrm{retract}}$: $\mathrm{Hom}(\mathcal{S}_{\mathrm{full}}, \mathbf{1}_{D+1})\times \mathrm{Hom}(\mathcal{S}_{\mathrm{full}}, \mathbf{1}_{D+1})\rightarrow \mathrm{Hom}(\mathcal{S}_{\mathrm{full}}, \mathbf{1}_{D+1})$. Meanwhile, the process of zipping additionally relies on a map $F:\mathrm{Hom}(\mathcal{S}_\mathrm{full},1)\rightarrow\mathrm{Hom}(\mathcal{S}_\mathrm{full}, 1)$ where $\mathcal{B}_{\mathrm{full}}=F(\mathcal{B}^{(\mathrm{retract})}_{\mathrm{full}})$ has additional degrees of freedom which is specified by the string construction. We leave a full exploration of this structure to future work.\footnote{Let us comment that it has recently been proposed that a suitable fusion $(D-1)$-category captures the categorical symmetry of a $D$-dimensional QFT (see e.g., \cite{Bartsch:2023pzl, Bhardwaj:2023wzd, Bartsch:2023wvv, Bhardwaj:2023ayw}). One can in principle still speak of the fusion $(D-1)$-category for the full theory $\mathcal{T}_{\mathrm{full}}$, but here we have observed the appearance of some additional structure as associated with a fusion $D$-category for the SymTFTs themselves.}

We now proceed to analyze the dressing of defects, and then turn to the dressing of symmetry operators. We turn to the stringy characterization of SymTrees in Section \ref{sec:TOPDOWN}.

\subsection{Heavy Defect Operators}\label{subsec:defects}

We begin by studying the heavy defects for a junction of SymTFTs, i.e., heavy the defects of our SymTree. The simplest non-trivial case is a trivalent junction $\mathcal{J}$ between the symmetry TFTs of the theories $\mathcal{T}_{1}$, $\mathcal{T}_{2}$ and $\mathcal{T}_{\mathrm{full}}$. We impose physical boundary conditions $\mathcal{B}^{(i)}_{\mathrm{phys}}$ for the $\mathcal{S}_{i}$ theories for $i=1,2$, and topological boundary conditions $\mathcal{B}_{\mathrm{top}}$ for theory $\mathcal{S}_{\mathrm{full}}$ (see figure \ref{fig:SymTree}).

\begin{figure}
\centering
\scalebox{0.6}{
\begin{tikzpicture}
\begin{pgfonlayer}{nodelayer}
		\node [style=none] (0) at (-8, 0) {};
		\node [style=none] (1) at (-8, 4) {};
		\node [style=none] (2) at (-4, 4) {};
		\node [style=none] (3) at (-4, 0) {};
		\node [style=none] (4) at (-0.25, 5.25) {};
		\node [style=none] (5) at (-1, 2.75) {};
		\node [style=none] (6) at (-0.25, 1.25) {};
		\node [style=none] (7) at (-1, -1.25) {};
		\node [style=none] (8) at (-1, 1) {};
		\node [style=none] (9) at (0.75, 4.5) {\Large $\mathcal{B}^{(2)}_\textnormal{phys}$};
		\node [style=none] (10) at (-0, -0.75) {\Large $\mathcal{B}^{(1)}_\textnormal{phys}$};
		\node [style=none] (11) at (-9, 2) {\Large $\mathcal{B}_\textnormal{top}$};
		\node [style=none] (12) at (-4.25, 4.75) {\Large $\mathcal{J}$};
		\node [style=none] (31) at (-3, -16) {};
		\node [style=none] (32) at (-3, -12) {};
		\node [style=none] (33) at (1, -12) {};
		\node [style=none] (34) at (1, -16) {};
		\node [style=none] (35) at (4.75, -10.75) {};
		\node [style=none] (36) at (4, -13.25) {};
		\node [style=none] (37) at (4.75, -14.75) {};
		\node [style=none] (38) at (4, -17.25) {};
		\node [style=none] (39) at (4, -15) {};
		\node [style=CirclePurple] (40) at (-4, 2) {};
		\node [style=CircleBlue] (42) at (-8, 2) {};
		\node [style=none] (46) at (2.5, -8) {};
		\node [style=none] (47) at (2.5, -4) {};
		\node [style=none] (48) at (6.5, -4) {};
		\node [style=none] (49) at (6.5, -8) {};
		\node [style=none] (50) at (10.25, -2.75) {};
		\node [style=none] (51) at (9.5, -5.25) {};
		\node [style=none] (52) at (10.25, -6.75) {};
		\node [style=none] (53) at (9.5, -9.25) {};
		\node [style=none] (54) at (9.5, -7) {};
		\node [style=CirclePurple] (55) at (6.5, -6) {};
		\node [style=CircleRed] (56) at (9.5, -7.25) {};
		\node [style=CircleRed] (57) at (10.25, -4.75) {};
		\node [style=none] (58) at (2.5, 0) {};
		\node [style=none] (59) at (2.5, 4) {};
		\node [style=none] (60) at (6.5, 4) {};
		\node [style=none] (61) at (6.5, 0) {};
		\node [style=none] (62) at (10.25, 5.25) {};
		\node [style=none] (63) at (9.5, 2.75) {};
		\node [style=none] (64) at (10.25, 1.25) {};
		\node [style=none] (65) at (9.5, -1.25) {};
		\node [style=none] (66) at (9.5, 1) {};
		\node [style=CirclePurple] (71) at (6.5, 2) {};
		\node [style=CircleRed] (72) at (9.5, 0.75) {};
		\node [style=CircleRed] (74) at (4, -15.25) {};
		\node [style=CircleRed] (75) at (4.75, -12.75) {};
		\node [style=CirclePurple] (76) at (1, -14) {};
		\node [style=CircleBlue] (77) at (-3, -14) {};
		\node [style=none] (78) at (-4, -1) {\Large(i)};
		\node [style=none] (79) at (6.5, -1) {\Large(ii)};
		\node [style=none] (80) at (-8, -8) {};
		\node [style=none] (81) at (-8, -4) {};
		\node [style=none] (82) at (-4, -4) {};
		\node [style=none] (83) at (-4, -8) {};
		\node [style=none] (84) at (-0.25, -2.75) {};
		\node [style=none] (85) at (-1, -5.25) {};
		\node [style=none] (86) at (-0.25, -6.75) {};
		\node [style=none] (87) at (-1, -9.25) {};
		\node [style=none] (88) at (-1, -7) {};
		\node [style=CirclePurple] (89) at (-4, -6) {};
		\node [style=CircleRed] (90) at (-1, -7.25) {};
		\node [style=CircleBlue] (91) at (-8, -6) {};
		\node [style=none] (92) at (-4, -9) {\Large(iii)};
		\node [style=none] (93) at (6.5, -9) {\Large(iv)};
		\node [style=none] (94) at (1, -17) {\Large(v)};
		\node [style=none] (95) at (1, -18) {};
		\node [style=none] (96) at (-6, 2.5) {\Large$D$};
		\node [style=none] (97) at (8.75, 0) {\Large $D^{(1)}$};
		\node [style=none] (98) at (8.75, -8) {\Large $D^{(1)}$};
		\node [style=none] (99) at (3.25, -16) {\Large $D^{(1)}$};
		\node [style=none] (100) at (-1.75, -8) {\Large $D^{(1)}$};
		\node [style=none] (101) at (9.5, -4.25) {\Large $D^{(2)}$};
		\node [style=none] (102) at (4, -12.25) {\Large $D^{(2)}$};
		\node [style=none] (103) at (-1, -13.5) {\Large $D$};
		\node [style=none] (104) at (-6, -5.5) {\Large $D$};
	\end{pgfonlayer}
	\begin{pgfonlayer}{edgelayer}
		\draw [style=BlueLine] (1.center) to (0.center);
		\draw [style=PurpleLine] (2.center) to (3.center);
		\draw [style=RedLine] (4.center) to (6.center);
		\draw [style=RedLine] (5.center) to (7.center);
		\draw [style=ThickLine] (1.center) to (2.center);
		\draw [style=ThickLine] (2.center) to (4.center);
		\draw [style=ThickLine] (2.center) to (5.center);
		\draw [style=ThickLine] (7.center) to (3.center);
		\draw [style=ThickLine] (3.center) to (0.center);
		\draw [style=ThickLine] (8.center) to (6.center);
		\draw [style=DottedLine] (3.center) to (8.center);
		\draw [style=BlueLine] (32.center) to (31.center);
		\draw [style=PurpleLine] (33.center) to (34.center);
		\draw [style=RedLine] (35.center) to (37.center);
		\draw [style=RedLine] (36.center) to (38.center);
		\draw [style=ThickLine] (32.center) to (33.center);
		\draw [style=ThickLine] (33.center) to (35.center);
		\draw [style=ThickLine] (33.center) to (36.center);
		\draw [style=ThickLine] (38.center) to (34.center);
		\draw [style=ThickLine] (34.center) to (31.center);
		\draw [style=ThickLine] (39.center) to (37.center);
		\draw [style=DottedLine] (34.center) to (39.center);
		\draw [style=BrownLine, snake it] (40) to (42);
		\draw [style=BlueLine] (47.center) to (46.center);
		\draw [style=PurpleLine] (48.center) to (49.center);
		\draw [style=RedLine] (50.center) to (52.center);
		\draw [style=RedLine] (51.center) to (53.center);
		\draw [style=ThickLine] (47.center) to (48.center);
		\draw [style=ThickLine] (48.center) to (50.center);
		\draw [style=ThickLine] (48.center) to (51.center);
		\draw [style=ThickLine] (53.center) to (49.center);
		\draw [style=ThickLine] (49.center) to (46.center);
		\draw [style=ThickLine] (54.center) to (52.center);
		\draw [style=DottedLine] (49.center) to (54.center);
		\draw [style=BrownLine, snake it] (56) to (55);
		\draw [style=BrownLine, snake it] (55) to (57);
		\draw [style=BlueLine] (59.center) to (58.center);
		\draw [style=PurpleLine] (60.center) to (61.center);
		\draw [style=RedLine] (62.center) to (64.center);
		\draw [style=RedLine] (63.center) to (65.center);
		\draw [style=ThickLine] (59.center) to (60.center);
		\draw [style=ThickLine] (60.center) to (62.center);
		\draw [style=ThickLine] (60.center) to (63.center);
		\draw [style=ThickLine] (65.center) to (61.center);
		\draw [style=ThickLine] (61.center) to (58.center);
		\draw [style=ThickLine] (66.center) to (64.center);
		\draw [style=DottedLine] (61.center) to (66.center);
		\draw [style=BrownLine, snake it] (72) to (71);
		\draw [style=BrownLine, snake it] (77) to (76);
		\draw [style=BrownLine, snake it] (76) to (74);
		\draw [style=BrownLine, snake it] (76) to (75);
		\draw [style=BlueLine] (81.center) to (80.center);
		\draw [style=PurpleLine] (82.center) to (83.center);
		\draw [style=RedLine] (84.center) to (86.center);
		\draw [style=RedLine] (85.center) to (87.center);
		\draw [style=ThickLine] (81.center) to (82.center);
		\draw [style=ThickLine] (82.center) to (84.center);
		\draw [style=ThickLine] (82.center) to (85.center);
		\draw [style=ThickLine] (87.center) to (83.center);
		\draw [style=ThickLine] (83.center) to (80.center);
		\draw [style=ThickLine] (88.center) to (86.center);
		\draw [style=DottedLine] (83.center) to (88.center);
		\draw [style=BrownLine, snake it] (90) to (89);
		\draw [style=BrownLine, snake it] (89) to (91);
	\end{pgfonlayer}
\end{tikzpicture}}
\caption{Sketch of possible defect configurations for a trivalent junction of symmetry TFTs. The red and purple dots denote the spacetime defects, the brown line marks the defect within the SymTree. The purple junction defect is said to \textit{dress} the red boundary defects.}
\label{fig:Defects}
\end{figure}
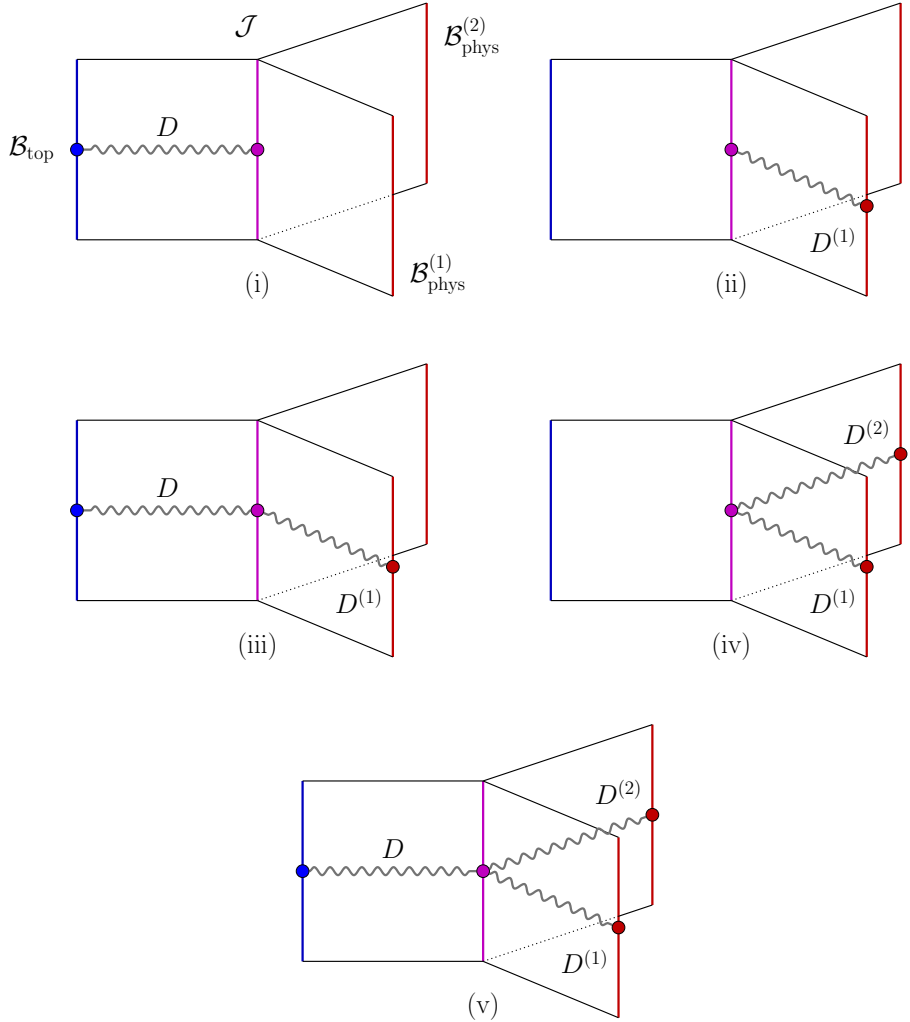

The topological boundary conditions determine which defects can end at the associated boundary, and sets the global form of the full multi-sector QFT. Consider a defect ending at $\mathcal{B}_{\textnormal{top}}$ and stretch it across $\mathcal{S}_{\mathrm{full}}$ to the junction $\mathcal{J}$. From here, three possibilities can occur:
\begin{enumerate}
\item The defect ends on the junction giving a defect of the junction theory. There is a single spacetime defect and it is not a defect of the relative theories $\mathcal{T}_{i}$. See subfigure (i) of figure \ref{fig:Defects}.
\item The defect can not end on the junction and continues on to the boundary $\mathcal{B}^{(i)}_\textnormal{phys}$ resulting overall in a defect of the junction theory and one of the relative theory. There are now two spacetime defects. We say the defect of the relative theory $\mathcal{B}^{(i)}_\textnormal{phys}$ is \textit{dressed} by the junction defect. See subfigure (iii) of figure \ref{fig:Defects}.
\item The defect does not end on the junction and fractionates into two defects which end on the two physical boundaries. We obtain a defect in the junction theory and the two relative theories $\mathcal{B}^{(i)}_\textnormal{phys}$. The latter pair is \textit{dressed} by the junction defect. See subfigure (v) of figure \ref{fig:Defects}.
\end{enumerate}
In addition to these three cases we also have the case
\begin{itemize}
\item The defect stretches from $\mathcal{B}^{(1)}_\textnormal{phys}$ to $\mathcal{B}^{(2)}_\textnormal{phys}$ or vice versa, passing through the junction and not attaching to the topological boundary. See subfigure (iv) of figure \ref{fig:Defects}.
\end{itemize}
In principle there could also exist defects which just stretch between the junction and the physical boundaries, see subfigure (ii) of figure \ref{fig:Defects}.

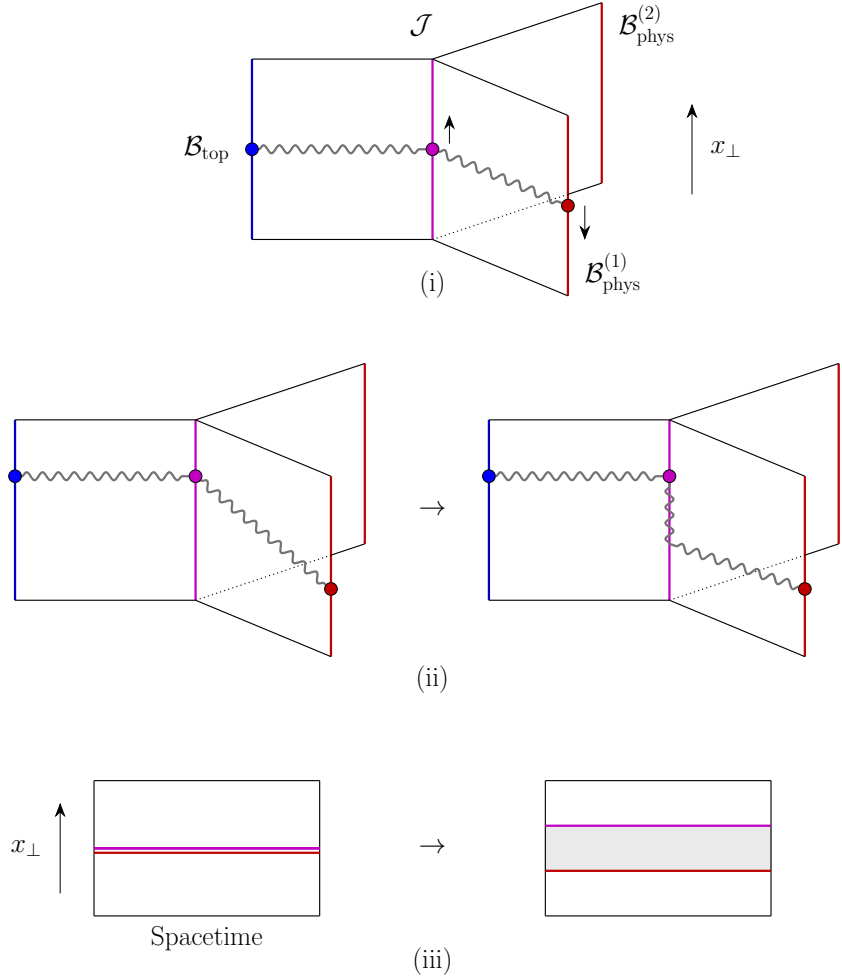
\begin{figure}
\centering
\scalebox{0.60}{
\begin{tikzpicture}
	\begin{pgfonlayer}{nodelayer}
		\node [style=none] (0) at (-10, -2) {};
		\node [style=none] (1) at (-10, 2) {};
		\node [style=none] (2) at (-6, 2) {};
		\node [style=none] (3) at (-6, -2) {};
		\node [style=none] (4) at (-2.25, 3.25) {};
		\node [style=none] (5) at (-3, 0.75) {};
		\node [style=none] (6) at (-2.25, -0.75) {};
		\node [style=none] (7) at (-3, -3.25) {};
		\node [style=none] (8) at (-3, -1) {};
		\node [style=CirclePurple] (13) at (-6, 0.75) {};
		\node [style=CircleBlue] (14) at (-10, 0.75) {};
		\node [style=CircleRed] (16) at (-3, -1.75) {};
		\node [style=none] (17) at (0.5, -2) {};
		\node [style=none] (18) at (0.5, 2) {};
		\node [style=none] (19) at (4.5, 2) {};
		\node [style=none] (20) at (4.5, -2) {};
		\node [style=none] (21) at (8.25, 3.25) {};
		\node [style=none] (22) at (7.5, 0.75) {};
		\node [style=none] (23) at (8.25, -0.75) {};
		\node [style=none] (24) at (7.5, -3.25) {};
		\node [style=none] (25) at (7.5, -1) {};
		\node [style=none] (28) at (-0.75, 0) {\Large $\rightarrow$};
		\node [style=CirclePurple] (30) at (4.5, 0.75) {};
		\node [style=CircleBlue] (31) at (0.5, 0.75) {};
		\node [style=none] (32) at (-0.75, -3.75) {\Large (ii)};
		\node [style=CircleRed] (33) at (7.5, -1.75) {};
		\node [style=none] (34) at (4.5, -0.75) {};
		\node [style=none] (35) at (-4.75, 6) {};
		\node [style=none] (36) at (-4.75, 10) {};
		\node [style=none] (37) at (-0.75, 10) {};
		\node [style=none] (38) at (-0.75, 6) {};
		\node [style=none] (39) at (3, 11.25) {};
		\node [style=none] (40) at (2.25, 8.75) {};
		\node [style=none] (41) at (3, 7.25) {};
		\node [style=none] (42) at (2.25, 4.75) {};
		\node [style=none] (43) at (2.25, 7) {};
		\node [style=none] (44) at (4, 10.75) {\Large $\mathcal{B}^{(2)}_\textnormal{phys}$};
		\node [style=none] (45) at (3.25, 5.25) {\Large $\mathcal{B}^{(1)}_\textnormal{phys}$};
		\node [style=none] (46) at (-5.75, 8) {\Large $\mathcal{B}_\textnormal{top}$};
		\node [style=none] (47) at (-1, 10.75) {\Large $\mathcal{J}$};
		\node [style=CirclePurple] (48) at (-0.75, 8) {};
		\node [style=CircleBlue] (49) at (-4.75, 8) {};
		\node [style=none] (50) at (-0.75, 5) {\Large (i)};
		\node [style=CircleRed] (51) at (2.25, 6.75) {};
		\node [style=none] (52) at (-0.375, 8.125) {};
		\node [style=none] (53) at (-0.375, 8.75) {};
		\node [style=none] (54) at (2.625, 6.75) {};
		\node [style=none] (55) at (2.625, 6) {};
		\node [style=none] (56) at (-8.25, -6) {};
		\node [style=none] (57) at (-3.25, -6) {};
		\node [style=none] (58) at (-8.25, -9) {};
		\node [style=none] (59) at (-3.25, -9) {};
		\node [style=none] (60) at (-5.75, -9.5) {\Large Spacetime};
		\node [style=none] (61) at (-8.25, -7.5) {};
		\node [style=none] (62) at (-3.25, -7.5) {};
		\node [style=none] (63) at (-0.75, -7.5) {\Large $\rightarrow$};
		\node [style=none] (64) at (1.75, -6) {};
		\node [style=none] (65) at (6.75, -6) {};
		\node [style=none] (66) at (1.75, -9) {};
		\node [style=none] (67) at (6.75, -9) {};
		\node [style=none] (69) at (1.75, -7) {};
		\node [style=none] (70) at (6.75, -7) {};
		\node [style=none] (71) at (5, 7) {};
		\node [style=none] (72) at (5, 9) {};
		\node [style=none] (73) at (5.75, 8) {\Large $x_\perp$};
		\node [style=none] (74) at (1.75, -8) {};
		\node [style=none] (75) at (6.75, -8) {};
		\node [style=none] (76) at (-9, -8.5) {};
		\node [style=none] (77) at (-9, -6.5) {};
		\node [style=none] (78) at (-9.75, -7.5) {\Large $x_\perp$};
		\node [style=none] (79) at (-0.75, -10) {\Large (iii)};
		\node [style=none] (80) at (-0.75, -11) {};
		\node [style=none] (81) at (-8.25, -7.6) {};
		\node [style=none] (82) at (-3.25, -7.6) {};
	\end{pgfonlayer}
	\begin{pgfonlayer}{edgelayer}
		\filldraw[fill={rgb,255: red,150; green,150; blue,150}, draw={rgb,255: red,200; green,200; blue,200}, fill opacity=0.2] (1.75, -7) -- (6.75, -7) -- (6.75, -8) -- (1.75, -8) -- cycle;
		\draw [style=BlueLine] (1.center) to (0.center);
		\draw [style=PurpleLine] (2.center) to (3.center);
		\draw [style=RedLine] (4.center) to (6.center);
		\draw [style=RedLine] (5.center) to (7.center);
		\draw [style=ThickLine] (1.center) to (2.center);
		\draw [style=ThickLine] (2.center) to (4.center);
		\draw [style=ThickLine] (2.center) to (5.center);
		\draw [style=ThickLine] (7.center) to (3.center);
		\draw [style=ThickLine] (3.center) to (0.center);
		\draw [style=ThickLine] (8.center) to (6.center);
		\draw [style=DottedLine] (3.center) to (8.center);
		\draw [style=BrownLine, snake it] (14) to (13);
		\draw [style=BrownLine, snake it] (13) to (16);
		\draw [style=BlueLine] (18.center) to (17.center);
		\draw [style=PurpleLine] (19.center) to (20.center);
		\draw [style=RedLine] (21.center) to (23.center);
		\draw [style=RedLine] (22.center) to (24.center);
		\draw [style=ThickLine] (18.center) to (19.center);
		\draw [style=ThickLine] (19.center) to (21.center);
		\draw [style=ThickLine] (19.center) to (22.center);
		\draw [style=ThickLine] (24.center) to (20.center);
		\draw [style=ThickLine] (20.center) to (17.center);
		\draw [style=ThickLine] (25.center) to (23.center);
		\draw [style=DottedLine] (20.center) to (25.center);
		\draw [style=BrownLine, snake it] (31) to (30);
		\draw [style=BrownLine, snake it] (33) to (34.center);
		\draw [style=BrownLine, snake it] (34.center) to (30);
		\draw [style=BlueLine] (36.center) to (35.center);
		\draw [style=PurpleLine] (37.center) to (38.center);
		\draw [style=RedLine] (39.center) to (41.center);
		\draw [style=RedLine] (40.center) to (42.center);
		\draw [style=ThickLine] (36.center) to (37.center);
		\draw [style=ThickLine] (37.center) to (39.center);
		\draw [style=ThickLine] (37.center) to (40.center);
		\draw [style=ThickLine] (42.center) to (38.center);
		\draw [style=ThickLine] (38.center) to (35.center);
		\draw [style=ThickLine] (43.center) to (41.center);
		\draw [style=DottedLine] (38.center) to (43.center);
		\draw [style=BrownLine, snake it] (49) to (48);
		\draw [style=BrownLine, snake it] (48) to (51);
		\draw [style=ArrowLineRight] (54.center) to (55.center);
		\draw [style=ArrowLineRight] (52.center) to (53.center);
		\draw [style=ThickLine] (56.center) to (58.center);
		\draw [style=ThickLine] (58.center) to (59.center);
		\draw [style=ThickLine] (59.center) to (57.center);
		\draw [style=ThickLine] (57.center) to (56.center);
		\draw [style=PurpleLine] (61.center) to (62.center);
		\draw [style=PurpleLine] (61.center) to (62.center);
		\draw [style=ThickLine] (64.center) to (66.center);
		\draw [style=ThickLine] (66.center) to (67.center);
		\draw [style=ThickLine] (67.center) to (65.center);
		\draw [style=ThickLine] (65.center) to (64.center);
		\draw [style=PurpleLine] (69.center) to (70.center);
		\draw [style=ArrowLineRight] (71.center) to (72.center);
		\draw [style=RedLine] (74.center) to (75.center);
		\draw [style=ArrowLineRight] (76.center) to (77.center);
		\draw [style=RedLine] (81.center) to (82.center);
		\draw [style=PurpleLine] (61.center) to (62.center);
	\end{pgfonlayer}
\end{tikzpicture}}
\caption{We sketch a deformation of the defect configuration depicted in (i). In (i) the purple and red defect are coincident in spacetime, as shown on the lefthand side in (iii). In (ii) we displace these along a spacetime direction $x_\perp$ and deforming the resulting defect configuration into a horizontal and vertical piece we find a topological operator bounded by the initial pair of defects, as depicted on the righthand side in (iii).}
\label{fig:spacetimeseparating}
\end{figure}

We have implicitly assumed that all spacetime defects are stacked, i.e., they have identical spacetime support. Let us now separate these defects. Consider for instance the setup of case 2, here we separate the junction defect from the physical defect. This results in a portion of the associated defect in the symmetry theory to run parallel to spacetime. We can localize this portion within the junction. In spacetime this portion realizes a topological operator, in one higher dimension, stretching between the separated spacetime defects (see figure \ref{fig:spacetimeseparating}).

Similar remarks hold for the other cases: whenever we separate spacetime defects a spacetime topological defect in one higher dimension emerges bounded by the initial defects. With this we can now introduce a notion of \textit{genuine} defects for the relative theories $\mathcal{B}^{(i)}_\textnormal{phys}$. We call defects genuine if they run between the topological boundary condition and a single physical boundary and non-genuine otherwise. Genuine defects are constructed from defects of a single physical boundary via dressings.
This definition is such that the defects depicted in subfigure (iv) of figure \ref{fig:Defects} are non-genuine. If we separate the defects of $\mathcal{B}^{(1)}_\textnormal{phys}$ and $\mathcal{B}^{(2)}_\textnormal{phys}$ in spacetime there will always be a topological operator running between these no matter the dressing. On the other hand, for example, the defect in subfigure (iii) of figure \ref{fig:Defects} is genuine.

\subsection{Topological Symmetry Operators}\label{subsec:sym operators}

We now study topological symmetry operators for our SymTree. To illustrate the main points, it again suffices to consider the trivalent junction comprised of SymTFTs $\mathcal{S}_{\mathrm{full}}$, $\mathcal{S}_{1}$ and $\mathcal{S}_2$. We assume that away from the junction we have imposed a topological boundary condition $ \mathcal{B}_{\mathrm{top}} $ for $\mathcal{S}_{\mathrm{full}}$, and physical boundary conditions $\mathcal{B}_{\mathrm{phys}}^{(i)}$ for the relative theories $\mathcal{T}_{i}$.

\begin{figure}
\centering
\scalebox{0.65}{
\begin{tikzpicture}
	\begin{pgfonlayer}{nodelayer}
		\node [style=none] (0) at (-9, 0) {};
		\node [style=none] (1) at (-9, 4) {};
		\node [style=none] (2) at (-5, 4) {};
		\node [style=none] (3) at (-5, 0) {};
		\node [style=none] (4) at (-1.25, 5.25) {};
		\node [style=none] (5) at (-2, 2.75) {};
		\node [style=none] (6) at (-1.25, 1.25) {};
		\node [style=none] (7) at (-2, -1.25) {};
		\node [style=none] (8) at (-2, 1) {};
		\node [style=none] (9) at (-0.25, 4.75) {\Large $\mathcal{B}^{(1)}_\textnormal{phys}$};
		\node [style=none] (10) at (-1, -0.75) {\Large $\mathcal{B}^{(2)}_\textnormal{phys}$};
		\node [style=none] (11) at (-10, 2) {\Large $\mathcal{B}_\textnormal{top}$};
		\node [style=none] (12) at (-5.25, 4.75) {\Large $\mathcal{J}$};
		\node [style=none] (13) at (-9, -16) {};
		\node [style=none] (14) at (-9, -12) {};
		\node [style=none] (15) at (-5, -12) {};
		\node [style=none] (16) at (-5, -16) {};
		\node [style=none] (17) at (-1.25, -10.75) {};
		\node [style=none] (18) at (-2, -13.25) {};
		\node [style=none] (19) at (-1.25, -14.75) {};
		\node [style=none] (20) at (-2, -17.25) {};
		\node [style=none] (21) at (-2, -15) {};
		\node [style=none] (24) at (1.5, 0) {};
		\node [style=none] (25) at (1.5, 4) {};
		\node [style=none] (26) at (5.5, 4) {};
		\node [style=none] (27) at (5.5, 0) {};
		\node [style=none] (28) at (9.25, 5.25) {};
		\node [style=none] (29) at (8.5, 2.75) {};
		\node [style=none] (30) at (9.25, 1.25) {};
		\node [style=none] (31) at (8.5, -1.25) {};
		\node [style=none] (32) at (8.5, 1) {};
		\node [style=none] (51) at (-5, -1) {\Large (i)};
		\node [style=none] (53) at (-9, -8) {};
		\node [style=none] (54) at (-9, -4) {};
		\node [style=none] (55) at (-5, -4) {};
		\node [style=none] (56) at (-5, -8) {};
		\node [style=none] (57) at (-1.25, -2.75) {};
		\node [style=none] (58) at (-2, -5.25) {};
		\node [style=none] (59) at (-1.25, -6.75) {};
		\node [style=none] (60) at (-2, -9.25) {};
		\node [style=none] (61) at (-2, -7) {};
		\node [style=none] (65) at (-5, -9) {\Large (iii)};
		\node [style=none] (66) at (5.5, -1) {\Large (ii)};
		\node [style=none] (67) at (-5, -17) {\Large (v)};
		\node [style=none] (68) at (0, -18) {};
		\node [style=none] (69) at (-7, 2.625) {\Large $\mathcal{U}$};
		\node [style=none] (72) at (-2.75, -14.125) {\Large $\mathcal{V}^{(1)}$};
		\node [style=none] (76) at (-7, -13.5) {\Large $\mathcal{V}$};
		\node [style=none] (77) at (-7, -5.5) {\Large $\mathcal{V}$};
		\node [style=GreenCircle] (78) at (-7, 2) {};
		\node [style=GreenCircle] (80) at (-7, -6) {};
		\node [style=CirclePurple] (81) at (-5, -6) {};
		\node [style=none] (82) at (7.5, 2.125) {\Large $\mathcal{V}^{(1)}$};
		\node [style=GreenCircle] (83) at (7, 1.5) {};
		\node [style=CirclePurple] (84) at (5.5, 2) {};
		\node [style=none] (85) at (1.5, -8) {};
		\node [style=none] (86) at (1.5, -4) {};
		\node [style=none] (87) at (5.5, -4) {};
		\node [style=none] (88) at (5.5, -8) {};
		\node [style=none] (89) at (9.25, -2.75) {};
		\node [style=none] (90) at (8.5, -5.25) {};
		\node [style=none] (91) at (9.25, -6.75) {};
		\node [style=none] (92) at (8.5, -9.25) {};
		\node [style=none] (93) at (8.5, -7) {};
		\node [style=none] (98) at (5.5, -9) {\Large (iv)};
		\node [style=none] (99) at (7.8, -6.5) {\Large $\mathcal{V}^{(1)}$};
		\node [style=none] (100) at (6.75, -5.125) {\Large $\mathcal{V}^{(2)}$};
		\node [style=GreenCircle] (102) at (-7, -14) {};
		\node [style=CirclePurple] (103) at (-5, -14) {};
		\node [style=GreenCircle] (104) at (-3.5, -14.5) {};
		\node [style=CirclePurple] (105) at (5.5, -6) {};
		\node [style=GreenCircle] (106) at (7, -6.5) {};
		\node [style=GreenCircle] (107) at (7.5, -5.5) {};
		\node [style=none] (108) at (1.5, -16) {};
		\node [style=none] (109) at (1.5, -12) {};
		\node [style=none] (110) at (5.5, -12) {};
		\node [style=none] (111) at (5.5, -16) {};
		\node [style=none] (112) at (9.25, -10.75) {};
		\node [style=none] (113) at (8.5, -13.25) {};
		\node [style=none] (114) at (9.25, -14.75) {};
		\node [style=none] (115) at (8.5, -17.25) {};
		\node [style=none] (116) at (8.5, -15) {};
		\node [style=none] (117) at (5.5, -17) {\Large (vi)};
		\node [style=none] (119) at (3.5, -13.5) {\Large $\mathcal{V}$};
		\node [style=GreenCircle] (120) at (3.5, -14) {};
		\node [style=CirclePurple] (121) at (5.5, -14) {};
		\node [style=GreenCircle] (122) at (7, -14.5) {};
		\node [style=GreenCircle] (123) at (7.5, -13.5) {};
		\node [style=none] (124) at (7.8, -14.5) {\Large $\mathcal{V}^{(1)}$};
		\node [style=none] (125) at (6.75, -13.125) {\Large $\mathcal{V}^{(2)}$};
	\end{pgfonlayer}
	\begin{pgfonlayer}{edgelayer}
		\draw [style=BlueLine] (1.center) to (0.center);
		\draw [style=PurpleLine] (2.center) to (3.center);
		\draw [style=RedLine] (4.center) to (6.center);
		\draw [style=RedLine] (5.center) to (7.center);
		\draw [style=ThickLine] (1.center) to (2.center);
		\draw [style=ThickLine] (2.center) to (4.center);
		\draw [style=ThickLine] (2.center) to (5.center);
		\draw [style=ThickLine] (7.center) to (3.center);
		\draw [style=ThickLine] (3.center) to (0.center);
		\draw [style=ThickLine] (8.center) to (6.center);
		\draw [style=DottedLine] (3.center) to (8.center);
		\draw [style=BlueLine] (14.center) to (13.center);
		\draw [style=PurpleLine] (15.center) to (16.center);
		\draw [style=RedLine] (17.center) to (19.center);
		\draw [style=RedLine] (18.center) to (20.center);
		\draw [style=ThickLine] (14.center) to (15.center);
		\draw [style=ThickLine] (15.center) to (17.center);
		\draw [style=ThickLine] (15.center) to (18.center);
		\draw [style=ThickLine] (20.center) to (16.center);
		\draw [style=ThickLine] (16.center) to (13.center);
		\draw [style=ThickLine] (21.center) to (19.center);
		\draw [style=DottedLine] (16.center) to (21.center);
		\draw [style=BlueLine] (25.center) to (24.center);
		\draw [style=PurpleLine] (26.center) to (27.center);
		\draw [style=RedLine] (28.center) to (30.center);
		\draw [style=RedLine] (29.center) to (31.center);
		\draw [style=ThickLine] (25.center) to (26.center);
		\draw [style=ThickLine] (26.center) to (28.center);
		\draw [style=ThickLine] (26.center) to (29.center);
		\draw [style=ThickLine] (31.center) to (27.center);
		\draw [style=ThickLine] (27.center) to (24.center);
		\draw [style=ThickLine] (32.center) to (30.center);
		\draw [style=DottedLine] (27.center) to (32.center);
		\draw [style=BlueLine] (54.center) to (53.center);
		\draw [style=PurpleLine] (55.center) to (56.center);
		\draw [style=RedLine] (57.center) to (59.center);
		\draw [style=RedLine] (58.center) to (60.center);
		\draw [style=ThickLine] (54.center) to (55.center);
		\draw [style=ThickLine] (55.center) to (57.center);
		\draw [style=ThickLine] (55.center) to (58.center);
		\draw [style=ThickLine] (60.center) to (56.center);
		\draw [style=ThickLine] (56.center) to (53.center);
		\draw [style=ThickLine] (61.center) to (59.center);
		\draw [style=DottedLine] (56.center) to (61.center);
		\draw [style=DashedLineGreen] (80) to (81);
		\draw [style=DashedLineGreen] (84) to (83);
		\draw [style=BlueLine] (86.center) to (85.center);
		\draw [style=PurpleLine] (87.center) to (88.center);
		\draw [style=RedLine] (89.center) to (91.center);
		\draw [style=RedLine] (90.center) to (92.center);
		\draw [style=ThickLine] (86.center) to (87.center);
		\draw [style=ThickLine] (87.center) to (89.center);
		\draw [style=ThickLine] (87.center) to (90.center);
		\draw [style=ThickLine] (92.center) to (88.center);
		\draw [style=ThickLine] (88.center) to (85.center);
		\draw [style=ThickLine] (93.center) to (91.center);
		\draw [style=DottedLine] (88.center) to (93.center);
		\draw [style=DashedLineGreen] (102) to (103);
		\draw [style=DashedLineGreen] (103) to (104);
		\draw [style=DashedLineGreen] (105) to (106);
		\draw [style=DashedLineGreen] (105) to (107);
		\draw [style=BlueLine] (109.center) to (108.center);
		\draw [style=PurpleLine] (110.center) to (111.center);
		\draw [style=RedLine] (112.center) to (114.center);
		\draw [style=RedLine] (113.center) to (115.center);
		\draw [style=ThickLine] (109.center) to (110.center);
		\draw [style=ThickLine] (110.center) to (112.center);
		\draw [style=ThickLine] (110.center) to (113.center);
		\draw [style=ThickLine] (115.center) to (111.center);
		\draw [style=ThickLine] (111.center) to (108.center);
		\draw [style=ThickLine] (116.center) to (114.center);
		\draw [style=DottedLine] (111.center) to (116.center);
		\draw [style=DashedLineGreen] (120) to (121);
		\draw [style=DashedLineGreen] (121) to (122);
		\draw [style=DashedLineGreen] (121) to (123);
	\end{pgfonlayer}
\end{tikzpicture}}
\caption{Sketch of different topological symmetry operator configurations for a trivalent junction of symmetry TFTs. We consider the initial configuration depicted in (i). Deforming it (partially) across/into the junction gives various equivalent configurations of topological operators presented in subfigures (ii) - (vi). The dashed lines indicate topological operators in one higher dimension. We denote non-genuine operators at their boundaries as $\mathcal{V},\mathcal{V}^{(1)},\mathcal{V}^{(2)}$, also represented by green dots. The purple dots again depict dressings. Both the dressings and the higher-dimensional operators can be trivial.}
\label{fig:SymOps}
\end{figure}
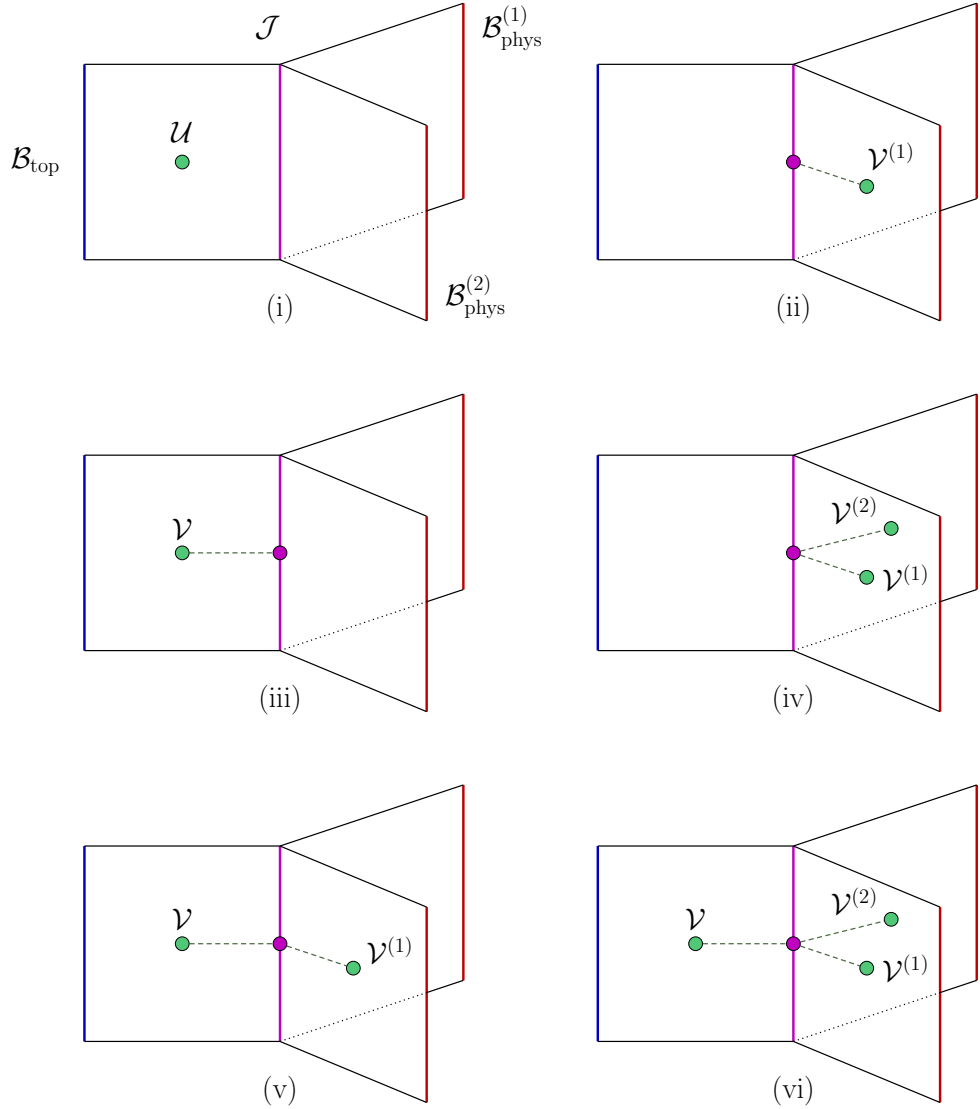

We begin by considering a symmetry operator located at the topological boundary condition for $\mathcal{S}_{\mathrm{full}}$. Our aim will be to understand the structure of this symmetry operator as we move it from the boundary $ \mathcal{B}_{\mathrm{top}} $ through the junction, and eventually on to either of the physical boundaries.\footnote{This subset of bulk operators of the SymTree is identified with the symmetry operators of a multi-sector QFT with string construction. In such settings topological symmetry operators can only be constructed in the asymptotic boundary which becomes $\mathcal{B}_{\rm top}$ in the SymTree.} With this our starting point is the configuration depicted in subfigure (i) of figure \ref{fig:SymOps}. As in our analysis of defect operators, we emphasize three generic possibilities:

\begin{enumerate}
\item The boundary conditions for $\mathcal{S}_{\mathrm{full}}$ set by the junction $\mathcal{J}$ are such that the topological symmetry operator can not be deformed across the junction.
\item The topological symmetry operator can be deformed across the junction to a topological symmetry operator in the slab with TFT $\mathcal{S}_{i}$. See subfigure (ii) of figure \ref{fig:SymOps}. The deformation across the junction comes at the cost of a dressing, i.e., an additional topological symmetry operator localized in the junction. This pair of symmetry operators can further be connected by a topological operator in one higher dimension.
\item The topological symmetry operator can be deformed across the junction to a collection of topological symmetry operator in the slabs $\mathcal{S}_{1}$ and $\mathcal{S}_2$. See subfigure (iv) of figure \ref{fig:SymOps}. Again, there may be a dressing and additional higher dimensional topological operators.
\end{enumerate}

Clearly there exist further configurations. For one we can, starting from (i), deform only a portion of the symmetry operators into/across the junction. This can give the configurations (iii), (v), (iv) in figure \ref{fig:SymOps}. We can also consider different starting points instead of (i), e.g., any of the configuration depicted in figure \ref{fig:SymOps} or configurations similar to (i) with a collection of operators $\mathcal{U}^{(i)}$ contained in the slabs attaching to physical boundaries. Given this large collection of symmetry operators the key point is that they are subject to an equivalence. Two configurations are equivalent precisely when they can be deformed into each other.

From this, the action of the symmetry operators, deformation equivalent to configuration (i) of figure \ref{fig:SymOps}, on heavy defects is now clear. For such operators the action on defects is given by considering the distinguished representative purely contained in the slab of the symmetry theory $\mathcal{S}_\textnormal{full}$. It acts in standard fashion via linking on the part of the defect which extends into that slab. The computation therefore fully restricts to $\mathcal{S}_\textnormal{full}$.

Similar to the spacetime deformation depicted in figure \ref{fig:spacetimeseparating} we can separate the various components contributing to symmetry operators, e.g., as depicted in figure \ref{fig:SymOps}. In the spacetime we have topological operators in one higher dimension which bound the individual components.



\section{Top Down Approach to SymTrees} \label{sec:TOPDOWN}

In the previous Section we presented a bottom up analysis of multi-sector QFTs and the SymTree which governs their symmetries. We now proceed from the top down, showing that for multi-sector QFTs with a stringy realization, this treelike structure directly descends from extra-dimensional geometry. Throughout, we shall work on spacetimes of the form $\mathbb{R}^{D-1,1} \times X$ where $X$ is taken to be a non-compact background which preserves some amount of supersymmetry in the $D$-dimensional spacetime.\footnote{The supersymmetry condition is more so that we have tractable examples to discuss rather than any intrinsic limitation.}

\begin{figure}
\centering
\scalebox{0.6}{
\begin{tikzpicture}
	\begin{pgfonlayer}{nodelayer}
		\node [style=none] (0) at (-1, -2) {};
		\node [style=none] (1) at (1, -2) {};
		\node [style=none] (2) at (0, 0) {};
		\node [style=none] (3) at (3, 2) {};
		\node [style=none] (4) at (-3, 2) {};
		\node [style=none] (5) at (0, 3.125) {\Large Asymptotic Boundary};
		\node [style=Star] (6) at (-1, -2) {};
		\node [style=Star] (7) at (1, -2) {};
		\node [style=none] (8) at (-2, -2) {\Large $\mathsf{Sing}_1$};
		\node [style=none] (9) at (2.1, -2) {\Large $\mathsf{Sing}_2$};
		\node [style=none] (10) at (2, 0) {};
		\node [style=none] (11) at (-2, 0) {};
		\node [style=none] (12) at (4.125, 0) {\Large Critical Slice};
		\node [style=none] (13) at (-5.375, 0) {};
		\node [style=none] (14) at (0, -3) {};
	\end{pgfonlayer}
	\begin{pgfonlayer}{edgelayer}
		\draw [style=ThickLine] (4.center) to (0.center);
		\draw [style=ThickLine] (0.center) to (2.center);
		\draw [style=ThickLine] (2.center) to (1.center);
		\draw [style=ThickLine] (1.center) to (3.center);
		\draw [style=ThickLine, bend left=15] (4.center) to (3.center);
		\draw [style=ThickLine, bend right=15] (4.center) to (3.center);
		\draw [style=DottedLine, bend right=15] (11.center) to (2.center);
		\draw [style=DottedLine, bend right=15] (2.center) to (10.center);
		\draw [style=DottedLine, bend left=15] (11.center) to (2.center);
		\draw [style=DottedLine, bend left=15, looseness=0.75] (2.center) to (10.center);
	\end{pgfonlayer}
\end{tikzpicture}}
\caption{Sketch of double throat internal geometry with two sets of localized degrees of freedom (red). Horizontal slices of constant radius are initially disjoint and then combine resulting in a connected asymptotic boundary.}
\label{fig:NewStyle}
\end{figure}
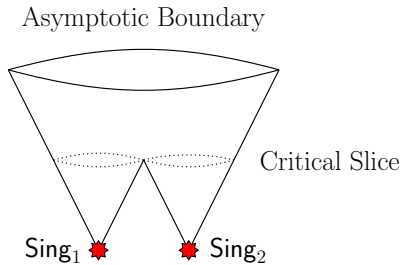

With this in mind, we will be interested in either a 10D (e.g., type IIA and IIB backgrounds) or 11D (i.e., M-theory) starting point in which we have QFTs which are localized at singularities. We refer to this higher-dimensional bulk theory as $\mathsf{BLK}$. In the extra-dimensional geometry such singularities can either arise from a singular metric profile (as in geometric engineering) or from branes probing the local geometry (which might also have metric curvature singularities). In principle these singularities need not be isolated: when gravity is decoupled there can be additional branes / singularities supported on non-compact cycles. For ease of exposition we shall mainly focus on cases where we do not have such ``flavor branes'' but the analysis we present naturally extends to these cases as well. We denote the singularities of $X$ by $\mathsf{Sing}=\sqcup_{i\in I\,}\mathsf{Sing}_i$ where $i\in I$ labels connected pairwise disjoint components of $\mathsf{Sing}$. We take $\mathsf{Sing}$ to have finitely many compact components and no non-compact components. In this case, there is a one to one correspondence
\be
\mathcal{B}_{\textnormal{phys}}^{(i)} \leftrightarrow \mathsf{Sing}_i
\ee
between relative QFTs and singular components. We shall often depict these geometries by putting ``$\infty$'' at the top of a figure, and the individual singularities / throats near the bottom. One should view this as a fattening up of the SymTree, but in which we have rotated the picture by 90 degrees so that the topological boundary conditions are now at the top. We do this in part to emphasize the top down nature of the construction, but also because it is easier to read off the relevant physical data in this presentation. See figure \ref{fig:NewStyle} for a depiction of such a top down geometry.

To illustrate how one reads off the symmetry TFT associated with such a geometry, suppose first that we have a single isolated singularity, and that our background $X$ has the form of a conical singularity $\mathrm{Cone}(\partial X)$ where $\partial X$ refers to the conformal boundary of $X$, and the relative QFT
is localized at the tip of the cone. As a point of notation, we shall introduce a radial coordinate $r$ and refer to $r = 0$ as the tip of the cone (where the QFT lives) and $r = \infty$ as the asymptotic boundary. From this starting point, we can consider branes which extend from $\partial X$ to the tip of the cone, giving rise to heavy defects \cite{DelZotto:2015isa, GarciaEtxebarria:2019caf, Albertini:2020mdx, Morrison:2020ool}. Branes purely wrapped in $\partial X$ give rise to symmetry operators, i.e., topological defects \cite{Apruzzi:2022rei, GarciaEtxebarria:2022vzq, Heckman:2022muc}. The global form of the relative QFT is specified by a choice of boundary conditions at $\partial X$ for the bulk supergravity fields. By inspection, the radial direction of the conical geometry suggestively resembles the extra dimension of a symmetry TFT.

Indeed, in reference \cite{Apruzzi:2021nmk} (see also \cite{Aharony:1998qu, Heckman:2017uxe}) it was noted that one can start from the topological terms of the bulk supergravity theory $\mathsf{BLK}$ and dimensionally reduce along the linking geometry $\partial X$ which might also be threaded by various supergravity fluxes (sourced by the branes at the tip of the cone), which we denote as $F$. This results in a $(D+1)$-dimensional TFT which captures some of the interactions terms of the SymTFT which we label as $\mathcal{S}(\partial X, {F})$, in the obvious notation.\footnote{Some of the terms of this SymTFT can be recovered by requiring appropriate braiding rules for extended operators in the associated SymTFT. These braiding rules follow from bulk kinetic terms, as shown in Appendix \ref{app:Kineticterms}.} This construction beautifully shows how SymTFTs arise from an extra-dimensional starting point. Moreover, the boundary states $\langle \mathcal{B}_{\mathrm{top}} \vert$ and $\vert \mathcal{B}_{\mathrm{phys}} \rangle$ are clearly manifest as the boundary conditions of the conical geometry at $r = \infty$ and $r = 0$, respectively.


Multi-sector QFTs naturally arise in backgrounds where $X$ supports multiple singularities. In what follows, we again assume that $X$ is asymptotically conical, i.e., we assume that there exists a coordinate $0 \leq r \leq \infty$ such that near $r = \infty$, we have a conical geometry $\mathrm{Cone}(\partial X)$. The different sectors are sequestered from each other because branes stretching between different singularities have a mass scale set by the size of this distance. In the corresponding effective field theory, this serves as a suppression scale for higher-dimension operators. Even so, there can still be non-trivial topological couplings between sectors, as captured by defects and symmetry operators.

The SymTFT for this multi-sector QFT arises from a similar procedure to that given in \cite{Apruzzi:2021nmk}: In principle, we simply need to perform a dimensional reduction on the linking geometry $\partial X$, and read off the corresponding $(D+1)$-dimensional SymTFT. Observe, however, that in this case the ``radial direction'' only makes sense near $r = \infty$. Indeed, as we proceed to the interior of the geometry we find additional structure as captured by the individual sectors of the model. Proceeding deep into the interior (i.e., for $r$ sufficiently small) we now allow for the geometry to fragment into other local conical geometries, each with their own localized singularity supporting a relative QFT. Indexing the collection of such localized singularities as $\mathsf{Sing}_{i}$ and their associated relative theories as $\mathcal{T}_i$, we have a local radial coordinate $r_i$ which points out from its corresponding singularity. For each such local patch we can again speak of a background $X_i$ and boundary $\partial X_i$. Consequently, we see that the ``full'' SymTFT $\mathcal{S}(\partial X, {F})$ is given by reduction on $\partial X$, and the SymTFT for each sector is instead captured by $\mathcal{S}(\partial X_i, {F}_i)\equiv \mathcal{S}_i$, in the obvious notation. Each such $\mathcal{S}_i$ admits physical boundary conditions $\vert \mathcal{B}_{\mathrm{phys}}^{(i)} \rangle$ at $r_i = 0$ (i.e., where the relative QFT lives), as well as topological boundary conditions $\langle \mathcal{B}_{\mathrm{top}}^{(i)} \vert$.

By inspection, we see that the resulting background $X$ resembles a tree. On the other hand, the SymTree picture has now been ``fattened up'' in the extra dimensions, so that sharp edges associated with junctions are now smoothed out / delocalized. We can project down to a tree again by integrating the topological terms of the bulk theory $\mathsf{BLK}$ over the linking geometry which is piecewise of the form $\partial X_{i}$, but additional care is needed in the fusion of multiple linking geometries, as well as the flux profiles. We return to this shortly, but for now it should be clear that the SymTree has an interpretation in the stringy extra-dimensional geometry.

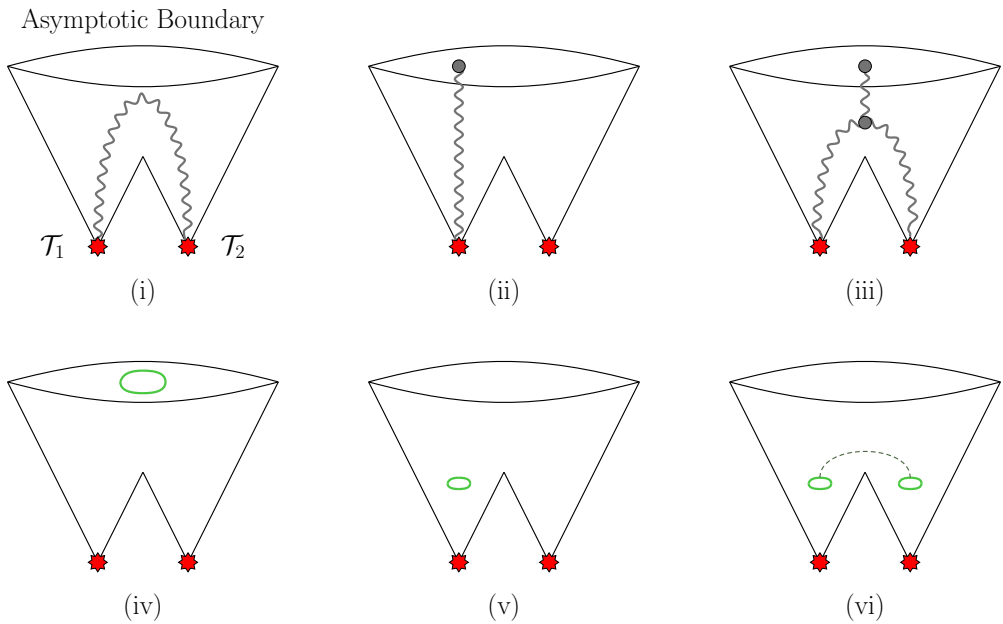
\begin{figure}
\centering

\scalebox{0.6}{
\begin{tikzpicture}
	\begin{pgfonlayer}{nodelayer}
		\node [style=none] (32) at (-9, 3) {};
		\node [style=none] (33) at (-7, 3) {};
		\node [style=none] (34) at (-8, 5) {};
		\node [style=none] (35) at (-5, 7) {};
		\node [style=none] (36) at (-11, 7) {};
		\node [style=none] (37) at (-1, 3) {};
		\node [style=none] (38) at (1, 3) {};
		\node [style=none] (39) at (0, 5) {};
		\node [style=none] (40) at (3, 7) {};
		\node [style=none] (41) at (-3, 7) {};
		\node [style=BrownCircle] (42) at (-1, 7) {};
		\node [style=none] (43) at (-8, 2) {\Large (i)};
		\node [style=none] (44) at (0, 2) {\Large (ii)};
		\node [style=none] (45) at (-8, 8) {\Large Asymptotic Boundary};
		\node [style=Star] (46) at (-9, 3) {};
		\node [style=Star] (47) at (-7, 3) {};
		\node [style=Star] (48) at (-1, 3) {};
		\node [style=Star] (49) at (1, 3) {};
		\node [style=none] (50) at (-9, -4) {};
		\node [style=none] (51) at (-7, -4) {};
		\node [style=none] (52) at (-8, -2) {};
		\node [style=none] (53) at (-5, 0) {};
		\node [style=none] (54) at (-11, 0) {};
		\node [style=none] (55) at (-8, -5) {\Large (iv)};
		\node [style=Star] (56) at (-9, -4) {};
		\node [style=Star] (57) at (-7, -4) {};
		\node [style=none] (59) at (-1, -4) {};
		\node [style=none] (60) at (1, -4) {};
		\node [style=none] (61) at (0, -2) {};
		\node [style=none] (62) at (3, 0) {};
		\node [style=none] (63) at (-3, 0) {};
		\node [style=none] (64) at (0, -5) {\Large (v)};
		\node [style=Star] (65) at (-1, -4) {};
		\node [style=Star] (66) at (1, -4) {};
		\node [style=none] (68) at (7, -4) {};
		\node [style=none] (69) at (9, -4) {};
		\node [style=none] (70) at (8, -2) {};
		\node [style=none] (71) at (11, 0) {};
		\node [style=none] (72) at (5, 0) {};
		\node [style=none] (73) at (8, -5) {\Large (vi)};
		\node [style=Star] (74) at (7, -4) {};
		\node [style=Star] (75) at (9, -4) {};
		\node [style=none] (78) at (-10, 3) {\Large $\mathcal{T}_1$};
		\node [style=none] (79) at (-6, 3) {\Large $\mathcal{T}_2$};
		\node [style=none] (80) at (7, 3) {};
		\node [style=none] (81) at (9, 3) {};
		\node [style=none] (82) at (8, 5) {};
		\node [style=none] (83) at (11, 7) {};
		\node [style=none] (84) at (5, 7) {};
		\node [style=BrownCircle] (85) at (8, 5.75) {};
		\node [style=none] (86) at (8, 2) {\Large (iii)};
		\node [style=Star] (87) at (7, 3) {};
		\node [style=Star] (88) at (9, 3) {};
		\node [style=BrownCircle] (89) at (8, 7) {};
		\node [style=none] (89) at (8, 7) {};
		\node [style=none] (90) at (-8.5, 0) {};
		\node [style=none] (91) at (-7.5, 0) {};
		\node [style=none] (92) at (-8, 0.25) {};
		\node [style=none] (93) at (-8, -0.25) {};
		\node [style=none] (94) at (-1.25, -2.25) {};
		\node [style=none] (95) at (-0.75, -2.25) {};
		\node [style=none] (96) at (-1, -2.125) {};
		\node [style=none] (97) at (-1, -2.375) {};
		\node [style=none] (98) at (6.75, -2.25) {};
		\node [style=none] (99) at (7.25, -2.25) {};
		\node [style=none] (100) at (7, -2.125) {};
		\node [style=none] (101) at (7, -2.375) {};
		\node [style=none] (102) at (8.75, -2.25) {};
		\node [style=none] (103) at (9.25, -2.25) {};
		\node [style=none] (104) at (9, -2.125) {};
		\node [style=none] (105) at (9, -2.375) {};
	\end{pgfonlayer}
	\begin{pgfonlayer}{edgelayer}
		\draw [style=ThickLine] (36.center) to (32.center);
		\draw [style=ThickLine] (32.center) to (34.center);
		\draw [style=ThickLine] (34.center) to (33.center);
		\draw [style=ThickLine] (33.center) to (35.center);
		\draw [style=ThickLine, bend left=15] (36.center) to (35.center);
		\draw [style=ThickLine, bend right=15] (36.center) to (35.center);
		\draw [style=BrownLine, snake it, bend left=90, looseness=5.50] (32.center) to (33.center);
		\draw [style=ThickLine] (41.center) to (37.center);
		\draw [style=ThickLine] (37.center) to (39.center);
		\draw [style=ThickLine] (39.center) to (38.center);
		\draw [style=ThickLine] (38.center) to (40.center);
		\draw [style=ThickLine, bend left=15] (41.center) to (40.center);
		\draw [style=ThickLine, bend right=15] (41.center) to (40.center);
		\draw [style=BrownLine, snake it] (42.center) to (37.center);
		\draw [style=ThickLine] (54.center) to (50.center);
		\draw [style=ThickLine] (50.center) to (52.center);
		\draw [style=ThickLine] (52.center) to (51.center);
		\draw [style=ThickLine] (51.center) to (53.center);
		\draw [style=ThickLine, bend left=15] (54.center) to (53.center);
		\draw [style=ThickLine, bend right=15] (54.center) to (53.center);
		\draw [style=ThickLine] (63.center) to (59.center);
		\draw [style=ThickLine] (59.center) to (61.center);
		\draw [style=ThickLine] (61.center) to (60.center);
		\draw [style=ThickLine] (60.center) to (62.center);
		\draw [style=ThickLine, bend left=15] (63.center) to (62.center);
		\draw [style=ThickLine, bend right=15] (63.center) to (62.center);
		\draw [style=ThickLine] (72.center) to (68.center);
		\draw [style=ThickLine] (68.center) to (70.center);
		\draw [style=ThickLine] (70.center) to (69.center);
		\draw [style=ThickLine] (69.center) to (71.center);
		\draw [style=ThickLine, bend left=15] (72.center) to (71.center);
		\draw [style=ThickLine, bend right=15] (72.center) to (71.center);
		\draw [style=ThickLine] (84.center) to (80.center);
		\draw [style=ThickLine] (80.center) to (82.center);
		\draw [style=ThickLine] (82.center) to (81.center);
		\draw [style=ThickLine] (81.center) to (83.center);
		\draw [style=ThickLine, bend left=15] (84.center) to (83.center);
		\draw [style=ThickLine, bend right=15] (84.center) to (83.center);
		\draw [style=BrownLine, snake it, in=90, out=180, looseness=0.75] (85.center) to (80.center);
		\draw [style=BrownLine, snake it] (85.center) to (89.center);
		\draw [style=BrownLine, snake it, in=90, out=0, looseness=0.75] (85.center) to (88);
		\draw [style=LineGreen, in=270, out=0] (93.center) to (91.center);
		\draw [style=LineGreen, in=360, out=90] (91.center) to (92.center);
		\draw [style=LineGreen, in=75, out=-180] (92.center) to (90.center);
		\draw [style=LineGreen, in=180, out=-90] (90.center) to (93.center);
		\draw [style=LineGreen, in=270, out=0] (97.center) to (95.center);
		\draw [style=LineGreen, in=360, out=90] (95.center) to (96.center);
		\draw [style=LineGreen, in=75, out=-180] (96.center) to (94.center);
		\draw [style=LineGreen, in=180, out=-90] (94.center) to (97.center);
		\draw [style=LineGreen, in=270, out=0] (101.center) to (99.center);
		\draw [style=LineGreen, in=360, out=90] (99.center) to (100.center);
		\draw [style=LineGreen, in=75, out=-180] (100.center) to (98.center);
		\draw [style=LineGreen, in=180, out=-90] (98.center) to (101.center);
		\draw [style=LineGreen, in=270, out=0] (105.center) to (103.center);
		\draw [style=LineGreen, in=360, out=90] (103.center) to (104.center);
		\draw [style=LineGreen, in=75, out=-180] (104.center) to (102.center);
		\draw [style=LineGreen, in=180, out=-90] (102.center) to (105.center);
		\draw [style=DashedLineGreen, bend left=90, looseness=1] (100.center) to (104.center);
	\end{pgfonlayer}
\end{tikzpicture}
}
\caption{Top row: Defects of a double throat geometry $X$ with two local sectors. Strings / branes either run between singularities (i) or between a singularity and the asymptotic boundary (ii) or between multiple singularities and the asymptotic boundary (iii). \newline
Bottom row: Symmetry Operators of a double throat geometry. The symmetry operators of the full theory are strings / branes wrapped in the asymptotic boundary (iv). These admit deformations into a single local model (v) or deformations into multiple local models (vi) joined by a possibly trivial string / brane configuration.}
\label{fig:Throats}
\end{figure}

The construction of heavy defects and topological symmetry operators proceeds much as in the case of a single singularity.
Symmetry operators result from (flux-)branes wrapped on asymptotic cycles in $\partial X$. These are at infinite distance from the QFT degrees of freedom which we assume to be localized in the bulk and hence result in operators interacting only topologically with the QFT. See \cite{Apruzzi:2022rei, GarciaEtxebarria:2022vzq, Heckman:2022muc, Cvetic:2023plv} for further details and \cite{Heckman:2022xgu, Dierigl:2023jdp, Bah:2023ymy, Apruzzi:2023uma, Cvetic:2023pgm} for applications. As such they engineer topological operators in the QFT.\footnote{Reading off the precise form of the generalized symmetries directly from the string background depends on the details of the geometry and fluxes. In the absence of fluxes, it is captured by a relative homology group, but when fluxes are present a suitable generalization of twisted K-theory must be used. For further discussion on the latter point, see Appendix A of reference \cite{Cvetic:2023plv}.} The novelty here is that as the heavy defects descend to different sectors of our system, they can fractionate and become dressed by operators of the smoothed out junctions, as well as defects of other sectors. Note also that a topological operator of the full system can, in an individual throat, end up being dressed by other defects, rendering it ``non-genuine''. See figure \ref{fig:Throats} for a depiction of how heavy defects and symmetry operators descend to individual sectors of the system.

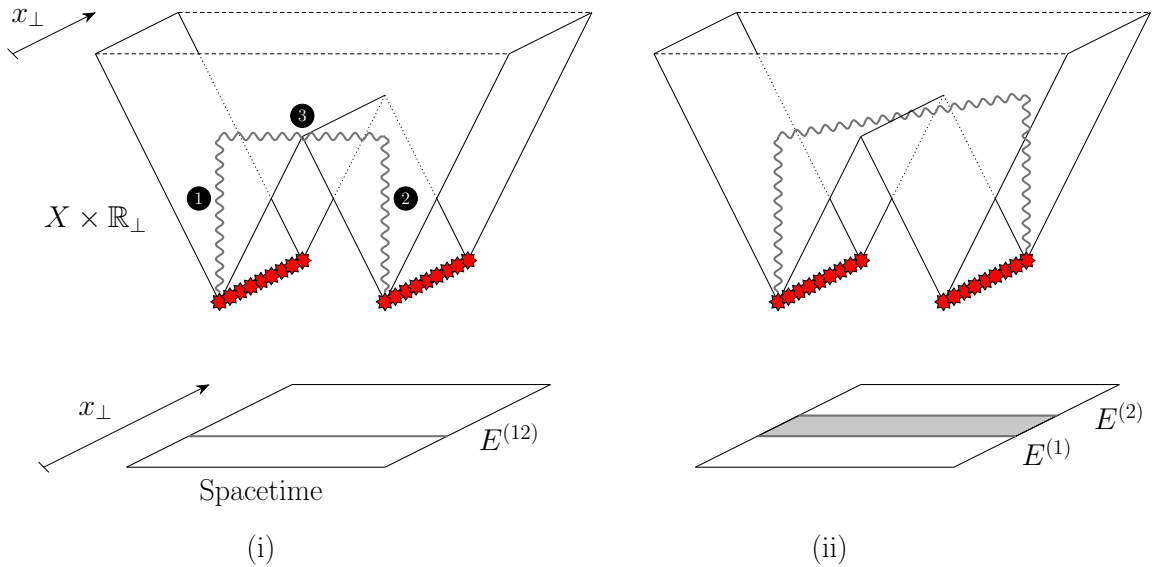
\begin{figure}
\centering

\scalebox{0.55}{
\begin{tikzpicture}
\begin{pgfonlayer}{nodelayer}
		\node [style=none] (0) at (-5, -2) {};
		\node [style=none] (1) at (-1, -2) {};
		\node [style=none] (2) at (-3, 2) {};
		\node [style=none] (4) at (-7, 2) {};
		\node [style=none] (5) at (-8, 4) {};
		\node [style=none] (6) at (2, 4) {};
		\node [style=none] (9) at (-3, -1) {};
		\node [style=none] (10) at (1, -1) {};
		\node [style=none] (11) at (-1, 3) {};
		\node [style=none] (13) at (-5.5, 4) {};
		\node [style=none] (14) at (-6, 5) {};
		\node [style=none] (15) at (4, 5) {};
		\node [style=none] (18) at (-3.75, 0.5) {};
		\node [style=none] (19) at (-2.25, 0.5) {};
		\node [style=none] (20) at (0.25, 0.5) {};
		\node [style=Star] (21) at (-3, -1) {};
		\node [style=Star] (22) at (1, -1) {};
		\node [style=Star] (23) at (-1, -2) {};
		\node [style=Star] (24) at (-5, -2) {};
		\node [style=none] (25) at (-5, 2) {};
		\node [style=none] (26) at (-1, 2) {};
		\node [style=none] (27) at (-7.25, -6) {};
		\node [style=none] (28) at (-3.25, -4) {};
		\node [style=none] (29) at (-1, -6) {};
		\node [style=none] (30) at (3, -4) {};
		\node [style=none] (31) at (-5.75, -5.25) {};
		\node [style=none] (32) at (0.5, -5.25) {};
		\node [style=none] (33) at (-5, -2.625) {};
		\node [style=none] (34) at (-1, -2.625) {};
		\node [style=Star] (37) at (-4.5, -1.75) {};
		\node [style=Star] (38) at (-4, -1.5) {};
		\node [style=Star] (39) at (-3.5, -1.25) {};
		\node [style=Star] (40) at (-0.5, -1.75) {};
		\node [style=Star] (41) at (0, -1.5) {};
		\node [style=Star] (42) at (0.5, -1.25) {};
		\node [style=Star] (43) at (-0.75, -1.875) {};
		\node [style=Star] (44) at (-0.25, -1.625) {};
		\node [style=Star] (45) at (0.25, -1.375) {};
		\node [style=Star] (46) at (-3.75, -1.375) {};
		\node [style=Star] (47) at (-4.25, -1.625) {};
		\node [style=Star] (48) at (-4.75, -1.875) {};
		\node [style=none] (49) at (-9.25, -6) {};
		\node [style=none] (50) at (-5.25, -4) {};
		\node [style=none] (51) at (-8, -4.625) {\LARGE $x_\perp$};
		\node [style=none] (52) at (-10, 4) {};
		\node [style=none] (53) at (-8, 5) {};
		\node [style=none] (54) at (-9.625, 4.875) {\LARGE $x_\perp$};
		\node [style=none] (55) at (-8, 0) {\LARGE $X\times \mathbb{R}_\perp$};
		\node [style=none] (56) at (-4, -6.625) {\LARGE Spacetime};
		\node [style=none] (57) at (1.75, -5.75) {};
		\node [style=none] (58) at (2, -5.25) {\LARGE $E^{(12)}$};
		\node [style=none] (59) at (8.5, -2) {};
		\node [style=none] (60) at (12.5, -2) {};
		\node [style=none] (61) at (10.5, 2) {};
		\node [style=none] (62) at (6.5, 2) {};
		\node [style=none] (63) at (5.5, 4) {};
		\node [style=none] (64) at (15.5, 4) {};
		\node [style=none] (65) at (10.5, -1) {};
		\node [style=none] (66) at (14.5, -1) {};
		\node [style=none] (67) at (12.5, 3) {};
		\node [style=none] (68) at (8, 4) {};
		\node [style=none] (69) at (7.5, 5) {};
		\node [style=none] (70) at (17.5, 5) {};
		\node [style=none] (71) at (9.75, 0.5) {};
		\node [style=none] (72) at (11.25, 0.5) {};
		\node [style=none] (73) at (13.75, 0.5) {};
		\node [style=Star] (74) at (10.5, -1) {};
		\node [style=Star] (75) at (14.5, -1) {};
		\node [style=Star] (76) at (12.5, -2) {};
		\node [style=Star] (77) at (8.5, -2) {};
		\node [style=none] (78) at (8.5, 2) {};
		\node [style=none] (79) at (12.5, 2) {};
		\node [style=none] (80) at (6.5, -6) {};
		\node [style=none] (81) at (10.5, -4) {};
		\node [style=none] (82) at (12.75, -6) {};
		\node [style=none] (83) at (16.75, -4) {};
		\node [style=none] (84) at (8, -5.25) {};
		\node [style=none] (85) at (14.25, -5.25) {};
		\node [style=none] (86) at (8.5, -2.625) {};
		\node [style=none] (87) at (12.5, -2.625) {};
		\node [style=Star] (90) at (9, -1.75) {};
		\node [style=Star] (91) at (9.5, -1.5) {};
		\node [style=Star] (92) at (10, -1.25) {};
		\node [style=Star] (93) at (13, -1.75) {};
		\node [style=Star] (94) at (13.5, -1.5) {};
		\node [style=Star] (95) at (14, -1.25) {};
		\node [style=Star] (96) at (12.75, -1.875) {};
		\node [style=Star] (97) at (13.25, -1.625) {};
		\node [style=Star] (98) at (13.75, -1.375) {};
		\node [style=Star] (99) at (9.75, -1.375) {};
		\node [style=Star] (100) at (9.25, -1.625) {};
		\node [style=Star] (101) at (8.75, -1.875) {};
		\node [style=none] (110) at (15.5, -5.75) {};
		\node [style=none] (111) at (15, -5.625) {\LARGE $E^{(1)}$};
		\node [style=none] (112) at (14.5, 3) {};
		\node [style=none] (113) at (9, -4.75) {};
		\node [style=none] (114) at (15.25, -4.75) {};
		\node [style=none] (115) at (16.75, -4.75) {\LARGE $E^{(2)}$};
		\node [style=none] (116) at (-4, -8) {\LARGE (i)};
		\node [style=none] (117) at (9.75, -8) {\LARGE (ii)};
		\node [style=Star] (118) at (-3.25, -1.125) {};
		\node [style=Star] (119) at (0.75, -1.125) {};
		\node [style=Star] (120) at (14.25, -1.125) {};
		\node [style=Star] (121) at (10.25, -1.125) {};
		\node [style=none] (122) at (-9.375, -5.875) {};
		\node [style=none] (123) at (-9.125, -6.125) {};
		\node [style=none] (124) at (-10.125, 4.125) {};
		\node [style=none] (125) at (-9.875, 3.875) {};
		\node [style=circle] (129) at (-3, 2.5) {};
		\node [style=circle] (130) at (-5.5, 0.5) {};
		\node [style=circle] (131) at (-0.5, 0.5) {};
		\node [style=none] (126) at (-3, 2.507) {\color{white}{3}};
		\node [style=none] (127) at (-5.5, 0.507) {\color{white}{1}};
		\node [style=none] (128) at (-0.5, 0.507) {\color{white}{2}};
	\end{pgfonlayer}
	\begin{pgfonlayer}{edgelayer}
		\filldraw[fill={rgb,255: red,200; green,200; blue,200}] (113) rectangle (85);
		\filldraw[fill={rgb,255: red,200; green,200; blue,200}] (9, -4.75) -- (8, -5.25) -- (14.25, -5.25) -- (15.25, -4.75) -- cycle;
		\draw [style=ThickLine] (5.center) to (0.center);
		\draw [style=ThickLine] (0.center) to (2.center);
		\draw [style=ThickLine] (2.center) to (1.center);
		\draw [style=ThickLine] (1.center) to (6.center);
		\draw [style=DashedLineThin] (5.center) to (6.center);
		\draw [style=DashedLineThin] (14.center) to (15.center);
		\draw [style=DottedLine] (13.center) to (18.center);
		\draw [style=DottedLine] (19.center) to (11.center);
		\draw [style=DottedLine] (11.center) to (20.center);
		\draw [style=ThickLine] (14.center) to (13.center);
		\draw [style=ThickLine] (20.center) to (10.center);
		\draw [style=ThickLine] (19.center) to (9.center);
		\draw [style=ThickLine] (9.center) to (18.center);
		\draw [style=ThickLine] (10.center) to (15.center);
		\draw (5.center) to (14.center);
		\draw (6.center) to (15.center);
		\draw (2.center) to (11.center);
		\draw [style=BrownLine, snake it] (24) to (25.center);
		\draw [style=BrownLine, snake it] (25.center) to (26.center);
		\draw [style=BrownLine, snake it] (26.center) to (23);
		\draw [style=ThickLine] (27.center) to (28.center);
		\draw [style=ThickLine] (28.center) to (30.center);
		\draw [style=ThickLine] (30.center) to (29.center);
		\draw [style=ThickLine] (29.center) to (27.center);
		\draw [style=BrownLine] (31.center) to (32.center);
		\draw [style=ArrowLineRight] (49.center) to (50.center);
		\draw [style=ArrowLineRight] (52.center) to (53.center);
		\draw [style=ThickLine] (63.center) to (59.center);
		\draw [style=ThickLine] (59.center) to (61.center);
		\draw [style=ThickLine] (61.center) to (60.center);
		\draw [style=ThickLine] (60.center) to (64.center);
		\draw [style=DashedLineThin] (63.center) to (64.center);
		\draw [style=DashedLineThin] (69.center) to (70.center);
		\draw [style=DottedLine] (68.center) to (71.center);
		\draw [style=DottedLine] (72.center) to (67.center);
		\draw [style=DottedLine] (67.center) to (73.center);
		\draw [style=ThickLine] (69.center) to (68.center);
		\draw [style=ThickLine] (73.center) to (66.center);
		\draw [style=ThickLine] (72.center) to (65.center);
		\draw [style=ThickLine] (65.center) to (71.center);
		\draw [style=ThickLine] (66.center) to (70.center);
		\draw (63.center) to (69.center);
		\draw (64.center) to (70.center);
		\draw [style=BrownLine, snake it] (77) to (78.center);
		\draw [style=ThickLine] (80.center) to (81.center);
		\draw [style=ThickLine] (81.center) to (83.center);
		\draw [style=ThickLine] (83.center) to (82.center);
		\draw [style=ThickLine] (82.center) to (80.center);
		\draw [style=BrownLine] (84.center) to (85.center);
		\draw [style=BrownLine, snake it] (112.center) to (75);
		\draw [style=BrownLine, snake it] (78.center) to (112.center);
		\draw (61.center) to (67.center);
		\draw [style=BrownLine] (113.center) to (114.center);
		\draw [style=ThickLine] (124.center) to (125.center);
		\draw [style=ThickLine] (122.center) to (123.center);
	\end{pgfonlayer}
\end{tikzpicture}
}
\caption{(i):\;String\,/\,brane running between two local models. Such an object admits a partition into three pieces, two of which are contained in a local model, and one which connect the two via the bulk of $X$. In the QFT spacetime the result is a pair of defects $E^{(12)}$.\newline
(ii):\;Deforming the configuration into a spacetime direction $x_\perp$ the bulk part of the string / brane gives a topological operator bounded by the defects. Individually the defects $E^{(i)}$ are non-genuine.}
\label{fig:Throats2}
\end{figure}

With this we can turn to the question of which operators constructed in figure \ref{fig:Throats} are genuine and non-genuine. For example, consider the brane configuration (i) in figure \ref{fig:Throats}. Deforming the locus along which a defect attaches to $\mathsf{Sing}$ in spacetime, as shown in figure \ref{fig:Throats2}, we find the component of the string / brane stretching between the local models gives rise to a topological operator. From the perspective of an individual throat the initial defect is non-genuine. Similar comments apply to other configurations displayed in figure \ref{fig:Throats}. Note that this discussion exactly parallels our ``bottom up'' analysis in Section \ref{sec:SYMTREE}.

\subsection{Projection to a SymTree} \label{sec:Filtrations}

Having spelled out the general strategy for extracting the SymTree of a multi-sector model directly from $X$, we now provide an algorithmic procedure for reading off this data. The discussion splits up into the contribution from the geometry of $X$, and if present, additional contributions from fluxes threading $\partial X$ as well as the individual branches $\partial X_{i}$. With this in mind, we first begin by explaining in more detail how the different ``boundary geometries'' $\partial X_{i}$ fuse together to form the SymTree, focusing in particular on the singular homology of these spaces and how they consistently glue together. To read off the SymTFT we will also extract the associated differential cohomology (following \cite{Belov:2006jd,Apruzzi:2021nmk}), but one might entertain a generalization such as differential K-theory. The contribution from fluxes follows a similar procedure: we find that along each segment of the resulting SymTree, we have a piecewise constant contribution from flux, but that this ``jumps'' across the junctions of the SymTree. This sort of jumping phenomena is indicative of additional degrees of freedom localized at the junction, precisely as expected on general grounds. Reduction of the bulk theory $\mathsf{BLK}$ topological terms then results in our SymTree theory.

\subsubsection{Filtrations and Trees} \label{sssec:FILTRUM}

We now turn to the treelike structure obtained by ``projecting down'' onto the radial direction of $X$. Recall that $X$ has singularities $\mathsf{Sing}=\sqcup_{i\in I} \mathsf{Sing}_i$ with finite index set $I$. Generically there is no unique treelike structure given $X$. Rather, a particular tree $\Upsilon$ is only specified once we have determined how to sweep out $X$ via radial shells, as specified by a filtration $\mathcal{F}_X$.
Much as in other stringy realizations of QFTs, ambiguities in reading off a specific SymTree from geometry amount to non-trivial dualities / associator moves of a SymTree.

Given a geometry $X$ we therefore also require the existence of a filtration $\mathcal{F}_X$ over the real half-line, parameterized by the ``radius'' $r\in [0,\infty)$ and with radial shells $\U_r$ and the indexed family of sets $ \{B_r\}_{r\geq 0}$ where $B_r=\cup_{s\;\!\leq\;\! r\;\!}U_s$ and $\partial B_r=U_r$. Among all possible filtrations we consider those with following favorable properties:
\begin{itemize}
\item The filtration is centered on the singularities of $X$; we impose $B_0=\mathsf{Sing}$. The filtration sweeps out the full geometry; we impose $B_\infty =X$.
\item The filtration describes a disjoint collection of local models at small radii. We impose
\be
0 < r \leq \epsilon\,: \quad  B_r\sim \sqcup_i \:\! \textnormal{Tube}(\mathsf{Sing}_i)
\ee
for some $\epsilon>0$ where Tube$(\mathsf{Sing}_i)$ is the tubular neighbourhood of $\mathsf{Sing}_i$ and no two tubes overlap. Tubes capture topological structure of a local model centered on $\mathsf{Sing}_i$.

\item The filtration is topologically piecewise constant. There only exist finitely many critical radii, denoted $r_k$ which we label as $r_k<r_l$ for $k<l$, such that balls wedged between the same critical radii are topologically equivalent:
\be
r_k< R_1 < R_2  <r_l\,: ~ ~  B_{R_2}\rightarrow B_{R_1}\,.
\ee
Here $\rightarrow$ denotes a deformation retraction from $B_{R_2}$ to $ B_{R_1}$. In particular the integral homology (and homotopy) groups are constant along the interval $(r_k,r_l)$, so we have
\be
H_n(B_{R_1})\cong H_n(B_{R_2})\,, \qquad H_n(U_{R_1})\cong H_n(U_{R_2})\,.
\ee
\item The filtration has one asymptotic boundary. The maximal radius $r_*=\textnormal{max}\{ r_k \}$  is such that $B_{r_*}$ is connected and there are deformation retractions $X\rightarrow B_r$ and $\partial X \rightarrow \partial B_{r}$ for all $r\geq r_*$.

\begin{figure}
\centering
\scalebox{0.8}{\begin{tikzpicture}
	\begin{pgfonlayer}{nodelayer}
		\node [style=none] (0) at (-2, -2) {};
		\node [style=none] (1) at (2, -2) {};
		\node [style=none] (2) at (0, -4) {};
		\node [style=none] (3) at (4, -4) {};
		\node [style=none] (4) at (0, 0) {};
		\node [style=none] (5) at (-4, -4) {};
		\node [style=CircleBlue] (6) at (0, 2) {};
		\node [style=none] (7) at (-0.125, 2) {};
		\node [style=none] (8) at (0.125, 2) {};
		\node [style=none] (9) at (-6, 2) {};
		\node [style=none] (10) at (-6, -4) {};
		\node [style=none] (11) at (-5.875, -4) {};
		\node [style=none] (12) at (-6.125, -4) {};
		\node [style=none] (13) at (-5.875, 2) {};
		\node [style=none] (14) at (-6.125, 2) {};
		\node [style=CirclePurple] (15) at (2, -2) {};
		\node [style=CirclePurple] (16) at (0, 0) {};
		\node [style=Star] (17) at (-4, -4) {};
		\node [style=Star] (18) at (0, -4) {};
		\node [style=Star] (19) at (4, -4) {};
		\node [style=none] (20) at (-7, -4) {$r=0$};
		\node [style=none] (21) at (-7, 2) {$r=\infty$};
		\node [style=none] (22) at (-6, -2) {};
		\node [style=none] (23) at (-6, 0) {};
		\node [style=none] (24) at (-7, -2) {$r_1$};
		\node [style=none] (25) at (-7, 0) {$r_2=r_*$};
		\node [style=none] (26) at (-3.925, -4.5) {$\mathsf{Sing}_1$};
		\node [style=none] (27) at (0.075, -4.5) {$\mathsf{Sing}_2$};
		\node [style=none] (28) at (4.075, -4.5) {$\mathsf{Sing}_3$};
		\node [style=none] (29) at (-4.25, -3) {};
		\node [style=none] (30) at (4.25, -3) {};
		\node [style=none] (31) at (-4.25, -1) {};
		\node [style=none] (32) at (4.25, -1) {};
		\node [style=none] (33) at (-4.25, 1) {};
		\node [style=none] (34) at (4.25, 1) {};
		\node [style=none] (35) at (5.225, 1) {$U_{r\;\!>\;\!r_2}$};
		\node [style=none] (36) at (5.5, -1) {$U_{r_2\;\!>\;\!r\;\!>\;\!r_1}$};
		\node [style=none] (37) at (5.45, -3) {$U_{r_1\;\!>\;\!r\;\!>\;\!0}$};
		\node [style=none] (38) at (-4.25, -2) {};
		\node [style=none] (39) at (4.25, -2) {};
		\node [style=none] (40) at (-4.25, 0) {};
		\node [style=none] (41) at (4.25, 0) {};
		\node [style=none] (42) at (5, 0) {$U_{r_2}$};
		\node [style=none] (43) at (5, -2) {$U_{r_1}$};
		\node [style=none] (44) at (0, 2.5) {Boundary Conditions};
		\node [style=Circle] (45) at (-6, -2) {};
		\node [style=Circle] (46) at (-6, 0) {};
		\node [style=none] (47) at (0, -5) {};
		\node [style=none] (48) at (0, 3) {Topological};
	\end{pgfonlayer}
	\begin{pgfonlayer}{edgelayer}
		\draw [style=ThickLine] (14.center) to (13.center);
		\draw [style=ThickLine] (12.center) to (11.center);
		\draw [style=ThickLine] (9.center) to (10.center);
		\draw [style=ThickLine] (5.center) to (4.center);
		\draw [style=ThickLine] (4.center) to (6.center);
		\draw [style=ThickLine] (7.center) to (8.center);
		\draw [style=ThickLine] (4.center) to (1.center);
		\draw [style=ThickLine] (1.center) to (2.center);
		\draw [style=ThickLine] (1.center) to (3.center);
		\draw [style=DottedLine] (33.center) to (34.center);
		\draw [style=DottedLine] (32.center) to (31.center);
		\draw [style=DottedLine] (29.center) to (30.center);
		\draw [style=DottedLine] (39.center) to (38.center);
		\draw [style=DottedLine] (41.center) to (40.center);
	\end{pgfonlayer}
\end{tikzpicture}
}
\caption{Radial filtration  $\{B_r\}_{r\geq 0} $ of the singular geometry $X$ with $\partial B_r=U_r$.}
\label{fig:ExampleFiltration}
\end{figure}
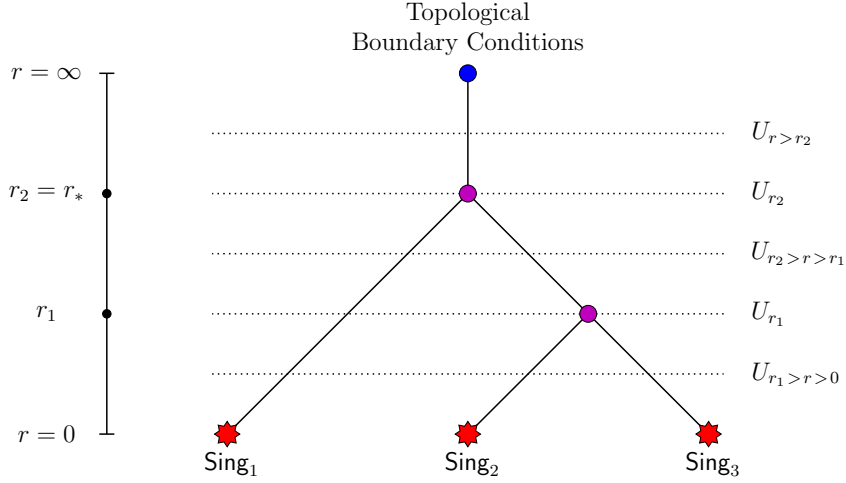

\end{itemize}

\noindent Now, given such a filtration $\mathcal{F}_X$ of the geometry $X$ we associate to it a SymTree, for a given choice of theory $\textnormal{IIA,\,IIB,\,M}$, by compactification of the topological terms of the respective supergravity theory $\mathsf{BLK}$ over the boundary $\partial B_r=U_r$. At each value of the radius $r$ we have $\pi_0(\partial B_r)$ connected components and on each component distinct topological fields are present. At the critical radii $\pi_0$ jumps and previously disjoint local models combine. Consequently, the symmetry TFT is defined on a tree $\Upsilon$ (see figure \ref{fig:ExampleFiltration}).

Generically, this tree has $\left|I\right|+1$ external nodes. Of these $\left|{I}\right|$ are located at $r=0$ and a single vertex is at $r=\infty$. The singularities $\mathsf{Sing}= \sqcup_{i\in I} \mathsf{Sing}_i$ specify relative theories setting enriched Neumann boundary conditions $\mathcal{B}_{\textnormal{phys}}^{(i)}$ at the $|{I}|$ vertices at $r=0$. Further, there is a topological boundary condition $\mathcal{B}_{\textnormal{top}}$ at the asymptotic node at infinity which determines the overall global form.

Internal nodes arise whenever the number of connected components of $\partial B_r=U_r$ change. At the first transition $r=r_1$ some local models centered on the individual connected singular loci $\mathsf{Sing}_i$ combine, such combined neighbourhoods then continue to grow, merging with similar neighborhoods at critical radii $r_k$ into larger neighbourhoods containing more and more components of $\mathsf{Sing}$. The number of connected components $\pi_0(\partial B_r)$ decreases with increasing radius $r$ and is locally constant away from critical radii (see figure \ref{fig:ExampleFiltration}).

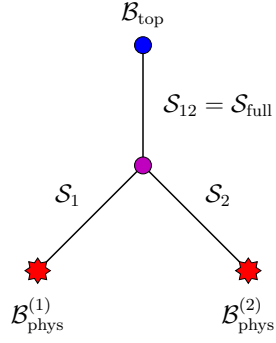
\begin{figure}
    \centering
    \scalebox{0.8}{
    \begin{tikzpicture}
	\begin{pgfonlayer}{nodelayer}
		\node [style=none] (0) at (0, 0) {};
		\node [style=CircleBlue] (4) at (0, 2) {};
		\node [style=none] (5) at (1.75, -1.75) {};
		\node [style=none] (6) at (-1.75, -1.75) {};
		\node [style=none] (7) at (0, 0) {};
		\node [style=none] (8) at (0, 0) {};
		\node [style=none] (9) at (-0.25, 2) {};
		\node [style=none] (10) at (0.25, 2) {};
		\node [style=Star] (11) at (-1.75, -1.75) {};
		\node [style=Star] (12) at (1.75, -1.75) {};
		\node [style=none] (13) at (-1.75, -2.5) {$\mathcal{B}_{{\textnormal{phys}}}^{(1)}$};
		\node [style=none] (14) at (1.75, -2.5) {$\mathcal{B}_{{\textnormal{phys}}}^{(2)}$};
		\node [style=none] (17) at (0, 2.5) {$\mathcal{B}_{{\textnormal{top}}}$};
		\node [style=none] (18) at (0, -3) {};
		\node [style=none] (19) at (1.25, -0.5) {$\mathcal{S}_2$};
		\node [style=none] (20) at (-1.25, -0.5) {$\mathcal{S}_1$};
		\node [style=none] (22) at (1.25, 1) {$\mathcal{S}_{12}=\mathcal{S}_{\textnormal{full}}$};
		\node [style=CirclePurple] (23) at (0, 0) {};
	\end{pgfonlayer}
	\begin{pgfonlayer}{edgelayer}
		\draw [style=ThickLine] (4.center) to (0.center);
		\draw [style=ThickLine] (8.center) to (5.center);
		\draw [style=ThickLine] (7.center) to (6.center);
	\end{pgfonlayer}
\end{tikzpicture}
    }
    \caption{Two relative theories $\mathcal{B}_{{\textnormal{phys}}}^{(i)}$ with symmetry topological field theory $\mathcal{S}_i$ which combine to $\mathcal{S}_{ij}$ terminated by topological boundary conditions $\mathcal{B}_{{\textnormal{top}}}$. From this fundamental junction more generic graphs can be built.}
    \label{fig:GluingSymTFT}
\end{figure}

\paragraph{Y-shaped Junctions and their Homology}\mbox{}\medskip \\  Generically there are $|{I}|-1$ internal trivalent vertices at which two previously disjoint local models combine. The tree $\Upsilon$ then parameterizes a collection of such combinations. It suffices to consider a single trivalent vertex as in figure \ref{fig:GluingSymTFT}. Such a junction of symmetry TFTs is supported on a Y-shaped tree. On the two legs at small radii we have the TFTs $\mathcal{S}_1$ and $\mathcal{S}_2$. At the internal vertex these attach to the TFT $\mathcal{S}_{12}$ describing symmetries at large radii. At the internal vertex additional fields can be localized and enter into the gluing conditions.

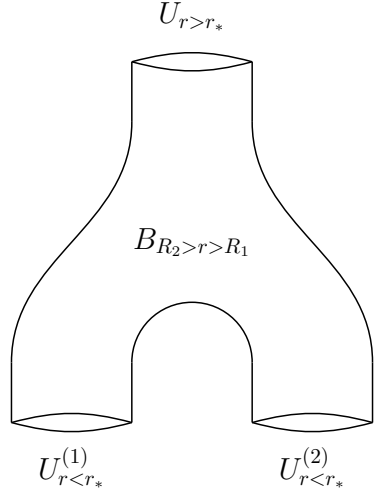
\begin{figure}
\centering
\scalebox{0.8}{
\begin{tikzpicture}
	\begin{pgfonlayer}{nodelayer}
		\node [style=none] (22) at (-3, -2) {};
		\node [style=none] (23) at (-1, -2) {};
		\node [style=none] (24) at (1, -2) {};
		\node [style=none] (25) at (3, -2) {};
		\node [style=none] (26) at (1, -1) {};
		\node [style=none] (27) at (-1, -1) {};
		\node [style=none] (28) at (-3, -1) {};
		\node [style=none] (29) at (3, -1) {};
		\node [style=none] (30) at (0, 0) {};
		\node [style=none] (32) at (2, 1) {};
		\node [style=none] (33) at (-2, 1) {};
		\node [style=none] (34) at (-1, 3) {};
		\node [style=none] (35) at (1, 3) {};
		\node [style=none] (36) at (-1, 4) {};
		\node [style=none] (37) at (1, 4) {};
		\node [style=none] (38) at (0, 4.75) {\large $U_{r>r_*}$};
		\node [style=none] (39) at (-2, -2.75) {\large $U^{(1)}_{r<r_*}$};
		\node [style=none] (40) at (2, -2.75) {\large $U^{(2)}_{r<r_*}$};
		\node [style=none] (41) at (0, 1) {\large$B_{R_2>r>R_1}$};
		\node [style=none] (42) at (0, 5.25) {};
		\node [style=none] (43) at (0, -3.25) {};
	\end{pgfonlayer}
	\begin{pgfonlayer}{edgelayer}
		\draw [style=ThickLine, bend left=15] (37.center) to (36.center);
		\draw [style=ThickLine] (36.center) to (34.center);
		\draw [style=ThickLine, in=90, out=-90] (34.center) to (28.center);
		\draw [style=ThickLine] (28.center) to (22.center);
		\draw [style=ThickLine] (23.center) to (27.center);
		\draw [style=ThickLine, in=-180, out=90] (27.center) to (30.center);
		\draw [style=ThickLine, in=90, out=0] (30.center) to (26.center);
		\draw [style=ThickLine] (26.center) to (24.center);
		\draw [style=ThickLine] (25.center) to (29.center);
		\draw [style=ThickLine, in=-90, out=90] (29.center) to (35.center);
		\draw [style=ThickLine] (37.center) to (35.center);
		\draw [style=ThickLine, bend right=15] (22.center) to (23.center);
		\draw [style=ThickLine, bend right=15] (24.center) to (25.center);
		\draw [style=ThickLine, bend left=15] (36.center) to (37.center);
		\draw [style=ThickLine, bend left=15] (24.center) to (25.center);
		\draw [style=ThickLine, bend left=15] (22.center) to (23.center);
	\end{pgfonlayer}
\end{tikzpicture}
}
\caption{Pair of pants describing the uplift of a SymTree junction.}
\label{fig:PairofPants}
\end{figure}

We now derive the gluing conditions at the junction from geometry. First, note that the topological fields on the legs of the Y-shaped graph derive via dimensional reduction over the radial slices $U_r$ and we therefore need to track the corresponding cycles of $U_{r<r_*}$ through $U_{r=r_*}$ to cycles of $U_{r\;\!>\;\!r_*}$. We reformulate this problem by noting that $U_{r\;\!>\;\!r_*}$ is the deformation retract of the pair of pants
\be \label{eq:PairOfPants}
B_{R_2\;\!>\;\!r\;\!>\;\! R_1}=\bigcup_{R_2\;\!>\;\!r\;\!>\;\! R_1} U_r\,, \qquad R_2>r_*>R_1\,,
\ee
which has boundary $\partial B_{R_2\;\!>\;\!r\;\!>\;\! R_1}=U_{R_2} \cup U_{R_1}$. Clearly there are embedding maps $U_{R_i}\hookrightarrow B_{R_2\;\!>\;\!r\;\!>\;\! R_1}$ and we denote their degree $n$ lift to homology by
\be\label{eq:EmbeddingMaps}  \ba
\jmath_n^{(r<r_*)}\,: H_n(U_{r<r_*})\cong H_n(U_{R_1}) ~\rightarrow~ H_n(B_{R_2\;\!>\;\!r\;\!>\;\! R_1}) \cong H_n(U_{r=r_*})\\[0.25em]
\jmath_n^{(r>r_*)}\,: H_n(U_{r>r_*})\cong H_n(U_{R_2}) ~\rightarrow~  H_n(B_{R_2\;\!>\;\!r\;\!>\;\! R_1}) \cong H_n(U_{r=r_*})\\
\ea\ee
which compare cycles in large radius shells with those of small radius shells by embedding them both into the critical shell (see figure \ref{fig:PairofPants}).

\newpage

Cycles of small and large radius shells are further put in relation by two Mayer-Vietoris long exact sequence, one for small radii and one for large radii. These are setup such that the mappings \eqref{eq:EmbeddingMaps} are maps of these sequences, offering a tool to compute them.

We begin by describing the small radius sequence. Denote the two connected components of $U_{r<r_*}$ as
\be
U_{r<r_*}= U^{(1)}_{r<r_*} \sqcup U^{(2)}_{r<r_*}\,,
\ee
where we suppress the index for notational purposes below. These two shells grow until they touch along some locus
\be
U^{(12)}_{r=r_*}=U^{(1)}_{r=r_*}\cap U^{(2)}_{r=r_*}
\ee
where the intersecting sets are such that we have deformation retractions $U^{(i)}_{r=r_*}\rightarrow U^{(i)}_{r<r_*}$. There we also have $U^{(1)}_{r=r_*}\cup U^{(2)}_{r=r_*}=U_{r=r_*}$ which is precisely the covering we use in formulating the Mayer-Vietoris sequence. The sequence
\be\label{eq:seq1}
\dots ~\rightarrow~ H_n(U^{(12)}_{r=r_*})   ~\xrightarrow[]{~\imath_n^{(r<r_*)}~} ~ H_n(U^{(1)}_{r=r_*})\oplus H_n(U^{(2)}_{r=r_*})~\xrightarrow[]{~\jmath_n^{(r<r_*)}~} ~ H_n(U_{r=r_*}) ~\rightarrow~\dots
\ee
then contains the map $\jmath_n^{(r<r_*)}$ which relates cycles (and therefore fields) at small radii to those of the critical shell. We label these maps by $r<r_*$ to denote that the covering of the critical slice is derived by approaching it from small radii.

\begin{figure}
\centering
\scalebox{0.8}{
\begin{tikzpicture}
	\begin{pgfonlayer}{nodelayer}
		\node [style=none] (0) at (-7, 0) {};
		\node [style=none] (1) at (-5, 0) {};
		\node [style=none] (2) at (-6, 1) {};
		\node [style=none] (3) at (-6, -1) {};
		\node [style=none] (4) at (-6, 0) {};
		\node [style=none] (5) at (-6, 1.5) {\large $U^{(1)}_{r=r_*}$};
		\node [style=none] (6) at (-4, 0) {};
		\node [style=none] (7) at (-2, 0) {};
		\node [style=none] (8) at (-3, 1) {};
		\node [style=none] (9) at (-3, -1) {};
		\node [style=none] (10) at (-3, 0) {};
		\node [style=none] (11) at (-3, 1.5) {\large $U^{(2)}_{r=r_*}$};
		\node [style=none] (12) at (-1, 0) {};
		\node [style=none] (13) at (0, 0) {};
		\node [style=none] (14) at (1, 0) {};
		\node [style=none] (15) at (3, 0) {};
		\node [style=none] (16) at (2, 1) {};
		\node [style=none] (17) at (2, -1) {};
		\node [style=none] (19) at (3, 1.5) {\large $U_{r=r_*}$};
		\node [style=none] (20) at (3, 0) {};
		\node [style=none] (21) at (5, 0) {};
		\node [style=none] (22) at (4, 1) {};
		\node [style=none] (23) at (4, -1) {};
		\node [style=none] (24) at (6, 0) {};
		\node [style=none] (25) at (7, 0) {};
		\node [style=none] (26) at (8, 0) {};
		\node [style=none] (28) at (9, 1) {};
		\node [style=none] (29) at (9, -1) {};
		\node [style=none] (30) at (9.75, 1.5) {\large $U_{r>r_*}$};
		\node [style=none] (32) at (11.5, 0) {};
		\node [style=none] (33) at (10.5, 1) {};
		\node [style=none] (34) at (10.5, -1) {};
		\node [style=none] (35) at (9.75, 0.5) {};
		\node [style=none] (36) at (9.75, -0.5) {};
		\node [style=none] (37) at (9.75, -1.25) {\large \color{red}${V}$};
	\end{pgfonlayer}
	\begin{pgfonlayer}{edgelayer}
		\draw [style=ThickLine, bend right=45] (3.center) to (1.center);
		\draw [style=ThickLine, bend right=45] (1.center) to (2.center);
		\draw [style=ThickLine, bend right=45] (2.center) to (0.center);
		\draw [style=ThickLine, bend right=45] (0.center) to (3.center);
		\draw [style=ThickLine, bend right=45] (9.center) to (7.center);
		\draw [style=ThickLine, bend right=45] (7.center) to (8.center);
		\draw [style=ThickLine, bend right=45] (8.center) to (6.center);
		\draw [style=ThickLine, bend right=45] (6.center) to (9.center);
		\draw [style=ArrowLineRight] (12.center) to (13.center);
		\draw [style=ThickLine, bend right=45] (17.center) to (15.center);
		\draw [style=ThickLine, bend right=45] (15.center) to (16.center);
		\draw [style=ThickLine, bend right=45] (16.center) to (14.center);
		\draw [style=ThickLine, bend right=45] (14.center) to (17.center);
		\draw [style=ThickLine, bend right=45] (23.center) to (21.center);
		\draw [style=ThickLine, bend right=45] (21.center) to (22.center);
		\draw [style=ThickLine, bend right=45] (22.center) to (20.center);
		\draw [style=ThickLine, bend right=45] (20.center) to (23.center);
		\draw [style=ArrowLineRight] (24.center) to (25.center);
		\draw [style=ThickLine, bend right=45] (28.center) to (26.center);
		\draw [style=ThickLine, bend right=45] (26.center) to (29.center);
		\draw [style=ThickLine, bend right=45] (34.center) to (32.center);
		\draw [style=ThickLine, bend right=45] (32.center) to (33.center);
		\draw [style=RedLine] (35.center) to (36.center);
		\draw [style=ThickLine, in=-180, out=90, looseness=0.75] (35.center) to (33.center);
		\draw [style=ThickLine, in=90, out=0, looseness=0.75] (28.center) to (35.center);
		\draw [style=ThickLine, in=0, out=-90, looseness=0.75] (36.center) to (29.center);
		\draw [style=ThickLine, in=180, out=-90, looseness=0.75] (36.center) to (34.center);
	\end{pgfonlayer}
\end{tikzpicture}
}
\caption{Movie growing the shells $U^{(i)}_{r=r_*}$, until they touch and subsequently overlap in $V$. On the righthand side $V$ and $U_{r>r_*}$ only overlap in $\partial V$. }
\label{fig:Covering2}
\end{figure}
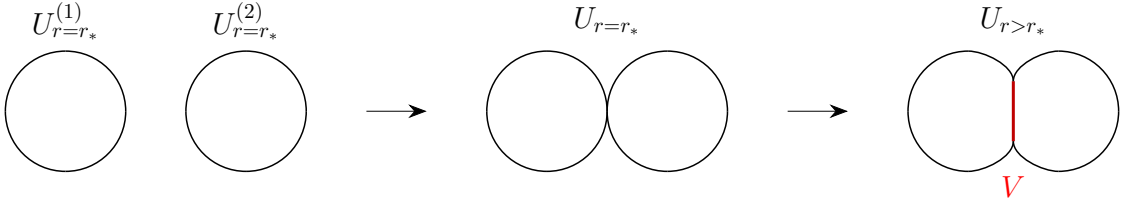


Let us next discuss the large radius sequence. Consider growing $U^{(i)}_{r<r_*}$ to the critical slice $U_{r=r_*}$ and then further, without changing the homotopy type, to what we will denote as $U^{(i)}_{r>r_*}$. The shells press up against each other and share
\be
V= U^{(1)}_{r>r_*}\cap U^{(2)}_{r>r_*}
\ee
which can be arranged such that the closures of $U^{(i)}_{r>r_*}\setminus V$ are not $U^{(i)}_{r>r_*}$ and $\partial V\neq \emptyset$ (see figure \ref{fig:Covering2}). Then the closure of $( U^{(1)}_{r>r_*}\cup U^{(2)}_{r>r_*})\setminus V$ is the shell $U_{r\;\!>\;\!r_*}$. Finally note that $U_{r\;\!>\;\!r_*}\cup V$ is deformation equivalent to $U_{r=r_*}$. The Mayer-Vietoris sequence of the union $U_{r\;\!>\;\!r_*}\cup V$ therefore is
\be\label{eq:seq2}
 \dots ~\rightarrow~ H_n(\partial V)   ~\xrightarrow[]{~\imath_n^{(r>r_*)}~} ~ H_n(U_{r\;\!>\;\!r_*})\oplus H_n(V)~\xrightarrow[]{~\jmath_n^{(r>r_*)} \oplus \,\jmath_n^{(V)}~} ~ H_n(U_{r=r_*}) ~\rightarrow~\dots
\ee
containing the map $\jmath_n^{(r>r_*)}$ which relates cycles at large radii to those of the critical shell.

Let us anticipate the geometric origin of an extension problem implicit in the above. First, note that $V$ can be viewed as a fattening of $U^{(12)}_{r=r_*}$. Conversely $V$ deformation retracts to $U^{(12)}_{r=r_*}$, and consequently we have the pair of embeddings
\be \label{eq:Ext1}
\partial V\hookrightarrow V\,, \qquad U^{(12)}_{r=r_*}\hookrightarrow V\,.
\ee
The cycles in $\partial V$ and $U^{(12)}_{r=r_*}$ can therefore be compared.
We can follow $\partial V$ through the deformation retraction $V\rightarrow U^{(12)}_{r=r_*}$
and therefore there exists a group homomorphism
\be \label{eq:Ext2}
\iota\,:\quad H_*(\partial V) \rightarrow H_*(U^{(12)}_{r=r_*})
\ee
which is neither injective nor surjective in general. Rather, generically, the image $\textnormal{Im}\,\iota$ is extended by elements of the homology group $H_*(U^{(12)}_{r=r_*})$ to its saturation $\overline{\textnormal{Im}\,\iota}$. The physical interpretation of this extension to the saturation is essentially the same as that given in reference \cite{Cvetic:2023pgm}; a $\U(1)$ symmetry of the ``bulk'' can, when pushed into one of the relative boundaries instead descend to a torsional symmetry generator in the boundary relative theory. This is in some sense just a consequence of having suitable objects which can partially screen the associated defects.

\paragraph{Y-shaped Junctions and Differential Cohomology}\mbox{}\medskip \\
Dimensional reduction of the $\mathsf{BLK}$ topological terms results in the SymTFT for a given branch of our tree. This reduction involves expanding the bulk fields in generators for differential cohomology classes for the internal geometry \cite{Apruzzi:2021nmk}. With this in mind, we now turn to an analysis of how the different differential cohomology groups fuse in the tree. Again, it suffices to consider the case of a Y-shaped junction.
More specifically, the symmetry TFT fields originate via expansions along generators of the differential cohomology groups $\breve{H}^*(U_{r})$.

The differential cohomology groups $\breve{H}^*(U_{r})$ sit in the short exact sequence\footnote{They also sit in $0 \rightarrow H^{p-1}(U_r,\mathbb{R}/\Z)\rightarrow\breve{H}^p(U_{r}) \rightarrow \Omega^p_\Z(U_r)\rightarrow0$. }
\be \ba \label{eq:DiffCohoRelations}
0~\rightarrow~\Omega^{p-1}(U_r)/\Omega^{p-1}_\Z(U_r)~\rightarrow~\breve{H}^p(U_{r})~\xrightarrow[]{~\pi~}~H^p(U_r,\Z)~\rightarrow~0\,,
\ea \ee
where $\Omega^p(U_r)$ (resp. $\Omega^p_\Z(U_r)$) denotes closed differential $p$-forms (resp. with integral periods) \cite{bar2014differential}. The groups $ H^p(U_r,\Z)$ are standard singular cohomology groups which we can relate via the universal coefficient theorem to the singular homology groups appearing in the Mayer-Vietoris sequences \eqref{eq:seq1} and \eqref{eq:seq2}.


The embeddings \eqref{eq:EmbeddingMaps} tracking the deformation of cycles at small and large radius through the critical slice corresponding to the junction, dualize to the restriction maps
\be\label{eq:EmbeddingMaps2}  \ba
\bar{\jmath}^{\,p}_{(r<r_*)}:& ~~  H^p(U_{r=r_*})    ~\rightarrow~  H^p(U_{R_1})\cong H^p(U_{r<r_*})\\[0.25em]
\bar{\jmath}^{\,p}_{(r>r_*)}:&~~   H^p(U_{r=r_*}) ~\rightarrow~  H^p(U_{R_2}) \cong H^p(U_{r>r_*})    \\
\ea\ee
on cocycles and subsequently lift to the restrictions $\breve{\jmath}^{\,p}_{(r<r_*)}, \breve{\jmath}^{\,p}_{(r>r_*)} $ on differential cohomology classes which we use momentarily to formulate the boundary conditions for bulk cocycles of the SymTree at the junction.


\paragraph{Flux Contributions}\mbox{}\medskip \\ So far, we have mainly concentrated on the background geometry $X$, but this can be threaded by fluxes in many cases. These fluxes emanate out from the localized singularities, e.g., in systems with a brane probe of geometry. Such branes source a supergravity flux $F(x)$, which is continuous away from the singularities of $X$.

Again, we first pick a favorable filtration $\mathcal{F}$ of the space $X$ which satisfies the properties listed out in subsection \ref{sec:Filtrations}. There then exist cycles $\Sigma_r$ within the shells $U_r$ such that
\be
n(r)\equiv \int_{\Sigma_r} F
\ee
is a piecewise constant function. Appropriately normalized $n(r)$ is a signed counting function, counting how many branes sourcing the flux $F$ are contained in the ball $B_r=\cup_{\:\!r\:\!'\leq\;\! r}U_{r'}$. Consequently $n(r)$ is constant on branches of the SymTree and jumps at the junctions.\footnote{This jumping phenomenon also signals the presence of additional light degrees of freedom, as dictated by anomaly inflow considerations. We expand on this point in specific examples.}

\paragraph{Junction Theory}\mbox{}\medskip \\
Finally, we come to the contributions localized at the junctions. As already mentioned, this analysis is somewhat more delicate since in the stringy construction, such ``jagged edges'' have already been smoothed out. In the case of brane probes of singularities, this is compounded by the fact that some modes (e.g., $\U(1)$'s) can end up being delocalized across the geometry, so projecting them onto the junction is a somewhat discontinuous process.

That being said, there is no obstacle in seeing how the different differential cohomology groups on branches fuse together at such a junction.
Indeed, the group $\breve{H}^p(U_{r=r_*})$ also specifies gluing conditions across the junction for dynamical fields of the symmetry TFTs attaching to the junction.  Whenever we have fields in the SymTFT $\mathcal{S}_{r<r_*}=\{ \mathcal{S}_1,\mathcal{S}_2\}$ and $\mathcal{S}_{r>r_*}=\mathcal{S}_{12}$, which result respectively via expansion along the classes
\be\label{eq:Image}
\breve{\jmath}_{\;\!r<r_*}(\breve{t}_{r=r_*})~~\textnormal{and}~~ \breve{\jmath}_{\;\!r>r_*}(\breve{t}_{r=r_*})
\ee
with common origin within the pair of pants / the critical slice (recall they are deformation equivalent), we have that their profiles necessarily glue along the junction. Equivalently, they restrict to the same value at the junction due to their common origin as modes along $\breve{t}_{r=r_*}$. This corresponds to imposing Dirichlet boundary conditions for two out of the three associated bulk fields at the junction. The overall glued profile is permitted to fluctuate, i.e., Neumann boundary conditions are imposed.


The fields not constrained by such a gluing condition are localized to the junction and characterize dynamical degrees of freedom
on the junction, such fields result via expansion along the classes
\be \label{eq:TheKernel}
\breve{\textnormal{K}}\textnormal{er}^{\;\!p}_{r=r_*}\equiv \textnormal{Ker}\,\breve{\jmath}_{\;\!r<r_*}^{\,p}\,\cap \textnormal{Ker}\,\breve{\jmath}_{\;\!r>r_*}^{\,p}\subset \breve{H}^p(U_{r=r_*})\,.
\ee
In many cases, especially when focusing on discrete symmetry structures, it will be sufficient to focus on the singular cohomology subgroup
\be \label{eq:TheKernel2}
{\textnormal{K}}\textnormal{er}^{\;\!p}_{r=r_*}\equiv \textnormal{Ker}\,\overline{\jmath}_{\;\!r<r_*}^{\,p}\,\cap \textnormal{Ker}\,\overline{\jmath}_{\;\!r>r_*}^{\,p}\subset {H}^p(U_{r=r_*})\,.
\ee
In order to completely determine the dynamics of the junction we however require information beyond topology. The modes characterized by \eqref{eq:TheKernel2} are dynamical only if this internal $p$-form profile is $L^2$-normalizable. When non-normalizable the corresponding field profiles are better viewed as Lagrange multipliers which enforce identifications between different SymTFT branches.

The embedding maps $\jmath^{(r>r_*)}_n,\jmath^{(r<r_*)}_n$ completely determine a Y-shaped SymTree up to this question of normalizability. Their domain and codomain determine the TFTs on the branches and the mappings themselves indicate, via dualization to cohomology and lifts to differential cohomology, how the bulk fields interact across the junction. Above we have interpreted the image \eqref{eq:Image} and kernel \eqref{eq:TheKernel} of the related maps in cohomology, it remains to interpret the cokernel.

Once the junction has been determined to support degrees of freedom, we may ask how to interpret the bulk SymTree fields as associated with backgrounds for these. We consider the cokernel of the mapping
\be
\breve{J}\equiv (\breve{\jmath}_{\;\!r<r_*}, \breve{\jmath}_{\;\!r>r_*})\,:\quad \breve{H}^*(U_{r=r_*})\rightarrow \breve{H}^*(U_{r<r_*}) \oplus \breve{H}^*(U_{r>r_*})
\ee
which correspond to bulk fields which are not fixed by gluing conditions.
In particular, denoting the composition of $\breve{J}$ with the projection
down to singular cohomology by $\overline{J}$, we get the projections:
\be \ba
\pi_{r<r_*}\,:\quad  \textnormal{coker}\,\bar{J}=\frac{  {H}^*(U_{r<r_*}) \oplus {H}^*(U_{r>r_*})}{ {H}^*(U_{r=r_*})} ~ \rightarrow~ \frac{ {H}^*(U_{r<r_*}) }{{H}^*(U_{r=r_*})} \\[0.5em]
 \pi_{r>r_*}\,:\quad   \textnormal{coker}\,\bar{J}=  \frac{  {H}^*(U_{r<r_*}) \oplus {H}^*(U_{r>r_*})}{ {H}^*(U_{r=r_*})} ~ \rightarrow~ \frac{ {H}^*(U_{r>r_*}) }{{H}^*(U_{r=r_*})} \\
\ea \ee
and consequently $ \textnormal{coker}\,\bar{J}$ can be viewed as an extension of the image of $\pi_{r<r_*}, \pi_{r>r_*}$ by its kernel. This is the relevant cohomological version of the extension described around line \eqref{eq:Ext1}. The quotients on the righthand side are associated with SymTree bulk fields which do not participate in gluing conditions and for which the junction imposes Neumann boundary conditions. The bulk fields descending from $ \textnormal{coker}\,\bar{J}$ relate to background fields for the junction degrees of freedom.

\section{Illustrative Example: Adjoint Higgsing of 7D SYM} \label{sec:7DSYM}

To illustrate the considerations spelled out in the previous sections, we now turn to some examples. Many of the key features are already present in the case of 7D Super Yang-Mills theory (7D SYM), and so we first treat this case in detail. An advantage of this case is that we have both an explicit stringy realization of this theory (and thus implicitly a UV completion) as well as a field theoretic characterization of the multi-sector system.

With this in mind, this Section is organized as follows. First, we review the case of a single 7D SYM theory, as well as the construction of heavy defects and topological symmetry operators. We also explain how adjoint Higgsing is captured by deformations of the associated M-theory background. With this in place, we next turn to adjoint Higgsing / geometric deformations which produces a multi-sector QFT at low energies.
We explicitly show how the SymTree theory arises in this context.

\subsection{Gauge Theory via Geometry}

Our starting point is 7D SYM, as realized by taking M-theory on the supersymmetric background:
\begin{equation}
\mathbb{R}^{6,1} \times X
\end{equation}
where $X = \mathbb{C}^2 / \Gamma$, and $\Gamma$ is a finite subgroup of $\SU(2)$ with group action dictated by the condition that we preserve 7D $\mathcal{N} = 1$ supersymmetry (16 real supercharges). There is an ADE classification of such finite subgroups, and these in turn specify the ADE type of the 7D SYM theory. As explained in \cite{Albertini:2020mdx} (see also \cite{DelZotto:2015isa, GarciaEtxebarria:2019caf, Morrison:2020ool}) the global form of the gauge group is fixed by a choice of boundary conditions on $\partial X = S^3 / \Gamma$. In gauge theoretic terms the center of the simply connected ADE Lie group is just the abelianization of $\Gamma$. This follows directly from the underlying geometry / gauge theory correspondence.

Indeed, in the relative QFT we can discuss the spectrum of Wilson lines (codimension 6) and 't Hooft defects (codimension 3) and a choice of global form fixes the spectrum in the absolute QFT. Wilson lines arise from M2-branes which wrap a torsional cycle in $H_1(S^3 / \Gamma)$ of the boundary lens space, and which sweep out the radial direction as well, i.e., they wrap $\mathrm{Cone}(\gamma)$ for $\gamma \in H_{1}(S^3 / \Gamma)$. Similarly, the 't Hooft defects arise from M5-branes which wrap $\mathrm{Cone}(\gamma)$ for $\gamma \in H_{1}(S^3 / \Gamma)$. As found in \cite{Apruzzi:2022rei, GarciaEtxebarria:2022vzq, Heckman:2022muc}, the associated topological symmetry operators which act on these defects arise from branes wrapped ``at infinity''. Indeed, the 1-form symmetry operator which acts on Wilson lines is generated by M5-branes wrapped on a boundary 1-cycle, and the 4-form symmetry operator which acts on 't Hooft defects is generated by M2-branes wrapped on a boundary 1-cycle. Specifying a consistent choice of boundary conditions then fixes an absolute theory.

The Symmetry TFT for this theory follows directly from the braiding relations for the various fields, and can be derived from dimensionally reducing the 11D kinetic term for the M-theory 3-form field.\footnote{For details on this see Appendix \ref{app:Kineticterms}.} Consider the special case where $\Gamma = \mathbb{Z}_N$.
Then, the 8D topological action for the SymTFT is:
\begin{equation}\label{eq:Sym7DTFT}
S_{\mathrm{8D}} = \frac{i}{2 \pi} N \int B_{2} \wedge d C_{5} + \dots\,,
\end{equation}
where the superscript indicates the form degree. Here $B_2$ and $C_5$ take values in $\U(1)$ and the overall coefficient of $N$ restrict their periods to take values in $\Z_N\subset\U(1)$. We have also dropped terms of the SymTFT associated with 2-form and $(-1)$-form symmetries.
These can also be extracted by dimensional reduction of the topological terms of 11D supergravity \cite{Apruzzi:2021nmk},
but for ease of exposition we focus on the 1-form and 4-form symmetries.
We specify physical boundary conditions on one end and topological boundary conditions on the other to fix the global form of the theory.

Similar considerations hold for more general choices of $\Gamma$; when the abelianization $\mathrm{Ab}(\Gamma)$ is a cyclic group (all cases other than $D_{2k}$) we simply take $N = \vert \mathrm{Ab}(\Gamma) \vert$, and one can likewise extract a similar expression when $\Gamma = D_{2k}$, where the abelianization is just $\mathbb{Z}_2 \times \mathbb{Z}_2$.

\subsection{Multi-Sector QFT via Adjoint Higgsing}

Starting from this theory, we can generate a multi-sector QFT via adjoint Higgsing. Geometrically, we start with a single singularity and then consider either a smoothing deformation or blowup of the singularity so that the resulting geometry has distinct singularities after the deformation. In field theory terms, we are switching on a background vacuum expectation value for some combination of the R-symmetry triplet of adjoint-valued scalars in the 7D $\mathcal{N} = 1$ vector multiplet. Of course, since all of these vacua are part of the same moduli space, there is a sense in which the original SymTFT still governs the structure of the spectrum of heavy defects and symmetry operators. On the other hand, there is clearly some approximate notion of the gauge theory corresponding to a single isolated singularity and its associated heavy defects and symmetry operators. Our aim will be to sharpen this correspondence.

The essential points are all captured by the case $\Gamma = \mathbb{Z}_N$ so in what follows we again focus on this special case. There are natural generalizations to the rest of the ADE series, albeit at the expense of a few more complications in writing out the explicit forms of blowups and smoothing deformations.

To begin, then, we recall that an $A_{N-1}$ singularity $X=\C^2/\Z_N$ can be presented as the singular hypersurface swept out by the locus:
\begin{equation}
x^2 + y^2 = z^N.
\end{equation}
Adjoint Higgsing amounts to a deformation or resolution of this singularity. Here, we focus on a complex deformation of the form:
\begin{equation}
x^2 + y^2 = \prod_{i=1}^{K} (z - t_i)^{N_i},
\end{equation}
where $N_1 + ... + N_K = N$, and $N_1t_1 + ... + N_Kt_K = 0$ (a tracelessness condition).
This corresponds to the breaking pattern:
\begin{equation}
\SU(N) \supset S(U(N_1) \times ... \times\U(N_K)),
\end{equation}
as triggered by a complex adjoint valued Higgs field of the form:\footnote{Two out of the three components of the $\SU(2)$ R-symmetry triplet are being switched on here. The third one in this choice of complex structure corresponds to a blowup mode.}
\begin{equation}
\langle \Phi \rangle = t_{1} \mathbf{1}_{N_1 \times N_1} \oplus ... \oplus t_{K} \mathbf{1}_{N_K \times N_K},
\end{equation}
where $\mathbf{1}_{M \times M}$ denotes the $M \times M$ identity.

After this Higgsing, we find multiple sectors at low energies, i.e., where we restrict all field ranges to be below the scales set by the $t_j$. Indeed, we have massive W-bosons as obtained from M2-branes which stretch between the separated singularities. This mass goes as:
\begin{equation}
M_{ij} \sim \vert t_i - t_j \vert.
\end{equation}
Further, in the vicinity of any individual singularity we have a geometry of the form $\mathbb{C}^2 / \mathbb{Z}_{N_j}$, and a corresponding 7D SYM theory with Lie algebra $\mathfrak{su}_{N_j}$. There are also $\mathfrak{u}(1)$ sectors which are delocalized / spread across the different singularities, and small fluctations about the values of the $t_j$ (as well as the accompanying resolution parameters) fill out R-symmetry triplets for the associated vector multiplets.

Focussing on just the non-abelian factors, we see a multi-sector QFT, but one in which there are still residual couplings to abelian sectors as well as additional TFT degrees of freedom. Our plan will be to extract the corresponding SymTree for this configuration.

\subsection{Extracting the SymTree}

By inspection, it is enough to focus on just the case where we Higgs the parent 7D SYM theory
to two non-abelian factors. Indeed, all other treelike structures can be obtained
by further Higgsing operations. In the case at hand, the Higgsed gauge group is:
\be \label{eq:quotient}
\SU(N) \supset \lb \SU(N_1)\times \SU(N_2)\times\U(1)\rb \Big /\,\mathbb{Z}_{L}\,,
\ee
with $N=N_1+N_2$ and where $L = \mathrm{lcm}(N_1,N_2)$ is the least common multiple of $N_1$ and $N_2$. We denote by $X^{\prime}$ the partial smoothing of $X$. The space $X'$ contains an $A_{N_1-1}$ and $A_{N_2-1}$ singularity at $z = t_1$ and $z = t_2$ respectively and a compact 2-cycle stretching between these. The adjoint fields of the SYM theory reorganize following the decomposition
\be\label{eq:adjointhiggs}\begin{aligned}
\mathfrak{su}_N~\rightarrow~ &\mathfrak{su}_{N_1}\oplus \mathfrak{su}_{N_2}\oplus \mathfrak{u}(1)\\[0.5em]
\textnormal{Ad}\big[\mathfrak{su}_N \big]~\rightarrow~ & \textnormal{Ad}\big[\mathfrak{su}_{N_1}\oplus \mathfrak{su}_{N_2}\oplus \mathfrak{u}(1) \big]  \oplus \big({\mathbf N}_1,\overline {\mathbf N}_2\big)_{N/g}\oplus \big(\:\!\overline {\mathbf N}_1, {\mathbf N}_2\big)_{-N/g} \\[0.5em]
\mathbf{N}~\rightarrow~&(\mathbf{N}_1,\mathbf{1})_{N_2/g}\oplus  (\mathbf{1},\mathbf{N}_2)_{-N_1/g}
\end{aligned}\ee
where $g=\textnormal{gcd}(N_1,N_2)$. Here the bifundamental fields are the massive W-bosons (in the off-diagonal blocks) and arise from M2-branes wrapped on the compact 2-cycle.

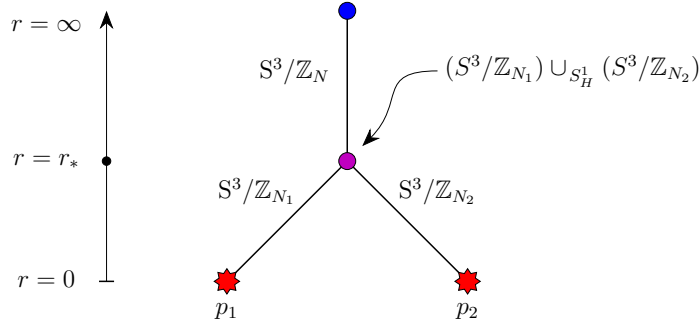
\begin{figure}
    \centering
    \scalebox{0.8}{\begin{tikzpicture}
	\begin{pgfonlayer}{nodelayer}
		\node [style=none] (0) at (0, 0) {};
		\node [style=none] (1) at (-2, -2) {};
		\node [style=none] (2) at (2, -2) {};
		\node [style=CircleBlue] (3) at (0, 2.5) {};
		\node [style=Star] (4) at (-2, -2) {};
		\node [style=Star] (5) at (2, -2) {};
		\node [style=none] (6) at (-2, -2.5) {$p_1$};
		\node [style=none] (7) at (2, -2.5) {$p_2$};
		\node [style=none] (8) at (-1.5, -0.5) {${\mathrm S}^3/\mathbb{Z}_{N_1}$};
		\node [style=none] (9) at (1.5, -0.5) {${\mathrm S}^3/\mathbb{Z}_{N_2}$};
		\node [style=none] (10) at (-0.875, 1.5) {${\mathrm S}^3/\mathbb{Z}_{N}$};
		\node [style=none] (11) at (0.25, 0.25) {};
		\node [style=none] (12) at (1.5, 1.5) {};
		\node [style=none] (13) at (3.75, 1.5) {$\lb S^3/\mathbb{Z}_{N_1}\rb\cup_{\;\!S^1_H}\lb S^3/\mathbb{Z}_{N_2}\rb$};
		\node [style=none] (14) at (-4, -2) {};
		\node [style=none] (15) at (-4, 2.5) {};
		\node [style=none] (16) at (-5, 0) {$r=r_*$};
		\node [style=none] (17) at (-5, -1.95) {$r=0$};
		\node [style=none] (18) at (-5, 2.25) {$r=\infty$};
        \node [style=none] (19) at (0, -3) {};
        \node [style=CirclePurple] (20) at (0, 0) {};
        \node [style=none] (21) at (-3.875, -2) {};
         \node [style=none] (22) at (-4.125, -2) {};
         \node [style=Circle] (23) at (-4, 0) {};
	\end{pgfonlayer}
	\begin{pgfonlayer}{edgelayer}
		\draw [style=ThickLine] (1.center) to (0.center);
		\draw [style=ThickLine] (0.center) to (2.center);
		\draw [style=ThickLine] (3.center) to (0.center);
		\draw [style=ArrowLineRight, in=45, out=180, looseness=1.50] (12.center) to (11.center);
		\draw [style=ArrowLineRight] (14.center) to (15.center);
		\draw [style=ThickLine] (21.center) to (22.center);
	\end{pgfonlayer}
\end{tikzpicture}}
    \caption{We sketch the partially smoothed geometry $X'$ as a fibration over a Y-shaped base. This picture presents a horizontal slice of figures similar to figure \ref{fig:SymTree}. }
    \label{fig:FibrationWithDisconnectedSlices}
\end{figure}

\paragraph{Filtration and Critical Slice }\mbox{}\medskip \\ We obtain the SymTree by first describing a convenient choice of filtration $\mathcal{F}_{X'}$ sweeping out the partial smoothing $X'$. The filtration has radial shells
\be \label{eq:HomoLens} \ba
 U_{r\;\!>\;\! r_*}&=S^3/\mathbb{Z}_{N_1+N_2}\,,\\[0.3em]
  U_{r= r_*}&=\lb S^3/\mathbb{Z}_{N_1}\rb\cup_{\;\!S^1_H}\lb S^3/\mathbb{Z}_{N_2}\rb\,, \\[0.3em]
  U_{r<r_*}&=(S^3/\mathbb{Z}_{N_1})\sqcup(S^3/\mathbb{Z}_{N_2})\,,
\ea \ee
with a single critical slice at $r=r_*$ (see figure \ref{fig:FibrationWithDisconnectedSlices}). Here $\cup_{\;\!S^1_H}$ denotes the gluing of the two lens spaces along one of their Hopf circles. Running the Mayer-Vietoris sequence we find the critical slice to be characterized by the homology groups
\be  \label{eq:criticalslicehomo}
H_n(\lb S^3/\mathbb{Z}_{N_1}\rb\cup_{\;\!S^1_H}\lb S^3/\mathbb{Z}_{N_2})\rb\cong  \begin{cases} \Z  & k=0 \\ \Z_g & k=1 \\  \Z \qquad\quad & k=2 \\ \Z^2   &k=3 \end{cases}
\ee
where $g=\textnormal{gcd}(N_1,N_2)$.
In Appendix \ref{app:FiltrationsALE} we identify the generators of \eqref{eq:criticalslicehomo}.

The filtration $\mathcal{F}_{X'}$ is motivated by the IIA dual setup consisting of two D6-brane stacks with common transverse space $\R^3$. We denote the location of the two stacks by $p_1,p_2\in \R^3$ and by $S^2_{1,r},S^2_{2,r}$ spheres of radius $r$ centered on these. A filtration $\mathcal{F}_{\:\!{\rm IIA}}$ of $\R^3$ is constructed by first growing $r$. In the process two 3-balls are swept out, they grow until they meet in a point. This results in a critical slice at radius $r=r_*$ which is the wedge sum
\be\label{eq:Wedgesum}
S^2_{12,r_*}= S^2_{1,r_*}\vee_p\, S^2_{2,r_*}\,.
\ee
The rest of $\mathcal{F}_{\:\!{\rm IIA}}$ follows by continuing to grow the spheres, taking the radial shells to be their `peanut' shaped exterior. The filtration $\mathcal{F}_{X'}$ is then the M-theory lift of $\mathcal{F}_{\:\!{\rm IIA}}$ via the Gibbons-Hawking ansatz, and the shells of $\mathcal{F}_{\:\!{\rm IIA}}$ are extended to $X'$ by including all M-theory circles projecting to these. In particular the point of kissing\footnote{Namely, an osculation.} $p$ in line \eqref{eq:Wedgesum} lifts to the circle $S^1_H$, and line \eqref{eq:Wedgesum} lifts to $U_{r=r_*}$.

\paragraph{Projection to SymTree}\mbox{}\medskip \\ We now reduce 11D supergravity on the radial shells $U_r$. We retain only topological data following the approach in \cite{Apruzzi:2021nmk}. This results in an 8D TFT action for each branch of the Y-shaped graph of figure \ref{fig:FibrationWithDisconnectedSlices} and a non-topological 7D action describing the junction degrees of freedom. These 7D and 8D modes derive from the 11D field strength $\breve{G}_4$ via KK reduction.

We discuss this reduction in detail in Appendices \ref{app:Kineticterms} and \ref{app:FiltrationsALE}. There we show that we can discuss discrete generalized symmetries in isolation of other structures\footnote{Of course the 8D TFTs on the branches of the SymTree are simply the SymTFTs of the $A_{N-1},A_{N_1-1},A_{N_2-1}$ 7D SYM theories. The string theory analysis for these also produces $(-1)$-form and 2-form symmetries. We defer a discussion of interaction terms near the junction to Appendix \ref{app:FiltrationsALE}.} occurring. The relevant discrete symmetries are the 1-form symmetries of the various SYM sectors, and their dual 4-form symmetries. The background fields for these symmetries are dynamical in the 8D TFTs and interact at the junction.

In terms of differential cohomology classes we are restricting our attention to the coefficients in the expansions
\be \ba\label{eq:DiffCoho}
(r>r_*)\,:\quad \breve{G}_4&=\breve{B}_{2}^{(r>r_*)}\star \breve{t}^{\;\!(r>r_*)}_2+\dots \\[0.3em]
(r=r_*)\,:\quad \breve{G}_4&=\breve{B}_{2}^{(r=r_*)}\star \breve{t}^{\;\!(r=r_*)}_2+\breve{F}_{2}\star \breve{u}_2+\dots \\[0.3em]
(r<r_*)\,:\quad \breve{G}_4 &= \breve{B}_{2}^{(r<r_*,1)}\star \breve{t}^{\;\!(r<r_*,1)}_{2}+ \breve{B}_{2}^{(r>r_*,2)}\star \breve{t}^{\;\!(r<r_*,2)}_{2}+\dots
\ea \ee
and omitting fields and interactions resulting from the ``...'' terms. Here, the generators $\breve{t}_2$ are torsional classes and are in correspondence with the torsional 1-cycles of the radial shells $U_r$. Their exponent indicates the SymTree branch they live on. The respective coefficient fields valued in $\U(1)$ with periods taking values in finite subgroups of $\U(1)$. We make this explicit by changing the normalization such that the fields now take values in $\Z_K$ where $K=N,N_1,N_2,g$. This rescaling is reflected in notation as
\be\label{eq:Renormalization}
\breve{B}_{2}^{(r>r_*)} =\frac{2\pi i}{N}\breve{B}_{2}^{(N)}\,, \quad \breve{B}_{2}^{(r=r_*)}  =\frac{2\pi i}{g}\breve{B}_{2}^{(g)}\,,\quad \breve{B}_{2}^{(r<r_*,i)}  =\frac{2\pi i}{N_i}\breve{B}_{2}^{(N_i)}\,,
\ee
where the exponent now keeps track of the order of the form with the index continuing to record its degree. The free generator $\breve{u}_2$ is in correspondence with the free 2-cycle in \eqref{eq:criticalslicehomo} which only exists at the critical radius $r=r_*$ and results in an abelian gauge field localized to the junction. The fields of the SymTree under consideration are thus
\be\label{eq:backgroundleg} \ba
(r> r_*) \,: &\quad B^{(N)}_{2} \\[0.1em]
(r=r_*) \,: &\quad B^{(g)}_{2}, A_{1}\\
(r<r_*)\,: &\quad B^{(N_1)}_{2}\,, B^{(N_2)}_{2}
\ea \ee
together with their magnetic duals. The discrete fields $B_{2}$ are background fields for the 1-form center symmetry.

We now determine the theory localized at the junction and the gluing conditions across the junction. Evaluating the Mayer-Vietoris sequences of lines \eqref{eq:seq1} and \eqref{eq:seq2} we find the intersection of kernels in line \eqref{eq:TheKernel2} yields:
\be \label{eq:guyatjunction}
{\rm Ker}^{(2)}_{r=r_*}=\Z\,.
\ee
This kernel characterizes the fields at the junction not arising as restriction of bulk fields. Since
the field content is supersymmetric, we conclude that the dynamical junction degrees of freedom organize into the following theory\footnote{This mode is understood to arise via Higgsing from which we infer that the $2$-form \eqref{eq:guyatjunction} is $L^2$-normalizable, i.e., the $\mathfrak{u}(1)$ theory is dynamical.}:
\be
\textnormal{7D $\mathcal{N}=1$ $\mathfrak{u}(1)$ vector multiplet}.
\ee
Further, line \eqref{eq:guyatjunction} implies that the junction field $B^{(g)}_{2}$ is eaten up by identifications. Indeed computing the images of the maps $\overline{\jmath}_{\;\!r<r_*}\,\overline{\jmath}_{\;\!r>r_*}$, defined in line \eqref{eq:EmbeddingMaps2}, we find
\be
\textnormal{Im}\,\overline{\jmath}_{\;\!r<r_*}^{(2)}\cong \Z_g\oplus \Z_g\,, \qquad \textnormal{Im}\,\overline{\jmath}_{\;\!r>r_*}^{(2)}\cong \Z_g\,,
\ee
implying that the low radius pair $B_{2}^{(N_1)}$ and $B_{2}^{(N_2)}$ glues via $B_{2}^{(g)}$ to the large radius field $B^{(N)}_{2}$. More precisely, at the critical slice we have the restrictions
\be
\frac{N_1}{g} B^{(N_1)}_{2}\Big|_{  {r= r_*}}=B^{(g)}_{2}\,, \qquad  \frac{N_2}{g} B^{(N_2)}_{2}\Big|_{  {r= r_*}}=B^{(g)}_{2}\,,  \qquad  \frac{N}{g}B^{(N)}_{2}\Big|_{ {r= r_*}}=B^{(g)}_{2}\,.
\ee
The background field $B^{(g)}_{2}$ is thus auxiliary and can be eliminated, straightforwardly implying the junction gluing condition
\be \label{eq:glue}
\frac{N_1}{g} B^{(N_1)}_{2}\Big|_{  {r= r_*}}= \frac{N_2}{g} B^{(N_2)}_{2}\Big|_{  {r= r_*}}= \frac{N}{g}B^{(N)}_{2}\Big|_{  {r= r_*}}
\ee
which is an equation with coefficients in $\Z_g$.

We turn to discuss how the abelian junction theory interacts with the three 8D TFTs. The point of view taken here is that the junction theory itself is relative and that the fields $B^{(N_1)}_{2},B^{(N_2)}_{2},B^{(N)}_{2}$ admit an interpretation as background fields for (a subgroup of) its 1-form center symmetry. Solving the extension problem in geometry we determine the relevant subgroup to be $\Z_{LN/g}\subset\U(1)$ which naturally arises in the extension
\be \label{eq:ext}
0~\rightarrow~\Z_{L}~\rightarrow~\Z_{LN/g}~\rightarrow~\Z_{N/g}~\rightarrow~0
\ee
where $L=\textnormal{lcm}(N_1,N_2)$. Let us denote 1-form symmetry backgrounds of the junction theory contained in this subgroup as $B^{({\ell},\U(1))}_{2}$ where we introduce $\ell=LN/g$ for convenience, similarly we introduce a background for the subgroup $\Z_L\subset\U(1)$. Studying the small radius Mayer-Vietoris sequence we find the identifications
\be \label{eq:Id1}
B^{(N_1)}_{2}=\frac{N_2}{g}B_{2}^{(L,\U(1))}\,, \qquad B^{(N_2)}_{2}=\frac{N_1}{g}B_{2}^{(L,\U(1))}
\ee
which fundamentally are identifications between center subgroups of $\U(1)$ and $\SU(N_i)$. Next, studying the large radius Mayer-Vietoris sequence we find the identifications
\be  \label{eq:Id2}
g B^{(N)}_{2}=L B_{2}^{(\ell,\U(1))}\,.
\ee
The righthand sides are related simply as $(N/g) B_{2}^{(\ell,\U(1))}=B_{2}^{(L,\U(1))}$. Note, that these are identifications and not gluing conditions, the restriction to the critical slice $|_{  {r= r_*}}$ is missing compared to \eqref{eq:glue}. One can check that the interaction of the 8D fields with and across the junction is now fully determined. They either glue across the junctions or enter the $\mathfrak{u}(1)$ theory. For discussion and details on geometrization of the above see Appendix \ref{app:FiltrationsALE}.

Overall the action of the SymTree now takes the form
\be \ba
S&=\sum_{\textnormal{branches}\,b} S_{\textnormal{8D}}^{(b)}+\sum_{\textnormal{internal\,nodes}\,n} S_{\:\!  \textnormal{7D}}^{(n)} \\[0.25em]
&=  S_{\textnormal{8D}}^{(1)}+S_{\textnormal{8D}}^{(2)}+S_{\textnormal{8D}}^{(12)}
+ S_{\textnormal{7D}}^{(\mathcal{J})},\label{eq:8daction}\ea \ee
where the final term is the junction theory, which includes a 7D $\mathcal{N} = 1$ vector multiplet $\mathfrak{u}(1)$, and also enforces the identifications for the 8D bulk modes. The first two 8D terms are supported at $r\in (0,r_*)$, the third term is supported along $r\in(r_*,\infty)$ and the fourth terms is in 7D located at $r=r_*$. The first three terms each correspond to a leg of the Y-shaped SymTree and are topological, explicitly
\be \ba
S^{(1)}_{\rm 8D}&=\frac{2\pi i}{N_1}  \int_{\;\!\R^{6,1}\times (0,r_*)} B^{(N_1)}_{2} \cup \delta C^{(N_1)}_{5}+\dots \\[0.25em]
S^{(2)}_{\rm 8D}&=\frac{2\pi i}{N_2}  \int_{\;\!\R^{6,1}\times (0,r_*)}  B^{(N_2)}_{2} \cup \delta C^{(N_2)}_{5}+\dots \\[0.25em]
S^{(12)}_{\rm 8D}&=\frac{2 \pi i}{N}  \int_{\;\!\R^{6,1}\times (r_*,\infty) } B^{(N)}_{2} \cup \delta C^{(N)}_{5} +\dots
\ea \ee
where the omitted terms include fields for the 2-form and $(-1)$-form symmetry derived from the 11D Chern-Simons term $\breve{G}_4^3$, see Appendix \ref{app:FiltrationsALE}. We have switched from $d$
to $\delta$ for our differentials from wedge product to cup products compared to \eqref{eq:Sym7DTFT} to emphasize the change from differential forms valued in $\U(1)$ to discrete cocycles.

We now comment on the difference between $\mathcal{B}_{\textnormal{phys}}^{(\textnormal{retract})}$, the boundary condition obtained from retracting the $\mathfrak{su}_{N_i}$ SymTree branches into the $\mathfrak{u}(1)$ junction, as in figure \ref{fig:Tuck}, and $\mathcal{B}_{\textnormal{phys}}^{(\textnormal{full})}$, the boundary condition associated the parent $\mathfrak{su}_{N_1+N_2}$ theory.  Recall from Section \ref{sec:SYMTREE}, that retraction consists of simply dimensionally reducing the two SymTFTs of the branches along the interval.
This means that the retracted theory includes the 7D gauge theories from each interval end which combine, with the junction degrees of freedom, to form a $\mathfrak{su}_{N_1}\oplus\mathfrak{su}_{N_2}\oplus\mathfrak{u}(1) $ gauge theory 
while $S^{(1)}_{\rm 8D}$ and $S^{(2)}_{\rm 8D}$ vanish. Interestingly, this boundary condition has defects which are localized within it. Upon contraction defects running between the $\mathfrak{su}_{N_i}$ boundaries (as in configuration (iv) of figure \ref{fig:Defects}) descend to defects in the engineered theory. The manner in which bulk fields reduce to background fields of the 7D theory is precisely given by lines \eqref{eq:glue} and \eqref{eq:Id1}. In comparison, $\mathcal{B}_{\textnormal{phys}}^{(\textnormal{full})}$ in this example is realized by physically moving the $A_{N_1-1}$ and $A_{N_2-1}$ singularities together to form a $A_{N_1+N_2-1}$ singularity. This fuses the boundary conditions of the branches of the SymTree to a 7D $\mathfrak{su}_{N_1+N_2}$ gauge theory, setting the standard physical boundary conditions for $\mathcal{S}_{\textnormal{full}}$.




\paragraph{Field Theory Interpretation}\mbox{}\medskip \\ We now discuss the above from a field theoretic perspective. This perspective relies heavily on understanding the origin of the junction via the Higgsing specified by line \eqref{eq:quotient}. More invariantly, we are relying on an understanding of how the physical edge modes fuse with the junction theory. In contrast the geometric analysis yields identical results without this additional input.

Our starting point is the Higgsing, which we repeat for convenience here:
\be \label{eq:higgs2}
\SU(N)~\rightarrow~\lb \SU(N_1)\times \SU(N_2)\times\U(1)\rb \Big /\,\mathbb{Z}_{L}\,.
\ee
Note that this deformation does not change the 1-form center symmetry of the theory. Indeed, screening arguments are insensitive to such deformations as is clear from the geometries $X,X'$ exhibiting the same boundaries and the partial smoothing $X'\rightarrow X$ introducing no 1-cycles.

We then may ask how to represent a 2-form background field configuration of the lefthand side via data of the righthand side. This is equivalent to asking how to represent a center element of the lefthand side as a combination of center elements on the righthand side modulo identification imposed by $\mathbb{Z}_{L}$.

Observe that the $\Z_g\subset \Z_N$ subgroup embeds into $\Z_{N_1}\times \Z_{N_2}$ without involving the $\U(1)$ which is the content of the gluing condition \eqref{eq:glue}. Therefore it remains to track the $\Z_{N/g}$ from left to right. This now necessarily involves the $\U(1)$ and turns on line \eqref{eq:Id2}. However no element of $\U(1)$ except the identity is a central subgroup of $\SU(N)$, we therefore need to compensate the $\U(1)$ profile by turning on a subgroup of $\Z_{N_1}\times \Z_{N_2}$ which is not central in $\SU(N)$. The subgroup $\Z_g\subset \Z_{N_1}\times \Z_{N_2}$ is central in $\SU(N)$, therefore we are turning on a subgroup of $(\Z_{N_1}\times \Z_{N_2})/\Z_g$, exactly as in line \eqref{eq:Id1}.

\subsection{Multi-Sector Defects and Symmetry Operators}

Let us next turn to the defects and symmetry operators of the multi-sector QFT. We have already reviewed how these arise in the UV parent theory, and in an electric polarization where we have the $\mathbb{Z}_N$ center symmetry, the absolute theory admits the following objects:
\be \ba\label{eq:SymDef}
\textnormal{Defect Operators:}\quad &\textnormal{M2-brane wrapped on Cone}(\gamma) \\
\textnormal{Symmetry Operators:}\quad  &\textnormal{M5-brane wrapped on }\gamma
\ea \ee
where $\gamma \in H_1(S^3/\Z_N)$ and the cone over $\gamma$ stretches to the singularity.

Consider next the SymTree generated by our smoothing deformation / adjoint Higgsing where we are left with $\mathbb{C}^2 / \mathbb{Z}_{N_{1}}$ and $\mathbb{C}^2 / \mathbb{Z}_{N_2}$ singularities. In addition to the localized non-abelian gauge groups, we also have a delocalized $\mathfrak{u}(1)$ sector (i.e., an $\mathcal{N} = 1$ vector multiplet) which we interpret as part of the junction of the SymTree.

Let us now turn to possible heavy defects and symmetry operators of the parent UV theory and how they are interpreted in the multi-sector QFT.
To do this, it is enough to study the fate of the various boundary cycles $\gamma \in H_{1}(S^3 / \mathbb{Z}_N)$ and what happens to them as we push them down to the junction and to the individual singularities. The Gibbons-Hawking ansatz tells us that the generator of $H_{1}(S^3 / \mathbb{Z}_N)$ pushes downwards into the local model of $\mathbb{C}^2/\Z_{N_i}$ without obstruction. However, the equivalence relations imposed on the free group generated by this chain is sensitive to the radial shell it is being considered in, giving different homology groups in degree 1. Similarly generators of $H_1(S^3/\Z_{N_i})$ can be deformed between the two singularities without obstruction. From here 2-cycles are constructed, which are simply traced out by these 1-cycles following these deformations.

Defects follow by wrapping M2-/M5-branes on the constructed 2-cycles, and returning to figure \ref{fig:Defects}, we see that fundamentally we are constructing defects of type (iii) and  (iv). Any defect of type (v) is a composite of these. Defects of type (ii) do not exist.

Defects of type (iii) and (iv) constructed in this way correspond to the representations in \eqref{eq:adjointhiggs} carrying $\U(1)$ charge. Dressing amounts to tensoring the non-abelian representations by the abelian ones, dressing the former by the latter. The $\U(1)$ charge is computed via an intersection. In Appendix \ref{app:FiltrationsALE} we show that the abelian generator is geometrized as (see the discussion leading to \eqref{eq:gluing} for notation)
\be
\Sigma\equiv \lb N_2\Sigma_1/g\rb\cup_{\;\!LS^1_H}  \lb -N_1\Sigma_2/g\rb .
\ee
In an electric frame, Wilson lines in the fundamental representation ${\bf N}_i$ are constructed by M2-branes wrapping the 1-cycle homology generator at ``true infinity'' $S^3 / \mathbb{Z}_N$ fibered to the $\C^2/\Z_{N_i}$ singularity. In the critical slice we find a $\U(1)$ line with charge $q_i=\gamma_i\cdot \Sigma$ where $\gamma_i$ is a representative for the 1-cycle from the $i$-th local model. We have $\gamma_i\cdot \Sigma_j=\delta_{ij}$ and therefore
\be
q_1=\frac{N_2}{g}\,, \qquad q_2=-\frac{N_1}{g}\,,
\ee
correctly reproducing \eqref{eq:adjointhiggs}. Taking orientations into account, the charge of the bifundamental is of course $q_1-q_2$. We have the dressing ${\bf N}_i\otimes {\bf 1}_{q_i}\equiv ({\bf N}_i)_{q_i}$ and $({\bf N}_1,{\bf N}_2) \otimes {\bf 1}_{q_1-q_2}\equiv({\bf N}_1,{\bf N}_2)_{q_1-q_2} $ which are the true defects of the system.

There also exist defects which need not be dressed. By the above analysis we found a subgroup $\Z_g\subset \Z_N$ to glue across the junction, to $\Z_g\subset \Z_{N_1}\times \Z_{N_2}$ and not involve the $\U(1)$. At the level of defects this describes a configuration of type (v). We begin with $N/g$ copies of the generator of $H_1(S^3/\Z_N)$, fiber these radially inwards, and after the critical slice we fiber $N_i/g$ of these to $\C^2/\Z_{N_i}$. The dressing line has charge $(N_1/g)q_1+(N_2/g)q_2=0$ and is trivial.

Similar considerations hold for the topological symmetry operators, with M5-branes now wrapped on the torsional cycles of $H_{1}(S^3 / \mathbb{Z}_N)$. We discussed this in general terms in figure \ref{fig:SymOps}. For M5-branes wrapped on $\gamma \in \mathbb{Z}_g \subset \mathbb{Z}_N$, the symmetry operator can be pushed down into the pair of throats according to $\Z_g\subset \Z_{N_1}\times \Z_{N_2}$ where they act on the corresponding lines. This corresponds to subfigure (iv) of figure \ref{fig:SymOps}. When $\gamma \notin \mathbb{Z}_g \subset \mathbb{Z}_N$, we have a further dressing by operators of the junction theory, as captured by the $\mathbb{Z}_L$ charge. This corresponds to subfigure (ii) of figure \ref{fig:SymOps}.

\section{More Multi-Sector QFTs via Moduli Space Flows} \label{sec:FLOW}

In the previous Section we focused on the special case of 7D SYM as realized by M-theory on an ADE singularity. Under adjoint Higgsing, i.e., a smoothing and / or blowup of the geometry, we arrived at a multi-sector QFT. Similar considerations apply to other QFTs which admit such geometric deformations. In this Section we present a few such examples which have either a similar geometric or field theoretic realization.

As a first class of examples, we consider the case of multi-sector 6D SCFTs realized via tensor branch deformations. Geometrically, these are quite similar to the case of 7D SYM, but with a different physical interpretation of the SymTree and associated defects and symmetry operators.
We illustrate how this works for brane probes of smooth and singular geometries. A pleasant feature of some brane probe theories is that, in a suitable large $N$ limit, they result in multi-throat AdS configurations.
This in turn provides us with a holographic description of the SymTree.

\subsection{6D SCFTs and their Compactifications}

Let us now turn to the SymTree for 6D SCFTs as realized in F-theory backgrounds \cite{Heckman:2013pva, DelZotto:2014hpa, Heckman:2015bfa} (see e.g., \cite{Heckman:2018jxk, Argyres:2022mnu} for reviews). In all these cases, the base of the F-theory model is of the form $\mathbb{C}^2 / \Gamma$ for $\Gamma$ a finite subgroup of $\U(2)$. A suitable elliptic fibration results in a non-compact elliptically fibered Calabi-Yau threefold which preserves (at least) $\mathcal{N} = (1,0)$ supersymmetry. In this case, we have stringlike surface operators defects from D3-branes wrapping cones of boundary one-cycles $\mathrm{Cone}(\gamma)$ with $\gamma \in H_{1}(S^3 / \Gamma) \simeq \mathrm{Ab}(\Gamma)$. Indeed, the 2-form symmetry for the relative theory is specified by the abelianization of $\Gamma$ (see reference \cite{DelZotto:2015isa}). There can in principle also be 0-form and 1-form symmetries, but these are model dependent so we defer an analysis of this structure to future work.\footnote{For some discussion of this, see e.g., references \cite{Apruzzi:2021mlh, Cvetic:2022imb, Hubner:2022kxr, Heckman:2022suy, Baume:2022cot, Lawrie:2023tdz}.} For ease of exposition, we also assume that $\Gamma$ is of generalized $A$-type, namely that it is always just a cyclic group $\mathbb{Z}_N$ with some group action induced from the action of $\U(2)$ on $\mathbb{C}^2$. All other choices were classified in \cite{DelZotto:2015isa} and result in quite similar conclusions.

Focusing, then, on just the 2-form symmetry, the SymTFT for this theory follows from dimensional reduction of the topological action associated with the chiral 4-form of type IIB string theory:\footnote{A proper derivation of the 7D SymTFT is a bit subtle because we are dealing with reduction of a chiral 4-form. Following the treatment in \cite{Heckman:2017uxe} as well as \cite{Belov:2006jd, Belov:2006xj, Apruzzi:2021nmk,  Apruzzi:2022dlm, Lawrie:2023tdz}, one starts from an 11D spacetime and a Chern-Simons-like action equipped with a Wu structure:
\begin{equation}
S_{\mathrm{11D}} = \frac{i}{4 \pi} \int C_5 \wedge dC_5.
\end{equation}
Then, treating the 10D spacetime as a boundary, we impose the condition $C_5 = \ast_{\mathrm{10D}} C_5$ as $C_5$ is an 11D extension of the self-dual 5-form RR flux in IIB. Following a similar analysis to that presented in Appendix \ref{app:Kineticterms}, we can then consider the reduction of the associated ``boundary kinetic term'' on the linking $S^3 / \mathbb{Z}_N$ to arrive at the 7D TFT action. For related discussions see e.g., references \cite{Heckman:2017uxe, Apruzzi:2022dlm}.}
\begin{equation}
S_{\mathrm{7D} } = \frac{i}{4 \pi} N \int C_3 \wedge d C_3 + ...,
\end{equation}
where the ``...'' refers to additional topological terms which captured contributions from possible 0-form and 1-form symmetries.

Now starting from this 6D SCFT, we can consider a tensor branch flow which results in the system breaking up into a multi-sector QFT. Geometrically this corresponds to blowing up some collection of the previously collapsed curves. Doing so, we can get local singularities of the form $\mathbb{C}^2 / \mathbb{Z}_{N_1}$ and $\mathbb{C}^2 / \mathbb{Z}_{N_2}$. It is now clear that the geometric structure of a SymTree found in the case of 7D SYM simply carries over since we again have boundary lens spaces which fuse together.\footnote{One might ask whether the difference between a finite subgroup of $\SU(2)$ versus $\U(2)$ plays a role here. At the level of topological structures, it does not appear to make much of a difference, although it can affect link-pairings between cycles.} Carrying on in this way, we can simply take our previous analysis of the SymTree action and make some small adjustments to reach the answer for our 7D theory.

Overall the action of the SymTree for a trivalent junction now takes the form
\be
\ba
S&=\sum_{\textnormal{branches}\,b} S_{\textnormal{7D}}^{(b)}+\sum_{\textnormal{internal\,nodes}\,n} S_{\:\!  \textnormal{6D}}^{(n)} \\[0.25em]
&=  S_{\textnormal{7D}}^{(1)}+S_{\textnormal{7D}}^{(2)}+S_{\textnormal{7D}}^{(12)}+S_{\textnormal{6D}}^{(\mathcal{J})}
\ea
\ee
where the final term is the junction theory, which includes a 6D $\mathcal{N} = (1,0)$ tensor multiplet, and also enforces the identifications for the 7D bulk modes. The first three terms each correspond to a leg of the Y-shaped SymTree:
\be \ba
S^{(1)}_{\rm 7D}&=\frac{\pi i}{ N_1} \int_{\;\!\R^{5,1}\times (0,r_*)} C_3^{(N_1)} \cup \delta C_3^{(N_1)}+\dots \\[0.25em]
S^{(2)}_{\rm 7D}&=\frac{\pi i}{N_2}  \int_{\;\!\R^{5,1}\times (0,r_*)}  C_3^{(N_2)} \cup \delta C_3^{(N_2)}+\dots \\[0.25em]
S^{(12)}_{\rm 7D}&=\frac{\pi i }{N}  \int_{\;\!\R^{5,1}\times (r_*,\infty) } C_3^{(N)} \cup \delta C_3^{(N)} +\dots.
\ea \ee
So, up to a few small rearrangements in the physical interpretation of various higher-form potentials, we see that we again reach precisely the same SymTree structure considered previously. Here we have again rescaled fields similar to \eqref{eq:Renormalization} as indicated by their raised index.

Similar considerations hold for compactifications of 6D SCFTs. For example, starting from the 6D $\mathcal{N} = (2,0)$ theory, compactification on a $T^2$ results in a 4D $\mathcal{N} = 4$ SYM theory, the global form of which depends on the compactification and boundary data. In this case, we can again extract a similar SymTree via adjoint Higgsing. We can also engineer various 4D $\mathcal{N} = 2$ SCFTs with a Coulomb branch moduli space by compactifying the 6D $\mathcal{N} = (2,0)$ theories on a genus $g > 1$ Riemann surface, as well as by compactifying 6D $\mathcal{N} = (1,0)$ theories on a $T^2$. In all these cases, we get multi-sector QFTs as dictated by geometric deformations of a single parent theory.

\subsection{Branes in Flat Space} \label{ssec:BRANES}

We can engineer much the same sort of theories starting from brane probes of geometry. For example, 4D $\mathcal{N} = 4$ SYM with an A-type gauge group follows from a stack of coincident D3-branes filling 4D Minkowski space and sitting at a common point of $\mathbb{C}^3$.

Focusing on the brane realization, the SymTFT is in this case obtained via dimensional reduction on the boundary $\partial \mathbb{C}^3 = S^5$ in the presence of the RR 5-form flux sourced by the D3-branes. Indeed, reduction of the type IIB term $F_5 \wedge B_2 \wedge F_3$ results in a 5D SymTFT action (see \cite{Aharony:1998qu}):
\begin{equation}
S_{\mathrm{5D}} = \frac{i}{2 \pi } N \int B_2 \wedge d C_2,
\end{equation}
when we have $N$ coincident D3-branes. Partitioning up the stacks into individual segments by adjoint Higgsing, we again see a treelike structure emerge.

Overall the action of the SymTree for a trivalent junction now takes the form
\be \ba
S&=\sum_{\textnormal{branches}\,b} S_{\textnormal{5D}}^{(b)}+\sum_{\textnormal{internal\,nodes}\,n} S_{\:\!  \textnormal{4D}}^{(n)} \\[0.25em]
&=  S_{\textnormal{5D}}^{(1)}+S_{\textnormal{5D}}^{(2)}+S_{\textnormal{5D}}^{(12)} + S_{\textnormal{4D}}^{(\mathcal{J})}
\ea \ee
where the final term is the junction theory, which includes a 4D $\mathcal{N} = 4$ $\mathfrak{u}(1)$ vector multiplet,
and also enforces the identifications for the 5D bulk modes. The first three terms each correspond to a leg of the Y-shaped SymTree:
\be \ba
S^{(1)}_{\rm 5D}&=\frac{2\pi i}{N_1}  \int_{\;\!\R^{3,1}\times (0,r_*)} B^{(N_1)}_{2} \cup \delta C^{(N_1)}_{2}+\dots \\[0.25em]
S^{(2)}_{\rm 5D}&=\frac{2 \pi i}{N_2} \int_{\;\!\R^{3,1}\times (0,r_*)}  B^{(N_2)}_{2} \cup \delta C^{(N_2)}_{2}+\dots \\[0.25em]
S^{(12)}_{\rm 5D}&=\frac{2 \pi i}{N}  \int_{\;\!\R^{3,1}\times (r_*,\infty) } B^{({N})}_{2} \cup \delta C^{({N})}_{2} +\dots.
\ea \ee
Here we have again rescaled fields similar to \eqref{eq:Renormalization} as indicated by their raised index.

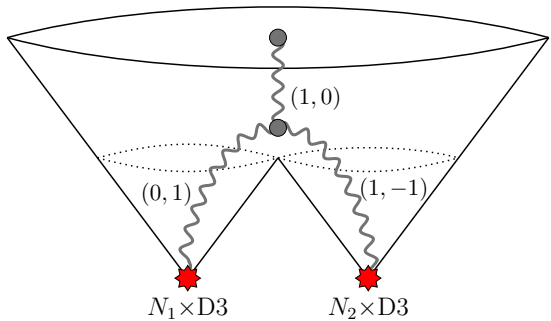
\begin{figure}
\centering
\scalebox{0.8}{
\begin{tikzpicture}
	\begin{pgfonlayer}{nodelayer}
		\node [style=none] (0) at (-1.5, -2) {};
		\node [style=none] (1) at (1.5, -2) {};
		\node [style=none] (2) at (0, 0) {};
		\node [style=none] (3) at (4.5, 2) {};
		\node [style=none] (4) at (-4.5, 2) {};
		\node [style=Star] (6) at (-1.5, -2) {};
		\node [style=Star] (7) at (1.5, -2) {};
		\node [style=none] (8) at (-1.5, -2.5) {$N_1\times$D3};
		\node [style=none] (10) at (3, 0) {};
		\node [style=none] (11) at (-3, 0) {};
		\node [style=none] (14) at (0, -3) {};
		\node [style=none] (15) at (1.5, -2.5) {$N_2\times$D3};
		\node [style=none] (16) at (0, 2) {};
		\node [style=none] (17) at (0, 0.5) {};
		\node [style=none] (18) at (1.5, -2) {};
		\node [style=none] (19) at (-1.5, -2) {};
		\node [style=BrownCircle] (20) at (0, 2) {};
		\node [style=none] (21) at (0.625, 1) {\small $(1,0)$};
		\node [style=none] (22) at (-1.85, -0.6) {\small $(0,1)$};
		\node [style=none] (23) at (1.925, -0.52) {\small $(1,-1)$};
		\node [style=BrownCircle] (25) at (0, 0.5) {};
	\end{pgfonlayer}
	\begin{pgfonlayer}{edgelayer}
		\draw [style=ThickLine] (4.center) to (0.center);
		\draw [style=ThickLine] (0.center) to (2.center);
		\draw [style=ThickLine] (2.center) to (1.center);
		\draw [style=ThickLine] (1.center) to (3.center);
		\draw [style=ThickLine, bend left=15, looseness=0.75] (4.center) to (3.center);
		\draw [style=ThickLine, bend right=15, looseness=0.75] (4.center) to (3.center);
		\draw [style=DottedLine, bend right=15] (11.center) to (2.center);
		\draw [style=DottedLine, bend right=15] (2.center) to (10.center);
		\draw [style=DottedLine, bend left=15] (11.center) to (2.center);
		\draw [style=DottedLine, bend left=15, looseness=0.75] (2.center) to (10.center);
		\draw [style=BrownLine, snake it, in=90, out=180, looseness=0.75] (17.center) to (19.center);
		\draw [style=BrownLine, snake it, in=90, out=0, looseness=0.75] (17.center) to (18.center);
		\draw [style=BrownLine, snake it] (16.center) to (17.center);
	\end{pgfonlayer}
\end{tikzpicture}
}
\caption{Depiction of line operators after adjoint Higgsing of 4D $\mathcal{N} = 4$ SYM with gauge group $\SU(N) = \SU(N_1 + N_2)$ to $S(U(N_1) \times\U(N_2))$. A non-genuine 't Hooft line $(0,1)$ and dyonic line $(1,-1)$ of two sectors of the multi-sector theory can fuse at the junction with an electric Wilson line $(1,0)$. These are realized by $(p,q)$-strings, i.e., bound states of $p$ F1-strings and $q$ D1-strings. This implements an explicit example of the general phenomena anticipated in figure \ref{fig:Throats}.}
\label{fig:DYONIC}
\end{figure}

One can also consider the explicit construction and analysis of defects and symmetry operators in this case. In the D3-brane case, the heavy defects of the relative theory are engineered via F1- and D1-strings which stretch from the boundary of $\mathbb{C}^3$ to the stacks of D3-branes.\footnote{The defect group is expected to be captured by a suitable generalization of twisted K-theory to RR fluxes (see Appendix A of \cite{Heckman:2022xgu}), but at the level of the SymTFT, this matters little.} Starting from an electric polarization of the parent $\SU(N)$ gauge theory, we observe that the electric Wilson lines persist in the individual throats. Additionally, we observe that we can also form string junctions which start as F1 strings at the boundary $S^5$, but which fragment to a D1-string and a dyonic $(1, -1)$-string on the other throats. These are associated with ``non-genuine'' line operators of the individual sectors. We see this explicitly in the decoupling limit because this non-genuine line ends on a topological surface operator constructed from everything else attached to the junction theory (which is now at infinity). See figure \ref{fig:DYONIC} for a depiction of this phenomenon.

\begin{figure}
\centering
\scalebox{0.8}{
\begin{tikzpicture}
	\begin{pgfonlayer}{nodelayer}
		\node [style=none] (0) at (-3.5, -2) {};
		\node [style=none] (1) at (-1.5, -2) {};
		\node [style=none] (2) at (-2.5, 0) {};
		\node [style=none] (3) at (5.5, 2) {};
		\node [style=none] (4) at (-5.5, 2) {};
		\node [style=none] (9) at (-0.5, 0) {};
		\node [style=none] (10) at (-0.5, -2) {};
		\node [style=none] (11) at (0.5, -2) {};
		\node [style=none] (12) at (1.5, -2) {};
		\node [style=none] (13) at (2.5, 0) {};
		\node [style=none] (14) at (3.5, -2) {};
		\node [style=none] (16) at (5.5, 2) {};
		\node [style=none] (17) at (0.5, 0) {};
		\node [style=none] (20) at (-4.25, -0.5) {};
		\node [style=none] (21) at (-2.75, -0.5) {};
		\node [style=none] (22) at (-2.25, -0.5) {};
		\node [style=none] (23) at (-0.75, -0.5) {};
		\node [style=none] (24) at (0.75, -0.5) {};
		\node [style=none] (25) at (2.25, -0.5) {};
		\node [style=none] (26) at (2.75, -0.5) {};
		\node [style=none] (27) at (4.25, -0.5) {};
		\node [style=Circle] (28) at (-3.5, -2) {};
		\node [style=Circle] (29) at (-1.5, -2) {};
		\node [style=Circle] (30) at (1.5, -2) {};
		\node [style=Circle] (31) at (3.5, -2) {};
		\node [style=none] (32) at (-3.5, -2.75) {AdS$_{5}^{(1)}$};
		\node [style=none] (33) at (-1.5, -2.75) {AdS$_{5}^{(2)}$};
		\node [style=none] (34) at (1.5, -2.75) {AdS$_{5}^{(K-1)}$};
		\node [style=none] (35) at (3.5, -2.75) {AdS$_{5}^{(K)}$};
		\node [style=none] (36) at (7, -2) {};
		\node [style=none] (37) at (7, 2) {};
		\node [style=Circle] (38) at (7, -2) {};
		\node [style=CirclePurple] (39) at (7, 0) {};
		\node [style=none] (40) at (7.75, 0) {$R_*$};
		\node [style=none] (41) at (-7.75, 2) {};
		\node [style=none] (42) at (0, -3.5) {};
	\end{pgfonlayer}
	\begin{pgfonlayer}{edgelayer}
		\draw [style=ThickLine] (4.center) to (0.center);
		\draw [style=ThickLine] (0.center) to (2.center);
		\draw [style=ThickLine] (2.center) to (1.center);
		\draw [style=ThickLine, bend left=15, looseness=0.50] (4.center) to (3.center);
		\draw [style=ThickLine, bend right=15, looseness=0.50] (4.center) to (3.center);
		\draw [style=ThickLine] (12.center) to (13.center);
		\draw [style=ThickLine] (13.center) to (14.center);
		\draw [style=ThickLine] (14.center) to (16.center);
		\draw [style=ThickLine] (17.center) to (12.center);
		\draw [style=DottedLine, bend right=15] (20.center) to (21.center);
		\draw [style=DottedLine, bend right=15] (22.center) to (23.center);
		\draw [style=DottedLine, bend right=15] (24.center) to (25.center);
		\draw [style=DottedLine, bend right=15] (26.center) to (27.center);
		\draw [style=DashedLineThin] (10.center) to (11.center);
		\draw [style=DottedLine, bend left=15] (20.center) to (21.center);
		\draw [style=DottedLine, bend left=15] (22.center) to (23.center);
		\draw [style=DottedLine, bend left=15] (24.center) to (25.center);
		\draw [style=DottedLine, bend left=15] (26.center) to (27.center);
		\draw [style=ThickLine] (29) to (9.center);
		\draw [style=ArrowLineRight] (36.center) to (37.center);
	\end{pgfonlayer}
\end{tikzpicture}}
\caption{Sequestered substacks of D3-branes are described in a near horizon limit by a multi-throat AdS configuration. Throats merge at the length scale $R_*$ characterizing the depth of the throats.}
\label{fig:Multithroat}
\end{figure}
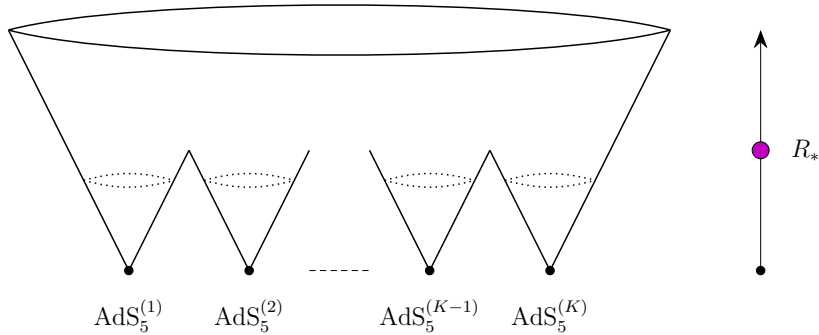

A pleasant feature of this setup is that the near horizon limit of an individual stack of $N$ D3-branes results in the semi-classical gravity dual $\mathrm{AdS}_5 \times S^5$ with $N$ units of self-dual 5-form flux threading the geometry \cite{Maldacena:1997re}. Partitioning up $N = N_1 + ... + N_K$, and assuming each $N_i$ is still sufficiently large to produce a semi-classical gravity dual on its own, we observe that we get a multi-throat configuration of AdS vacua, as in figure \ref{fig:Multithroat}. Clearly, we still retain the structure of a SymTree, although here, the radial direction of the AdS throats corresponds to the radial direction of the SymTree. Down a given throat, we have a geometry of the form $\mathrm{AdS}_5^{(i)} \times S^5_{(i)}$ threaded by $N_i$ units of RR 5-form flux. The merger between the different throats results in a jump in the level of the associated bulk 5D topological term. Additional degrees of freedom are localized at these special radial slices, and these are just the locations of the junction theory (in the SymTree). Of course, in the holographic dual the 5-form flux varies smoothly over the 10D geometry; the jumping occurs because of reduction on the linking $S^5$'s.

\paragraph{The View from a Single Stack}\mbox{}\medskip \\ Although we have emphasized a ``democratic perspective'' for how to realize the SymTree in terms of branes probing an extra-dimensional geometry, it is also interesting to consider the view from a single stack, where we continuously enlarge the size of shells including the branes. Doing so, we begin down in the deep IR of a single AdS throat with a fixed value of $N_1$ for the amount of brane flux. As we go further into the UV, we encounter a domain wall (another stack) and the total amount of flux jumps to $N_1 + N_2$. At the domain wall, we have localized degrees of freedom which absorb the anomaly inflow generated by the mismatch in topological terms on the two sides of the interface. Continuing in this fashion, we again build up the local structure of a SymTree.

\subsubsection{Wilson Line Dressing}\label{ssec:WLineDressing}

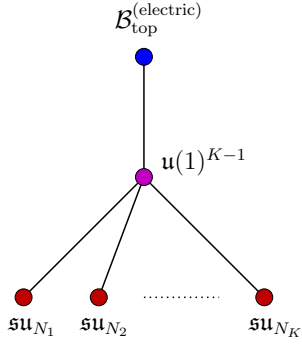
\begin{figure}
    \centering
    \scalebox{0.8}{
    \begin{tikzpicture}
	\begin{pgfonlayer}{nodelayer}
		\node [style=none] (0) at (0, 2) {};
		\node [style=none] (1) at (0, 0) {};
		\node [style=none] (2) at (-2, -2) {};
		\node [style=none] (3) at (-0.75, -2) {};
		\node [style=none] (4) at (0, -2) {};
		\node [style=none] (7) at (2, -2) {};
		\node [style=none] (8) at (-1.85, -2.5) {$\mathfrak{su}_{N_1}$};
		\node [style=none] (9) at (-0.65, -2.5) {$\mathfrak{su}_{N_2}$};
		\node [style=none] (10) at (2.2, -2.5) {$\mathfrak{su}_{N_K}$};
		\node [style=none] (11) at (1.25, -2) {};
		\node [style=CircleBlue] (12) at (0, 2) {};
		\node [style=CircleRed] (13) at (-2, -2) {};
		\node [style=CircleRed] (14) at (-0.75, -2) {};
		\node [style=CircleRed] (15) at (2, -2) {};
		\node [style=none] (16) at (1, 0.25) {$\mathfrak{u}(1)^{K-1}$};
		\node [style=none] (17) at (0.25, 2.625) {$\mathcal{B}_{\textnormal{top}}^{(\textnormal{electric})}$};
        \node [style=CirclePurple] (18) at (0, 0) {};
	\end{pgfonlayer}
	\begin{pgfonlayer}{edgelayer}
		\draw [style=ThickLine] (2.center) to (1.center);
		\draw [style=ThickLine] (1.center) to (0.center);
		\draw [style=ThickLine] (1.center) to (7.center);
		\draw [style=ThickLine] (3.center) to (1.center);
		\draw [style=DottedLine] (4.center) to (11.center);
	\end{pgfonlayer}
\end{tikzpicture}
    }
    \caption{SymTree with topological boundary condition for $SU(N_1+N_2+\dots+N_K)$ global form, and junction and tree $\Upsilon$ describing
    the Higgsing $SU(N_1+N_2+\dots+N_K) \supset S(U(N_1)\times\U(N_2)\times\dots\times\U(N_K))$. Node theories are 4D SYM with indicated gauge algebra.}
    \label{fig:LargeNAvgTree}
\end{figure}

It is instructive to consider in more detail how heavy defects end up being dressed by additional degrees of freedom localized near the junction of the SymTree. To illustrate, we again focus on the case of 4D $\mathcal{N} = 4$ SYM with gauge group $\SU(N) = \SU(N_1+ ... + N_K)$, i.e., the electric polarization of the relative $\mathfrak{su}_{N}$ theory and consider adjoint Higgsing to the subgroup $S(U(N_1) \times ... \times\U(N_K))$. The relative $\mathfrak{su}_{N_i}$ theories specify physical boundary conditions of the SymTree, while the $\mathfrak{u}(1)$ factors sit at the junction(s). See figure \ref{fig:LargeNAvgTree} for a depiction of the case with a single multi-valent junction.

We can rephrase the above in terms of $\U(N_i)$ gauge connections which we decompose as:
\begin{equation}
\mathcal{A}_i = A_i + \frac{a_i}{N_i} \mathbf{1}_{N_i \times N_i},
\end{equation}
which embed in the $\SU(N)$ connection in the obvious way.\footnote{$a_i$ is the trace of the U$(N_i)$ gauge connection in the fundamental representation.} In particular, we have the overall trace constraint:
\begin{equation}\label{eq:traceconstraint}
N_1a_1 + ... + N_Ka_K = 0.
\end{equation}

Suppose we now attempt to construct a Wilson line in an irreducible representation $\mathbf{R}_i$ of the $\mathfrak{su}_{N_i}$ throat.\footnote{In stringy terms, we get the fundamental representation of $\mathfrak{su}_{N_i}$ from an F1-string which descends to the stack of D3-branes. Higher-dimensional representations are obtained by merging such lines, i.e., via fusion of these fundamental lines.} On its own, this does not really make sense because the global form of the $\SU(N)$ gauge group has an electric $\mathbb{Z}_{N}^{(1)}$ symmetry, whereas the $\mathfrak{su}_{N_i}$ gauge theory has electric defects acted on by $\mathbb{Z}_{N_i}^{(1)}$. We can dress the ``naive'' Wilson line for $\SU(N_i)$ gauge theory by an overall $\U(1)_i$ line:
\begin{equation}\label{eq:Dressing}
\mathcal{W}_{\mathbf{R}_i , q_i} = \mathcal{W}_{\mathbf{R}_i}^{\mathrm{naive}} \exp\left( - i q(\mathbf{R}_i) \int a_i \right),
\end{equation}
where $q(\mathbf{R}_i)$ is simply the $N_i$-ality of the representation:
\begin{equation}
q(\mathbf{R}_i) = \frac{\# \mathrm{boxes}(\mathbf{R}_i)}{N_i},
\end{equation}
in the obvious notation.

Returning to line \eqref{eq:traceconstraint}, we can of course also project down to the $K-1$ independent $\U(1)$ vector potentials, e.g., matching back onto \eqref{eq:adjointhiggs}. Consider for example the case $K=2$ with $\mathbf{R}_i$ the fundamental representation of $\mathfrak{su}_{N_i}$. The dressing of line \eqref{eq:Dressing} then corresponds to the representation $(\mathbf{R}_i)_1$ where the added $\U(1)$ gauge field is that of the abelian factor in
\be \label{eq:smth}
\U(N_i)=\frac{\SU(N_i)\times\U(1)_i}{\Z_{N_i}}\,.
\ee
We now reduce to a single abelian factor as in
\be \label{eq:Higgs54}
S(\U(N_1)\times \U(N_2))=\SU(N_1)\times \SU(N_2)\times\U(1)/\Z_{L},
\ee
where $L$ is the least common multiple $\text{lcm}(N_1,N_2)$.
Denoting this $U(1)$ potential as $a$, it is related to the $U(1)_i$ vector potentials as
\begin{equation}
    a_1=\frac{N_2}{g}a, \quad \quad \quad a_2=-\frac{N_1}{g}a
\end{equation}
which is consistent with the constraint \eqref{eq:traceconstraint} as well as the charge assignment in the final line of \eqref{eq:adjointhiggs}.

From manipulations like the above we now see that the individual 1-form symmetries across the different relative $\mathfrak{su}_{N_i}$ theories are correlated. Indeed, precisely because all of these Wilson lines descend from a parent $\SU(N)$ gauge theory, observe that the electric $\mathbb{Z}_N^{(1)}$ 1-form symmetry naturally acts on each of the $\mathcal{W}_{\mathbf{R}_i , q_i}$'s. Suppose next that we fix a Young diagram $\mathcal{Y}$ such that this can be interpreted as a representation $\mathbf{R}^{\mathcal{Y}}_i$ of each $\mathfrak{su}_{N_i}$ factor.\footnote{Of course the notation here is a bit redundant once we specify $\mathbf{R}_{i}$. It is more to emphasize the point that it is all dictated by the Young diagram.} We can then build an operator which transforms under the parent 1-form symmetry:
\begin{equation}\label{eq:Wdressed}
\mathbb{W}_{\mathcal{Y}} = \sum_{i = 1}^{K} \mathcal{W}_{\mathbf{R}^{\mathcal{Y}}_i , q_i}
\end{equation}
where $\mathbf{R}^{\mathcal{Y}}_i$ denotes the representation of $\mathfrak{su}_{N_i}$ with the same Young diagram. The line operator $\mathbb{W}_{\mathcal{Y}}$ has $N$-ality charge dictated by the number of boxes. This makes the transformation of the Wilson lines consistent in each individual throat, and the dressing by the $\U(1)$ line operators ensures that there is a common phase rotation for each summand.

Since we have labelled our Wilson line operators by a choice of Young diagram $\mathcal{Y}$, it is natural to ask what happens when this Young diagram has a sufficient number of boxes which anti-symmetrize indices in the representation so that $\mathcal{Y}$ does not specify a genuine representation for some choice of $N_i$. In this case, we take this to mean that the original $\mathcal{W}_{\mathbf{R}_i^{\mathcal{Y}}}$ of a given $\mathfrak{su}_{N_i}$ theory has actually broken up into a product of Wilson lines labelled by smaller Young diagrams. This is in accord with how we would treat heavy probe quarks in the corresponding representation.

\subsection{Branes at Singularities}

\begin{figure}
\centering
\scalebox{0.67}{
\begin{tikzpicture}
	\begin{pgfonlayer}{nodelayer}
		\node [style=none] (0) at (-1, 3) {};
		\node [style=none] (1) at (-1, 2) {};
		\node [style=none] (2) at (-0.5, 2.5) {};
		\node [style=none] (3) at (-1.5, 2.5) {};
		\node [style=none] (4) at (2.5, 1.5) {};
		\node [style=none] (5) at (2.5, 0.5) {};
		\node [style=none] (6) at (3, 1) {};
		\node [style=none] (7) at (2, 1) {};
		\node [style=none] (8) at (1, -2) {};
		\node [style=none] (9) at (1, -3) {};
		\node [style=none] (10) at (1.5, -2.5) {};
		\node [style=none] (11) at (0.5, -2.5) {};
		\node [style=none] (16) at (1, 3) {};
		\node [style=none] (17) at (1, 2) {};
		\node [style=none] (18) at (1.5, 2.5) {};
		\node [style=none] (19) at (0.5, 2.5) {};
		\node [style=none] (20) at (2.5, -0.5) {};
		\node [style=none] (21) at (2.5, -1.5) {};
		\node [style=none] (22) at (3, -1) {};
		\node [style=none] (23) at (2, -1) {};
		\node [style=none] (24) at (-1, -2) {};
		\node [style=none] (25) at (-1, -3) {};
		\node [style=none] (26) at (-0.5, -2.5) {};
		\node [style=none] (27) at (-1.5, -2.5) {};
		\node [style=none] (28) at (-2.5, 1.75) {};
		\node [style=none] (29) at (-2.5, -1.75) {};
		\node [style=none] (30) at (1.375, 2.125) {};
		\node [style=none] (31) at (2.125, 1.375) {};
		\node [style=none] (32) at (2.125, -1.375) {};
		\node [style=none] (33) at (1.375, -2.125) {};
		\node [style=none] (112) at (14.25, 0) {\Large $+$};
		\node [style=none] (113) at (3.5, 0) {};
		\node [style=none] (114) at (6.5, 0) {};
		\node [style=none] (115) at (5, 0.5) {$L=L_1+L_2$};
		\node [style=none] (115) at (5, -0.5) {$N=N_1+N_2$};
		\node [style=none] (116) at (-1, 2.5) {$N$};
		\node [style=none] (117) at (1, 2.5) {$N$};
		\node [style=none] (118) at (2.5, 1) {$N$};
		\node [style=none] (119) at (2.5, -1) {$N$};
		\node [style=none] (120) at (1, -2.5) {$N$};
		\node [style=none] (121) at (-1, -2.5) {$N$};
		\node [style=none] (122) at (-1.75, 0) {$L$ total};
		\node [style=none] (123) at (9.5, 3) {};
		\node [style=none] (124) at (9.5, 2) {};
		\node [style=none] (125) at (10, 2.5) {};
		\node [style=none] (126) at (9, 2.5) {};
		\node [style=none] (127) at (13, 1.5) {};
		\node [style=none] (128) at (13, 0.5) {};
		\node [style=none] (129) at (13.5, 1) {};
		\node [style=none] (130) at (12.5, 1) {};
		\node [style=none] (131) at (11.5, -2) {};
		\node [style=none] (132) at (11.5, -3) {};
		\node [style=none] (133) at (12, -2.5) {};
		\node [style=none] (134) at (11, -2.5) {};
		\node [style=none] (135) at (11.5, 3) {};
		\node [style=none] (136) at (11.5, 2) {};
		\node [style=none] (137) at (12, 2.5) {};
		\node [style=none] (138) at (11, 2.5) {};
		\node [style=none] (139) at (13, -0.5) {};
		\node [style=none] (140) at (13, -1.5) {};
		\node [style=none] (141) at (13.5, -1) {};
		\node [style=none] (142) at (12.5, -1) {};
		\node [style=none] (143) at (9.5, -2) {};
		\node [style=none] (144) at (9.5, -3) {};
		\node [style=none] (145) at (10, -2.5) {};
		\node [style=none] (146) at (9, -2.5) {};
		\node [style=none] (147) at (8, 1.75) {};
		\node [style=none] (148) at (8, -1.75) {};
		\node [style=none] (149) at (11.875, 2.125) {};
		\node [style=none] (150) at (12.625, 1.375) {};
		\node [style=none] (151) at (12.625, -1.375) {};
		\node [style=none] (152) at (11.875, -2.125) {};
		\node [style=none] (153) at (9.5, 2.5) {$N_1$};
		\node [style=none] (154) at (11.5, 2.5) {$N_1$};
		\node [style=none] (155) at (13, 1) {$N_1$};
		\node [style=none] (156) at (13, -1) {$N_1$};
		\node [style=none] (157) at (11.5, -2.5) {$N_1$};
		\node [style=none] (158) at (9.5, -2.5) {$N_1$};
		\node [style=none] (159) at (8.75, 0) {$L_1$ total};
		\node [style=none] (160) at (17.25, 3) {};
		\node [style=none] (161) at (17.25, 2) {};
		\node [style=none] (162) at (17.75, 2.5) {};
		\node [style=none] (163) at (16.75, 2.5) {};
		\node [style=none] (164) at (20.75, 1.5) {};
		\node [style=none] (165) at (20.75, 0.5) {};
		\node [style=none] (166) at (21.25, 1) {};
		\node [style=none] (167) at (20.25, 1) {};
		\node [style=none] (168) at (19.25, -2) {};
		\node [style=none] (169) at (19.25, -3) {};
		\node [style=none] (170) at (19.75, -2.5) {};
		\node [style=none] (171) at (18.75, -2.5) {};
		\node [style=none] (172) at (19.25, 3) {};
		\node [style=none] (173) at (19.25, 2) {};
		\node [style=none] (174) at (19.75, 2.5) {};
		\node [style=none] (175) at (18.75, 2.5) {};
		\node [style=none] (176) at (20.75, -0.5) {};
		\node [style=none] (177) at (20.75, -1.5) {};
		\node [style=none] (178) at (21.25, -1) {};
		\node [style=none] (179) at (20.25, -1) {};
		\node [style=none] (180) at (17.25, -2) {};
		\node [style=none] (181) at (17.25, -3) {};
		\node [style=none] (182) at (17.75, -2.5) {};
		\node [style=none] (183) at (16.75, -2.5) {};
		\node [style=none] (184) at (15.75, 1.75) {};
		\node [style=none] (185) at (15.75, -1.75) {};
		\node [style=none] (186) at (19.625, 2.125) {};
		\node [style=none] (187) at (20.375, 1.375) {};
		\node [style=none] (188) at (20.375, -1.375) {};
		\node [style=none] (189) at (19.625, -2.125) {};
		\node [style=none] (190) at (17.25, 2.5) {$N_2$};
		\node [style=none] (191) at (19.25, 2.5) {$N_2$};
		\node [style=none] (192) at (20.75, 1) {$N_2$};
		\node [style=none] (193) at (20.75, -1) {$N_2$};
		\node [style=none] (194) at (19.25, -2.5) {$N_2$};
		\node [style=none] (195) at (17.25, -2.5) {$N_2$};
		\node [style=none] (196) at (16.5, 0) {$L_2$ total};
		\node [style=none] (197) at (10.5, -3.5) {};
	\end{pgfonlayer}
	\begin{pgfonlayer}{edgelayer}
		\draw [style=ThickLine, bend right=45] (1.center) to (2.center);
		\draw [style=ThickLine, bend right=45] (2.center) to (0.center);
		\draw [style=ThickLine, bend right=45] (0.center) to (3.center);
		\draw [style=ThickLine, bend right=45] (3.center) to (1.center);
		\draw [style=ThickLine, bend right=45] (5.center) to (6.center);
		\draw [style=ThickLine, bend right=45] (6.center) to (4.center);
		\draw [style=ThickLine, bend right=45] (4.center) to (7.center);
		\draw [style=ThickLine, bend right=45] (7.center) to (5.center);
		\draw [style=ThickLine, bend right=45] (9.center) to (10.center);
		\draw [style=ThickLine, bend right=45] (10.center) to (8.center);
		\draw [style=ThickLine, bend right=45] (8.center) to (11.center);
		\draw [style=ThickLine, bend right=45] (11.center) to (9.center);
		\draw [style=ThickLine, bend right=45] (17.center) to (18.center);
		\draw [style=ThickLine, bend right=45] (18.center) to (16.center);
		\draw [style=ThickLine, bend right=45] (16.center) to (19.center);
		\draw [style=ThickLine, bend right=45] (19.center) to (17.center);
		\draw [style=ThickLine, bend right=45] (21.center) to (22.center);
		\draw [style=ThickLine, bend right=45] (22.center) to (20.center);
		\draw [style=ThickLine, bend right=45] (20.center) to (23.center);
		\draw [style=ThickLine, bend right=45] (23.center) to (21.center);
		\draw [style=ThickLine, bend right=45] (25.center) to (26.center);
		\draw [style=ThickLine, bend right=45] (26.center) to (24.center);
		\draw [style=ThickLine, bend right=45] (24.center) to (27.center);
		\draw [style=ThickLine, bend right=45] (27.center) to (25.center);
		\draw [style=DottedLine, bend right=45] (28.center) to (29.center);
		\draw [style=ThickLine] (2.center) to (19.center);
		\draw [style=ThickLine] (5.center) to (20.center);
		\draw [style=ThickLine] (11.center) to (26.center);
		\draw [style=ThickLine] (30.center) to (31.center);
		\draw [style=ThickLine] (32.center) to (33.center);
		\draw [style=ArrowLineRight] (113.center) to (114.center);
		\draw [style=ThickLine, bend right=45] (124.center) to (125.center);
		\draw [style=ThickLine, bend right=45] (125.center) to (123.center);
		\draw [style=ThickLine, bend right=45] (123.center) to (126.center);
		\draw [style=ThickLine, bend right=45] (126.center) to (124.center);
		\draw [style=ThickLine, bend right=45] (128.center) to (129.center);
		\draw [style=ThickLine, bend right=45] (129.center) to (127.center);
		\draw [style=ThickLine, bend right=45] (127.center) to (130.center);
		\draw [style=ThickLine, bend right=45] (130.center) to (128.center);
		\draw [style=ThickLine, bend right=45] (132.center) to (133.center);
		\draw [style=ThickLine, bend right=45] (133.center) to (131.center);
		\draw [style=ThickLine, bend right=45] (131.center) to (134.center);
		\draw [style=ThickLine, bend right=45] (134.center) to (132.center);
		\draw [style=ThickLine, bend right=45] (136.center) to (137.center);
		\draw [style=ThickLine, bend right=45] (137.center) to (135.center);
		\draw [style=ThickLine, bend right=45] (135.center) to (138.center);
		\draw [style=ThickLine, bend right=45] (138.center) to (136.center);
		\draw [style=ThickLine, bend right=45] (140.center) to (141.center);
		\draw [style=ThickLine, bend right=45] (141.center) to (139.center);
		\draw [style=ThickLine, bend right=45] (139.center) to (142.center);
		\draw [style=ThickLine, bend right=45] (142.center) to (140.center);
		\draw [style=ThickLine, bend right=45] (144.center) to (145.center);
		\draw [style=ThickLine, bend right=45] (145.center) to (143.center);
		\draw [style=ThickLine, bend right=45] (143.center) to (146.center);
		\draw [style=ThickLine, bend right=45] (146.center) to (144.center);
		\draw [style=DottedLine, bend right=45] (147.center) to (148.center);
		\draw [style=ThickLine] (125.center) to (138.center);
		\draw [style=ThickLine] (128.center) to (139.center);
		\draw [style=ThickLine] (134.center) to (145.center);
		\draw [style=ThickLine] (149.center) to (150.center);
		\draw [style=ThickLine] (151.center) to (152.center);
		\draw [style=ThickLine, bend right=45] (161.center) to (162.center);
		\draw [style=ThickLine, bend right=45] (162.center) to (160.center);
		\draw [style=ThickLine, bend right=45] (160.center) to (163.center);
		\draw [style=ThickLine, bend right=45] (163.center) to (161.center);
		\draw [style=ThickLine, bend right=45] (165.center) to (166.center);
		\draw [style=ThickLine, bend right=45] (166.center) to (164.center);
		\draw [style=ThickLine, bend right=45] (164.center) to (167.center);
		\draw [style=ThickLine, bend right=45] (167.center) to (165.center);
		\draw [style=ThickLine, bend right=45] (169.center) to (170.center);
		\draw [style=ThickLine, bend right=45] (170.center) to (168.center);
		\draw [style=ThickLine, bend right=45] (168.center) to (171.center);
		\draw [style=ThickLine, bend right=45] (171.center) to (169.center);
		\draw [style=ThickLine, bend right=45] (173.center) to (174.center);
		\draw [style=ThickLine, bend right=45] (174.center) to (172.center);
		\draw [style=ThickLine, bend right=45] (172.center) to (175.center);
		\draw [style=ThickLine, bend right=45] (175.center) to (173.center);
		\draw [style=ThickLine, bend right=45] (177.center) to (178.center);
		\draw [style=ThickLine, bend right=45] (178.center) to (176.center);
		\draw [style=ThickLine, bend right=45] (176.center) to (179.center);
		\draw [style=ThickLine, bend right=45] (179.center) to (177.center);
		\draw [style=ThickLine, bend right=45] (181.center) to (182.center);
		\draw [style=ThickLine, bend right=45] (182.center) to (180.center);
		\draw [style=ThickLine, bend right=45] (180.center) to (183.center);
		\draw [style=ThickLine, bend right=45] (183.center) to (181.center);
		\draw [style=DottedLine, bend right=45] (184.center) to (185.center);
		\draw [style=ThickLine] (162.center) to (175.center);
		\draw [style=ThickLine] (165.center) to (176.center);
		\draw [style=ThickLine] (171.center) to (182.center);
		\draw [style=ThickLine] (186.center) to (187.center);
		\draw [style=ThickLine] (188.center) to (189.center);
	\end{pgfonlayer}
\end{tikzpicture}}
\caption{Deformation of a necklace quiver with $L$ nodes to a pair of quivers with $L_1,L_2$ nodes respectively. The quivers are (locally) engineered by a stack of $N$ (respectively $N_i$) D3-branes probing $\mathbb{C}^2/\Z_L$ (respectively $\mathbb{C}^2/\Z_{L_i}$).}
\label{fig:Necklace}
\end{figure}
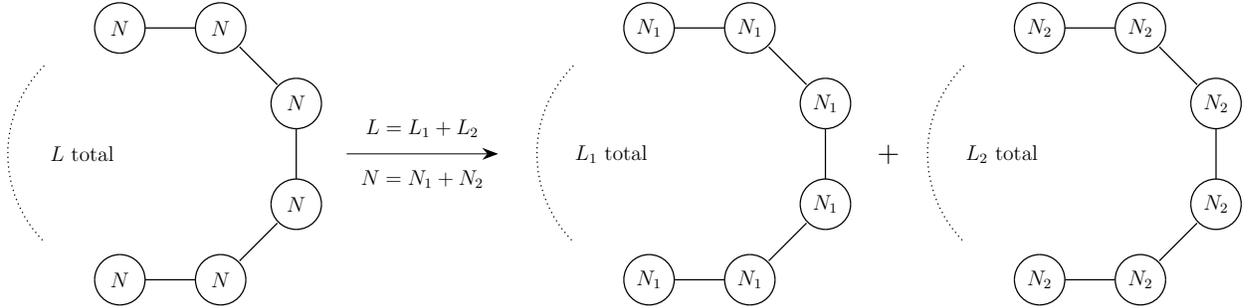

As another class of examples, we now turn to multi-sector QFTs realized by brane probes of singularities. More precisely, we assume that we have a single stack of branes probing a Calabi-Yau singularity, where we can geometrically deform both the singularity, as well as the stacks of branes. From the perspective of the Calabi-Yau singularity, we are dealing with a higher-dimensional QFT which has a collection of defects, as realized by the probe branes. For a different approach to bulk modes coupled to dynamical SCFT edge modes realized via geometry, see reference \cite{Acharya:2023bth}.

There are many ways in which such systems can arise. For example, we can take the 7D SYM theories as realized by M-theory on $\mathbb{C}^2 / \mathbb{Z}_L$, and introduce a stack of $N$ M5-branes probing the singularity. This results in a 6D conformal matter theory.\footnote{For various properties of conformal matter, see e.g., references \cite{DelZotto:2014hpa, Heckman:2014qba, Baume:2020ure, Heckman:2020otd, Baume:2022cot}.} Introducing a smoothing deformation, the original A-type singularity breaks up into a collection of A-type singularities which are each locally of the form $\mathbb{C}^2 / \mathbb{Z}_{L_i}$ with $L = L_1 + ... + L_K$. We can also partition up the stacks of M5-branes as $N = N_1 + ... + N_K$.
Similar considerations hold for a stack of $N$ D3-branes probing a $\mathbb{C}^2 / \mathbb{Z}_L$ singularity where we can perform a similar partitioning of the geometry (see figure \ref{fig:Necklace}). To keep our discussion general, we assume we are dealing with a local Calabi-Yau singularity $X$ which admits smoothing deformations to local singularities $\{X_j \}_{j = 1}^{M}$ and a stack of $N$ branes which we partition up as $N = N_1 + ... + N_K$, i.e., we assume there are no obstructions in the moduli space (as generated by a potential).

As a general comment, it can happen that the brane probe specifies a QFT $\mathcal{T}_{\mathrm{brane}}$
in dimension $D$ but should be viewed as a defect of a $D^{\prime} \geq D$-dimensional QFT $\mathcal{T}_{\mathrm{geom}}$
engineered by a geometric singularity.\footnote{Of course, it could happen that $D^{\prime} = D$, in which case the subtleties which we now discuss do not arise.} From the perspective of the SymTFT $\mathcal{S}_{\mathrm{brane}}$ for the brane probe theory, the theory $\mathcal{T}_{\mathrm{geom}}$ actually fills out the higher-dimensional bulk. This occurs, for example, in configurations such as M5-brane probes of ADE singularities, where the holographic dual is (in the large number of M5's limit) $\mathrm{AdS}_7 \times S^4 / \Gamma_{ADE}$ with the 7D SYM sector filling all of $\mathrm{AdS}_7$. This issue is not unique to SymTrees but generically arises in the stringy realization of a SymTFT in the first place. Strictly speaking, then, the SymTFT is no longer purely a TFT since we still have gapless degrees of freedom in the bulk. This arises in many string constructions, see e.g., reference \cite{Acharya:2023bth} for some recent examples along these lines. In this case, then, the SymTree will generically have branches which might include both gapped as well as gapless degrees of freedom.

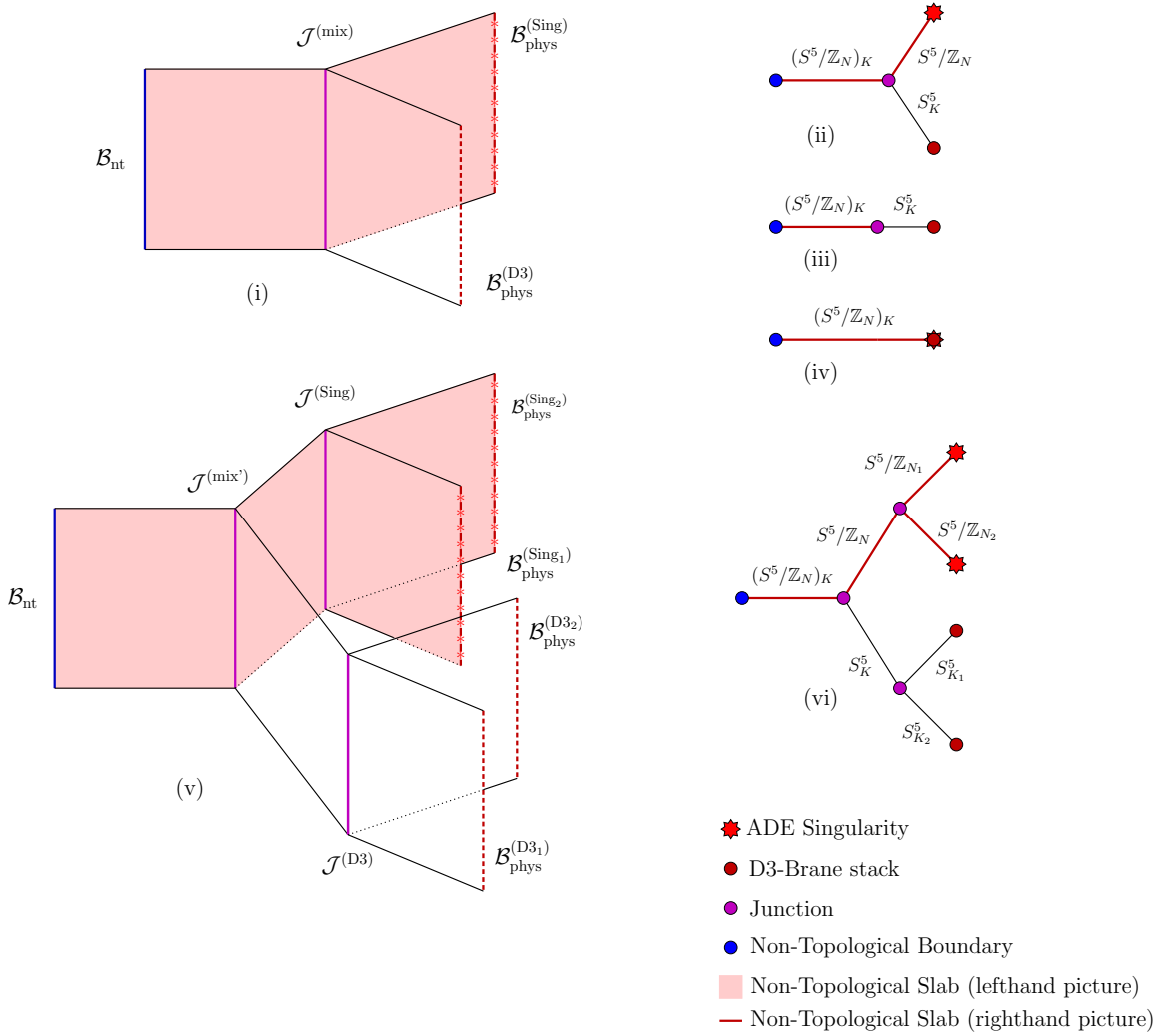
\begin{figure}
\centering
\scalebox{0.6}{
\begin{tikzpicture}
\begin{pgfonlayer}{nodelayer}
\node [style=none] (0) at (-2, -1.75) {};
\node [style=none] (1) at (-2, 2.25) {};
\node [style=none] (2) at (0, 4) {};
\node [style=none] (3) at (0, 0) {};
\node [style=none] (4) at (3.75, 5.25) {};
\node [style=none] (5) at (3, 2.75) {};
\node [style=none] (6) at (3.75, 1.25) {};
\node [style=none] (7) at (3, -1.25) {};
\node [style=none] (8) at (3, 1) {};
\node [style=none] (9) at (-6, -1.75) {};
\node [style=none] (10) at (-6, 2.25) {};
\node [style=none] (11) at (-2, 2.25) {};
\node [style=none] (12) at (-2, -1.75) {};
\node [style=none] (13) at (0.5, -1) {};
\node [style=none] (14) at (0.5, -5) {};
\node [style=none] (20) at (4.75, 4.5) {$\mathcal{B}_{\textnormal{phys}}^{(\textnormal{Sing}_2)}$};
\node [style=none] (23) at (-6.7, 0.25) {\large $\mathcal{B}_{\textnormal{nt}}$};
\node [style=none] (24) at (0.5, -1) {};
\node [style=none] (25) at (0.5, -5) {};
\node [style=none] (26) at (4.25, 0.25) {};
\node [style=none] (27) at (3.5, -2.25) {};
\node [style=none] (28) at (4.25, -3.75) {};
\node [style=none] (29) at (3.5, -6.25) {};
\node [style=none] (30) at (3.5, -4) {};
\node [style=none] (33) at (9.25, 0.25) {};
\node [style=none] (34) at (11.5, 0.25) {};
\node [style=none] (35) at (12.75, 2.25) {};
\node [style=none] (36) at (14, 3.5) {};
\node [style=none] (37) at (14, 1) {};
\node [style=none] (38) at (14, -0.475) {};
\node [style=none] (39) at (14, -3) {};
\node [style=none] (40) at (12.75, -1.75) {};
\node [style=CirclePurple] (41) at (11.5, 0.25) {};
\node [style=CirclePurple] (42) at (12.75, 2.25) {};
\node [style=CirclePurple] (43) at (12.75, -1.75) {};
\node [style=Star] (44) at (14, 3.5) {};
\node [style=Star] (45) at (14, 1) {};
\node [style=CircleRed] (46) at (14, -0.475) {};
\node [style=CircleRed] (47) at (14, -3) {};
\node [style=CircleBlue] (48) at (9.25, 0.25) {};
\node [style=none] (49) at (-3, -4) {\large (v)};
\node [style=none] (50) at (11, -2) {\large (vi)};
\node [style=none] (51) at (-4, 8) {};
\node [style=none] (52) at (-4, 12) {};
\node [style=none] (53) at (0, 12) {};
\node [style=none] (54) at (0, 8) {};
\node [style=none] (55) at (3.75, 13.25) {};
\node [style=none] (56) at (3, 10.75) {};
\node [style=none] (57) at (3.75, 9.25) {};
\node [style=none] (58) at (3, 6.75) {};
\node [style=none] (59) at (3, 9) {};
\node [style=none] (60) at (10, 11.75) {};
\node [style=none] (61) at (12.5, 11.75) {};
\node [style=none] (62) at (13.5, 13.25) {};
\node [style=none] (63) at (13.5, 10.25) {};
\node [style=CirclePurple] (64) at (12.5, 11.75) {};
\node [style=CircleBlue] (67) at (10, 11.75) {};
\node [style=none] (68) at (11, 10.5) {\large (ii)};
\node [style=Star] (69) at (13.5, 13.25) {};
\node [style=CircleRed] (70) at (13.5, 10.25) {};
\node [style=none] (71) at (10, 8.5) {};
\node [style=none] (72) at (12.25, 8.5) {};
\node [style=none] (74) at (13.5, 8.5) {};
\node [style=CirclePurple] (75) at (12.25, 8.5) {};
\node [style=CircleBlue] (76) at (10, 8.5) {};
\node [style=none] (77) at (11, 7.75) {\large (iii)};
\node [style=CircleRed] (79) at (13.5, 8.5) {};
\node [style=none] (80) at (1.5, -0.625) {};
\node [style=none] (82) at (-1.5, 7) {\large (i)};
\node [style=none] (83) at (11.25, 12.25) {$(S^5/\Z_N)_K$};
\node [style=none] (84) at (13.75, 12.25) {$S^5/\Z_N$};
\node [style=none] (85) at (13.4, 11.25) {$S^5_K$};
\node [style=none] (86) at (12.875, 9) {$S^5_K$};
\node [style=none] (87) at (11.1, 9) {$(S^5/\Z_N)_K$};
\node [style=none] (88) at (11.5, 1.625) {$S^5/\Z_{N}$};
\node [style=none] (89) at (12.625, 3.25) {$S^5/\Z_{N_1}$};
\node [style=none] (90) at (14.25, 1.75) {$S^5/\Z_{N_2}$};
\node [style=none] (91) at (13.125, -2.75) {$S^5_{K_2}$};
\node [style=none] (92) at (13.875, -1.35) {$S^5_{K_1}$};
\node [style=none] (93) at (11.875, -1.25) {$S^5_{K}$};
\node [style=none] (95) at (10.375, 0.75) {$(S^5/\Z_{N})_K$};
\node [style=none] (96) at (4.75, 12.75) {\large$\mathcal{B}_{\textnormal{phys}}^{\textnormal{(Sing)}}$};
\node [style=none] (97) at (4.1, 7.25) {\large$\mathcal{B}_{\textnormal{phys}}^{\textnormal{(D3)}}$};
\node [style=none] (98) at (-4.75, 10) {\large$\mathcal{B}_{\textnormal{nt}}$};
\node [style=none] (99) at (4.75, 1) {\large $\mathcal{B}_{\textnormal{phys}}^{(\textnormal{Sing}_1)}$};
\node [style=none] (100) at (4.375, -5.5) {\large $\mathcal{B}_{\textnormal{phys}}^{(\textnormal{D3}_1)}$};
\node [style=none] (101) at (5.125, -0.5) {\large $\mathcal{B}_{\textnormal{phys}}^{(\textnormal{D3}_2)}$};
\node [style=none] (102) at (0, 12.75) {\large $\mathcal{J}^{\textnormal{(mix)}}$};
\node [style=none] (103) at (0, 4.75) {\large $\mathcal{J}^{\textnormal{(Sing)}}$};
\node [style=none] (104) at (-2.375, 2.875) {\large $\mathcal{J}^{(\textnormal{mix'})}$};
\node [style=none] (105) at (0.5, -5.625) {\large $\mathcal{J}^{(\textnormal{D3})}$};
\node [style=Star] (106) at (9, -4.875) {};
\node [style=CircleRed] (107) at (9, -5.75) {};
\node [style=none] (108) at (11.15, -4.875) {\large ADE Singularity};
\node [style=none] (109) at (11.075, -5.75) {\large D3-Brane stack};
\node [style=none] (110) at (5, -8) {};
\node [style=CirclePurple] (111) at (9, -6.625) {};
\node [style=CircleBlue] (112) at (9, -7.5) {};
\node [style=none] (113) at (10.35, -6.625) {\large Junction};
\node [style=none] (114) at (12.35, -7.5) {\large Non-Topological Boundary};
\node [style=none] (115) at (10, 6) {};
\node [style=none] (116) at (12.25, 6) {};
\node [style=none] (117) at (13.5, 6) {};
\node [style=CircleBlue] (119) at (10, 6) {};
\node [style=Star] (123) at (13.5, 6) {};
\node [style=CircleRed] (120) at (13.5, 6) {};
\node [style=none] (122) at (11.75, 6.5) {$(S^5/\Z_N)_K$};
\node [style=none] (124) at (11, 5.25) {\large (iv)};
\node [style=none] (125) at (8.75, -8.125) {};
\node [style=none] (126) at (9.25, -8.125) {};
\node [style=none] (127) at (9.25, -8.625) {};
\node [style=none] (128) at (8.75, -8.625) {};
\node [style=none] (129) at (13.76, -8.375) {\large Non-Topological Slab (lefthand picture)};
\node [style=none] (130) at (8.75, -9.125) {};
\node [style=none] (131) at (9.25, -9.125) {};
\node [style=none] (132) at (13.92, -9.125) {\large Non-Topological Slab (righthand picture)};
\end{pgfonlayer}
\begin{pgfonlayer}{edgelayer}
\filldraw[fill=red!20, draw=red!20]  (3.75, 13.25) -- (0, 12) -- (0, 8) -- (3.75, 9.25) -- cycle;
\filldraw[fill=red!20, draw=red!20]  (-4, 8) -- (-4, 12) -- (0, 12) -- (0, 8) -- cycle;
\filldraw[fill=red!20, draw=red!20]  (0, 4) -- (3.75, 5.25) --  (3.75, 1.25) -- (0, 0) -- cycle;
\filldraw[fill=red!20, draw=red!20]  (0, 0) -- (0, 4) --  (-2, 2.25) -- (-2, -1.75)  -- cycle;
\filldraw[fill=red!20, draw=red!20]  (0, 0) -- (0, 4) --  (3, 2.75) --  (3, -1.25)  -- cycle;
\filldraw[fill=red!20, draw=red!20]  (-6, -1.75) -- (-6, 2.25) --  (-2, 2.25) -- (-2, -1.75)  -- cycle;
\filldraw[fill=red!20, draw=red!20]  (8.75, -8.125) -- (9.25, -8.125) --  (9.25, -8.625)  -- (8.75, -8.625)  -- cycle;
\draw [style=PurpleLine] (2.center) to (3.center);
\draw [style=RedLine, dashstar] (4.center) to (6.center);
\draw [style=RedLine, dashstar] (5.center) to (7.center);
\draw [style=ThickLine] (1.center) to (2.center);
\draw [style=ThickLine] (2.center) to (4.center);
\draw [style=ThickLine] (2.center) to (5.center);
\draw [style=ThickLine] (8.center) to (6.center);
\draw [style=DottedLine] (3.center) to (8.center);
\draw [style=BlueLine] (10.center) to (9.center);
\draw [style=PurpleLine] (11.center) to (12.center);
\draw [style=ThickLine] (10.center) to (11.center);
\draw [style=ThickLine] (12.center) to (9.center);
\draw [style=ThickLine] (11.center) to (13.center);
\draw [style=ThickLine] (14.center) to (12.center);
\draw [style=DottedLine] (12.center) to (3.center);
\draw [style=PurpleLine] (13.center) to (14.center);
\draw [style=PurpleLine] (24.center) to (25.center);
\draw [style=DashedRed] (26.center) to (28.center);
\draw [style=DashedRed] (27.center) to (29.center);
\draw [style=ThickLine] (24.center) to (26.center);
\draw [style=ThickLine] (24.center) to (27.center);
\draw [style=ThickLine] (29.center) to (25.center);
\draw [style=ThickLine] (30.center) to (28.center);
\draw [style=DottedLine] (25.center) to (30.center);
\draw [style=RedLine] (33.center) to (34.center);
\draw [style=RedLine] (34.center) to (35.center);
\draw [style=RedLine] (35.center) to (36.center);
\draw [style=RedLine] (35.center) to (37.center);
\draw [style=ThickLine] (38.center) to (40.center);
\draw [style=ThickLine] (40.center) to (39.center);
\draw [style=ThickLine] (40.center) to (34.center);
\draw [style=BlueLine] (52.center) to (51.center);
\draw [style=PurpleLine] (53.center) to (54.center);
\draw [style=RedLine, dashstar] (55.center) to (57.center);
\draw [style=DashedRed] (56.center) to (58.center);
\draw [style=ThickLine] (52.center) to (53.center);
\draw [style=ThickLine] (53.center) to (55.center);
\draw [style=ThickLine] (53.center) to (56.center);
\draw [style=ThickLine] (58.center) to (54.center);
\draw [style=ThickLine] (54.center) to (51.center);
\draw [style=ThickLine] (59.center) to (57.center);
\draw [style=DottedLine] (54.center) to (59.center);
\draw [style=RedLine] (60.center) to (61.center);
\draw [style=RedLine] (61.center) to (62.center);
\draw [style=ThickLine] (63.center) to (61.center);
\draw [style=RedLine] (71.center) to (72.center);
\draw [style=ThickLine] (74.center) to (72.center);
\draw [style=DottedLine] (80.center) to (7.center);
\draw [style=ThickLine] (3.center) to (80.center);
\draw [style=RedLine] (115.center) to (116.center);
\draw [style=RedLine] (117.center) to (116.center);
\draw [style=RedLine] (130.center) to (131.center);
\end{pgfonlayer}
\end{tikzpicture}}
\caption{SymTrees for brane probes of singularities. (i): stack of $K$ D3-branes probing an $\mathbb{C}^2/\Z_N$ singularity. In the extra dimensional geometry the D3-branes and the singularity are separated. (ii): cross-section of this configuration. (iii): we contract a branch / center the filtration on the D3-brane stack. Moving radially outwards from the D3-brane stack the linking $S^5_K$ with $K$ units of D3 flux sweeps over the singularity and is folded to an $(S^5/\Z_N)_K$. (iv): we contract another branch pushing the brane stack into the singularity. This alters the physical boundary condition and the massless edge modes are now organized into a quiver gauge theory. The shaded slabs / red cross-section signify, from the perspective of 4D edge modes / 5D slabs, that the ADE locus is non-compact. It stretches to infinity of the respective filtration and alters the boundary conditions.\newline
(v): the SymTree of two stacks of D3-branes probing a partial resolution of a $\mathbb{C}^2/\Z_N$ singularity. The internal dimensions contain a pair $\C^2/\Z_{N_i}$ singularities and $K_i$ D3-brane, where $i=1,2$, and which are all separated. In (vi) we show the cross-section and label branches by the geometry of the corresponding radial shells and their D3-brane flux. Note, there is a purely geometric junction $\mathcal{J}^{(\textnormal{Sing})}$ and junction purely characterized by adjoint Higgsing of a brane stack $\mathcal{J}^{(\textnormal{D3})}$. }
\label{fig:ChristmasTree}
\end{figure}

With these subtleties addressed, let us now turn to the structure of the SymTree in this setting. First of all, we have the ``true infinity'' which consists of the boundary $\partial X$ as threaded by some units of fluxes (as sourced by the probe branes). On the other side of the tree, we have physical boundary conditions, as obtained both from the stacks of branes and singularities. Indeed, we can begin by separating the singularities from the branes, and they specify different sorts of boundary conditions. For example, the singularities are generically a higher-dimensional relative QFT and the branes specify a lower-dimensional ``defect'' QFT. Starting near one such boundary condition
we can build up ever larger radial slices around any one of the singularities or stacks of branes. Eventually, they touch, and we produce a SymTree again. In the cases just mentioned, we have three generic situations for the local structure of the SymTree:
\begin{itemize}
\item Trivalent junction from fusing two stacks of branes
\item Trivalent junction from fusing two singularities
\item Trivalent junction from fusing a brane with a singularity
\end{itemize}
We have already dealt with the first two cases, where we also dealt with the junction theory. This leaves us with the fusion of a brane probe with a singularity. In this case, we analyze the junction by pushing the singularity up into the bulk of the SymTree. This results in a single branch which begins with a brane stack physical boundary condition, and which then transitions to the singularity. Prior to reaching the singularity, the local geometry of the brane stack is simply $\mathbb{R}^n$ with a boundary $S^{n-1}$. Once we cross the singularity, however, the boundary jumps to $\partial X_{j}$. This can therefore be treated as a single SymTFT, and reduction on the boundary in the presence of a flux proceeds in precisely the same way as already discussed. See figure \ref{fig:ChristmasTree} for a depiction of a hybrid SymTree with both branes and singularities.

A special multi-sector QFT of interest is where we focus on the worldvolume degrees of freedom of the branes, ignoring the contributions from the higher-dimensional QFT generated by the singularity. To analyze this case, we start from the SymTFT for $N_j$ branes at the $j^{th}$ singularity and pull the singularity off the physical boundary condition. Doing this for all the sectors, we can now fuse together the singularities first. Thus, this reduces to one of the cases previously considered.

\section{Isolated Multi-Sector QFTs} \label{sec:ISOLATED}

The operating theme in many of the previous examples has been to start from a single parent UV theory, and to then initiate a flow on moduli space which, at low energies, results in a multi-sector QFT. In this Section we consider a class of multi-sector QFTs which arise simply from geometries with multiple singularities, but in which the collision of such singularities is always at infinite distance in moduli space. Consequently, we do not have a single sector parent boundary condition $\mathcal{B}_{\textnormal{phys}}^{(\textnormal{full})}$, however $\mathcal{B}_{\textnormal{phys}}^{(\textnormal{retract})}$ is still well-defined as in figure \ref{fig:Tuck}. To illustrate, we focus on some examples of isolated orbifold singularities where the underlying QFTs are coupled by purely topological terms. We begin by revisiting 7D SYM theories, as well as 6D SCFTs, and then turn to an example with 5D SCFTs.

As a general comment, a common feature we shall encounter is the appearance of a putative $\U(1)$ symmetry factor as associated with the local motion of the isolated singularities. In a compact model we would indeed get a normalizable mode associated with this deformation, but in the decompactified limit, this mode is log-normalizable, i.e., its kinetic term is proportional to $\log \mathrm{Vol}X$ (in Planck units). To maintain continuity with our discussion of the junction theories in earlier sections, we shall therefore continue to include this $U(1)$ factor, but it is important to bear in mind that in the limit decoupled from gravity, it does not contribute a normalizable mode in the SymTree. However, once we couple to gravity, this mode ``comes to life'' so we include it in what follows.

\subsection{Revisiting Multi-Sector 7D Models}

To generate isolated multi-sector 7D SYM sectors, we return to M-theory, but now introduce ADE singularities such that the product group $G_1 \times ... \times G_K$ cannot be obtained from adjoint Higgsing from a single simple gauge group factor. This occurs in many situations, e.g., when all $G_j$ are $E$-type gauge groups.

One way to geometrically engineer such examples is to start in F-theory, where we realize such gauge group factors in terms of a suitable $\mathrm{SL}(2,\mathbb{Z})$ monodromy transformation around a codimension two singularity of a 7-brane, as marked by a distinguished point in the complex line $\mathbb{C}$.\footnote{For example, in terms of the standard $A$,$B$,$C$ non-perturbative 7-branes of reference \cite{Gaberdiel:1997ud}, the $E_N$ series is realized as $A^{N-1} B C^2$.} In this way, the entire ADE series of Lie algebras dictates a specific monodromy structure.
Far away from the 7-brane, we characterize this in terms of an $\mathrm{SL}(2,\mathbb{Z})$ duality bundle on the boundary $S^1 = \partial \mathbb{C}$, i.e., $\theta \rightarrow \theta + 2 \pi$ means acting by $M_{\rm 7-brane}$, the monodromy matrix for the 7-brane in question. Note that this also implicitly specifies a $T^2$ bundle over the boundary $S^1$ \cite{Cvetic:2021sxm, Hubner:2022kxr, Debray:2023yrs}.

Suppose now that we have multiple codimension two singularities, as captured by monodromy matrices $M_1, ... ,M_K$. The structure of the F-theory geometry is just as before, but now the ``true infinity'' involves a monodromy given by the product $M_1 ... M_K$. The fusion of junctions and the corresponding change in the bundles at each step is also captured by products of such monodromy matrices.

With the F-theory characterization in place, we pass to M-theory by further compactification on a circle, in which case the F-theory torus now becomes part of the target space geometry. Everything we have just specified in terms of monodromy matrices is directly specified by a choice of $T^2$ bundle over an $S^1$. We also see that when the radial shells surrounding two distinct singularities just touch, there is an additional free generator in the homology group of the boundary 3-manifold. This is the $\U(1)$ of the junction theory. 

Observe that in the above discussion, we did not assume that we could push the singularities on top of one another. Of course, we can specialize to the case of adjoint Higgsing, e.g., when all the $M_j$ commute and each specifies a monodromy of the form $\tau \rightarrow \tau + N_j$ as we get for an $\mathfrak{su}_{N_j}$ gauge algebra. In this case, we can merge all the singularities at finite distance in moduli space.

On the other hand, we can also consider cases where merging the singularities is at infinite distance in moduli space. This case gives us an example of a multi-sector QFT where each sector is fully isolated from its neighbors.\footnote{In a compact model where this modulus is normalizable (instead of log-normalizable as in the presence situation), one can push the singularities but this results in a higher-dimensional theory. For further discussion, see reference \cite{Lee:2021usk}.}

We now turn to the construction of the SymTree in these cases.

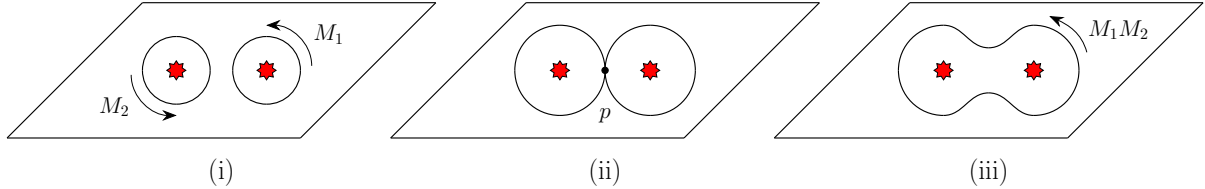
\begin{figure}
\scalebox{0.6}{
\begin{tikzpicture}
\begin{pgfonlayer}{nodelayer}
\node [style=none] (0) at (-4.75, -1.5) {};
\node [style=none] (1) at (-1.75, 1.5) {};
\node [style=none] (2) at (4.75, 1.5) {};
\node [style=none] (3) at (1.75, -1.5) {};
\node [style=Star] (4) at (-1, 0) {};
\node [style=none] (5) at (-1, 0.75) {};
\node [style=none] (6) at (-0.25, 0) {};
\node [style=none] (7) at (-1.75, 0) {};
\node [style=none] (8) at (-1, -0.75) {};
\node [style=Star] (9) at (1, 0) {};
\node [style=none] (10) at (1, 0.75) {};
\node [style=none] (11) at (1.75, 0) {};
\node [style=none] (12) at (0.25, 0) {};
\node [style=none] (13) at (1, -0.75) {};
\node [style=none] (14) at (3.75, -1.5) {};
\node [style=none] (15) at (6.75, 1.5) {};
\node [style=none] (16) at (13.25, 1.5) {};
\node [style=none] (17) at (10.25, -1.5) {};
\node [style=Star] (18) at (7.5, 0) {};
\node [style=none] (19) at (7.5, 1) {};
\node [style=none] (20) at (8.5, 0) {};
\node [style=none] (21) at (6.5, 0) {};
\node [style=none] (22) at (7.5, -1) {};
\node [style=Star] (23) at (9.5, 0) {};
\node [style=none] (24) at (9.5, 1) {};
\node [style=none] (25) at (10.5, 0) {};
\node [style=none] (26) at (8.5, 0) {};
\node [style=none] (27) at (9.5, -1) {};
\node [style=none] (28) at (12.25, -1.5) {};
\node [style=none] (29) at (15.25, 1.5) {};
\node [style=none] (30) at (21.75, 1.5) {};
\node [style=none] (31) at (18.75, -1.5) {};
\node [style=Star] (32) at (16, 0) {};
\node [style=none] (33) at (16, 1) {};
\node [style=none] (35) at (15, 0) {};
\node [style=none] (36) at (16, -1) {};
\node [style=Star] (37) at (18, 0) {};
\node [style=none] (38) at (18, 1) {};
\node [style=none] (39) at (19, 0) {};
\node [style=none] (41) at (18, -1) {};
\node [style=none] (42) at (17, 0.5) {};
\node [style=none] (43) at (17, -0.5) {};
\node [style=none] (44) at (0, -2.25) {\Large (i)};
\node [style=none] (45) at (8.5, -2.25) {\Large (ii)};
\node [style=none] (46) at (17, -2.25) {\Large (iii)};
\node [style=none] (47) at (2, 0.1) {};
\node [style=none] (48) at (1, 1) {};
\node [style=none] (49) at (-2, -0.1) {};
\node [style=none] (50) at (-1, -1) {};
\node [style=none] (51) at (19.175, 0.4) {};
\node [style=none] (52) at (18.35, 1.2) {};
\node [style=none] (53) at (2.375, 0.8) {\large $M_1$};
\node [style=none] (54) at (-2.375, -0.8) {\large $M_2$};
\node [style=none] (55) at (19.875, 0.9) {\large $M_1M_2$};
\node [style=Circle] (56) at (8.5, 0) {};
\node [style=none] (57) at (8.5, -1) {\large $p$};
\node [style=none] (58) at (8.5, -2.75) {};
\end{pgfonlayer}
\begin{pgfonlayer}{edgelayer}
\draw [style=ThickLine] (1.center) to (2.center);
\draw [style=ThickLine] (2.center) to (3.center);
\draw [style=ThickLine] (3.center) to (0.center);
\draw [style=ThickLine] (0.center) to (1.center);
\draw [style=ThickLine, in=-90, out=0] (8.center) to (6.center);
\draw [style=ThickLine, in=0, out=90] (6.center) to (5.center);
\draw [style=ThickLine, in=90, out=180] (5.center) to (7.center);
\draw [style=ThickLine, in=180, out=-90] (7.center) to (8.center);
\draw [style=ThickLine, in=-90, out=0] (13.center) to (11.center);
\draw [style=ThickLine, in=0, out=90] (11.center) to (10.center);
\draw [style=ThickLine, in=90, out=180] (10.center) to (12.center);
\draw [style=ThickLine, in=180, out=-90] (12.center) to (13.center);
\draw [style=ThickLine] (15.center) to (16.center);
\draw [style=ThickLine] (16.center) to (17.center);
\draw [style=ThickLine] (17.center) to (14.center);
\draw [style=ThickLine] (14.center) to (15.center);
\draw [style=ThickLine, in=-90, out=0] (22.center) to (20.center);
\draw [style=ThickLine, in=0, out=90] (20.center) to (19.center);
\draw [style=ThickLine, in=90, out=180] (19.center) to (21.center);
\draw [style=ThickLine, in=180, out=-90] (21.center) to (22.center);
\draw [style=ThickLine, in=-90, out=0] (27.center) to (25.center);
\draw [style=ThickLine, in=0, out=90] (25.center) to (24.center);
\draw [style=ThickLine, in=90, out=180] (24.center) to (26.center);
\draw [style=ThickLine, in=180, out=-90] (26.center) to (27.center);
\draw [style=ThickLine] (29.center) to (30.center);
\draw [style=ThickLine] (30.center) to (31.center);
\draw [style=ThickLine] (31.center) to (28.center);
\draw [style=ThickLine] (28.center) to (29.center);
\draw [style=ThickLine, in=90, out=180] (33.center) to (35.center);
\draw [style=ThickLine, in=180, out=-90] (35.center) to (36.center);
\draw [style=ThickLine, in=-90, out=0] (41.center) to (39.center);
\draw [style=ThickLine, in=0, out=90] (39.center) to (38.center);
\draw [style=ThickLine, in=180, out=0] (36.center) to (43.center);
\draw [style=ThickLine, in=180, out=0] (43.center) to (41.center);
\draw [style=ThickLine, in=180, out=0] (42.center) to (38.center);
\draw [style=ThickLine, in=0, out=-180] (42.center) to (33.center);
\draw [style=ArrowLineRight, in=180, out=-90] (49.center) to (50.center);
\draw [style=ArrowLineRight, in=0, out=90] (47.center) to (48.center);
\draw [style=ArrowLineRight, in=-25, out=105] (51.center) to (52.center);
\end{pgfonlayer}
\end{tikzpicture}}
\caption{We depict the base of the local K3 $\pi:X\rightarrow B=\mathbb{C}$. The preimage under $\pi$ of a base filtration $\mathcal{F}_B$ gives a filtration $\mathcal{F}_X$. The critical slice projects to a figure eight. The filtration for the base lifts to the full geometry, as determined by the $\mathrm{SL}(2,\Z)$ monodromy matrices $M_1,M_2$.}
\label{Fig:Elliptic}
\end{figure}

\paragraph{Filtration and Critical Slice }\mbox{}\medskip \\ The filtration of $\mathcal{F}_X$ is constructed using the elliptic fibration structure $\pi : X\rightarrow B$, with base $B = \mathbb{C}$, via the lift of a filtration $\mathcal{F}_B$ of the base. Any base filtration $\mathcal{F}_B$ lifts to a filtration of $X$ by replacing the radial shells of $\mathcal{F}_B$ by their preimage with respect to $\pi$. Now, the ADE singularities of $X$ project to two points $p_i\in \C$ and we simply pick $\mathcal{F}_B$ to be swept out, at small radius by two circles $S^1_{i}$ centered on $p_i$. These then kiss at a point $p$, resulting in a figure eight, and then merge into a single circle $S_{12}^1$ at large radius. See figure \ref{Fig:Elliptic} where small to large radii are depicted as (i) $\rightarrow$ (ii) $\rightarrow$ (iii). The point $p$ lifts to a generic torus fiber $\pi^{-1}(p)=T^2_p$.

The radial shells of $\mathcal{F}_{X}$ are
\be \label{eq:HomoElliptic} \ba
 U_{r\;\!>\;\! r_*}&=\Sigma_3^{M_1M_2}\,,\\[0.3em]
  U_{r= r_*}&=\lb \Sigma_3^{M_1} \rb\cup_{\;\!T^2}\lb \Sigma_3^{M_2}\rb\,, \\[0.3em]
  U_{r<r_*}&=\Sigma_3^{M_1}\sqcup \Sigma_3^{M_2}\,,
\ea \ee
where $\Sigma_3^M$ denotes the three-manifold constructed by fibering a two-torus over a circle with monodromy twist $M$. Here we will specify $M$ by an element of $\mathrm{SL}(2,\Z)$ which gives the monodromy action on $H_1(T^2)$. The elliptic singularity, via Kodaira's classification, fixes the conjugacy class of $M$. When considering multiple elliptic singularities there may not exist a $\mathrm{SL}(2,\Z)$ frame in which all monodromy matrices take Kodaira's canonical form simultaneously, as often happens in the context of K3 surfaces. Monodromy matrices are compared by traversing closed loops, starting and ending at one common point in $B$, linking a singularity. Using the monodromy data the homology groups of $\Sigma_3^M$ are
\be  \label{eq:noncriticalslicehomo}
H_n\!\lb \Sigma_3^M\rb\cong  \begin{cases} \Z  & k=0 \\ \Z\oplus \textnormal{coker}(M-1) & k=1 \\  \Z \oplus \textnormal{coker}(M-1)^\wedge\qquad\quad & k=2 \\ \Z   &k=3 \end{cases}
\ee
where we introduced the shorthand notation
\be
\textnormal{coker}\,(M-1)^\wedge= \textnormal{Hom}(\textnormal{coker}\,(M-1),\Z)\cong \textnormal{coker}(M-1)/\textnormal{Tor}\,\textnormal{coker}\,(M-1)
\ee
for a free group. These groups are invariant with respect to $\mathrm{SL}(2,\Z)$ conjugation of $M$.
The homology groups of the critical slice are:
\be  \label{eq:criticalslicehomoEll}
H_n \lb \lb \Sigma_3^{M_1} \rb\cup_{\;\!T^2}\lb \Sigma_3^{M_2}\rb \rb \cong  \begin{cases} \Z  & k=0 \\ \Z^2\oplus \lbb \Z^2 /\,\! \textnormal{Im}(M_1-1,M_2-1) \rbb & k=1 \\  \Z \oplus \textnormal{coker}(M_1-1)^\wedge \oplus \textnormal{coker}(M_2-1)^\wedge\oplus  F \qquad\quad & k=2 \\ \Z^2   &k=3 \end{cases}
\ee
where with the second entry in degree one we are denoting a quotient both by the image of $M_1-1$ and $M_2-1$. This quotient is only invariant with respect to simultaneous $\mathrm{SL}(2,\Z)$ transformations of $M_1$ and $M_2$. Further, depending on the case, we have $F=0,\Z,\Z^2$. The rank of $F$, denoted $|F|=\textnormal{rank}\,F$, counts the number of two-cycles stretching between the ADE singularities which depends both on the pair of elliptic singularities and their relative orientation in $\mathrm{SL}(2,\Z)$. We discuss the computation of these homology groups and their generators in Appendix \ref{app:SEQISOLATED}.

\paragraph{Projection to the SymTree}\mbox{}\medskip \\ We now reduce 11D supergravity on the radial shells $U_r$.
The computation follows the same steps as laid out in Section \ref{sec:TOPDOWN} and Appendix \ref{app:SEQISOLATED}.

Again, we will focus on the subset of the generalized symmetries originating from the center symmetries of the gauge theory sector. More precisely, we focus on the discrete 1-from symmetries and their dual 4-form symmetries. We defer the discussion of a $(-1)$-form $\U(1)$ symmetry, 2-form $\U(1)$ instanton symmetry, KK $\U(1)$ 0-form symmetry, associated with the zero section, and a $\U(1)$ 1-form symmetry of a decoupled abelian vector multiplet associated with the fiber class. The latter is fully decoupled, the fiber class does not intersect any compact curves and is not even topologically coupled \cite{Cvetic:2023pgm}. For further discussion on these see \cite{Cvetic:2021sxm, Hubner:2022kxr}.

In terms of geometry these restrictions amount to focusing on 1-cycles and 2-cycles with one leg the elliptic fiber, these are
\be \ba \label{eq:singlelegs}
H_1\!\lb \Sigma_3^{M_i}\rb &\cong  \textnormal{coker}(M_i-1)\oplus \dots  \\[0.25em]
H_1 \lb \lb \Sigma_3^{M_1} \rb\cup_{\;\!T^2}\lb \Sigma_3^{M_2}\rb \rb &\cong  \Z^2 /\,\! \textnormal{Im}(M_1-1,M_2-1)\oplus \dots\\[0.25em]
H_2 \lb \lb \Sigma_3^{M_1} \rb\cup_{\;\!T^2}\lb \Sigma_3^{M_2}\rb \rb &\cong F\oplus \dots
\ea \ee
which, lifting to differential cohomology, translates to restricting our attention to the coefficients in the expansions\footnote{The lowered indices denote form degrees, the raised indices specify, in order, the radial shells the fields are localized to, if these have multiple connected components which component, and finally and index running over the relevant homology generators for that component. We further attach the index ``$\U(1)$" for background fields for continuous symmetries, as obtained via expansion along free classes.}
\be \ba\label{eq:DiffCohoElliptic}
(r>r_*)\,:\quad \breve{G}_4&= \sum_{j\in J}  \breve{B}^{(r>r_*,j)}_{2}\star \breve{t}^{\;\!(r>r_*,j)}_{2}+ \sum_{j\in J'}  \breve{H}^{(r>r_*,j)}_{2,\U(1)}\star \breve{u}^{\;\!(r>r_*,j)}_{2}+\dots \\[0.3em]
(r=r_*)\,:\quad \breve{G}_4&= \sum_{j\in J_*}\breve{B}^{(r=r_*,j)}_{2}\star \breve{t}^{\;\!(r=r_*)}_{2}+ \sum_{k\in K}\breve{F}^{(r=r_*,k)}_{2,\U(1)}\star \breve{u}_2^{(r=r_*,k)}+\dots \\[0.3em]
(r<r_*)\,:\quad \breve{G}_4 &= \sum_{j\in J_1}\breve{B}^{(r<r_*,1,j)}_{2}\star \breve{t}^{\;\!(r<r_*,1,j)}_{2}+ \sum_{j\in J_2} \breve{B}^{(r<r_*,2,j)}_{2}\star \breve{t}^{\;\!(r<r_*,2,j)}_{2}\\&\;\!+\sum_{j\in J_1'}\breve{H}_{2,\U(1)}^{(r<r_*,1,j)}\star \breve{u}^{\;\!(r<r_*,1,j)}_{2}+ \sum_{j\in J_2'} \breve{H}_{2,\U(1)}^{(r<r_*,2,j)}\star \breve{u}^{\;\!(r<r_*,2,j)}_{2}+\dots
\ea \ee
where $J_1,J_2,J_*,J\in\{\emptyset,\{0\},\{0,1\}\}$ which count the number of torsional groups in the prime decomposition of the finitely generated abelian groups computed as the torsion contributions above. Similarly the primed index sets describe free contributions. We have $|K|=\textnormal{rank}\,F$. The geometrization of these groups is analogous to the discussion near line \eqref{eq:DiffCoho}. The fields of the SymTree under consideration are thus, projecting back down to cohomology,
\be\label{eq:backgroundlegElliptic} \ba
(r> r_*) \,: &\quad  {B}_{2}^{(r>r_*,j)},H_{2,\U(1)}^{(r>r_*,\ell)} && j\in J, \ell \in J'\\[0.1em]
(r=r_*) \,: &\quad B_{2}^{(r=r_*,j)}, A_{1,\U(1)}^{(r=r_*,k)} \quad && j\in J_*,k\in K \\
(r<r_*)\,: &\quad B^{(r<r_*,1,j)}_{2}\,, B^{(r<r_*,2,\ell)}_{2},H_{2,\U(1)}^{(r<r_*,1,r)},H_{2,\U(1)}^{(r<r_*,2,s)}\quad &&j\in J_1, \ell \in J_2, r\in J_1', s \in J_2'
\ea \ee
together with their magnetic duals. The fields on the SymTree branches are background fields for the discrete and continous symmetry of the 7D theory. The $H_2$ fields are backgrounds for the decoupled abelian vector multiplets mentioned above.

The elliptic fiber has two 1-cycles and the geometric origin of these legs thus restricts the number of fields under consideration to two on each SymTree branch
\be
|J_i|+|J_i'|\leq 2\,, \qquad |J|+|J'|\leq 2\,.
\ee
There are also up to two $\U(1)$ log-normalizable vector multiplets localized at the junction, i.e., by supersymmetry the junction theory is
\be\label{eq:JunctionDOF}
|F| \textnormal{ 7D $\mathcal{N}=1$ $\mathfrak{u}(1)$ log-normalizable vector multiplets}
\ee
Additionally, the junction fields $B_{2}^{(r=r_*,j)}$ are eaten up by identifications. Indeed the part of the maps $\overline{\jmath}_{\;\!r<r_*}\,\overline{\jmath}_{\;\!r>r_*}$, defined in \eqref{eq:EmbeddingMaps2}, relevant for
the above fields are the restriction maps associated via duality to the homology embedding maps ${\jmath}_{\;\!r<r_*}\,{\jmath}_{\;\!r>r_*}$ defined by
\be \ba
{\jmath}_{\;\!r<r_*}^{(i)}\,:&~ \Z^2/ \textnormal{Im}(M_i-1)~\rightarrow~  \Z^2/ \textnormal{Im}(M_1-1,M_2-1) \\
{\jmath}_{\;\!r>r_*}\,:& ~\Z^2/ \textnormal{Im}(M_1M_2-1)~\rightarrow~  \Z^2/ \textnormal{Im}(M_1-1,M_2-1)
\ea \ee
with ${\jmath}_{\;\!r<r_*}={\jmath}_{\;\!r<r_*}^{(1)}+{\jmath}_{\;\!r<r_*}^{(2)}$. 
Because the mappings ${\jmath}_{\;\!r<r_*}^{(i)}$ impose further identifications, the dual of these maps have vanishing kernel. With this all of the junction field $B_{2}^{(r=r_*,j)}$ arises as restrictions from the discrete field localized on the small radius branches, i.e., they are eaten up completely by gluing conditions. This of course matches with the junction degrees of freedom being solely those of line \eqref{eq:JunctionDOF}.

Instead of writing out the gluing conditions explicitly, as in \eqref{eq:glue} and \eqref{eq:Id1}, we now consider a representative examples for the cases $|F|=1$. The case $|F|=2$ as for example engineered by two elliptic singularities both of Kodaira type $I_0^*$ is rather subtle.\footnote{In the context of an F-theory compactification, the monodromy around a pair of $I_{0}^{\ast}$ singularities results in a $(-1)^F$ transformation (see e.g., \cite{Dierigl:2022reg}). This monodromy transformation is present in $\mathrm{Mp}(2,\mathbb{Z})$, the Spin cover of $\mathrm{SL}(2,\mathbb{Z})$ (see \cite{Pantev:2016nze}) as well as the $\mathsf{Pin}^{+}$ cover of $GL(2,\mathbb{Z})$ (see \cite{Tachikawa:2018njr}). For further discussion on some of the physical implications of these finer duality structures, see e.g., \cite{Dierigl:2022reg, Debray:2023yrs, Debray:2021vob}.} The case $|F|=0$ has topological junctions at which only gluing conditions are formulated.

\newpage
\paragraph{Case $|F|=1$ with $\mathfrak{u}(1)$ Junction}\mbox{}\medskip \\ Consider a pair of elliptic singularities both of Kodaira type IV$^*$ engineering two $\mathfrak{e}_6$ gauge theory sectors. The relevant monodromies are
\be
M\equiv M_1=M_2=\lb \begin{array}{cc} -1 & -1 \\ 1 & 0 \end{array}\rb\,, \qquad M_1M_2=\lb \begin{array}{cc} 0 & 1 \\ -1 & -1 \end{array}\rb
\ee
which determine the fields of the SymTree:
\be\label{eq:backgroundlegElliptic2} \ba
(r> r_*) \,: &\quad B_{2}^{(r>r_*)} \\[0.1em]
(r=r_*) \,: &\quad B_2^{(r=r_*)}, A_{1}^{(r=r_*)}  \\
(r<r_*)\,: &\quad B_{{2}}^{(r<r_*,1)}\,, B_{{2}}^{(r<r_*,2)}
\ea \ee
where all the $B$-fields are valued in $\Z_3\subset\U(1)$. The gluing conditions across the junctions are
\be
B_{{2}}^{(r<r_*,1)}\Big|_{r=r_*}=B_{{2}}^{(r<r_*,2)}\Big|_{r=r_*}=- B_{2}^{(r>r_*)}\Big|_{r=r_*}= B_{{*,j}}^{(2)}\,
\ee
where the minus sign is due to $M^{-1}=M^2$. There is no mixing between $B$-fields and the $\mathfrak{u}(1)$ junction sector, i.e., none of the $B$-fields serve as background fields for the $\mathfrak{u}(1)$ junction. These gluing conditions reflect the gauge group, in an electric frame,
\be
G=\frac{E_6\times E_6\times\U(1)}{\Z_3}
\ee
where $E_6$ is the simply connected Lie group with Lie algebra $\mathfrak{e}_6$. We also find massive bifundamentals $ ({\bf 27}, \overline{\bf 27})_2\oplus (\overline{\bf 27}, {\bf 27})_{-2} $ which is compatible with the overall observed symmetries. Again, we comment that in the decompactified limit, this $U(1)$ factor is log-normalizable, i.e., it is not really part of the SymTree junction.

\subsection{6D SCFTs}

As another class of multi-sector QFTs, consider $\mathcal{N} = (2,0)$ SCFTs as realized by type IIB string theory on an ADE singularity. Such singularities are modelled as $\mathbb{C}^2 / \Gamma$ for $\Gamma$ a finite subgroup of $\SU(2)$, but we can also reach the same sort of structures from non-compact elliptically fibered Calabi-Yau twofolds. Indeed, the \textit{same} geometry introduced in our analysis of isolated 7D sectors
works equally well in this case as well. For each individual 6D SCFT, we have a 7D SymTFT. In these cases, the junction theory again contains a log-normalizable tensor multiplet (as in the non-isolated case). Again, there is a topological coupling between the different 6D SCFTs as captured by the bulk 3-form potential of the SymTFT branches.

We can perform a similar analysis in the more general case of 6D SCFTs as realized by F-theory on a non-compact elliptically fibered Calabi-Yau threefold with multiple canonical singularities. Indeed, the main requirement here is that we start from independent contracting configurations of curves which cannot be combined into a single configuration of curves. We remark that this happens rather frequently in explicit 6D supergravity models realized in F-theory.

\subsection{5D SCFTs}

Consider next the case of isolated multi-sector 5D SCFTs. In M-theory we get examples of 5D SCFTs by working on the background $\mathbb{R}^{4,1} \times X$ for $X$ a Calabi-Yau threefold with a canonical singularity. Some aspects of the geometry, as well as higher-form symmetries for these cases were studied e.g., in references \cite{Albertini:2020mdx, Morrison:2020ool, Tian:2021cif, DelZotto:2022fnw, Cvetic:2022imb, Acharya:2023bth}. In general, we can consider a Calabi-Yau threefold which has multiple isolated canonical singularities that cannot be merged. These furnish examples of multi-sector models with isolated 5D SCFTs which only couple via topological terms.

To illustrate these considerations in more detail, we focus on multi-sector models where each sector is just the $E_0$ Seiberg SCFT \cite{Seiberg:1996bd, Morrison:1996xf}. Geometrically, the $E_0$ SCFT is realized via M-theory on $\mathbb{C}^3 / \mathbb{Z}_3$. The boundary geometry is the generalized lens space $S^5 / \mathbb{Z}_3$. The model has a $\mathbb{Z}_3$ 1-form symmetry with symmetry operators obtained from M5-branes wrapped on torsional 3-cycles at the boundary geometry. This symmetry links / acts on M2-branes which stretch from the singularity out to the torsional one-cycles of the boundary, i.e., $\mathrm{Cone}(\gamma)$ for $\gamma \in H_{1}(S^5 / \mathbb{Z}_3)$.
The relevant homology groups in this case are:
\be \label{eq:Lens}
H_k(S^5/\Z_3)\cong \begin{cases} \Z  & k=0 \\  \Z_3 & k=1 \\ 0 & k=2 \\ \Z_3 \qquad\quad &k=3 \\0  & k=4\\ \Z & k=5
\end{cases}
\ee

To produce a collection of $E_0$ SCFTs, it suffices to compactify one of the complex directions of our original model. With this in mind, we consider the quotient space $X = (T^2 \times \mathbb{C} \times \mathbb{C}) / \mathbb{Z}_3$, where we fix the complex structure of the $T^2$ to be $\tau = \exp(2 \pi i / 6)$. Each holomorphic factor has a local coordinate $z_{i}$, and the group action is simply
$(z_1,z_2,z_3) \rightarrow (\omega z_1, \omega z_2, \omega z_3)$ where $\omega^3=1$. This results in three codimension six singularities, each of which has the form $\mathbb{C}^3 / \mathbb{Z}_3$. In this case, the asymptotic geometry for the full system is $\partial X = (T^2 \times S^3) / \mathbb{Z}_3$. We can view this as a specific $\mathrm{SL}(2, \mathbb{Z})$ bundle over the lens space $S^3 / \mathbb{Z}_3$, or as a lens space bundle over the quotient space $T^2 / \mathbb{Z}_3$. While we focus on this case, similar considerations hold for related spaces such as $(T^2 \times T^2 \times \mathbb{C}) / \mathbb{Z}_3$. The fully compactified case $T^6 / \mathbb{Z}_3$ (i.e., 5D SCFTs coupled to gravity) and their higher-form symmetries was studied in \cite{Cvetic:2023pgm}.

The 5D field theory obtained from M-theory on $(T^2 \times \mathbb{C}^2) / \mathbb{Z}_3$ contains three $E_0$ SCFT sectors. Additionally, because we have a compact $T^2$ factor which also has finite volume at the conformal boundary of $X$, reduction of the M-theory 3-form potential on this 2-cycle results in a continuous $\mathfrak{u}(1)$ 0-form gauge symmetry. Observe that since the $E_0$ theories do not have a continuous $0$-form symmetry,\footnote{Geometrically, this is a consequence of $H_3(S^5 / \mathbb{Z}_3) = \mathbb{Z}_3$ being pure torsion. There is of course also a continuous $\SU(2)$ R-symmetry, but that is not relevant for the present discussion.} this gauge symmetry can only couple via massive modes / topological terms to the $E_0$ theories. The decoupling limit for the model corresponds to sending the volume of the $T^2$ factor to infinite size.

\begin{figure}
\centering
\scalebox{0.6}{
\begin{tikzpicture}
	\begin{pgfonlayer}{nodelayer}
		\node [style=CirclePurple] (0) at (-3.5, 0) {};
		\node [style=CircleBlue] (1) at (-3.5, 3) {};
		\node [style=CircleRed] (2) at (-5.5, -2.5) {};
		\node [style=CircleRed] (3) at (-3.5, -2.5) {};
		\node [style=CircleRed] (4) at (-1.5, -2.5) {};
		\node [style=none] (16) at (0, -4) {};
		\node [style=CircleRed] (19) at (2, 1) {};
		\node [style=CirclePurple] (20) at (2, 2) {};
		\node [style=CircleBlue] (21) at (2, 3) {};
		\node [style=none] (22) at (5.25, 3) {\Large Topological Boundary};
		\node [style=none] (23) at (5.0625, 1) {\Large Physical Boundaries};
		\node [style=none] (24) at (3.625, 2) {\Large Junction};
		\node [style=none] (25) at (-2.75, 1.5) {\Large $\partial X$};
		\node [style=CircleRed] (27) at (2, -1) {};
		\node [style=none] (28) at (5, -1) {};
		\node [style=none] (29) at (5.5, -1) {};
		\node [style=none] (30) at (3.5, -1.5) {\Large $S^5/\Z_3$};
		\node [style=none] (31) at (-2.5, 0.125) {\Large $X^\circ$};
	\end{pgfonlayer}
	\begin{pgfonlayer}{edgelayer}
		\draw [style=ThickLine] (2) to (0);
		\draw [style=ThickLine] (0) to (1);
		\draw [style=ThickLine] (0) to (3);
		\draw [style=ThickLine] (0) to (4);
		\draw [style=ThickLine] (27) to (28.center);
		\draw [style=DottedLine] (28.center) to (29.center);
	\end{pgfonlayer}
\end{tikzpicture}}
\caption{SymTree derived from the filtration $\mathcal{F}_X$ of the orbifold $X=(T^2\times \mathbb{C}^2)/\Z_3$.
We give the topological models for the radial shells at the legs and junctions. The junction valency is 4.}
\label{fig:SymTrees5D}
\end{figure}
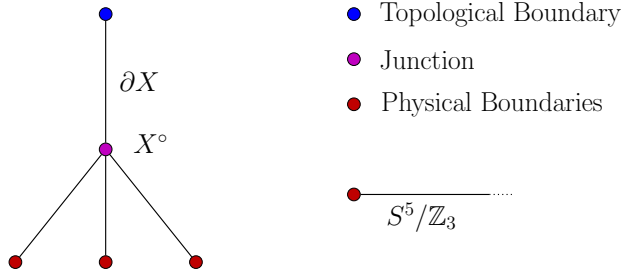

\paragraph{Filtration and Critical Slice}\mbox{}\medskip \\ We now determine the SymTree by first describing a convenient choice of filtration $\mathcal{F}_{X}$ sweeping out $X = (T^2 \times \mathbb{C}^2) / \mathbb{Z}_3$.
The filtration is again constructed by growing tubular neighborhoods of the singularities, this results in the 5D radial shells
\be \label{eq:Homo} \ba
 U_{r\;\!>\;\! r_*}&=\partial X\,,\\[0.3em]
  U_{r= r_*}&=X^\circ \big|_{\textnormal{retract}}\,, \\[0.3em]
  U_{r<r_*}&=(S^5/\mathbb{Z}_{3})\sqcup(S^5/\mathbb{Z}_{3})\sqcup (S^5/\mathbb{Z}_{3})\,, \\[0.25em]
\ea \ee
with a single critical slice at $r=r_*$. Here $X^\circ$ is the total space $X$ with the three singularities excised and $X^\circ |_{\textnormal{retract}}$ denotes the deformation retraction of $X^\circ$ to 5D. The space $X^\circ$ is a topological model for three-legged pants\footnote{That is, three legs and one waist. The cross-section of these is $S^5/\Z_3$ and $\partial X$ respectively. See figure \ref{fig:TripletofPants} in Appendix \ref{app:SEQISOLATED}.}. Via various dualities in algebraic topology, the critical slice are computed to have the homology groups:
\be  \label{eq:criticalslicehomo5D}
H_n(X^\circ)\cong H_n\!\lb X^\circ \big|_{\textnormal{retract}}\rb \cong \begin{cases} \Z  & k=0 \\ \Z_3^2  & k=1 \\  \Z   \qquad\quad & k=2 \\  \Z_3^3  &k=3 \\0 & k=4\\ \Z^3 & k=5 \end{cases}
\ee
See subfigure (i) in figure \ref{fig:SymTrees5D} for a sketch of the filtration. The asymptotic boundary is smooth and admits two fibrations $X\rightarrow T^2/\Z_3$ and $X\rightarrow S^3/\Z_3$ and from here we compute
\be \label{eq:AsympBdry}
H_k(\partial X)\cong \begin{cases} \Z  & k=0 \\  \Z_3^2 & k=1 \\ \Z & k=2 \\ \Z\oplus \Z_3^2 \qquad\quad &k=3 \\0  & k=4\\ \Z & k=5
\end{cases}
\ee
The three-legged pants $X^\circ$ runs between three copies of $S^5/\Z_3$ and one copy of $\partial X$. In Appendix \ref{app:FiltrationsALE} we give additional details and identify generators. The homology groups $H_k(\partial X)$ determine the defect and symmetry operators, as in references \cite{Apruzzi:2022rei, GarciaEtxebarria:2022vzq, Heckman:2022muc, Cvetic:2023plv}. Returning to our discussion near line (\ref{eq:Lens}), we see that we can again speak of a 1-form symmetry operator as realized by M5-branes wrapping a torsional generator of $H_{3}(\partial X)$. Similar considerations hold for the other symmetry operators, as well as the asymptotic profile of defects in the SymTFT which become defects of the relative QFTs localized at singularities.

But compared with the case of line (\ref{eq:Lens}), we also see the appearance of free generators in $H_2(\partial X)$ and $H_3(\partial X)$.
We already anticipated the appearance of such free factors; they are associated with the presence of a $\U(1)$ 0-form gauge symmetry in the 5D theory. This $\U(1)$ field does not directly couple to the $E_0$ SCFTs because these theories do not have a continuous global $\U(1)$ symmetry. On the other hand, we clearly have massive M2-branes stretched between the different sectors, as captured by elements of $H_2( X)$. Proceeding to one of our 5D SCFT sectors, we integrate out these massive M2-branes; their remnant consists of line defects of the individual $E_0$ theory. In the local $E_0$ sectors, the remnant of the $\U(1)$ gauge symmetry is a global $\mathbb{Z}_3$ 1-form symmetry which acts on these lines.

\paragraph{Projection to SymTree}\mbox{}\medskip \\ We now determine the SymTree by reducing 11D supergravity on the radial shells $U_r$. This resembles the steps resulting in \eqref{eq:backgroundleg}, hence we will be brief and only highlight differences to the previous cases.

We focus on the SymTree fields (and their duals)
\be\label{eq:backgroundleg5D}
\ba
\eqref{eq:Lens}\quad\rightarrow\quad(r<r_*)\,: &\quad B^{{(r<r_*,i)}}_{2}\,,B^{{(r<r_*,j)}}_{0}\,, \dots \\
\eqref{eq:criticalslicehomo5D}\quad\rightarrow\quad(r=r_*) \,:  &\quad  B^{{(r=r_*,a)}}_{2}\,, B^{{(r=r_*,i)}}_{0}\,,  F_{2,\U(1)}^{(r=r_*)} \,,\dots \\
\eqref{eq:AsympBdry} \quad\rightarrow\quad (r> r_*) \,: &\quad  B^{{(r>r_*,a)}}_{2}\,,B^{{(r>r_*,b)}}_{0}\,,H_{2,\U(1)}^{(r>r_*)} \,,H_{1,\U(1)}^{(r>r_*)}\,,\dots
\ea \ee
derived via KK reduction from $\breve{G}_4$. Here $i,j=1,2,3$ run over the three generalized lens spaces at low radii and $a,b=1,2$. The $B$-fields have discrete periods taking values in $\Z_3$, all other fields take values in $\U(1)$. The indices $a,b$ can be thought to label differences of labels $i,j$ and we sometimes write $a_{ij},b_{ij}$. This notation is to indicate that, if the fields labelled by $i$ are associated with cycles $\sigma_i$, then those lablled by $a_{ij}$ are associated with $\sigma_i-\sigma_j$. See Appendix \ref{app:SEQISOLATED} for explicit discussion.


We begin by determining the how junction fields glue to fields living on the branches of the SymTree. From the Mayer-Vietoris sequences \eqref{eq:seq1} and \eqref{eq:seq2} we compute, transitioning to cohomology,
\be\ba
H_{2,\U(1)}^{(r>r_*)} \Big|_{  {r= r_*}}&= F_{2,\U(1)}^{(r=r_*)} \\
H_{1,\U(1)}^{(r>r_*)}\Big|_{  {r= r_*}}&=0
\ea\ee
for fields valued in $\U(1)$ and associated with free generators. From the second relation we conclude that the corresponding 0-form symmetry does not couple to the $E_0$ sectors. From the first relation we conclude that there is no isolated junction theory. All junction fields arise as restrictions from fields on the branches of the SymTree.

For torsional fields we compute
\be\ba
\lb B^{{(r<r_*,i)}}_{2}-B^{{(r<r_*,j)}}_{2}\rb\Big|_{  {r= r_*}}=B^{{(r>r_*,a_{ij})}}_{2}\Big|_{  {r= r_*}}&= B^{{(r=r_*,a_{ij})}}_{2}\\ B^{{(r<r_*,i)}}_{0}\Big|_{  {r= r_*}}&= B^{{(r=r_*,i)}}_{0}\\
B^{{(r>r_*,a_{ij})}}_{0}\Big|_{  {r= r_*}}  &= B^{{(r=r_*,i)}}_{0}-B^{{(r=r_*,j)}}_{0}
\ea \ee
and find these to be fully eaten up by gluing conditions. These result follow by geometrizing the various homology group generators as in appendix \ref{app:SEQISOLATED}. The geometry $X\rightarrow T^2/\Z_3$ clearly has a compact 2-cycle (the zero section), however, the corresponding metric modulus is not normalizable.
Overall the action of the SymTree now takes the form
\be \ba
S&=\sum_{\textnormal{branches}\,b} S_{\textnormal{6D}}^{(b)}+\sum_{\textnormal{internal nodes}\,n} S_{\:\!  \textnormal{5D}}^{(n)} \\[0.25em]
&=  S_{\textnormal{6D}}^{(1)}+S_{\textnormal{6D}}^{(2)}+S_{\textnormal{6D}}^{(3)}+S_{\textnormal{6D}}^{(123)}
+ S_{\:\! \textnormal{5D},\;\!\mathcal{J}},
\ea \ee
where the 5D junction action simply enforces the gluing conditions between the different 6D bulk fields.
The first four terms each correspond to a leg of the Y-shaped SymTree and are topological:
\be \ba
S^{(i)}_{\rm 6D}&=\frac{i}{2 \pi} 3 \int_{\;\!\R^{4,1}\times (0,r_*)} B^{{(r<r_*,i)}}_{2} \wedge dB^{{(r<r_*,i)}}_{3}+\frac{1}{3}B^{{(r<r_*,i)}}_{2} \wedge B^{{(r<r_*,i)}}_{2}  \wedge B^{{(r<r_*,i)}}_{2} +\dots \\[0.25em]
S^{(123)}_{\rm 6D}&=\frac{i}{2 \pi} 3 \sum_{a=1,2}\int_{\;\!\R^{4,1}\times (r_*,\infty) }B^{{(r>r_*,a)}}_{2} \wedge dB^{{(r>r_*,a)}}_{3}  +... \\[0.25em]
&~~~~\;\!+\frac{i}{2 \pi} \int_{\;\!\R^{4,1}\times (r_*,\infty) }H_{2,\U(1)}^{(r>r_*)} \wedge dH_{3,\U(1)}^{(r>r_*)}+H_{1,\U(1)}^{(r>r_*)} \wedge dH_{4,\U(1)}^{(r>r_*)}+ ...
\ea \ee
where the ``...'' involves bulk fields for the 2-form and $(-1)$-form symmetry derived from the 11D Chern-Simons term $\breve{G}_4^3$.
See reference \cite{Apruzzi:2021nmk} for additional discussion.

\section{Non-Supersymmetric Example} \label{sec:NONSUSY}

So far, we have mainly focused on examples which are also supersymmetric.
This is mainly so that we can maintain technical control over the construction,
and also so that we can match to known string constructions, which are often implicitly supersymmetric.

That being said, the general structure of SymTFTs applies more broadly and does not really rely on supersymmetry at all. With this in mind, we now present a non-supersymmetric example which illustrates much of the same structure found in the supersymmetric setting.

Along these lines, we consider 4D $\SU(N)$ gauge theory with matter given by a complex adjoint-valued scalar $\phi$.
We shall be interested in a model in which $\phi$ has a potential
energy density $V(\phi, \phi^{\dag})$ which leads to Higgsing of the $\SU(N)$ gauge theory to a gauge group of the form
\begin{equation}\label{eq:GIR}
G = \frac{\SU(N_1) \times \SU(N_2) \times\U(1)}{\mathbb{Z}_{L}},
\end{equation}
where $L = \mathrm{lcm}(N_1, N_2) = N_1 N_2 / \mathrm{gcd}(N_1, N_2)$. To achieve this, we assume the vacuum expectation value of $\phi$ is of the form:\footnote{In a supersymmetric theory this can be arranged by a suitable choice of superpotential. Recall that the physical potential in a supersymmetric theory is of the schematic form $V = \vert \partial W / \partial \phi \vert^2$, with superpotential $W(\phi) = a \mathrm{Tr} \phi^2 + b \mathrm{Tr} \phi^3 + \lambda \mathrm{Tr} \phi$, and where the $\lambda$ serves as a Lagrange multiplier enforcing the tracelessness constraint. $W$ also implicitly defines a function of a single complex variable, and so we can enforce the desired choice of critical points by demanding (by abuse of notation) $W^{\prime}(z) = (z - v_1)(z - v_2)$ with $N_1 v_1 + N_2 v_2 = 0$. In the non-supersymmetric setting, additional tuning and / or higher order terms are typically necessary to achieve this breaking pattern.}
\begin{equation}
\langle \phi \rangle = \mathrm{diag}(\underbrace{v_{1},...,v_{1}}_{N_1},\underbrace{v_{2},...,v_{2}}_{N_2}),
\,\,\, \text{with} \,\,\, N_1 v_1 + N_2 v_2 = 0.
\end{equation}
Below the characteristic energy scale set by this vacuum expectation value we reach the expected gauge group of line (\ref{eq:GIR}). Similar considerations hold for breaking patterns which involve additional $\mathfrak{su}_{N_i}$ factors, so we leave this extension implicit in what follows.

Now, after adjoint Higgsing, we observe that the two $\mathfrak{su}_{N_i}$ factors have different beta functions, and so the gauge coupling for the gauge group factor with more colors will run to strong coupling faster.\footnote{Recall that in pure $\SU(N)$ gauge theory, the one-loop running of $\alpha = g^2 / 4 \pi $ is $d \alpha^{-1} / dt = b / 2 \pi$ with $b = \frac{11N}{3}$.} Without loss of generality, we assume $N_1 \geq N_2$, and thus that $\Lambda_1 \geq \Lambda_2$ for the associated strong coupling scales.\footnote{Of course, we are also implicitly assuming that $v_1,v_2\gg \Lambda_1$.} We would like to understand now how the topological coupling imposed by the $\mathbb{Z}_L$ quotient in \eqref{eq:GIR} affects this model both for an observer at intermediate energy scales $\Lambda_1>E>\Lambda_2$ and in the deep IR $\Lambda_2>E$ where both factors are confined.

First, note that the junction theory connecting the two sectors is again given by the $\U(1)$ gauge theory with the same topological couplings as in Section \ref{ssec:BRANES}. Also similar to the supersymmetric moduli space flow example, the IR electric 1-form symmetry below the energy scales $\Lambda_1$ and $\Lambda_2$ is
\begin{equation}\label{eq:1FS}
  \textnormal{IR Electric 1-form Symmetry:} \quad \quad (\mathbb{Z}_g\times\U(1))^{(1)}
\end{equation}
where $g:=\mathrm{gcd}(N_1,N_2)$. The $\U(1)^{(1)}$ factor is of less interest to us since the $\U(1)$ photon in $G$ remains gapless and spontaneously breaks the $\U(1)^{(1)}$.\footnote{Also, the sense in which $\U(1)^{(1)}$ is an IR symmetry is that it is broken in the UV explicitly due to the gauge covariant derivative of $\SU(N)$, i.e., the conservation equation $d*F_{U(1)}=0$ implicit from the effective Lagrangian below the Higgsing scales receives corrections to the righthand side as it is realized in the equations of motion $D_{A_{\SU(N)}}*F_{\SU(N)}=0$ at high energies.} On the other hand, the $(\mathbb{Z}_g)^{(1)}$ factor is retained from the UV to the IR. This simply follows from the fact that we have an electric area law for the Wilson lines.

\section{Large $M$ Averaging and Multi-Sector Models} \label{sec:LARGEN}

Having presented a number of examples of multi-sector QFTs, as well as their associated SymTrees,
we now explain how we can use this same formalism to study large $M$\footnote{$M$ here is associated to the central charge of the CFT, commonly referred to as ``large $c$'' or ``large $N$'' in the literature.} averaging in CFTs with a gravity dual.
Large $M$ averaging was recently discussed in references \cite{Schlenker:2022dyo, Chandra:2022bqq} as a way to provide an approximate characterization of chaotic dynamics in holographic systems, especially for observables above the large\footnote{In 3D gravity, the black hole threshold is sharp, but in $D > 3$, we can also have small black holes.} black hole threshold, i.e., for operators
with dimension $\Delta \gtrsim M$. Our aim here will be study the structure of higher-form symmetries and whether it is compatible with such a large $M$ averaging procedure.

To keep our discussion concrete, we focus on the case of 4D $\mathcal{N} = 4$ SYM theory with gauge group $\SU(M)$, i.e., the electric polarization of the relative $\mathfrak{su}_{M}$ QFT. In that context, the electric Wilson lines provide order parameters for the confinement / deconfinement transition. Indeed, as found in \cite{Witten:1998zw}, putting the boundary theory on $S^1 \times S^3$, i.e., at finite temperature, the breaking of the center symmetry directly tracks with the Hawking Page transition \cite{Hawking:1982dh}. As such, Wilson line observables, and thus higher-form symmetries are directly sensitive to states near the black hole threshold. Thus, we expect that it should be possible to make sense of large $M$ averaging and higher-form symmetries.

At first glance, we meet with a puzzle: what does it mean to have a $\mathbb{Z}_{M}^{(1)}$ symmetry if we are going to average over $M$? At a pragmatic level, one might wish to assert that only self-averaging observables $\mathcal{O}_{\mathrm{self}}$ need to be considered, and that Wilson lines should be excluded from such considerations. But then it is unclear how to actually calculate correlation functions which involve both the $\mathcal{O}_{\mathrm{self}}$'s and the Wilson lines.

Our aim will be to reverse engineer a prescription in gauge theory which does allow for higher-form symmetries, even in the presence of large $M$ averaging. The main idea will be to use a similar proposal to that given in \cite{Heckman:2021vzx} where we directly build a multi-sector ensemble of QFTs. Each sector will be a relative $\mathfrak{su}_{N_i}$ theory with $N_i = M + \varepsilon_i$ for $\varepsilon_i$ an integer much smaller than $M$.  Projecting onto a diagonal subset of operators $\mathbb{O}$, we show that connected correlators for local operators exhibits large $M$ ensemble averaging. Moreover, by dressing ``naive'' Wilson lines of each relative theory, we show how to produce a diagonal subset of Wilson lines $\mathbb{W}$ which all transform under a common $\mathbb{Z}_{M}^{(1)}$ 1-form symmetry.

The rest of this Section is organized as follows. We begin by reviewing the top down construction of ensemble averaging proposed in \cite{Heckman:2021vzx} and explain how we can use it to implement large $M$ averaging for local observables. With this in place, we then show how to dress Wilson lines of the individual sectors of such a system so that extended operators correctly transform under a common 1-form symmetry.

As a general comment, though we couch our discussion in terms of stringy terms, there is clearly a bottom up prescription available where we simply consider a large number of replica theories with different values of $M$. Dressing the Wilson lines via the SymTree then yields precisely the same prescription.

\subsection{Artisanal Ensembles}

To study higher-form symmetries in CFTs with large $M$ averaging, we first review the proposal of \cite{Heckman:2021vzx} which engineers ``by hand'' an ensemble average with respect to parametric families of QFTs. In such artisinal ensembles, the main idea is to consider a multi-sector QFT with similar field content in each sector. After reviewing how this works when averaging over the marginal parameters of a CFT,\footnote{One can generalize this to cover more general parameters of a QFT, a feature which can be read off from the associated brane constructions.} we show that the same considerations extend to ensemble averaging in $M$ for large $M$ QFTs. Again, we emphasize that this procedure reverse engineers the same low energy behavior as that of self-averaging observables but can deviate from this result at short distances / high energies.

We begin by briefly reviewing how we can use a multi-sector QFT to engineer an ensemble average. Consider a multi-sector QFT comprised of decoupled CFTs which we label as $\mathcal{T}_k$, where the index $k = 1,...,K$ runs over all the sectors. We assume for now that the CFTs have the same operator content, but possibly different values of marginal parameters which we specify as $\vec{\lambda}_k$. With this in mind, suppose we now introduce a local operator $\mathcal{O}_k$ for one such sector. We can use a connection on the moduli space to construct its parallel transported version on the other copies of the multi-sector QFT. Doing so, we can speak of the operator obtained from the linear combination:
\begin{equation}\label{eq:BOLDERBIGGER}
\mathbb{O} = \mathcal{O}_1 + ... + \mathcal{O}_K.
\end{equation}
For connected correlators, we observe that there is a pleasant factorization of the associated correlation functions for the $\mathbb{O}$'s. Indeed, we have normalized connected correlation functions of the form:
\begin{equation}\label{eq:boldOObservable}
    \langle \bO^{(1)} \dotsm \bO^{(n)} \rangle_\text{conn,norm} \approx \frac{1}{K} \sum_{1 \le k \le K} \langle\cO_{k}^{(1)} \dotsm \cO_{k}^{(n)}\rangle_{\mathcal{T}_k}\,,
\end{equation}
namely, the correlation function breaks up into a sum over the distinct sectors. It is important to emphasize that we only discuss connected correlators here. Additionally, the normalization factor $1 / K$ reflects a normalization of the identity operator for the full system, and also ensures that the large $K$ limit is well-defined.

To see how this results in ensemble averaging, note that the $k$-th CFT sector has its corresponding set of parameters $\vec{\lambda}_k$. Therefore, we can define a discrete probability distribution over the parameter values $\vec{\lambda}$ with density
\begin{equation}\label{eq:deltaDistribution}
    p_\text{disc}(\vec{\lambda}) = \frac{K(\vec{\lambda})}{K}\,,
\end{equation}
where $K(\vec{\lambda})$ counts the number of brane stacks with parameters $\vec{\lambda}$.

In the context of stringy realizations of such ensembles, there is a natural sense in which we can always ``smooth out'' these discretized distributions to continuous probability distributions. As noted in \cite{Heckman:2021vzx}, we can realize this ensemble by taking brane configurations probing an extra-dimensional geometry. In this case, we still have a multi-sector QFT but one in which there are mixing terms between the sectors specified by irrelevant operators. The geometry of the extra dimensions specifies values of the parameters in the worldvolume theory. Further, since the branes have finite tension, the branes are not strictly localized at a point in the transverse direction, but are instead ``spread out'' over a characteristic length scale.\footnote{For a D$p$-brane, this scale is characterized by $l_\text{min} \sim \left(\frac{1}{T_p}\right)^\frac{1}{p}$ with $T_p \sim \left(g_s l_s^{p + 1}\right)^{-1}$, where $g_s$ is the string coupling and $l_s$ is the fundamental string length.} Approximating this spread as uniform, we smooth out the probability distribution (\ref{eq:deltaDistribution}) from a sum of $\delta$-functions (or ``comb'') into a sum of step functions, which can be further regarded as a ``binned'' approximation for a continuous distribution $p_\text{smooth}$. The resulting distribution can be expressed as a histogram function via an indicator $I_{\vec{\lambda}^\prime, \vec{\epsilon}_{\vec{\lambda}^\prime}}(\lambda)$,
\begin{equation}
    p_\text{smooth}(\vec{\lambda}) = \int_{\vec{\lambda}^\prime} \diff\vec{\lambda}^\prime \; I_{\vec{\lambda}^\prime, \vec{\epsilon}_{\vec{\lambda}^\prime}}(\vec{\lambda}) \frac{K(\vec{\lambda}^\prime)}{K}\,,
\end{equation}
where $\vec{\epsilon}_{\vec{\lambda}^\prime}$ is the appropriate vector of widths of the window centered on $\vec{\lambda}^{\prime}$ and $I_{\vec{\lambda}^\prime, \vec{\epsilon}_{\vec{\lambda}^\prime}}(\vec{\lambda})$ has unit area. Using this distribution, we can rewrite the correlation function (\ref{eq:BOLDERBIGGER}) as
\begin{equation}\label{eq:binnedApproxObservable}
    \langle\bO^{(1)} \dotsm \bO^{(n)}\rangle_\text{normalized} \approx \frac{1}{K} \sum_{1 \le k \le K} \langle\cO_k^{(1)} \dotsm \cO_k^{(n)}\rangle_{\cH_k} \approx \int \diff\vec{\lambda} \; p_{\text{smooth}}(\vec{\lambda}) \langle\cO^{(1)} \dotsm \cO^{(m)}\rangle_{\vec{\lambda}}\,,
\end{equation}
which mimics an ensemble averaged computation for the observable $\overline{\langle\cO^{(1)} \dotsm \cO^{(m)}\rangle}$.

In order to have a holographic interpretation of this ensemble averaging, we now assume that each individual CFT sector has its own semi-classical AdS dual, with the same value of the cosmological constant. It is indeed possible to engineer an ensemble of local CFT sectors with the same dual cosmological constant but different parameters, as demonstrated with concrete examples in reference \cite{Heckman:2021vzx}. Note that the operator $\bO$ given in (\ref{eq:BOLDERBIGGER}) is a sum over operators, each of which enjoys the same field content in its respective local CFT, and as such describes the collective motion of many copies $\cO_i$ of this operator. Therefore, although in fact the full system contains $K$ distinct AdS throats, when we restrict our attention to the set of observables of the form $\bO$, one only reconstructs a single AdS dual according to the GKP dictionary \cite{Gubser:1998bc}.

At low energies, we thus see that this top down ensemble averaging produces a probability distribution which can in principle match to the one which might be prescribed by other holographic considerations. The approximation can break down in various ways, both in terms of short distance limits, but also entropically by sampling sufficiently many times from the ``true distribution'' generated by a single chaotic system and that of the top down reverse engineered system.

Our discussion so far has focused on the case of averaging over the marginal parameters of a CFT. An interesting generalization of this proposal is to consider the consequences of also permitting a variation in the number of degrees of freedom in a large $M$ CFT, i.e., to allow for large $M$ averaging. Strictly speaking, the operator content of each CFT with a different value of $M$ is distinct. Even so, there is a clear notion of varying $M$, especially in large $M$ gauge theories. For example, in an $\SU(M)$ gauge theory we can label (possibly gauge non-invariant) operators by a representation of $\SU(M)$, as specified by a Young diagram $\mathcal{Y}$. The Young diagram is independent of $M$ insofar as we restrict our attention to representations where the number of anti-symmetrizing indices is ``small'' compared to $M$, and even when this is not the case, the dependence on $M$ is relatively mild (we simply pass to multi-particle states). Indeed, this is also quite natural in the context of brane constructions of large $M$ parametric families of branes with an AdS dual. Otherwise, the very notion of having a semi-classical gravity dual with $G_{\mathrm{Newton}}$ scaling as a power of $M$ would make little sense to begin with!

Proceeding in this fashion, then, we can now enlarge our notion of our ensemble of relative theories $\mathcal{T}_{i}$ to possibly include a variation in the marginal parameters as well as in the value of the $N_i$. Again, this is quite natural in the context of string constructions where the value of $N_i$ is really just the asymptotic value of a flux quanta sourced by a stack of branes, i.e., it is simply the background value of a flux operator in the gravity dual.

Indeed, we can implement this sort of ensemble both at the formal level of a multi-sector QFT, as well as in the context of explicit brane constructions. We actually encountered such systems in Section \ref{sec:FLOW} where we studied brane probes of singularities. In the near horizon limit, this results in a multi-throat configuration, and we can tune the marginal couplings as well as the parameter $N_i$ in each stack to even maintain the same value of the cosmological constant in each throat.

In the case of building an ensemble over $N_i$, provided all $N_i$ are of the form $N_i = M + \varepsilon_i$ for $\varepsilon_i$ an integer far smaller than $M$, the precise form of the distribution matters little. Indeed, in this case we achieve a reasonable approximation even using the uniform distribution and all other choices require sampling a large number of times from the $N_i$. As such, the choice of distribution is relatively insensitive to the particular brane configuration and we indeed find a ``preferred'' distribution for our average over $M$.\footnote{For some recent discussion on non-perturbative effects which distinguish the choice of distribution, see e.g., reference \cite{Johnson:2019eik}.}

\begin{figure}
    \centering
    \scalebox{0.8}{
\begin{tikzpicture}
	\begin{pgfonlayer}{nodelayer}
		\node [style=none] (0) at (-7, 2) {};
		\node [style=none] (1) at (-5.5, 0) {};
		\node [style=none] (2) at (-4, 2) {};
		\node [style=none] (4) at (-3.75, 2) {};
		\node [style=none] (5) at (-2.25, 0) {};
		\node [style=none] (6) at (-0.75, 2) {};
		\node [style=none] (10) at (0.75, 2) {};
		\node [style=none] (11) at (2.25, 0) {};
		\node [style=none] (12) at (3.75, 2) {};
		\node [style=none] (13) at (-0.5, 2) {};
		\node [style=none] (14) at (0.5, 2) {};
		\node [style=none] (15) at (-5.5, 2.5) {$\mathcal{T}_{1}$};
		\node [style=none] (16) at (-2.25, 2.5) {$\mathcal{T}_{i}$};
		\node [style=none] (17) at (2.25, 2.5) {$\mathcal{T}_{K}$};
		\node [style=none] (18) at (4.25, 1) {};
		\node [style=none] (19) at (6.75, 1) {};
		\node [style=none] (20) at (5.5, 1.375) {ensemble};
		\node [style=none] (21) at (5.5, 0.625) {average};
		\node [style=none] (22) at (7.25, 2) {};
		\node [style=none] (23) at (10, 0) {};
		\node [style=none] (24) at (12.75, 2) {};
		\node [style=none] (25) at (10, 2.5) {$\overline{\mathcal{T}}$};
		\node [style=none] (26) at (-5.5, 1.25) {$\mathcal{O}_{1}$};
		\node [style=none] (27) at (-2.25, 1.25) {$\mathcal{O}_{i}$};
		\node [style=none] (28) at (2.25, 1.25) {$\mathcal{O}_{K}$};
		\node [style=Circle] (29) at (-5.5, 0.75) {};
		\node [style=Circle] (30) at (-2.25, 0.75) {};
		\node [style=Circle] (31) at (2.25, 0.75) {};
		\node [style=none] (32) at (10, 1.25) {$\mathbb{O}$};
		\node [style=Circle] (33) at (10, 0.75) {};
	\end{pgfonlayer}
	\begin{pgfonlayer}{edgelayer}
		\draw [style=ThickLine] (0.center) to (1.center);
		\draw [style=ThickLine] (1.center) to (2.center);
		\draw [style=ThickLine] (4.center) to (5.center);
		\draw [style=ThickLine] (5.center) to (6.center);
		\draw [style=ThickLine] (10.center) to (11.center);
		\draw [style=ThickLine] (11.center) to (12.center);
		\draw [style=DottedLine] (13.center) to (14.center);
		\draw [style=RedLine] (0.center) to (2.center);
		\draw [style=RedLine] (4.center) to (6.center);
		\draw [style=RedLine] (10.center) to (12.center);
		\draw [style=ThickLine] (22.center) to (23.center);
		\draw [style=ThickLine] (23.center) to (24.center);
		\draw [style=RedLine] (22.center) to (24.center);
		\draw [style=ArrowLineRight] (18.center) to (19.center);
	\end{pgfonlayer}
\end{tikzpicture}
    }
    \caption{On the left, we depict the collection of separated 4D $\mathcal{N}=4$ $\mathfrak{su}_{N_i}$ SYM theories that we label by $\mathcal{T}_{i}$ where $N_i=M+\epsilon_i$ and a sum of local operators in each of these sectors. The wedges represent the $\mathrm{AdS}_5\times S^5$ dual spacetimes for each sector. On the right, we illustrate the averaged operator $\mathbb{O}$ in the ensemble averaged theory $\overline{\mathcal{T}}$.}
    \label{fig:LargeNAvg}
\end{figure}
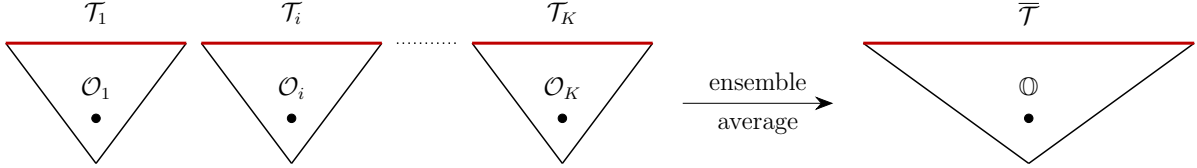

Proceeding as before, we also face no immediate obstacles in building local operators $\mathbb{O} = \mathcal{O}_1 + ... + \mathcal{O}_K$. Their normalized connected correlators again exhibit an ensemble average which now includes an average over $M$, see figure \ref{fig:LargeNAvg} for a depiction. We comment here that in the proposal of \cite{Schlenker:2022dyo}, the expectation is that up to non-perturbative corrections of order $\exp(-M)$, there is no ensemble averaging at all for operators with scaling dimension below the black hole threshold. In most of the cases we know of where we can implement a large $M$ average via branes in string theory, the typical situation is a $D > 2$ CFT, and the gravity dual also supports small black holes. As such, large $M$ averaging should be present (even if smaller) in all these cases. For further discussion on this point, see Appendix \ref{app:ENSEMBLE}.

For large $M$ averaging over extended operators, however, we face an additional complication because these are often sensitive to the arithmetic properties of the individual $N_i$ in each sector of our multi-sector QFT. For example, the Wilson lines of an $\SU(N_i)$ gauge theory are charged under the electric $\mathbb{Z}_{N_i}^{(1)}$ 1-form symmetry. To make sense of Wilson line operators, we can thus entertain two general possibilities: either we demand that a putative $\mathbb{W}_{\mathcal{Y}}$ (as specified by a choice of Young diagram / representation) has a well-defined charge under an electric 1-form symmetry, or we forfeit the existence of a 1-form symmetry in the large $M$ average. The latter possibility would be a pity because it would seem to also require abandoning the beautiful connection between bulk gravitational dynamics and center symmetry breaking in the gauge theory dual found in \cite{Witten:1998zw}. So, we shall instead proceed by constructing a suitable $\mathbb{W}_{\mathcal{Y}}$ which has a well-defined charge under the $\mathbb{Z}_{M}^{(1)}$ 1-form symmetry.

In fact, we have already presented the main elements of this construction in Section \ref{ssec:WLineDressing} where we considered the case of $N = N_1 + ... + N_K$ branes probing an extra-dimensional geometry. As we observed there, we can start with the Wilson line of the relative $\mathfrak{su}_{N_j}$ theory and then dress it by $\U(1)$ factors of the SymTree as in equation \eqref{eq:Wdressed}. Doing so, we produce an operator which has a well-defined charge under the $\mathbb{Z}_{N}^{(1)}$ 1-form symmetry of the $\SU(N)$ gauge theory. Similar considerations hold for other choices of polarization of the relative $\mathfrak{su}_{N}$ theory. To get the specific case of a Wilson line $\mathbb{W}_{\mathcal{Y}}$ charged under a $\mathbb{Z}_{M}^{(1)}$ 1-form symmetry we now specialize further by setting $N = L M$ and work in the polarization where the absolute theory has gauge group $\SU( L M ) / \mathbb{Z}_{L}$. This theory has an electric $\mathbb{Z}_{M}^{(1)}$ 1-form symmetry (isomorphic to the center of the gauge group), and as such, the $\mathbb{W}$ constructed in this way has a well-defined charge under the $\mathbb{Z}_{M}^{(1)}$ 1-form symmetry. A consequence of this is that in the diagonal theory with the ensemble operators such as $\mathbb{O}$ and $\mathbb{W}_{\mathcal{Y}}$, we can still speak of our 1-form symmetries, which matches to expectation from the bulk gravity dual. Again, let us emphasize that here we are interested in studying large $M$ averaging in its own right, and whether we can make sense of gauge / gravity dual in that setting. We expect, however, that these considerations connect with the analysis in \cite{Schlenker:2022dyo, Chandra:2022bqq}.

\section{Conclusions} \label{sec:CONC}

Much of the topological structure of global symmetries in a $D$-dimensional
QFT is captured by a bulk $(D+1)$-dimensional field theory with suitable
boundary conditions imposed to fix the global form of the QFT. In this paper
we have studied the case of a $D$-dimensional multi-sector QFT.
Each individual sector is associated to a SymTFT,
but these can form junctions, leading to topological mixing between the sectors.
Topological operators and defects of a given sector must then be dressed by additional
operators associated with modes localized at the (possibly non-topological)
junctions of different SymTFTs. We have illustrated these general
considerations in the context of various QFTs realized via geometry
and branes probing singularities. We have also presented some non-supersymmetric examples.
We also used this construction to study generalized symmetries in
holographic large $M$ ensemble averaging. In the remainder of this section
we discuss some potential avenues for future investigation.

A general feature of SymTree theories is the appearance of multiple boundaries. In this work we have focused on the appearance
of multiple physical boundaries, which covers the appearance of multi-sector QFTs. One can also entertain additional topological boundary conditions. This leads to a further generalization in the global structure of a QFT, as influenced by the presence of a junction of SymTFTs.
It would be interesting to study the structure of such theories, for example, their partition functions.

One of the general themes in recent work is the appearance of various
higher-categorical symmetries which capture the topological structure of such
QFTs. In most cases considered to date, heavy use has been made of the bulk
SymTFT associated with such a QFT. Given what we have observed here, one can sometimes have additional substructure
as captured by a SymTree. We have sketched some aspects of the higher-categorical structure
which enters here, but it would be interesting to formalize this further.

The structure of the SymTree resembles that of a tree-level Feynman diagram. Continuing with this analogy, it is natural to also
consider SymTrees which include closed loops as well. It would also be interesting to investigate the sense in which
there might be a ``meta-theory'' with scattering amplitudes associated with such diagrams, perhaps along the lines sketched in reference \cite{Vafa:2012fi}.

It is natural to study the fate of these categorical structures once we
switch on gravity. For example, in reference \cite{Cvetic:2023pgm}, it was noted that the heavy
defects and topological symmetry operators of individual sectors inevitably
become correlated in such systems. One might expect that including the
effects of gravity leads to additional constraints on multi-sector models
which are only topologically coupled when gravity is switched off. Studying
such constraints would likely be quite informative.

While we have presented a prescription for making sense of higher-form symmetries in a large $M$ averaging prescription,
one of the important features of reference \cite{Schlenker:2022dyo} is that this really ought to be viewed as taking place in a single
large $M$ gauge theory. It would nevertheless be interesting to see whether a more explicit map between the ensemble of theories considered here and possible replicas connected by wormhole configurations can also be constructed.

\newpage

\section*{Acknowledgements}

We thank M. Del Zotto, M. Dodelson, K. Jensen, Z. Komargodski, and H.Y. Zhang for helpful discussions.
ET would like to thank IFT Madrid and the CERN Theory Department for their hospitality during the completion of this work. XY thanks the UPenn Theory Group for their hospitality during part of this work. FB, JJH, MH, ET, and XY would like to thank the 2023 Simons Physics Summer Workshop for hospitality during part of this work. JJH and MH would like to thank the 2023 NORDITA program on Categorical Aspects of Symmetries for hospitality during part of this work. MH would like to thank the Aspen Center for Physics for hospitality during part of this work. JJH would like to thank the Yau Mathematical Sciences Center for hospitality during part of this work.
The work of FB is supported by the German Research Foundation through a German-Israeli Project Cooperation (DIP) grant
``Holography and the Swampland'', by the Swiss National Science Foundation (SNSF), grant
number P400P2 194341, and in part by the Deutsche Forschungsgemeinschaft under Germany’s Excellence
Strategy EXC 2121 Quantum Universe 390833306. The work of JJH, ET and AT is supported by DOE (HEP) Award DE-SC0013528.
The work of JJH, and MH is supported in part by a University Research Foundation grant at the University of Pennsylvania.
The work of JJH and MH is supported in part by BSF grant 2022100.
The work of MH is also supported by the Simons Foundation Collaboration grant
\#724069 on ``Special Holonomy in Geometry, Analysis and Physics''. The work of ET is supported in part
by the ERC Starting Grant QGuide-101042568 - StG 2021. The work of XY was supported in part by
and NYU James Arthur Graduate Associate Fellowship. The work of XY is supported by NSF grant PHY-2014086.


\appendix

\addtocontents{toc}{\protect\pagebreak}

\section{Coupling QFTs to TFTs \`{a} la Kapustin-Seiberg} \label{app:KapustinSeiberg}

Couplings between QFTs and TFTs were explored previously in \cite{Kapustin:2014gua}.
In this Appendix we review this construction and compare it to our discussion of SymTrees.

We focus on a 4D example with the QFT given by an $\SU(N)$ gauge theory and the TFT given by $\Z_N$ topological gauge theory $(\Z_N)_p$ with discrete $\theta$-parameter $p$. Once coupled, the system describes a $(\SU(N)/\Z_N)_p$ gauge theory with a discrete $\theta$-parameter $p$ turned on \cite{Kapustin:2014gua}.

In the SymTFT framework changing from $\SU(N)$ to $(\SU(N)/\Z_N)_p$ amounts to changing the topological boundary conditions of the associated 5D topological field theory, perhaps together with adding an SPT. Clearly this does not add any physical degrees of freedom and formulating the QFT/TFT coupling of \cite{Kapustin:2014gua} via a Y-shaped SymTree therefore, if possible, must therefore involve two topological boundary conditions and one physical boundary conditions, the latter supporting the relative $\mathfrak{su}_N$ theory.

To make this explicit let us discuss deforming the respective SymTFT into a SymTree. We begin by considering the SymTFT
\be
S_{\textnormal{5D}}=\frac{2\pi i}{N}\int B^{(N)}_2\cup \delta C^{(N)}_2
\ee
where fields now take values in $\Z_N$. We place this TFT on the slab $M_4\times [0,1]$ with one physical and topological boundary condition.

For concreteness we now also restrict to the case $N=2$. In this case the the physical boundary condition is determined by an edge mode $\mathfrak{su}_{2}$ theory:
 \be
\ket{ \mathcal{T}_{\mathfrak{su}_{2}}  }=\sum_{d}Z_{G_0}[d]\ket{G_0,d} \,,
\ee
where $G=\SU(2),\SO(3)_+,\SO(3)_-$ is one of the global forms of the gauge algebra $\mathfrak{su}_{2}$ and the subscript $p=0,1$ in $G_p$ labels a stacked SPT. The background 2-form fields for the 1-form symmetry of $G$ are denoted $d$, they are associated with SymTFT fields $B^{(2)}_2,C^{(2)}_2,B^{(2)}_2+C^{(2)}_2$ respectively, and $Z_{G_0}[d]$ the partition function of gauge theory with gauge group $G$ with background $d$ turned on. We refer to \cite{Kaidi:2022cpf, Heckman:2022xgu} for further details.

The topological Dirichlet and Neumann boundary conditions $\mathcal{B}_{\textnormal{top}}$ we consider are:
\be\ba
\textnormal{Dirichlet:}\quad &\bra{G_p,D} \\
\textnormal{Neumann:}\quad &\overline{\bra{G_p,D}} =\sum_d \exp\lb i\int d\cup D\rb\bra{G_p,d}
\ea \ee
where the overline denotes that the Neumann boundary condition is conjugate, via Fourier transformation, to the respective Dirichlet boundary conditions. Here $D$ is the Dirichlet boundary profile imposed on the relevant 2-form background. Upon contracting the SymTFT slab the partition functions compute as
\be
\langle {G_p,D \:\! | \:\! \mathcal{T}_{\mathfrak{su}_{2}}} \rangle = Z_{G_p}[D]\,, \qquad \overline{ \langle G_p,D \:\! |} \:\! \mathcal{T}_{\mathfrak{su}_{2}} \rangle = Z_{G_p/\:\!\Z_2}[D]\,.
\ee
Note also that via gauging of 1-form symmetries one finds the Fourier pairs
\be\ba
\exp\lb  i\pi \int D^{(1)}\cup D^{(2)}\rb&=\langle \SU(2)_0,D^{(1)} \, | \,\SO(3)_{+,0},D^{(2)} \rangle\\&=\langle\SO(3)_{+,1},D^{(1)} \, | \,\SO(3)_{-,1},D^{(2)} \rangle \\[0.6em] &=\langle \SU(2)_1,D^{(1)} \, | \,\SO(3)_{-,0},D^{(2)} \rangle
\ea \ee

In \cite{Heckman:2022xgu} it was shown that there is simple SymmetryTFT bulk operator for each Fourier pair, which maps one boundary condition of a Fourier pair onto the other and vice versa. For the discussion at hand the relevant Fourier operator is
\be \ba
P&=\exp\lb i \pi \int\frac{ \mathcal{P}\!\lb B_{2}+C_{2}\rb}{2} \rb\\ &=\exp\lb i \pi \int\frac{ \mathcal{P}\!\lb B_{2}\rb}{2}+\frac{\mathcal{P}\!\lb C_{2}\rb}{2}+B_{2}\cup C_{2} \rb\,.
\ea \ee
Here we denote the Pontryagin square of a 2-form $B_{2}$ as $\mathcal{P}(B_{2})$. This operator maps topological boundary conditions as
\be \ba
P\:\! \ket{\SU(2)_0,D} &=\ket{SO(3)_{+,0},D} \\[0.2em]
P\:\! \ket{\SU(2)_1,D}&=\ket{SO(3)_{+,1},D}\\[0.2em]
P\:\! \ket{SO(3)_{-,0},D} &=\ket{SO(3)_{-,1},D}\\[0.2em]
\ea \ee
and satisfies $P\circ P=1$. In particular, from the first line, we have the identity
\be
P\:\! \ket{\SU(2)_0,D} =\sum_d  \exp\lb \pi i\int d\cup D \rb \ket{\SU(2)_0,d} \,.
\ee

We now turn to phrase the coupling of the $\SU(2)$ theory to the discrete $(\Z_2)_p$ theory in the SymTFT framework, resulting in an $SO(3)_+$ theory. First, we note that we can express the $SO(3)_{+,0}$ gauge theory partition function as
\be
Z_{SO(3)_{+,0}}[D]=\bra{\SU(2)_0,D}\, {P}\,\ket{\mathcal{T}_{\mathfrak{su}_{2}}}
\ee
where acting with ${P}$ to the left we simply produce the boundary condition $\bra{SO(3)_{+,0},D}$. Acting on the right we obtain a new physical boundary condition $\ket{{P}\mathcal{T}_{\mathfrak{su}_{2}}}$ which is
\be \ba
\ket{{P}\mathcal{T}_{\mathfrak{su}_{2}}}&=\sum_{d_1}Z_{\SU(2)_0}[d_1] \sum_{d_2} \exp\lb i\pi \int d_1\cup d_2\rb \ket{\SU(2)_0,d_2}  \\
&=\sum_{d_2} \lbb \sum_{d_1}Z_{\SU(2)_0}[d_1] \exp\lb i\pi \int d_1\cup d_2\rb \rbb \ket{\SU(2)_0,d_2}\,.
\ea \ee
See figure \ref{fig:SymTFT}. Now note that the argument of the exponential is the action for a $(\Z_2)_0$ topological gauge theory coupled to a background $d_2$. The sum over the common $d_1$ is interpreted as a gauging and we write
\be
\ket{{P}\mathcal{T}_{\mathfrak{su}_{2}}}=\sum_d Z_{\SU(2)_0 | (\Z_2)_p} [d]\ket{\SU(2)_0,d}
\ee
where $\SU(2)_0 | (\Z_2)_p$ denotes the coupled system QFT/TFT in \cite{Kapustin:2014gua}. It is immediate that we have $ Z_{\SU(2)_0 | (\Z_2)_p} [d]= Z_{SO(3)_{+,0}} [d]$.

In the difference between $\ket{{P}\mathcal{T}_{\mathfrak{su}_{2}}}$ and $\ket{\mathcal{T}_{\mathfrak{su}_{2}}}$ it is thus crucial to keep track of the TFT basis $\{ \ket{G_p,D} \}$ in which the relative boundary condition is expanded. Equivalently, the coupling a TFT to a QFT in the framework of \cite{Kapustin:2014gua} can be phrased in SymTFT language as a manipulation of the physical boundary condition: once expanded in a TFT basis the coefficients are permuted against the basis elements.

\begin{figure}
\centering
\scalebox{0.8}{
\begin{tikzpicture}
	\begin{pgfonlayer}{nodelayer}
		\node [style=none] (0) at (-8, 0) {};
		\node [style=none] (1) at (-4, 0) {};
		\node [style=Star] (2) at (-4, 0) {};
		\node [style=CircleBlue] (3) at (-8, 0) {};
		\node [style=none] (4) at (-1.5, 0) {};
		\node [style=none] (5) at (2.5, 0) {};
		\node [style=Star] (6) at (2.5, 0) {};
		\node [style=CircleBlue] (7) at (-1.5, 0) {};
		\node [style=none] (8) at (-2.75, 0) {$=$};
		\node [style=CirclePurple] (9) at (1, 0) {};
		\node [style=none] (10) at (5, 0) {};
		\node [style=none] (11) at (9, 0) {};
		\node [style=Star] (12) at (9, 0) {};
		\node [style=CircleBlue] (13) at (5, 0) {};
		\node [style=none] (14) at (3.75, 0) {$=$};
		\node [style=none] (15) at (1, 0.625) {${P}$};
		\node [style=none] (16) at (-7.5, 0.625) {$\bra{SO(3)_{+,0},D}$};
		\node [style=none] (17) at (-4, 0.625) {$\ket{\mathcal{T}_{\mathfrak{su}_{2}}}$};
		\node [style=none] (18) at (-1, 0.625) {$\bra{\SU(2)_{0},D}$};
		\node [style=none] (19) at (2.5, 0.625) {$\ket{\mathcal{T}_{\mathfrak{su}_{2}}}$};
		\node [style=none] (20) at (5.5, 0.625) {$\bra{\SU(2)_{0},D}$};
		\node [style=none] (21) at (9, 0.625) {$\ket{{P}\mathcal{T}_{\mathfrak{su}_{2}}}$};
		\node [style=none] (22) at (0.5, -0.5) {};
		\node [style=CirclePurpleSmall] (23) at (9, 0) {};
	\end{pgfonlayer}
	\begin{pgfonlayer}{edgelayer}
		\draw [style=ThickLine] (0.center) to (1.center);
		\draw [style=ThickLine] (4.center) to (5.center);
		\draw [style=ThickLine] (10.center) to (11.center);
	\end{pgfonlayer}
\end{tikzpicture}}
\caption{In the SymTFT frame work changing from $\SU(2)$ to $\SU(2)/\Z_2=SO(3)_+$ gauge theory can be formulated as a change in topological boundary condition. Equivalently, we can realize this as an insertion of the Fourier operator ${P}$. This operator can the be collided with the physical boundary condition giving a notion of coupling the relative physical boundary to a TFT.}
\label{fig:SymTFT}
\end{figure}

We now formulate the discussion above using SymTrees. The point of our discussion will be that the topological couplings we have described throughout this paper are different from those analyzed by Kapustin and Seiberg in \cite{Kapustin:2014gua}. Naively, one might have thought that the coupling described there can be recast as a Y-shaped SymTree with external nodes associated to the $\SU(2)$ theory and topological $(\Z_2)_p$ theory and the topological boundary condition.

\begin{figure}
\centering
\scalebox{0.7}{
\begin{tikzpicture}
	\begin{pgfonlayer}{nodelayer}
		\node [style=CirclePurple] (0) at (0, 0) {};
		\node [style=CircleBlue] (1) at (2, 2) {};
		\node [style=CircleBlue] (2) at (-2.5, 0) {};
		\node [style=CircleRed] (3) at (2, -2) {};
		\node [style=none] (4) at (2, 2.5) { ${P}$};
		\node [style=none] (5) at (2, -2.5) { $\mathcal{T}_{\mathfrak{su}_{2}}$};
		\node [style=none] (6) at (-2.5, 0.5) { $\mathcal{B}_{\textnormal{top}}$};
		\node [style=CirclePurple] (7) at (7.5, 0) {};
		\node [style=CircleBlue] (9) at (5, 0) {};
		\node [style=CircleRed] (10) at (9.5, -2) {};
		\node [style=none] (11) at (9.5, -2.5) { $\mathcal{T}_{\mathfrak{su}_{2}}$};
		\node [style=none] (12) at (5, 0.5) { $\mathcal{B}_{\textnormal{top}}$};
		\node [style=none] (13) at (8, 0.5) { ${P}$};
		\node [style=none] (19) at (0, -3.5) {\Large (i)};
		\node [style=none] (20) at (7.5, -3.5) {\Large (ii)};
		\node [style=none] (22) at (7.5, -4) {};
		\node [style=none] (23) at (7.5, -0.5) {};
		\node [style=none] (24) at (6.5, -0.5) {};
		\node [style=none] (25) at (2, 1.5) {};
		\node [style=none] (26) at (1.25, 0.75) {};
		\node [style=none] (27) at (-0.5, 0.5) {Gluing};
	\end{pgfonlayer}
	\begin{pgfonlayer}{edgelayer}
		\draw [style=ThickLine] (0) to (1);
		\draw [style=ThickLine] (0) to (3);
		\draw [style=ThickLine] (0) to (2);
		\draw [style=ThickLine] (7) to (10);
		\draw [style=ThickLine] (7) to (9);
		\draw [style=ArrowLineRight] (25.center) to (26.center);
	\end{pgfonlayer}
\end{tikzpicture}}
\caption{SymTree with two topological boundary conditions realizing the polarization change $SU(2)$ to $SO(3)_+$.}
\label{fig:SymTreeKapustinSeiberg}
\end{figure}
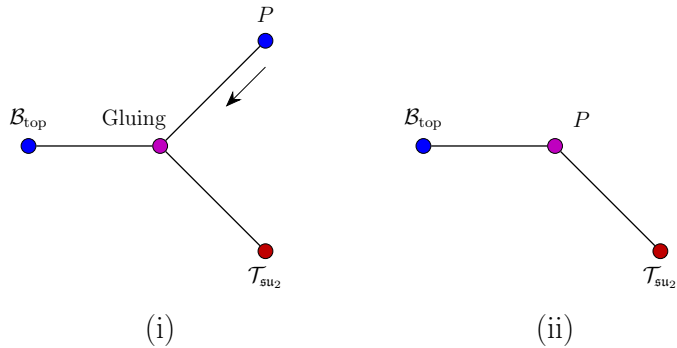

Starting from the central configuration in figure \ref{fig:SymTFT}, the SymTree is constructed by extracting a third edge which is glued trivially at the introduced trivalent junction (see figure \ref{fig:SymTreeKapustinSeiberg}). More precisely, the three edges $e$ we have the fields $B_{2}^{(2,e)}$ and at the junction they are all set pairwise equal, similarly for $C_{2}^{(2,e)}$. Figure \ref{fig:SymTreeKapustinSeiberg} then shows the deformation of this configuration back to the one appearing in figure \ref{fig:SymTFT}. The resulting edge supports the action
\be
S_{P}= \pi i \int\frac{ \mathcal{P}\!\lb B_{2}+C_{2}\rb}{2}
\ee
and upon retracting the third leg we revert to the configuration shown in figure \ref{fig:SymTFT} in the central subfigure.

\section{Single Derivative Terms in SymTFT Action}\label{app:Kineticterms}
The goal of this Appendix will be to derive the leading term in \eqref{eq:Sym7DTFT} from reducing the 11D M-theory kinetic term for the 3-form potential which we reproduce here
\begin{equation}\label{eq:Sym7DTFTREPEAT}
S_{\mathrm{8D}} = \frac{i}{2 \pi} N \int_{8D} B_{2} \wedge d C_{5}\,.
\end{equation}
This appears in the 8D topological action of the SymTFT of 7D $\mathfrak{su}_{N}$ SYM, as engineered from M-theory on $\mathbb{C}^2/\mathbb{Z}_N$.\footnote{It is straightforward to generalize to arbitrary $\Gamma_{ADE}\subset SU(2)$ but we will stick with $\Gamma=\mathbb{Z}_N$ for ease of exposition.} The key effect of line \eqref{eq:Sym7DTFTREPEAT} is that it creates a braiding algebra between the electric 1-form and magnetic 4-form symmetry operators, or equivalently, signifies a mixed 't Hooft anomaly between these symmetries.

We will perform this dimensional reduction in a similar fashion to \cite{Camara:2011jg} which represents torsional cocycles by non-harmonic differential forms.\footnote{For a more systematic derivation of similar BF-type terms in SymTFTs from reducing string/M-theory actions, see the upcoming work \cite{InakiSaghar}.} In particular, we can represent the generator of $\mathbb{Z}_N=H^2(S^3/\Gamma, \mathbb{Z})$ by a pair $(\alpha_2, \beta_1)$ where the 2-form $\alpha_2$ and 1-form $\beta_1$ obey
\begin{equation}\label{eq:torsionform}
  N\alpha_2=d\beta_1, \quad \quad \quad d^\dagger \beta_1=0.
\end{equation}
The M-theory fluxes can then be expanded along $(\alpha_2, \beta_1)$ as
\begin{align}
 & G_4=(dA_1+NB_2)\wedge \alpha_2+dB_2\wedge \beta_1 \\
  & G_7=(dA_4+NC_5)\wedge \alpha_2+dC_5\wedge \beta_1
\end{align}
This expands in the 11D kinetic term as
\begin{align}
 -2\pi i S_{\mathrm{11D}}&=\frac{1}{2}\int_{11D} G_4\wedge G_7 \\
  &=\frac{1}{2}\left(\int_{S^3/\mathbb{Z}_N}\alpha_2\wedge \beta_1 \right) \left( \int_{8D}dA_1\wedge dC_5+NB_2\wedge dC_5-dB_2\wedge dA_4-NdB_2\wedge C_5 \right)
\end{align}
where the ``8D" directions are the directions of the SymTFT are $\mathbb{R}_{+}\times M_7$, and the minus signs result from anticommuting $\beta_1$ through 5-forms. The two-derivative terms in the above expansion are not topological as $*_{8D}dC_5=dA_1$ and $*_{8D}dA_4=dB_2$ so these will not be of concern to us here.\footnote{At long distances the topological term dominates.} On the other hand, the single derivative terms are topological and after integrating by parts we have
\begin{equation}
  S_{\mathrm{8D}} = \frac{i}{2 \pi}\left(\int_{S^3/\mathbb{Z}_N}\alpha_2\wedge \beta_1 \right) \left(N \int_{8D} B_{2} \wedge d C_{5}\right)\,.
\end{equation}
The term $\int_{S^3/\mathbb{Z}_N}\alpha_2\wedge \beta_1$ is the cohomological version of the linking pairing of 1-cycles on $S^3/\mathbb{Z}_N$ and normalizing this integral to be $\equiv 1 \; \mathrm{mod}\; N$ reproduces line \eqref{eq:Sym7DTFTREPEAT}.

\section{SymTrees from ALE Spaces} \label{app:FiltrationsALE}

In this Appendix we discuss topological features of the filtration $\mathcal{F}_{X'}$ which sweeps out
\be \label{eq:PartialRes2}
X'\,:\quad x^2+y^2=(z-z_1)^{N_1}(z-z_2)^{N_2}\,,
\ee
as introduced in Section \ref{sec:FLOW}. We discuss the homology groups of the radial slices $U_r$ and their relationship across the critical slice as determined by the small radius (line \eqref{eq:seq1}) and large radius (line \eqref{eq:seq2}) Mayer-Vietoris sequences. Further, we dualize and lift to differential cohomology as relevant in the reduction of the 11D supergravity terms. We extend the SymTree analysis of Section \ref{sec:FLOW} and include generalized symmetries indicated by the ``...'' in the expansion of line \eqref{eq:DiffCoho}.

\subsection{Generators of $H_n(U_{r=r_*})$}
\begin{figure}
\centering
\scalebox{0.9}{
\begin{tikzpicture}
\begin{pgfonlayer}{nodelayer}
\node [style=none] (2) at (-2, 0) {};
\node [style=none] (3) at (2, 0) {};
\node [style=none] (4) at (-2, 0) {};
\node [style=none] (5) at (2, 0) {};
\node [style=none] (6) at (0, 3.5) {};
\node [style=none] (7) at (0, 2.75) {};
\node [style=none] (8) at (0, 4.25) {$S^3/\Z_{N_i}$};
\node [style=none] (9) at (0, -2.75) {$S^2$};
\node [style=none] (10) at (-2.75, 0) {$\partial \Sigma_i$};
\node [style=none] (11) at (0, -1) {$\Sigma_i$};
\node [style=none] (12) at (0, -3.25) {};
\end{pgfonlayer}
\begin{pgfonlayer}{edgelayer}
\draw [style=ThickLine, bend left=90, looseness=1.75] (2.center) to (3.center);
\draw [style=RedLine, bend right=15] (4.center) to (5.center);
\draw [style=DottedRed, bend left=15] (4.center) to (5.center);
\draw [style=RedLine, bend right=90, looseness=1.75] (4.center) to (5.center);
\draw [style=ArrowLineRight] (6.center) to (7.center);
\end{pgfonlayer}
\end{tikzpicture}}
\caption{Sketch of the Hopf-fibration $S^3/\Z_{N_i}\rightarrow S^2$ and the bounding chain $\Sigma_i$ within it. The Euler class $N_i \textnormal{vol}_{S^2}$ of this circle fibration characterizes the obstruction to the existence of a section. Consider attempting to construct such a section, as depicted, by starting at the south pole of $S^2$ and growing a disk inside of $S^3/\Z_{N_i}$, projecting to $S^2$ as shown. Upon reaching the North pole the boundary $\partial \Sigma_i$ does not close, rather it winds $N_i$ times around the Hopf fiber $S^1_H$. With this $\Sigma_i$ is a chain bounding $N_i$ copies of $S^1_H$. }
\label{fig:ChainVisual}
\end{figure}
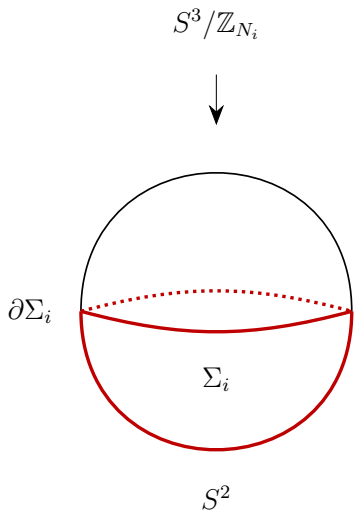

We begin by identifying the generators of the integral homology groups of the critical slice
\be
U_{r=r_*}=\lb S^3/\mathbb{Z}_{N_1}\rb\cup_{\;\!S^1_H}\lb S^3/\mathbb{Z}_{N_2}\rb
\ee
which are listed in line \eqref{eq:criticalslicehomo}. The bottom and top degree generators are clear. The generator in degree one is the common Hopf circle $S^1_H$. Next note, within each lens space there exists a chain $\Sigma_i$ such that
\be
\partial \Sigma_i = N_i S^1_H\,.
\ee
As a consequence $S^1_H$ is torsional, representing a class of order $g=\textnormal{gcd}(N_1,N_2)$. We move on to discuss the generator in degree two. For this note that we can glue multiples of the two chains $\Sigma_i$ to a 2-cycle of the critical slice:
\be \label{eq:gluing}
\Sigma\equiv \lb N_2\Sigma_1/g\rb\cup_{\;\!LS^1_H}  \lb -N_1\Sigma_2/g\rb .
\ee
Here $L=\textnormal{lcm}(N_1,N_2)$ and the sign is required for closure $\partial \Sigma=0$. To see that $\Sigma$ represents a free class we now compare this 2-cycle to the generator $e$ of $H_2(X')\cong\Z$. The cross-section of $\Sigma$ is $LS^1_H$ while the cross-section of $e$ is $S^1_H$. From this we conclude that mapping $\Sigma$ into $H_2(X')$ we have
\be \label{eq:Mult}
\Sigma=Le
\ee
or $\Sigma=0$. Clearly the embedding is not trivial. From the IIA dual we see that $\Sigma$, projected to $\R^{3}$, links both D6-brane stacks. See figure \ref{fig:ChainVisual} where we illustrate the 2-chains $\Sigma_i$ within the Hopf fibration  for $S^3/\Z_{N_i}$. Upon gluing this construction back to back as indicated in line \eqref{eq:gluing} we indeed find line \eqref{eq:Mult} holds.

\subsection{Small and Large Radius Mayer-Vietoris Sequences}

Next we determine how cycles contained within small\,/\,large radii slices deform to those of the critical slice. This data is carried by the maps within the small\,/\,large radius Mayer-Vietoris sequences whose associated coverings we now describe and which we then compute.

\begin{figure}
\centering
\scalebox{0.6}{
\begin{tikzpicture}
\begin{pgfonlayer}{nodelayer}
\node [style=none] (0) at (0, 1) {};
\node [style=none] (1) at (0, -1) {};
\node [style=none] (2) at (1.5, 2.5) {};
\node [style=none] (3) at (1.5, -2.5) {};
\node [style=none] (4) at (-1.5, 2.5) {};
\node [style=none] (5) at (-1.5, -2.5) {};
\node [style=none] (6) at (-1.5, -3) {\Large $S^2_1$};
\node [style=none] (7) at (1.5, -3) {\Large$S^2_2$};
\node [style=none] (8) at (-1.5, 3) {};
\node [style=none] (9) at (-1.5, 3.75) {};
\node [style=none] (10) at (1.5, 3) {};
\node [style=none] (11) at (1.5, 3.75) {};
\node [style=none] (13) at (1.5, 4.25) {\Large$S^3/\Z_{N_2}$};
\node [style=none] (14) at (-1.5, 4.25) {\Large$S^3/\Z_{N_1}$};
\node [style=none] (15) at (0.625, 0) {\Large$D^2$};
\node [style=none] (16) at (0, -3.5) {};
\node [style=Star] (17) at (2, 0) {};
\node [style=Star] (18) at (-2, 0) {};
\node [style=none] (19) at (-2, -0.75) {\Large $A_{N_1-1}$};
\node [style=none] (20) at (2, -0.75) {\Large$A_{N_2-1}$};
\end{pgfonlayer}
\begin{pgfonlayer}{edgelayer}
\draw [style=ThickLine, in=90, out=0] (4.center) to (0.center);
\draw [style=ThickLine, in=180, out=90] (0.center) to (2.center);
\draw [style=ThickLine, bend left=90, looseness=1.50] (2.center) to (3.center);
\draw [style=ThickLine, in=270, out=180] (3.center) to (1.center);
\draw [style=ThickLine, in=0, out=-90] (1.center) to (5.center);
\draw [style=ThickLine, bend right=270, looseness=1.50] (5.center) to (4.center);
\draw [style=BlueLine] (0.center) to (1.center);
\draw [style=ArrowLineRight] (9.center) to (8.center);
\draw [style=ArrowLineRight] (11.center) to (10.center);
\end{pgfonlayer}
\end{tikzpicture}}
\caption{Depict the covering of the large radius Mayer-Vietoris sequence with respect to the M-theory circle fibration. In $\R^3$, the IIA dual to $X'$, we have two spheres $S^2_i$ touching along a two-disk $D^2$ marked blue. The preimage of this disk and its complement are the large radius covering. The boundary of the preimage of the disk, which is the intersection of the two covering sets, is the circle $S^1_H$ fibered over the boundary of the disk.}
\label{fig:LargeRadius}
\end{figure}
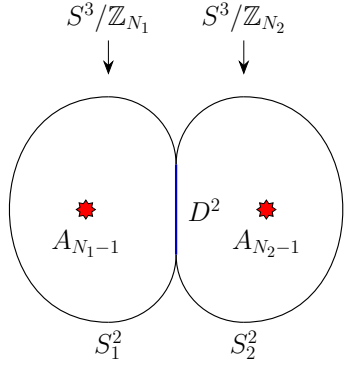

The covering of the critical slice associated with small radii is then given by the two patches  $S^3/\mathbb{Z}_{N_1},S^3/\mathbb{Z}_{N_2}$ which intersect in ${S}^1_H$. The large radii covering has patches $S^3/\mathbb{Z}_{N}$ and the tubular neighbourhood
\be
{S}^1_H~\hookrightarrow~T({S}^1_H)~\rightarrow~D^2
\ee which is a solid torus fibered over a disk $D^2$. These patches intersect along the torus ${\rm T}^2=\partial\;\! T({S}^1_H)$ which is a circle fibration over the circle $\partial D^2$ (see figure \ref{fig:LargeRadius}).

With respect to these decompositions the long exact Mayer-Vietoris sequences are
\begin{equation}\label{eq:MV1}
    \begin{array}{c||ccccccc}
&   H_n(S^1_H)  &  & H_n(S^3/\Z_{N_1}) \oplus H_n(S^3/\Z_{N_2}) &  &  H_n(U_{r=r_*}) &   &   \\[0.4em] \hline \hline \\[-0.9em]
      (n=3) ~&   0  & \rightarrow & \mathbb{Z}\oplus \mathbb{Z}& \rightarrow &  \mathbb{Z}^2 & \rightarrow & \\
         (n=2)~ &  0  &\rightarrow &0\oplus 0 & \rightarrow& \mathbb{Z} &\rightarrow & \\
       (n=1)~ &    \mathbb{Z}  &\rightarrow & \mathbb{Z}_{N_1}\oplus  \mathbb{Z}_{N_2}&\rightarrow & \mathbb{Z}_g &\rightarrow & \\
        (n=0) ~&   \mathbb{Z}  &\rightarrow & \mathbb{Z}\oplus\mathbb{Z} &\rightarrow & \mathbb{Z}  & \rightarrow & 0
    \end{array}
\end{equation}
at small radii and
\begin{equation}\label{eq:MV2}
    \begin{array}{c||ccccccc}
&   H_n({\rm T}^2)  &  & H_n(T(S^1_H)) \oplus H_n(S^3/\Z_{N}) &  &  H_n(U_{r=r_*}) &   &   \\[0.4em] \hline \hline \\[-0.9em]
      (n=3) ~&   0  & \rightarrow & \,0\oplus \mathbb{Z}& \rightarrow &  \mathbb{Z}^2 & \rightarrow  & \\
         (n=2)~ &  \mathbb{Z}  &\rightarrow & 0\oplus 0 & \rightarrow& \mathbb{Z} &\rightarrow &  \\
       (n=1)~ &   \mathbb{Z}^2  &\rightarrow & \;~\:\! \mathbb{Z}_{}\oplus  \mathbb{Z}_{N}&\rightarrow & \mathbb{Z}_g &\rightarrow &  \\
        (n=0) ~&   \mathbb{Z}  &\rightarrow &\mathbb{Z}\oplus\mathbb{Z} &\rightarrow & \mathbb{Z}  & \rightarrow & 0
    \end{array}
\end{equation}
 at large radii. We remark that the most relevant part of the above sequences is summarized, respectively, in the following two exact subsequences
\begin{align}\label{eq:ES}
    \begin{array}{ccccccccccc}
       0  & \rightarrow & \mathbb{Z}&  \xrightarrow[]{~\partial_2^{\tiny(r<r_*)}~}  &  \mathbb{Z} & \xrightarrow[]{~\imath_1^{\tiny(r<r_*)}~}  &  \mathbb{Z}_{N_1}\oplus \mathbb{Z}_{N_2} & \xrightarrow[]{~\jmath_1^{\tiny(r<r_*)}~}  & \mathbb{Z}_{g}  &\rightarrow  &0\\0  & \rightarrow & \mathbb{Z}&  \xrightarrow[]{~\partial_2^{\tiny(r>r_*)}~}  &  \mathbb{Z} & \xrightarrow[]{~\imath_1^{\tiny(r>r_*)}~}  &  \mathbb{Z}_{N_1+N_2}& \xrightarrow[]{~\jmath_1^{\tiny(r>r_*)}~}   & \mathbb{Z}_{g}  &\rightarrow&0
    \end{array}
\end{align}
where the index $p$ marks maps mapping from a domain of $p$-cycles and the exponent labels the Mayer-Vietoris sequence the subsequence was extracted from.

We now discuss the maps featuring in these subsequences. First, recall that the boundary map in the Mayer-Vietoris sequence is defined by cutting a cycle along the intersection of the two covering components and then considering one of the resulting halves. The initial cycle is then mapped to the boundary of one of its `halves'. With this the map $\partial_2^{\tiny(r<r_*)}$ maps $\Sigma$ onto its cross-section and, as the codomain of $\partial_2^{\tiny(r<r_*)}$ is generated by $S^1_H$ and considering line \eqref{eq:Mult}, therefore is multiplication by $L$.

In order to characterize $\partial_2^{\tiny(r>r_*)}$ let us consider the torus ${\rm T}^2=\partial\:\! T({S}^1_H)$ and denote its one cycles by $S^1_H$ and $\partial D^2=\beta_1$. We therefore have $\partial \Sigma_1=N_1 S^1_H +\beta_1$ and $\partial \Sigma_2=N_2S^1_H -\beta_1$. The two halves of $\Sigma$ overlap in $\mathrm{T}^2$ and therefore
\be
\partial_2^{\tiny(r>r_*)}\Sigma=(N_2/g)\partial \Sigma_1-(N_1/g)\partial \Sigma_2=(N/g)\beta_1\,.
\ee
Similar consideration result in noting that $\beta_1$ generates the codomain of $\partial_2^{\tiny(r>r_*)}$ which is therefore multiplication by $N/g=(N_1+N_2)/g$.

All remaining $n=1$ homology groups are generated by the obvious Hopf circles and in obvious, yet slightly redundant, notation we rewrite \eqref{eq:ES} as
\be\begin{aligned}\label{eq:ES2}
    \begin{array}{ccccccccccc}
       0  & \rightarrow & \langle \Sigma \rangle &  \xrightarrow[]{~1\;\!\mapsto\;\! L~}  &  \langle {\rm \alpha} \rangle  & \xrightarrow[]{~1;\!\mapsto\;\! (1,-1)~}  &  \langle (\mathrm{S}_H^1)_1\rangle\oplus\langle (\mathrm{S}_H^1)_2\rangle & \xrightarrow[]{~\textnormal{mod}\,(1,-1)~}  & \langle \mathrm{S}_H^1\rangle  &\rightarrow  &0\\0  & \rightarrow & \langle \Sigma \rangle &  \xrightarrow[]{~1\;\!\mapsto\;\! N/g~}  &  \langle \beta \rangle  & \xrightarrow[]{~1\;\!\mapsto\;\! g~}  &  \langle  (\mathrm{S}_H^1)_{12} \rangle & \xrightarrow[]{~\textnormal{mod}\,g~}   & \langle \mathrm{S}_H^1\rangle &\rightarrow&0\\[0.5em]
 \end{array}
 \end{aligned} \ee
 where $\alpha=S^1_H$.

\subsection{The Extension Problem}
\label{sec:GeoExt}
Now we turn to an extension problem, discussed around lines \eqref{eq:Ext1} and \eqref{eq:Ext2}. From figure \ref{fig:ChainVisual} and related discussion it follows that $\beta$ is a multiple of $\alpha$. Also note (see Section \ref{sec:FLOW}) that we have identified the $\U(1)$ localized to the critical slice as\footnote{As elsewhere, here $G^\vee=\textnormal{Hom}(G,\U(1))$ denotes the Pontryagin dual of an abelian group $G$.}
\be
\langle \Sigma \rangle ^\vee =\Z^\vee=\U(1)\,,
\ee
while various 1-form symmetry background fields on the edges of the SymTree attaching to the junction are related to the homology groups that $\alpha,\beta$ are mapped into. The relevance of the refinement of $\langle \beta \rangle$ into $\langle \alpha \rangle $ lies in noting that in the coupling of the junction $\U(1)$ to the edges of the SymTree only runs via the subgroup
\be
\Z_{LN/g} \subset\U(1)\,.
\ee
More precisely, as we explain later, we are permitted to interpret the fields on the SymTree edges as background fields for the relative $\U(1)$ junction theory which take values in this $ \Z_{LN/g} $ subgroup. Also the preferred nesting of groups is
\be \label{eq:extension}
0~\rightarrow~\Z_{L}~\rightarrow~\Z_{LN/g}~\rightarrow~\Z_{N/g}~\rightarrow~0\,,
\ee
where Pontryagin duality has flipped the arrow, i.e., $\langle \beta \rangle^\vee $ is a refinement of $\langle \alpha \rangle^\vee$ and the preferred subgroup of $\Z_{LN/g}\subset\U(1)$ is $\Z_{L}\subset\U(1)$.

\subsection{Differential Cohomology Uplift}

Homology groups of internal M-theory dimensions carry geometric intuition. The reduction of 11D supergravity however proceeds via expansions in differential cohomology classes. We now discuss how to move from homology to differential cohomology and discuss the reduction of the topological 11D SUGRA Chern-Simons terms on the radial shells.

First, note that, while the critical slice $U_{r=r_*}$ is not a manifold, it is a finite CW complex. We hence have, via the universal coefficient theorem, the cohomology groups
\be  \label{eq:criticalslicecohomo}
H^n(\lb S^3/\mathbb{Z}_{N_1}\rb\cup_{\;\!S^1_H}\lb S^3/\mathbb{Z}_{N_2})\rb\cong  \begin{cases} \Z\cong \langle 1_0 \rangle  & k=0 \\ 0& k=1 \\  \Z\oplus \Z_g \cong \langle u_2 \rangle \oplus \langle t_2 \rangle  \qquad\quad & k=2 \\ \Z^2 \cong \langle \textnormal{vol}^{(1)}_3,\textnormal{vol}^{(2)}_3  \rangle    &k=3 \end{cases}
\ee
which are lifted to the differential cohomology classes
\be
\breve{1}_0, \breve{u}_2, \breve{t}_2, \breve{\textnormal{v}}\textnormal{ol}_3^{(1)},\breve{\textnormal{v}}\textnormal{ol}_3^{(2)}
\ee
which with respect to the projection $\pi : \breve{H}^*\rightarrow H^*$ of the short exact sequence of line \eqref{eq:DiffCohoRelations} satisfy
\be
\pi(\breve{1}_0)=1_0\,, \quad  \pi( \breve{u}_2)=u_2\,, \quad \pi( \breve{t}_2)=t_2\,,\quad \pi(\breve{\textnormal{v}}\textnormal{ol}_3^{(i)} )=\textnormal{v}\textnormal{ol}_3^{(i)}
\ee
where $i=1,2$. Similarly, for lens spaces, we have
\be  \label{eq:criticalslicecohomo2}
{H}^n\! \lb  S^3/\mathbb{Z}_{K}\rb \cong  \begin{cases} \Z \cong \langle {1}_0 \rangle  & k=0 \\ 0& k=1 \\  \Z_K  \cong  \langle {t}_2 \rangle \qquad\quad & k=2 \\ \Z \cong \langle  {\textnormal{v}}\textnormal{ol}_3 \rangle  &k=3 \end{cases}
\ee
where $K=N$ for $r>r_*$ and $K=N_1,N_2$ for $r<r_*$ along the three SymTree edges. These are lifted
analogously to differential cohomology as in line \eqref{eq:DiffCohoRelations}.

Whenever it is necessary to resolve the redundancy in the notation of lines \eqref{eq:criticalslicecohomo} and \eqref{eq:criticalslicecohomo2} we append a raised label clarifying which of the SymTree edges we are referring to, for example
\be
\breve{t}_2^{\,(r<r_*),1},\breve{t}_2^{\,(r<r_*),2},\breve{t}_2^{\,(r>r_*)}
\ee
refer to the differential cohomology classes associated with the two small radius edges $r<r_*$ and large radius edge $r>r_*$ of the Y-shaped graph $\Upsilon$ supporting the SymTree.

All differential cohomology classes relevant in the KK reduction are the uplift of integral singular cohomology classes and as such are related to and across the critical slice via mappings which are dual to those appearing in lines \eqref{eq:MV1} and \eqref{eq:MV2}, i.e., the mappings appearing in the respective Mayer-Vietoris cohomology sequences. The boundary map $\partial$ dualizes to the coboundary map. All other maps are embeddings and dualize to restrictions for cocycles.

\subsection{11D SUGRA Reduction}

Next we turn to the KK reduction of the topological 11D SUGRA terms along the differential cohomology classes lifted from lines \eqref{eq:criticalslicecohomo} and \eqref{eq:criticalslicecohomo2}. First, we determine the bulk fields of the SymTree. This extends line \eqref{eq:DiffCoho}. Then we determine the SymTree action governing their interactions. It is in the latter part where the uplift to differential cohomology bears fruit in the computation of anomaly coefficients.

The fields on a branch of the SymTree are determined by KK reduction of the field strength $\breve{G}_4$ over the associated radial shell. For the lens space shells we have
\be\label{eq:G4expansion}
\breve{G}_4=\breve{H}^{(\mathrm{inst})}_{4}\!\star \breve{1}_0+\breve{B}_{2}^{(K)}\!\star \breve{t}_2+\breve{H}_{1}^{}\star \breve{\textnormal{v}}\textnormal{ol}_3+\dots
\ee
and the coefficients are the SymTree fields on the 8D branches of the SymTree. Here we have suppressed an additional label of the fields denoting the branch of the SymTree these live on, we shall add it as a raised index later. We also normalize such that $G_4$ has integral periods. For the small radius branches which attach to the physical boundaries supporting 7D SYM edge modes, one has the interpretations:
\begin{itemize}
\item  $\breve{H}^{\mathrm{inst}}_{4}$\,: continuous 4-form field strength, associated with a 3-form $\U(1)$ SymTree gauge potential $C_{3}^{\mathrm{inst}}$ and restricting to the background of a 2-form symmetry on the physical boundaries which is there interpreted as 2-form instanton symmetry.
\item $\breve{H}_{1}^{}$\,: continuous 1-form field strength, associated with a 0-form $\U(1)$ SymTree gauge potential $C_{0}^{\mathrm{inst}}$. The associated parameter is
\be
\theta = \int_{S^3/\Z_K} C_3
\ee
where $C_3$ is the 11D SUGRA 3-form potential. We return to the physical interpretation of this parameter after discussing the other fields.
\item $\breve{B}_{2}^{(K)}$\,: discrete 2-form gauge field with $\mathbb{Z}_K$ values. This field restricts on the boundary to the background potential for a 1-form symmetry on the physical boundaries which is there interpreted as the center symmetry.
\end{itemize}
The ``...'' omissions refer to an expansion along classes of $\Omega^{p-1}/ \Omega^{p-1}_\Z$, which lie in the kernel of the projection $\pi:\breve{H}^p\rightarrow H^p$ resulting in $\U(1)$ valued fields in 8D which do not couple to background fields for the discrete symmetries of the 7D SYM relative theories.

The one outlier in this discussion is the appearance of a continuous parameter coming from $\breve{H}_{1}^{}$, and its associated 0-form potential in the bulk 8D theory. This would suggest the appearance of an 8D topological term of the schematic form $\theta \breve{H}^{\mathrm{inst}}_{4} \cup \breve{H}^{\mathrm{inst}}_{4}$. Restricting to the 7D worldvolume, this would descend to a Chern-Simons-like theory of the form:
\begin{equation}
S_{7D,CS} = \frac{i}{4 \pi} \theta \int {C_3} \wedge d C_3,
\end{equation}
for the background 3-form associated with the 2-form symmetry. For $\theta \notin \mathbb{Z}$,
this would result in a theory with an improperly quantized level, i.e., it cannot be defined
independent of the 8D bulk.

Now we turn to discuss the TFT interactions between these fields which are determined via reduction of the 11D supergravity Chern-Simons term
\be
\frac{2\pi i}{6} \int \breve{G}_4\star  \breve{G}_4 \star  \breve{G}_4
\ee
over the lens space shells. Inserting the expansions of line \eqref{eq:G4expansion} this results in (see reference \cite{Apruzzi:2021nmk}):
\be
\breve{S}_{\:\!\mathcal{I},K}^{(\textnormal{anomaly})}=\pi i \int_{M_7\times \mathcal{I}}  \breve{H}^{(\mathrm{inst})}_{4}\star \breve{H}^{(\mathrm{inst})}_{4}\star\breve{H}_{1}^{}- 2\pi i  \frac{K-1}{2K}\int_{M_7\times \mathcal{I}}  \breve{H}^{(\mathrm{inst})}_{4}\star\breve{B}_{2}^{(K)}\star\breve{B}_{2}^{(K)}\,,
\ee
with spacetime $M_7$ and interval $\mathcal{I}=(r,r_*)$ for the case $K=N_1,N_2$ or $\mathcal{I}=(r_*,\infty)$ for the case $K=N$, specifying mixed anomalies. Along each branch of the SymTree we also have the one-derivative action
\be
S_{\:\!\mathcal{I},K}^{(0)}=  \frac{2\pi i}{K}  \int_{M_7\times \mathcal{I}} B_{2}^{(K)} \cup \delta C_{5}^{(K)}\,,
\ee
whose derivation is discussed in Appendix \ref{app:Kineticterms}.
The TFT action associated with one edge $e$ of the SymTree with internal radial shells $S^3/\Z_K$ is then
\be
{S}_{\:\!\mathcal{I},K}=S_{\:\!\mathcal{I},K}^{(0)}+S_{\:\!\mathcal{I},K}^{(\textnormal{anomaly})}+\dots
\ee
where we have projected back down to integral singular cohomology, replacing the star product $\star$ with the cup product $\cup$ in the process. When truncating to the indicated two terms we are describing the TFT associated with discrete symmetry structures in a background of continuous symmetry structures. Of course the full TFT does not distinguish between discrete and continuous, however we postpone details of this to future work.

Overall the full SymTree action takes the form
\be
S_{\Upsilon}={S}_{(0,r_*),N_1}+{S}_{(0,r_*),N_2}+{S}_{(r_*,\infty),N}+{S}_{(r=r_*),\:\!\mathcal{J}}\,.
\ee
where the last term describes a possibly non-topological relative junction theory which we are yet to determine. In addition to the junction action we are now required to supplement the overall action with boundary conditions for the bulk fields at the junction. The junction degrees of freedom crucially enter these boundary conditions, we therefore determine these first and then solve for the boundary conditions via geometry.

First, we determine the fields localized to the junction. For this, similar to line \eqref{eq:G4expansion},
we expand $\breve{G}_4$ in the classes of line \eqref{eq:criticalslicecohomo} of the critical slice resulting in
\be\label{eq:CriticalSliceFields}
\breve{G}_4\Big |_{r=r_*}=\breve{H}_{4}^{(r=r_*,\mathrm{inst})}\star \breve{1}_0+\breve{B}^{(r=r_*)}_{2}\star \breve{t}_2+\breve{F}^{(r=r_*)}_{2,\U(1)}\star \breve{u}_2+\breve{H}^{(r=r_*,1)}_{1}\star \breve{\textnormal{v}}\textnormal{ol}_3^{(1)}+\breve{H}_{1}^{(r=r_*,2)}\star \breve{\textnormal{v}}\textnormal{ol}_3^{(2)}+\dots\,.
\ee
The first set of boundary conditions derives by determining which of the coefficient fields arises as restrictions from fields on the edges of the SymTree to the junction. For this we need to relate the internal legs of differential cohomology class across the junction. In homology this amounts to studying how the dual cycles of the edges embed into the critical slice of the junction. For example the mappings ${\jmath}^{\,(r<r_*)}_1,{\jmath}^{\,(r>r_*)}_2$ of line \eqref{eq:ES} immediately gives the gluing conditions of line \eqref{eq:glue}. Similar embeddings straightforwardly give
\be
\breve{H}_{4}^{(r<r_*,1,\mathrm{inst})}\Big |_{r=r_*}=\breve{H}_{4}^{(r<r_*,2,\mathrm{inst})}\Big |_{r=r_*}=\breve{H}_{4}^{(r>r_*,\mathrm{inst})}\Big |_{r=r_*}=\breve{H}_{4}^{(r=r_*,\mathrm{inst})} \,,
\ee
\be
\breve{H}^{(r<r_*,1)}_{1}\Big |_{r=r_*}=\breve{H}_{1}^{(r=r_*,1)}\,, \quad
\breve{H}^{(r<r_*,2)}_{1}\Big |_{r=r_*}=\breve{H}_{1}^{(r=r_*,2)}\,, \quad
\breve{H}_{1}^{(r>r_*)}\Big |_{r=r_*}=\breve{H}_{1}^{(r=r_*,1)}=\breve{H}_{1}^{(r=r_*,2)}
 \ee
and we note that all coefficient fields of line \eqref{eq:CriticalSliceFields} are thus fixed by bulk fields of the SymTree except for the $\U(1)$ 2-form background fields. In order the raised indices here give the radii the fields live on, the connected component of this radial slice if it is disconnected and additional physical qualifiers. Next, following the discussion in Section \ref{sec:TOPDOWN}, we have the images
\be\ba
\textnormal{Im}\, \overline{\jmath}_{\;\!r<r_*}^{(0)}&\cong \Z\oplus \Z\,,\qquad && \textnormal{Im}\,{\overline{\jmath}}_{\;\!r>r_*}^{(0)} \cong \Z\,, \\
\textnormal{Im}\, \overline{\jmath}_{\;\!r<r_*}^{(2)}&\cong \Z_g\subset \Z_{N_1}\oplus \Z_{N_2}\,, \qquad && \textnormal{Im}\,\overline{\jmath}_{\;\!r>r_*}^{(2)}\cong \Z_g\subset \Z_N\,,\\
\textnormal{Im}\, \overline{\jmath}_{\;\!r<r_*}^{(3)}&\cong \Z\oplus \Z\,, \qquad && \textnormal{Im}\,\overline{\jmath}_{\;\!r>r_*}^{(3)}\cong \Z\,,
\ea \ee
with trivial cokernels in degree 0 and 3. Here notation is such that ${\jmath}_{(p)},\overline{\jmath}^{(p)},\breve{\jmath}^{\;\!(p)}$ respecitvely denote the embedding map in homology in the large radius and small radius Mayer-Vietoris sequences (which we distinguish by an additional lowered or raised index), the restriction map in the cohomology version of these sequences and the uplift of the latter to differential cohomology. We further have
\be
\textnormal{Ker}\, \overline{\jmath}_{\;\!r<r_*}^{(2)}\cap \textnormal{Ker}\,{\overline \jmath}_{\;\!r>r_*}^{(2)}\cong \Z\,,
\ee
generated by $u_2$ identifying $\breve{F}^{(r=r_*)}_{2,\U(1)}$ as a junction degree of freedom, i.e., it does not arise as restrictions of external bulk fields and is free to fluctuate.

Having identified the junction edge modes, we now view these as a relative theory with respect to the three SymTree edges attaching to the junction. We are thus required to give an interpretation of the bulk fields as background fields for the junction edge modes.
This results in lines \eqref{eq:Id1} and \eqref{eq:Id2}.

To proceed note that in restricting all the 2-form fields $B_2^{(K)}$, valued in $\Z_K$ and supported on the three SymTree edges, to the junction we have set boundary conditions for a subgroup
\be
\Z_g^2\subset \Z_{N_1}\oplus \Z_{N_2}\oplus \Z_N
\ee
via the gluing conditions of line \eqref{eq:glue}. The three $B_2^{(K)}$ fields are only fixed relative to each other, hence only $\Z_g^2$ is eaten up by the conditions, rather than $\Z_g^3$. The extension problem discussed in Appendix \ref{sec:GeoExt} now implies that the remaining B-field profiles in
\be \label{eq:quotientnlg}
 \frac{\Z_{N_1}\oplus \Z_{N_2}\oplus \Z_N}{\Z_g^2}
\ee
are to be interpreted as 2-form fields associated with $\U(1)$ backgrounds in the junction taking values in $\Z_{NL/g}$, the central entry in line \eqref{eq:extension}. The quotient of line \eqref{eq:quotientnlg} is precisely $\Z_{NL/g}=\Z_{N_1N_2N/g^2}$. The quotient of line \eqref{eq:quotientnlg} also clearly gives map from bulk fields to background field for the $\U(1)$ junction theory, it is simply the quotient map itself
\be
 Q\,: \qquad \Z_{N_1}\oplus \Z_{N_2}\oplus \Z_N\rightarrow \frac{\Z_{N_1}\oplus \Z_{N_2}\oplus \Z_N}{\Z_g^2}\,.
\ee
Working out the quotient for elements $(n_1,n_2,n)\in  \Z_{N_1}\oplus \Z_{N_2}\oplus \Z_N$ with one non-vanishing entry we find lines \eqref{eq:Id1} and \eqref{eq:Id2}.

\section{Sequences for Isolated Multi-Sector QFTs} \label{app:SEQISOLATED}

In this Appendix we supply additional details on the isolated multi-sector QFTs analyzed in Section \ref{sec:ISOLATED}.

\subsection{Sequences for 7D Models}

We now discuss the filtration $\mathcal{F}_X$ with radial shells of Section \ref{sssec:FILTRUM} in greater detail. Recall that we are considering elliptic local K3 surfaces $X\rightarrow B$ with base $B=\mathbb{C}$. As such we will repeatedly encounter three-manifolds $\Sigma_3^M$ which are smooth torus bundles
\be
T^2 \hookrightarrow \Sigma_3^M \rightarrow S^1
\ee
over a circle subject to a monodromy twist. For our purposes only the action of this monodromy on the $n$-th homology lattice of the torus fiber will be relevant, and we denote it by $M_n$. The homology groups of such a three-manifold derive from the exact sequence
\be
0~\rightarrow~\textnormal{Coker}(M^{(n)}-1)~\rightarrow~H_n(\Sigma_3^M)~\rightarrow~\textnormal{Ker}(M^{(n-1)}-1)~\rightarrow~0\,.
\ee
The monodromies we consider are such that $M^{(0)}=M^{(2)}=1$ and $M^{(1)}\equiv M\in \mathrm{SL}(2,\Z)$. From this sequence we derive the homology groups \eqref{eq:noncriticalslicehomo}. Let us discuss the generators of these groups. Top and bottom homology classes are clear. Consider $H_1(\Sigma_3^M)\cong \Z \oplus \textnormal{Coker}(M-1)$. The factor of $\Z$ admits a simple representative, the base circle $S^1$. Now consider a 1-cycle $\gamma_1\in H_1(T^2)$ of the elliptic fiber. Now trace out a 2-chain by transporting it once around the circle base transforming as
\be\label{eq:Monodromy}
\gamma_1\rightarrow M\gamma_1=(M-1)\gamma_1+\gamma_1
\ee
considering orientations the initial and final copied of $\gamma_1$ cancel against each other, the 2-chain has boundary $(M-1)\gamma_1$. This explains the contribution $ \textnormal{Coker}(M-1)$. The homology generators in degree 2 follow by duality, they are the fiber class and any eigenvector of $M$ fibered over the base circle.

Let us next compute the homology groups of the critical slice
\be
  U_{r= r_*}=\lb \Sigma_3^{M_1} \rb\cup_{\;\!T^2}\lb \Sigma_3^{M_2}\rb\,.
\ee
Again top and bottom homology classes are clear. Note that $U_{r= r_*}$ is fibered over a figure eight $S^1\vee S^1$ and hence the factor of $\Z^2$ in
\be
H_1(U_{r= r_*})=\Z^2\oplus [\Z^2 /\textnormal{Im}(M_1-1,M_2-1)]
\ee
are represented by the two base circles. Constructing 2-chains by transporting 1-cycles of the elliptic fiber around these circles as above then gives the quotient contribution. The degree two cycles\footnote{Given an abelian group $G$ we define a dual group as $G^\wedge = \textnormal{Hom}(G,\Z)$.}
\be
H_2(U_{r= r_*})= \Z \oplus \textnormal{coker}(M_1-1)^\wedge \oplus \textnormal{coker}(M_2-1)^\wedge\oplus  F
\ee
are respectively the fiber, monodromy eigen-1-cycles fibered over any one of the base circles and a case dependent contribution $F=0,\Z,\Z^2$ of a fiber 1-cycle fibered over the full figure eight base. From these considerations the restriction to line \eqref{eq:singlelegs} parameterizing the gauge theory data is clear.

Next we determine how cycles contained within small\,/\,large radii slices deform to those of the critical slice, studying the small\,/\,large radius Mayer-Vietoris sequences whose associated coverings we now describe and which we then compute.

The covering of the critical slice associated with small radii is then given by the two patches  $\Sigma_3^{M_1},\Sigma_3^{M_2}$ which intersect in $T^2$. The large radii covering has patches $\Sigma_3^{M_1M_2}$ and the cylinder $I\times T^2$. Growing the base circles the touch along a point and then an interval, the latter cylinder are simply all fibers projecting to the interval, see figure \ref{fig:Covering2}. These patches intersect along 2 tori ${\rm T}^2\sqcup {\rm T}^2$.

With respect to these decompositions the long exact Mayer-Vietoris sequences are
\begin{equation}\label{eq:MV1Ell}
    \begin{array}{c||ccccccc}
&   H_n(T^2)  &  & H_n(\Sigma_3^{M_1}) \oplus H_n(\Sigma_3^{M_2}) &  &  H_n(U_{r=r_*}) &   &   \\[0.4em] \hline \hline \\[-0.9em]
      (n=3) ~&   0  & \rightarrow & \mathbb{Z}\oplus \mathbb{Z}& \rightarrow &  \mathbb{Z}^2 & \rightarrow & \\
         (n=2)~ &  \mathbb{Z}  &\rightarrow &\Z\oplus C_1^\wedge \oplus \Z\oplus C_2^\wedge & \rightarrow& \Z \oplus C_{1}^\wedge\oplus C_2^\wedge \oplus F&\rightarrow & \\
       (n=1)~ &    \mathbb{Z}^2  &\rightarrow &\Z\oplus C_1 \oplus  \Z\oplus C_2 &\rightarrow & \Z^2 \oplus C_{1,2}&\rightarrow & \\
        (n=0) ~&   \mathbb{Z}  &\rightarrow & \mathbb{Z}\oplus\mathbb{Z} &\rightarrow & \mathbb{Z}  & \rightarrow & 0
    \end{array}
\end{equation}
at small radii, where we abbreviated $C_i=\textnormal{Coker}(M_i-1)$ and $C_{1,2}=\Z^2/\textnormal{Im}(M_1-1,M_2-1)$, and
\begin{equation}\label{eq:MV2Ell}
    \begin{array}{c||ccccccc}
&   H_n({\rm T}^2\sqcup T^2)  &  & H_n(T^2) \oplus H_n(\Sigma_3^{M_1M_2}) &  &  H_n(U_{r=r_*}) &   &   \\[0.4em] \hline \hline \\[-0.9em]
      (n=3) ~&   0  & \rightarrow & \,0\oplus \mathbb{Z}& \rightarrow &  \mathbb{Z}^2 & \rightarrow  & \\
         (n=2)~ &  \mathbb{Z}  \oplus \Z &\rightarrow & \Z \oplus \Z \oplus C_{12}^\wedge & \rightarrow&\Z \oplus C_{1}^\wedge\oplus C_2^\wedge \oplus F&\rightarrow &  \\
       (n=1)~ &   \mathbb{Z}^2 \oplus \mathbb{Z}^2  &\rightarrow & \;~\:\! \mathbb{Z}_{}^2\oplus  \mathbb{Z}\oplus C_{12}&\rightarrow &  \Z^2 \oplus C_{1,2}&\rightarrow &  \\
        (n=0) ~&   \mathbb{Z} \oplus \mathbb{Z}  &\rightarrow &\mathbb{Z}\oplus\mathbb{Z} &\rightarrow & \mathbb{Z}  & \rightarrow & 0\\[0.3em]
    \end{array}
\end{equation}
 at large radii, where we abbreviated $C_{12}=\textnormal{Coker}(M_1M_2-1)$. We remark that the most relevant part of the above sequences is summarized, respectively, in the following two exact subsequences
\begin{align}\label{eq:ESEll}
 \scalebox{0.95}{$
    \begin{array}{ccccccccccccc}
  & &      0  & \rightarrow &F &  \xrightarrow[]{~\partial_2^{\tiny(r<r_*)}~}  &  \mathbb{Z}^2 & \xrightarrow[]{~\imath_1^{\tiny(r<r_*)}~}  &  C_1\oplus C_2& \xrightarrow[]{~\jmath_1^{\tiny(r<r_*)}~}  & C_{1,2} &\rightarrow  &0\\ 0 & \rightarrow & C_{12}^\wedge  &  \xrightarrow[]{~\jmath_2^{\tiny(r>r_*)}~}  & C_1^\wedge \oplus C_2^\wedge \oplus F &  \xrightarrow[]{~\partial_2^{\tiny(r>r_*)}~}  &  \mathbb{Z}^2 & \xrightarrow[]{~\imath_1^{\tiny(r>r_*)}~}  & C_{12} & \xrightarrow[]{~\jmath_1^{\tiny(r>r_*)}~}   & C_{1,2}  &\rightarrow&0 \\[0.3em]
    \end{array}$}
\end{align}
where the index $p$ marks maps mapping from a domain of $p$-cycles and the exponent labels the Mayer-Vietoris sequence the subsequence was extracted from. These subsequences follow by explicit considerations from having identified the generators above. The sequence \eqref{eq:ESEll} should be compared  \eqref{eq:ES2} which forms the starting point for the SymTree discussion laid out in Appendix \ref{sec:GeoExt}.

Instead of laying out this very similar analysis we demonstrate how to make direct contact with previous results. As expected this can be achieved by considering two mutually local singularities of Kodaira type $I_{N_1},I_{N_2}$, respectively with monodromy matrices
\be
M_1 =\lb \begin{array}{cc} 1 & N_1  \\ 0  & 1 \end{array}
\rb \,, \qquad M_2 =\lb \begin{array}{cc} 1 & N_2  \\ 0  & 1 \end{array}
\rb\,.
\ee
This gives
\be\label{eq:Example} \ba
C_i&\cong \Z \oplus \Z_{N_i}\,, && C_{12}\cong \Z \oplus \Z_{N_1+N_2}\,, && C_{1,2} \cong \Z\oplus \Z_{g} \\
C_i^\wedge &\cong \Z \,, && C_{12}^\wedge \cong  \Z \,, && C_{1,2}^\wedge \cong \Z\,.
\ea \ee
Further there is a single 1-cycle of the fiber which pinches at both singularities and hence we expect $F\cong \Z$, corresponding to the compact 2-cycle obtained by fibering that 1-cycle between the singularities. However, the generator of $F$ is constructed from the 2-cycles constructed around \eqref{eq:Monodromy} in much the same way as \eqref{eq:gluing}, this reproduces \eqref{eq:Mult}. Inserting \eqref{eq:Example} into \eqref{eq:ESEll}, we find an exact subsequence $0\rightarrow \Z\rightarrow\Z^2\rightarrow\Z\rightarrow 0$ which we can remove in both sequences as well as an additional subsequence  $0\rightarrow \Z\rightarrow\Z\rightarrow 0$ in the second sequence. Once these trivial parts are cut we reproduce \eqref{eq:ES2}.

\subsection{Homology groups for $X = (T^2\times \C^2) / \Z_3$}

In this Appendix we expand on the homology computations for the geometry $X=T^2\times \C^2/\Z_3$ and in particular derive the homology groups lines \eqref{eq:criticalslicehomo5D} and \eqref{eq:AsympBdry} which are relevant for the small and large radius Mayer-Vietoris sequences with respect to the filtration \eqref{eq:Homo}. The SymTree for this case is depicted in figure \ref{fig:SymTrees5D}.

Let us first discuss the homology groups of $\partial X= (T^2\times S^3)/\Z_3$ given in \eqref{eq:AsympBdry} and identify their generators. The homology groups are determined by considering the two fibrations
\be
\pi_1\,:~\partial X \rightarrow T^2/\Z_3\,, \qquad \pi_2\,:~\partial X \rightarrow S^3/\Z_3\,,
\ee
which have generic fiber $S^3,T^2$ respectively. The boundary is smooth and hence its homology groups are organized by Poincar\`e duality. First, consider the fibration $\pi_1$, which has three exceptional fibers $(S^3/\Z_3)_i$ with $i=1,2,3$ which project to the three fixed points of $T^2/\Z_3$. For these we have
\be
3 (S^3/\Z_3)_i=S^3
\ee
where the lefthand side denotes the generic fiber of $\pi_1$. These three-cycles generate the third homlogy group of $\partial X$ and taking the above equivalence into account are isomorphic to $\Z\oplus \Z_3^2$. There are no four-cycles and Poincar\`e duality completely fixes the homology groups of $\partial X$. The one-cycles $\gamma_i$, generating copy of $\Z_3^2$, are given by the uplift of the three 1-cycles which link exactly one marked point on $T^2/\Z_3$. They satisfy the homology relations
\be\label{eq:sumtozero}
\gamma_1+\gamma_2+\gamma_3=0\,, \qquad 3\gamma_i=0\,.
\ee
Finally the 2-cycle is simply the fiber class of the fibration $\pi_2$, i.e., a copy of $T^2$.

\begin{figure}
\centering
\scalebox{0.8}{
\begin{tikzpicture}
\begin{pgfonlayer}{nodelayer}
		\node [style=none] (0) at (-1, -1) {};
		\node [style=none] (1) at (1, -1) {};
		\node [style=none] (2) at (2, -1) {};
		\node [style=none] (3) at (4, -1) {};
		\node [style=none] (4) at (-2, -1) {};
		\node [style=none] (5) at (-4, -1) {};
		\node [style=none] (6) at (-1, 3) {};
		\node [style=none] (7) at (1, 3) {};
		\node [style=none] (8) at (0, 3.75) {$\partial X$};
		\node [style=none] (9) at (-3, -1.75) {$(S^5/\Z_3)_1$};
		\node [style=none] (10) at (0, -1.75) {$(S^5/\Z_3)_2$};
		\node [style=none] (11) at (3, -1.75) {$(S^5/\Z_3)_3$};
		\node [style=none] (12) at (0, 1) {$X^\circ$};
		\node [style=none] (13) at (3, -2.25) {};
	\end{pgfonlayer}
	\begin{pgfonlayer}{edgelayer}
		\draw [style=ThickLine, bend right=90, looseness=0.50] (2.center) to (3.center);
		\draw [style=ThickLine, bend left=90, looseness=0.50] (1.center) to (0.center);
		\draw [style=ThickLine, bend left=90, looseness=0.50] (4.center) to (5.center);
		\draw [style=ThickLine, bend left=90, looseness=0.50] (5.center) to (4.center);
		\draw [style=ThickLine, bend left=90, looseness=0.50] (0.center) to (1.center);
		\draw [style=ThickLine, bend left=90, looseness=0.50] (2.center) to (3.center);
		\draw [style=ThickLine, bend left=90, looseness=0.50] (7.center) to (6.center);
		\draw [style=ThickLine, bend left=90, looseness=0.50] (6.center) to (7.center);
		\draw [style=ThickLine, in=90, out=-90] (6.center) to (5.center);
		\draw [style=ThickLine, bend left=90, looseness=3.50] (4.center) to (0.center);
		\draw [style=ThickLine, bend left=90, looseness=3.50] (1.center) to (2.center);
		\draw [style=ThickLine, in=-90, out=90] (3.center) to (7.center);
	\end{pgfonlayer}
\end{tikzpicture}}
\caption{Sketch of the geometry $X^\circ$.}
\label{fig:TripletofPants}
\end{figure}
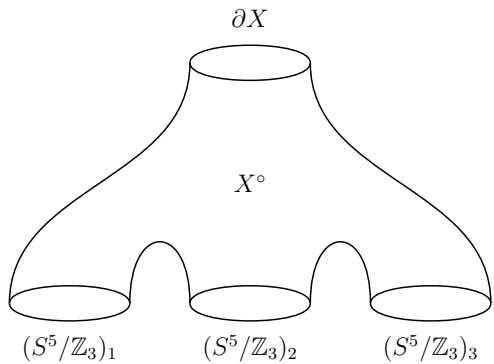

We can now compute the homology groups of $X^\circ$. We proceed via Poincar\`e-Lefschetz duality together with excision which establish the isomorphisms
\be
H_n(X^\circ)\cong H^{6-n}(X^\circ, \partial (X^\circ))=H^{6-n}(X, \partial X \cup \{x_1,x_2,x_3\})
\ee
where $x_i$ denote the location of the three codimension-6 singularities. We then compute the lefthand side using the long exact sequence in relative homology resulting in line \eqref{eq:criticalslicehomo5D}. The only not straightforward map in this computation is the restriction map
\be
R_2\,: ~H^2(X)\cong \Z~\rightarrow~\Z\cong H^2(\partial X\cup \{ x_1,x_2,x_3\})
\ee
which is multiplication by 3. We can now further geometrize the homology groups of line \eqref{eq:criticalslicehomo5D}. For this think of $X^\circ$ as three-legged pants with cross-section $(S^5/\Z_3)_i$ and one branch with cross-section $\partial X$ (see figure \ref{fig:TripletofPants}). The generators in degree 1 are again the $\gamma_i$ subject to the same relation \eqref{eq:sumtozero}. Given $\gamma_i$ there exists a deformation into the boundary component $(S^5/\Z_3)_i$ such that $\gamma_i$ generates $H_1((S^5/\Z_3)_i))\cong \Z_3$. The generator in degree two is again the $T^2$ fiber class. The generators in degree three are now all three lens spaces $(S^3/\Z_3)_i$. These however now generate $\Z_3^3$ because any such lens space is homologous to the $(S^3/\Z_3)_i\subset (S^5/\Z_3)_i$ making it clear that all class generated by these are torsional. Finally the top degree class follows because there are four boundary components which sum to zero in homology, the trivializing chain is $X^\circ$ itself.

With these identifications of generators the small and large radius Mayer-Vietoris sequences follow straightforwardly.

\section{Holography and Ensemble Averaging} \label{app:ENSEMBLE}

In this Appendix we briefly review some aspects of ensemble averaging in the context of the AdS/CFT correspondence.
There are by now many papers in the literature with differing viewpoints on the underlying reason that such averaging occurs,
so we mainly focus on the features most salient to our discussion in Section \ref{sec:LARGEN}.

To large extent, ensemble averaging in holography is expected due to a factorization puzzle which occurs in comparing the partition functions of causally disconnected boundary CFTs which are joined by a bulk wormhole configuration \cite{Witten:1999xp, Maldacena:2004rf}.

Consider a system in which the boundary has $n$ connected components $\Sigma = \Sigma_1 \sqcup \Sigma_2 \sqcup \dotsb \sqcup \Sigma_n$. The gravitational path integral must naively sum over all possible bulk manifolds with this conformal boundary. Near each boundary, we need to specify the asymptotic profile for the fields of the CFT, e.g., the moduli / parameters of the theory. We denote these as $\phi \vert_{\partial \Sigma_i} = J_i$ for $i = 1,...,n$. According to the standard holographic dictionary \cite{Maldacena:1997re, Gubser:1998bc, Witten:1998qj},
the path integral results in a connected correlation function:
\begin{equation}
    \langle Z[J_1] \dotsm Z[J_n]\rangle\,,
\end{equation}
which cannot generically be factored into a product of correlation functions for the individual components of the boundary. For example, with $n = 2$,
\begin{equation}
    \langle Z[J_1] Z[J_2]\rangle \neq \langle Z[J_1]\rangle \langle Z[J_2]\rangle\,,
\end{equation}
where $Z[J_1]$ and $Z[J_2]$ can be regarded as CFT partition functions over the two boundary components $\Sigma_1$ and $\Sigma_2$, respectively.

From the point of view of the effective theory and holography, the proposal is to interpret this non-factorization as due to the contribution of wormhole configurations in the gravitational path integral, which corresponds to the ensemble averaging of boundary theories. There are indeed low-dimensional examples demonstrating this behavior, e.g., the duality between 2D JT gravity and 1D random matrix theory (see e.g., \cite{Stanford:2019vob,Saad:2019lba,Chandra:2022bqq} and references therein), as well as the duality between semi-classical 3D gravity and 2D CFT ensemble averaging (see e.g., references \cite{Schlenker:2022dyo, Chandra:2022bqq}).

At present, a full understanding of the mechanism underlying ensemble averaging remains an open question. One interpretation is that this ensemble averaging is ``real'' in the sense that the gravitational theory really is dual to an ensemble of CFTs; this possibility is closely related to having a high-dimensional Hilbert space for baby universes \cite{Marolf:2020xie},\footnote{See, however, \cite{McNamara:2020uza, Heckman:2021vzx}.} which in turn would also appear to require treating gravity as an open system. On the other hand, a more conservative interpretation is that even within a single fixed CFT, any attempt to describe black hole physics will necessarily require some sort of chaotic dynamics. As such, one should expect averaging to be a generic feature of states which are sufficiently ``complex''. This would be in line with Wigner's approach to nuclear theory which makes use of a probabilistic ensemble of Hamiltonians \cite{Wigner:1955} to model the structure of large nuclei. Such an approach would also be in line with the general contours of the eigenstate thermalization hypothesis (see \cite{Deutsch:1991, Srednicki:1994}).

Some aspects of this more conservative interpretation were recently sharpened in the specific context of 3D gravitational theories in reference \cite{Schlenker:2022dyo} (see also \cite{Chandra:2022bqq}), but it is expected that some of these considerations apply more broadly. In reference \cite{Schlenker:2022dyo}, Schlenker and Witten studied the factorization puzzle by separating observables into those below and above the black hole threshold. Here, ``below the black hole threshold'' means, in the CFT dual, states with scaling dimensions above the ground state by only a fixed amount, i.e., above some $\Delta_{BH}$, in the large $M$ limit. It is claimed that observables below the black hole threshold do not demonstrate ensemble averaging while, by contrast, black hole states are responsible for the ensemble averaging behavior of the gravitational path integral.

This proposal is based on two statements: that black hole physics is chaotic, and that the Hilbert space $\cH_{M}$ describing black hole states does not have a large $M$ limit. To see this, note that the black hole entropy $\mathcal{S}_{BH}$ at a fixed temperature grows as a power of $M$, e.g., $\mathcal{S}_{BH} \propto M^2$ if the boundary CFT is a 4D large $M$ gauge theory. Then, in the limit of large $M$, if one changes $M$ to $M+1$, the dimension of $\cH_M$ grows by exponential factor as
\begin{equation}
    e^{M^2} \rightarrow e^{(M + 1)^2}\,.
\end{equation}
Therefore, it is very likely that the Hilbert space and the corresponding Hamiltonian of black hole states do not have a large $M$ limit.
The Hamiltonian $H_{M}$ of black hole states at given $M$ is then a pseudorandom matrix\footnote{A pseudorandom matrix is one generated by a deterministic causal algorithm, but one in which it cannot be distinguished from a truly random matrix by any pre-determined statistical test for randomness.} For neighboring values of $M$, each $H_{M}$ can be regarded as an independent draw from a random matrix ensemble.\footnote{An unfortunate feature of some of the literature on holographic ensemble averaging is that the notion of averaging is sometimes different across different papers, e.g., it sometimes refers to a quenched, and sometimes to an annealed average. This shows up in the present discussion because even though we are averaging over $M$, we are still treating the system as drawn from a single class of Hamiltonian operators \`{a} la Wigner.}

To better understand how this affects the computation of observables, let us focus on an arbitrary observable $\cO_{M}$ depending only on $H_{M}$. In random matrix theory, $\cO_{M}$ may be a ``self-averaging'' function, meaning that it has almost the same value for almost any draw from the ensemble. In this case, $\langle\cO_{M}\rangle$ will be a smooth function of $M$, with small $e^{-S}$ corrections reflecting the fact that self-averaging functions of a random matrix can differ slightly from draw to draw. If $\cO_{M}$ is not self-averaging, it will be an erratic function of $M$ whose expectation value $\langle\cO_M\rangle$ cannot be simply computed approximately. However, the gravitational path integral always produces a smooth function of $M$ by typically summing over the contributions of saddle points.\footnote{This holds even when classical solutions are not available. See \cite{Cotler:2020ugk,Maxfield:2020ale} for examples.} From the random matrix theory point of view, even for non-self-averaging observables, there still exists an averaged value within the ensemble, which is possibly computed using the gravitational path integral. In this sense, $\langle \cO_{M} \rangle$ as derived from the gravitational path integral should represent an averaged result over nearby values of large $M$. The reason underlying this property of the gravitational path integral is related to the coarse-graining nature of the semi-classical gravity (see e.g.,~\cite{Chandra:2022fwi}). We must emphasize that the Schlenker--Witten proposal by no means claims that the ensemble averaging is always over microscopically well-defined CFTs, which is in contrast to the procedure given in e.g., references \cite{Maloney:2020nni, Ashwinkumar:2021kav, Benjamin:2021wzr, Romaidis:2021hnh, Collier:2022emf, Romaidis:2023zpx, Ashwinkumar:2023jtz}.

Another subtlety with the Schlenker--Witten proposal is how it works in $\AdS_{D + 1}$/$\CFT_D$ when the CFT spacetime dimension is $D > 2$. Recall that in $(D + 1)$-dimensional gravity, the $\AdS_{D + 1}$ black hole solutions can be separated as
\begin{equation}
\begin{aligned}
    &\text{small black holes with} \; \rho_+ < \sqrt{\frac{D - 2}{D}} \; L_{\AdS}\,, \\
    &\text{large black holes with} \; \rho_+ > \sqrt{\frac{D - 2}{D}} \; L_{\AdS}\,,
\end{aligned}
\end{equation}
where $\rho_+$ is the location of the horizon and $L_{\AdS}$ is the length scale of $\AdS_{D + 1}$. In $\AdS_3$, there is \emph{no} small black hole and thus there exists a sharp black hole threshold scale to distinguish states with energies below and above it. When $D > 2$, however, there exist small black holes, so it is not clear in what sense one can determine whether states are sub-threshold or not. This would seem to suggest that at least in higher-dimensional CFTs, large $M$ ensemble averaging would need to be entertained even in considering correlators for low dimension operators.

Another question is how to implement large $M$ averaging when dealing with extended operators which transform non-trivially under higher-form symmetries. Such operators are often directly sensitive to the topological sector of the bulk gravitational dual, and in particular quantities such as $M$ itself. This occurs, for example, in the 5D topological term:
\begin{equation}
S_{5D} = \frac{i}{2 \pi} M \int B_2 \wedge dC_2.
\end{equation}
One of the aims of Section \ref{sec:LARGEN} is to construct extended operators which
still admit large $M$ averaging even whilst still retaining a higher-form symmetry.

\newpage

\bibliographystyle{utphys}
\bibliography{InfinityWars}

\end{document}